\documentclass[12pt]{article}
\usepackage[utf8]{inputenc}
\usepackage{amsmath,setspace,geometry}
\usepackage{amsthm}
\usepackage{amsfonts}
\usepackage[shortlabels]{enumitem}
\usepackage{rotating}
\usepackage{pdflscape}
\usepackage{graphicx}
\usepackage{bbm}
\usepackage{comment}
\usepackage[dvipsnames]{xcolor}
\usepackage{hyperref}
\hypersetup{colorlinks=true, linkcolor= BrickRed, citecolor = BrickRed, filecolor = BrickRed, urlcolor = BrickRed, hypertexnames = true}
\usepackage[]{natbib} 
\bibpunct[:]{(}{)}{,}{a}{}{,}
\usepackage[english]{babel}
\usepackage{float}
\usepackage{caption}
\usepackage{subcaption}
\usepackage{booktabs}
\usepackage{pdfpages}
\usepackage{threeparttable}
\usepackage{lscape}
\usepackage{bm}

\usepackage{multirow}
\usepackage{adjustbox}
\newtheorem{proposition}{Proposition}

\begin{document}

\title{Just After Minimum Wage Hikes:\\ Short-Run Labor-Demand Response and Reallocation}
\author{Hayato Kanayama\thanks{\href{mailto:ha13yato@toki.waseda.jp}{ha13yato@toki.waseda.jp}, Graduate School of Economics, Waseda University.}, Sho Miyaji\thanks{\href{mailto:sho.miyaji@yale.edu}{sho.miyaji@yale.edu}, Graduate School of Economics, Yale University}, Suguru Otani\thanks{\href{mailto:suguru.otani@e.u-tokyo.ac.jp}{suguru.otani@e.u-tokyo.ac.jp}, Market Design Center, Department of Economics, The University of Tokyo\\
 We thank Yu Awaya, Jun Goto, Takuya Hasebe, Yuichiro Kamada, Ryo Kambayashi, Michihiro Kandori, Daiji Kawaguchi, Keisuke Kawata, Kohei Kawaguchi, Fuhito Kojima, Ayako Kondo, Seongman Moon, Yuko Mori, Shunya Noda, Hiroko Okudaira, Hideo Owan, Shinpei Sano, Hitoshi Shigeoka, Uta Sch{\"o}nberg, Mari Tanaka, and Ken Yamada for their advice, and seminar and conference participants at the University of Tokyo, Kochi University of Technology, Sophia University, Doshisha University, the 100th Annual Conference of the Western Economic Association International, the 20th Applied Econometrics Conference, EALE 2026, the Workshop on the Economics of Human Resource Allocation, and the Kansai and Tokyo Labor Study Group for their comments. An earlier version was titled ``Who Bears the Cost? High-Frequency Evidence on Minimum Wage Effects and Non-wage Amenity Adjustment in Spot Labor Markets.'' We are grateful to Kazuki Takaishi and Takahide Kimura for sharing the data. This work was supported by JST ERATO (JPMJER2301) and JSPS KAKENHI (25K16620). } }

 \date{
First version: May 7, 2025; current version: \today
}
\maketitle

\begin{abstract}
    How labor markets adjust immediately after minimum wage hikes remains an open, policy-relevant question. This paper studies short-run minimum-wage effects in Japan's spot labor market using Timee data and a wage-bin difference-in-differences design. We find a normalized net employment-share change of $-0.010$, with similar estimates for vacancies and total working hours. Weekly and daily estimates show that these effects emerge immediately after implementation and then stabilize. Effects are more negative where the minimum-wage bite is higher and in restaurant work. Net share effects are negative for midnight shifts and postings offering transportation reimbursement. Gender-targeted and beginner-targeted postings show no significant net change. Decomposing vacancies by employers' posting histories, we find that incumbent employers cut postings below the new floor while employers that did not post before the revision post just above it. Cross-employer reallocation, a multi-year process in standard labor markets, thus begins immediately in spot gig work. 

\end{abstract}


\section{Introduction}

The gig economy—comprising temporary, contract-based, and freelance online work—has expanded rapidly \citep{boeri2020solo}. An increasing share of freelance workers is now matched with customers through online platforms.\footnote{For instance, \cite{kassi2018online} report that the demand for online gig work grew by approximately 21\% between 2016 and 2018, based on an analysis of open projects and tasks on various platforms.} Because gigs range from on-demand shifts to project-based contracts, gig work represents one of the most adjustable forms of labor demand. Yet, evidence on minimum-wage effects in this setting remains limited, with a few notable exceptions \citep{glasner2023minimum}. In particular, for alternative work arrangements \citep{mas2020alternative}—and especially for spot labor markets in which posted jobs require minimal skills and worker turnover is high—we still know little about the extent of employment declines, how quickly they occur, and which jobs contract in the short run. Clarifying the magnitude, timing, and composition of these short-run employment changes and the resulting reallocation is essential for designing timely, well-targeted policy support and for linking immediate adjustments to longer-run reallocation \citep{dustmann2022reallocation}.

We address these questions using confidential job-posting and contract-level data from \textit{Timee}, a leading Japanese on-site spot-work platform. Spot work platforms have expanded worldwide, reflecting a broader shift toward flexible, on-demand labor markets.\footnote{Examples include Taskmo (India), TROOPERS (Singapore), Geupgu (South Korea), QuikShift (Australia), Coople (Switzerland/UK), and Instawork (United States). Related on-demand platforms also include TaskRabbit (United States; task-based on-site services), Wonolo (United States), and GigSmart (United States).} We exploit the platform's data to disentangle labor demand from employment by studying vacancies and matches side by side and to trace short-run dynamics around minimum-wage changes at daily, weekly, and monthly frequencies. We further use non-wage amenities and job-description content to characterize job types and compare group-specific employment responses. These features provide a high-frequency view of how labor demand and job composition adjust in the immediate aftermath of a wage-floor increase.

Spot labor markets exhibit three features that make them ideal for studying short-run minimum-wage effects: limited labor heterogeneity, short-term employment contracts, and wage posting. First, they attract workers seeking jobs that require minimal skills and are therefore especially exposed to minimum-wage changes. The resulting matches are relatively homogeneous, making these markets close to the textbook homogeneous-labor benchmark. Second, short-term contracts allow labor demand to adjust rapidly. Firms hire for discrete, temporary tasks rather than permanent positions, allowing us to observe hiring responses with limited employment-adjustment frictions. Our spot-work data capture an extreme case of short-term labor-demand adjustment, a phenomenon rarely observed in the empirical literature, which often focuses instead on the labor supply of taxi and Uber drivers \citep{angrist2021uber,buchholz2023rethinking,kuroda2023exploring}. Finally, wage posting precludes bargaining---a central element in traditional models of long-term employment contracts---allowing us to focus on models grounded in monopsonistic or oligopsonistic labor-market structures. Our estimates characterize short-run responses in a flexible employment setting.\footnote{The worker-side evidence below is consistent with a labor-demand response but does not separately identify labor supply. Elasticities for narrowly defined demographic groups, such as less-educated young men in \citet{Kawaguchi_Mori2021}, are not directly comparable to our aggregate estimate for affected spot postings.} We construct a newsvendor model of spot labor demand based on \citet{mas2026newsvendor} to interpret the vacancy response in Appendix \ref{sec:newsvendor}: a binding wage floor lowers committed vacancies and raises the productivity cutoff for active postings, and the same mechanism generates compositional selection when productivity is correlated with posting attributes. These short-run responses are complementary to, rather than in conflict with, longer-run evidence on minimum-wage elasticities, reallocation, and firm exit, and they clarify the earliest and most flexible margin through which minimum-wage shocks propagate \citep{dube2024minimum}.

Our data offer three advantages for studying short-run adjustments in spot labor markets. First, the data are recorded at the event level with associated click logs, allowing us to track employers' job-posting activity and workers' application activity immediately before and after the implementation of the new minimum wage. This granularity is substantially finer than the month-by-state level used in \citet{sabia2009effects,sabia2009identifying,brochu2025minimum} and \citet{melo2025minimum} or the quarter-by-state level in \citet{addison2009minimum} and \citet{Cengiz2019}, enabling a more detailed view of temporal adjustments. Second, because establishments can readily modify job descriptions at the time of posting and the data provide clearly observable information on fixed-amount transportation reimbursements as adjustable non-wage amenities, we can use this amenity and description information to characterize job types and compare their short-run responses. Third, since spot labor demand is not tied to long-term employment relationships, it is less affected by reputation concerns or switching costs, making this an ideal setting for identifying immediate behavioral responses to policy changes.

From a macro perspective, Timee's role in part-time spot employment has grown markedly in recent years. As of 2024, more than 7 million workers were registered on the platform, more than 10\% of Japan's labor force of roughly 70 million, and roughly 1 million were active at the end of 2023. While the number of active users has increased gradually, the number of vacancies has grown sharply, especially since 2022, leading to a rise in labor-market tightness on the platform. Hiring counts show a substantial upward trend beginning in 2020, reflecting a pronounced increase in successful matches. The job-finding rate has risen above 1.0, indicating that each active worker matches with multiple spot vacancies per month. Meanwhile, the worker-finding rate has remained stable at approximately 0.8, that is, about 0.8 hires per posting. These trends highlight the platform's growing role in facilitating spot employment matching.

As an illustrative example, we first use granular platform data from Tokyo to examine how the 2023 statutory minimum-wage increase reshaped employment patterns at the intensive margin of job characteristics. This uniquely detailed dataset allows us to track daily, weekly, and monthly shifts in the employment distribution across wage levels, posted working hours, and transportation reimbursements. Descriptive evidence reveals a clear upward shift in the wage distribution around the new minimum-wage threshold, consistent with firms adjusting wages to comply with the policy. By contrast, the overall distributions of working hours and transportation reimbursements remain largely stable. These fine-grained patterns highlight the value of high-frequency, contract-level data in capturing immediate and heterogeneous responses to labor-market regulation, and they motivate a causal analysis that isolates the impact of the minimum-wage change from broader trends.

Given these patterns, we employ a difference-in-differences (DID) framework inspired by \cite{Cengiz2019} and \cite{dustmann2022reallocation}, which leverages variation in exposure to the minimum wage increase across wage bins. Rather than identifying treated workers based on demographic or job characteristics, we exploit the wage distribution itself, comparing outcomes in bins near the new minimum wage threshold with those in the upper tail of the distribution, which are assumed to be unaffected. We define treatment and control groups based on their relative position to the minimum wage in each prefecture, enabling a consistent bin-level comparison over time. This granular approach is well suited to our setting, as it fully utilizes the spot-work platform data aggregated at the prefecture-by-month-by-wage-bin level. By estimating dynamic effects through interactions of relative time and exposure group indicators, we trace the evolution of treatment effects before and after the policy change. To summarize these effects across the distribution, we implement the ``excess jobs minus missing jobs'' decomposition following \cite{Cengiz2019} and scale the net share response by minimum-wage and average-wage changes.

Our estimates show a shift in employment shares from below to above the new minimum wage. Relative to the September lower-tier employment share, missing jobs are $-28.0\%$, excess jobs are $27.0\%$, and Net is approximately $-1.0\%$. The corresponding share-based elasticity with respect to the minimum wage is $-0.211$. These measures summarize relative changes in the observed wage distribution. The net share effect is most negative in December and January, documenting a rapid distributional response in the spot market.

The vacancy and employment estimates are consistent with a labor-demand response. Vacancy postings decline sharply below the new minimum wage, while gains of similar size appear just above it, consistent with adjustment through vacancy posting. The normalized employment, vacancy, and total-hours estimates are similar. On the worker side, realized matches follow similar trends across gender and worker-status groups (e.g., students and stay-at-home spouses). These worker-group series do not separately identify labor supply. Using weekly and daily data, we also find that these effects appear immediately after implementation.

Next, we use our distributional DID approach to characterize heterogeneity in minimum-wage effects across prefectures, occupations, time-of-day categories, and job attributes. Exploiting prefecture-level variation in the minimum-wage bite, we find more negative net share estimates in prefectures with higher Kaitz indices. Across occupations, the net employment effect is negative for restaurant work, while the other occupations, including other low-wage occupations such as light work and retail, show no significant net change. Across time-of-day categories, net share estimates are negative for midnight and evening shifts, most strongly for midnight shifts.

Moving beyond broad occupational categories, we estimate employment responses separately by posting attributes. Net estimates are significantly negative for postings offering transportation reimbursement and for experienced-worker-targeted postings, whereas female-, male-, and beginner-targeted postings show no significant net change. The newsvendor framework generates compositional selection when posting productivity is correlated with posting attributes; our data do not measure posting productivity.

Taken together, these estimates describe responses within distinct job types. The larger reallocation occurs across employers: decomposing the vacancy effects by employers' posting histories, we find that incumbent employers cut postings below the new floor and recreate few of them just above it, while employers that did not post in the six months before the revision account for most of the excess jobs. Consistent with the reallocation mechanisms that \citet{dustmann2022reallocation} document in the years following the German reform, our evidence shows that reallocation across employers manifests immediately along the most flexible margin---spot labor demand---where adjustments occur first through the entry and exit of posting employers and the contraction of particular job types before propagating to longer-run changes in firm composition and employment structure.

\subsection{Related Literature}

Our paper contributes to three related literatures: minimum wage effects in platform-mediated spot labor markets, the reallocation of labor across employers and job types in response to minimum-wage hikes, and the short-run dynamics of job destruction and creation around the wage floor.

First, we contribute to the emerging literature on alternative work arrangements and online labor markets \citep{mas2020alternative, katz2019rise} by studying the employment effects of minimum wage increases in spot labor markets. Existing studies on platform-mediated work largely focus on remote gig work involving relatively high-skill tasks, such as programming or design \citep{glasner2023minimum}. In contrast, we study on-site spot work—short-term jobs with minimal skill requirements that require physical presence and are accessible to a broad segment of the workforce—which has expanded rapidly worldwide as part of a broader shift toward flexible, on-demand labor arrangements.

These features imply that spot labor markets differ fundamentally in their exposure to minimum-wage policies. Jobs are frequently posted at or near the minimum wage, and both workers and employers are directly affected by statutory wage floors, leaving limited scope for spillover effects to higher wages. Using comprehensive platform data, we find negative net employment-share effects, concentrated near the minimum wage.\footnote{A large literature examines the employment effects of minimum wages, documenting substantial heterogeneity across settings, demographic groups, and sectors. While many studies focus on specific age groups, such as teenagers or young workers \citep{card1992minimum, neumark1992employment, allegretto2011minimum, Neumark2014, allegretto2017credible}, others examine employment effects within particular industries, most notably low-wage sectors such as restaurants \citep{katz1992effect, card1994minimum, Dube_etal2010}. More recent work emphasizes cases in which employment effects are small or statistically insignificant \citep{clemens2021firms, Manning_2021, neumark2022myth}. Methodological debates and evidence on sectoral heterogeneity and spillovers are surveyed in \citet{dube2024minimum}.} \citet{lee1999wage} and \citet{autor2016contribution} study how minimum wages shape the wage distribution, while \citet{Cengiz2019} links changes in that distribution, including wage spillovers above the minimum wage, to employment effects. Our share-based estimates complement the employment-level evidence in \citet{Cengiz2019} by tracing distributional adjustment in spot work.\footnote{Related platform-based studies examine minimum-wage effects on vacancies rather than employment outcomes, including evidence from online and flexible labor platforms \citep{adams2020flexible, melo2025minimum}.}


Second, our paper contributes to the literature on minimum-wage-induced reallocation by documenting group-specific employment responses within a rapidly adjusting spot labor market \citep{dustmann2022reallocation}. While much of the minimum-wage literature emphasizes net employment effects, an equally important question is \emph{which} jobs contract, \emph{which} expand, and how employment is reallocated across locations, occupations, and job types. Using a distributional DID design on prefecture-by-month-by-wage-bin cells, we quantify missing and excess jobs and decompose these changes in four ways: by prefecture exposure to the minimum wage hike (measured by the Kaitz index), by occupation, by posting-based job characteristics, and by employers' posting histories. The estimates are more negative in heavily exposed regions, and restaurant work shows a significant negative net estimate, while the other occupations show no significant net change.

In this reallocation perspective, posting characteristics such as transportation reimbursement (non-wage amenities) and description-based targeting (gender and skill) serve as markers of job type rather than merely adjustment margins.\footnote{Recently, a growing body of literature has investigated various margins of adjustment for employers, such as price pass-through \citep{aaronson2007product, harasztosi2019pays, leung2021minimum}, reductions in non-wage compensation \citep{simon2004minimum, clemens2018minimum}, decreases in firm revenue \citep{draca2011minimum, bell2018minimum}, and firm entry and exit \citep{draca2011minimum, mayneris2018improving, aaronson2018industry, harasztosi2019pays, luca2019survival, dustmann2022reallocation, chava2023does}.
See \cite{clemens2021firms} and \cite{dube2024minimum} for comprehensive reviews. Job advertisement content has also been shown to shape labor-market segmentation and self-selection: titles and descriptions predict wages and applications \citep{marinescu2020opening} and often carry explicit or implicit gender or skill signals \citep{brenvcivc2010employers, kuhn2013gender, kuhn2020gender, helleseter2020age}.} The group-specific net estimates are significantly negative for postings offering transportation reimbursement, restaurant work, midnight and evening shifts, and experienced-worker-targeted postings. Female-, male-, and beginner-targeted postings show no significant net change. Decomposing vacancies by employers' posting histories, we find that most excess jobs just above the new floor are posted by employers that did not post before the revision, while incumbent employers cut postings below the floor and recreate few of them. These results complement findings on reallocation in standard employment settings \citep{dustmann2022reallocation} by highlighting that, in on-site spot markets, reallocation occurs more rapidly through the entry and exit of posting employers.



Third, we provide evidence on the very short-run \emph{speed} and \emph{composition} of job losses and job creation following minimum wage hikes. Building on the distributional approach of \citet{Cengiz2019} and the wage-bin design in \citet{dustmann2022reallocation}, we track month-by-month (and, in robustness checks, higher-frequency) movements in missing and excess jobs around the new wage floor.\footnote{Related applications of this wage-bin approach include \citet{gopalan2021state} and \citet{brochu2025minimum}.} This design allows us to identify which postings disappear immediately below the new minimum, which reappear just above it, and how these patterns vary across regions, occupations, and advertised job characteristics.

Relatedly, our findings contribute to the literature on minimum wages and employment flows. Existing studies show that minimum-wage increases can reduce both worker accessions and separations, so substantial changes in gross employment flows may occur even when employment stocks change little \citep{portugal2006disentangling, dube2016minimum}.
Our setting isolates an upstream margin of the accession response. 
Because each spot shift must be separately posted and matched, while the conventional retention margin associated with continuing employment relationships is limited, a reduction in hiring demand translates more directly into lower realized employment. 
By observing vacancies and realized matches separately at high frequency, we document similar immediate changes in their wage distributions. This mechanism is consistent with our newsvendor interpretation, in which a higher wage floor reduces firms’ optimal commitment to flexible labor under demand uncertainty.

Studying the short run is policy-relevant because firms may face liquidity constraints and adjustment frictions at the time of implementation; consequently, immediate job destruction may reflect temporary financial stress rather than long-run inefficiency \citep{doh2024economic}. Despite the significance of these immediate responses, there is limited evidence on firm adjustments at daily, weekly, or monthly frequencies, with only a few exceptions \citep{sabia2009effects,sabia2009identifying,brochu2025minimum, melo2025minimum}. 
By revealing the near-immediate sorting of postings around the new wage floor—where compliance is rapid and spillovers to higher wage bins are limited—we clarify how minimum wage policy reshapes the spot job distribution in the months following a reform. This perspective complements the longer-run literature on capital–labor substitution, firm exit, and reallocation \citep{Neumark_Wascher2008, dustmann2022reallocation} by identifying the initial margins and employer groups through which longer-run adjustments unfold.

The rest of the paper is organized as follows. Section~\ref{sec:background} outlines the spot-work context and the Japanese minimum-wage system. Section~\ref{sec:data} introduces the data and presents motivating patterns. Section~\ref{sec:empirical_strategy} describes the identification strategy. Section~\ref{sec:results} reports the main results. Section~\ref{sec:heterogeneity_analysis_across_prefecture_occupation} examines heterogeneity across prefectures, occupations, and posting attributes. Section~\ref{sec:conclusion} concludes.

\section{Background}\label{sec:background}

\subsection{Non-regular Work and Spot Work in Japan}
\begin{figure}[!ht]
  \begin{center}
  \subfloat[Hires, Labor force survey]{\includegraphics[width = 0.6\textwidth]
  {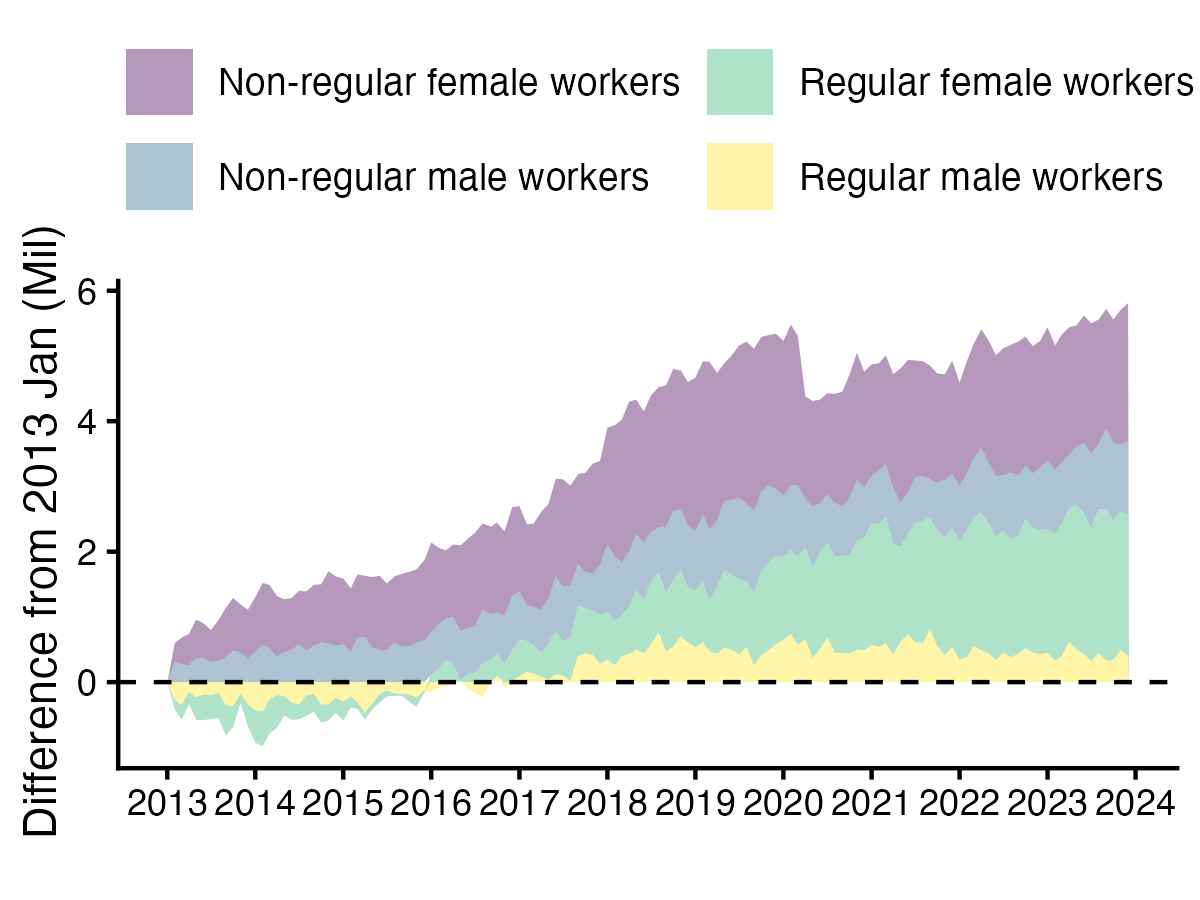}}\\
  \subfloat[Hires, Spot work platform]{\includegraphics[width = 0.6\textwidth]
  {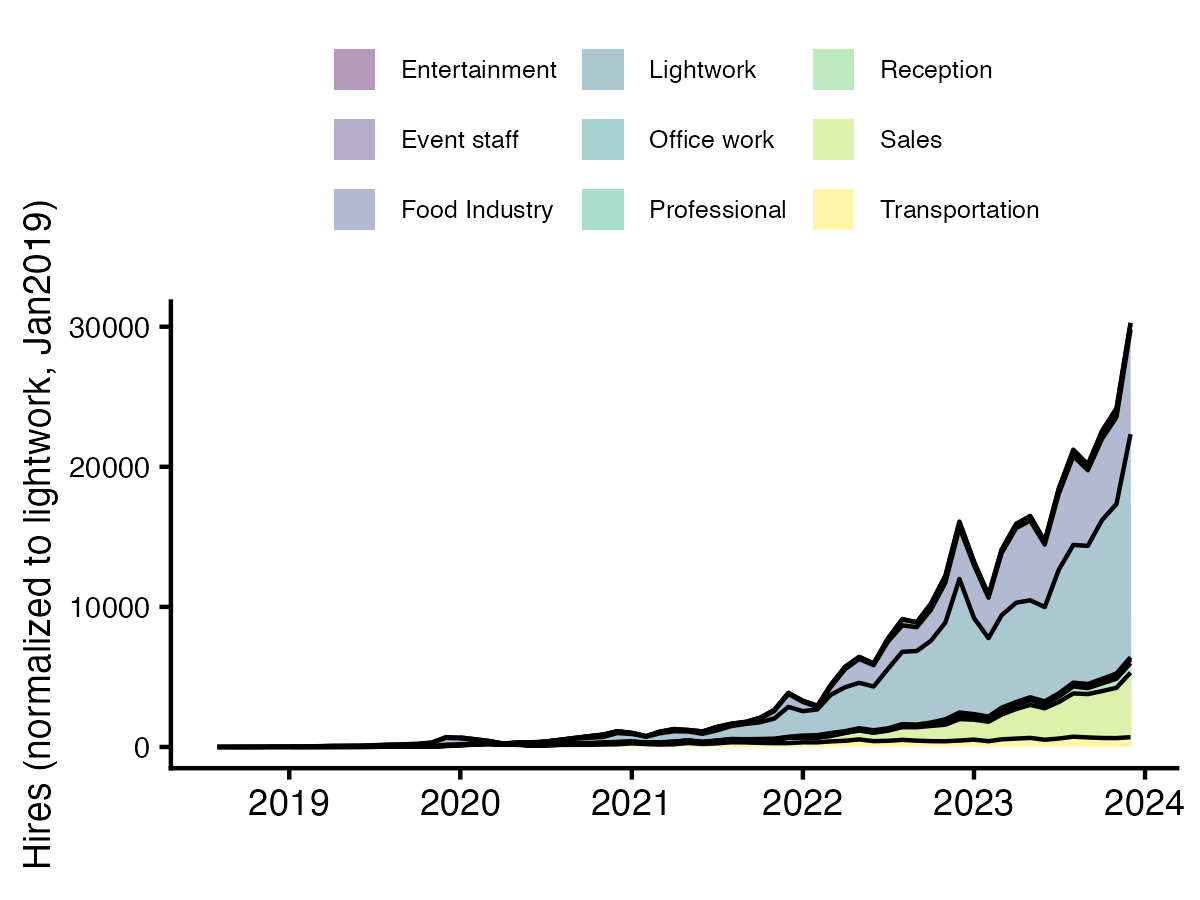}}
  \caption{Employment Changes: National vs. Spot Work Platform}
  \label{fg:labor_force} 
  \end{center}
  \footnotesize
  Note: Panel (a) depicts the time series of job changes in Japan, standardized to January 2013. The colored areas indicate changes within specific worker groups. This graph is based on data from the Economic Survey by the Japanese Cabinet Office (see details at \url{https://www5.cao.go.jp/keizai3/2024/0802wp-keizai/setsumei-e2024.pdf}, accessed September 19, 2024). Panel (b) reports monthly platform-side hires by job category, each divided by the number of Lightwork hires in January 2019. Panel (b) is computed from the Timee contract-level data described in Section \ref{sec:data}. See the survey of \cite{miyamoto2025macroeconomic} for an overview of labor markets in Japan.
\end{figure}

Figure \ref{fg:labor_force} shows that both non-regular work and the importance of spot labor markets have grown substantially in Japan between 2013 and 2024. Panel (a) documents the rising prevalence of non-regular workers. Using platform data, Panel (b) shows that spot-work hiring spans a wide range of job categories, with Lightwork, Food Industry, and Sales accounting for the largest shares and the remaining categories contributing smaller shares. Together, these patterns suggest a growing reliance on spot-work platforms for flexible employment opportunities and broader acceptance of on-demand work arrangements in Japan.

\subsection{Spot Work Platform}

\begin{figure}[!ht]
  \begin{center}
  \subfloat[User $U$, Vacancy $V$, and Tightness ($\frac{V}{U}$)]{\includegraphics[width = 0.33\textwidth]
  {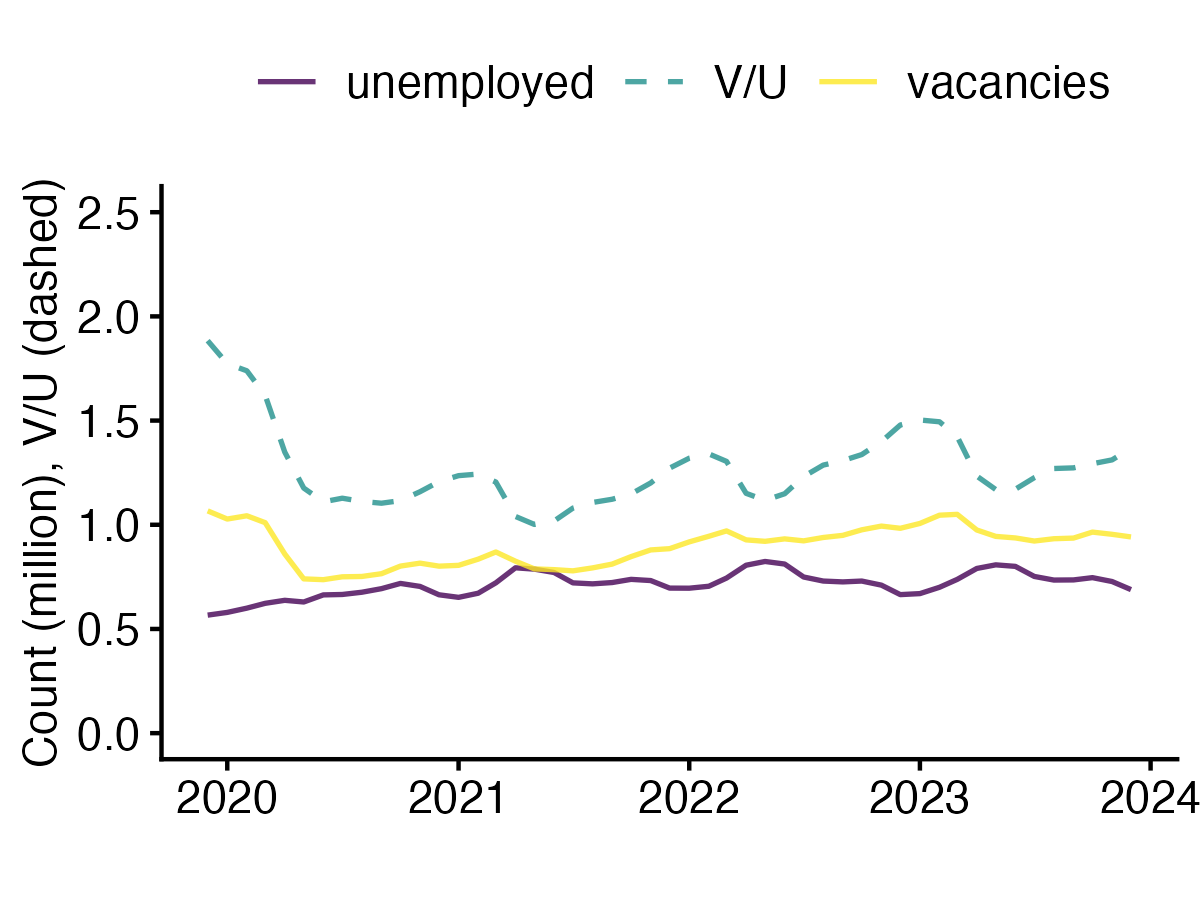}\includegraphics[width = 0.33\textwidth]
  {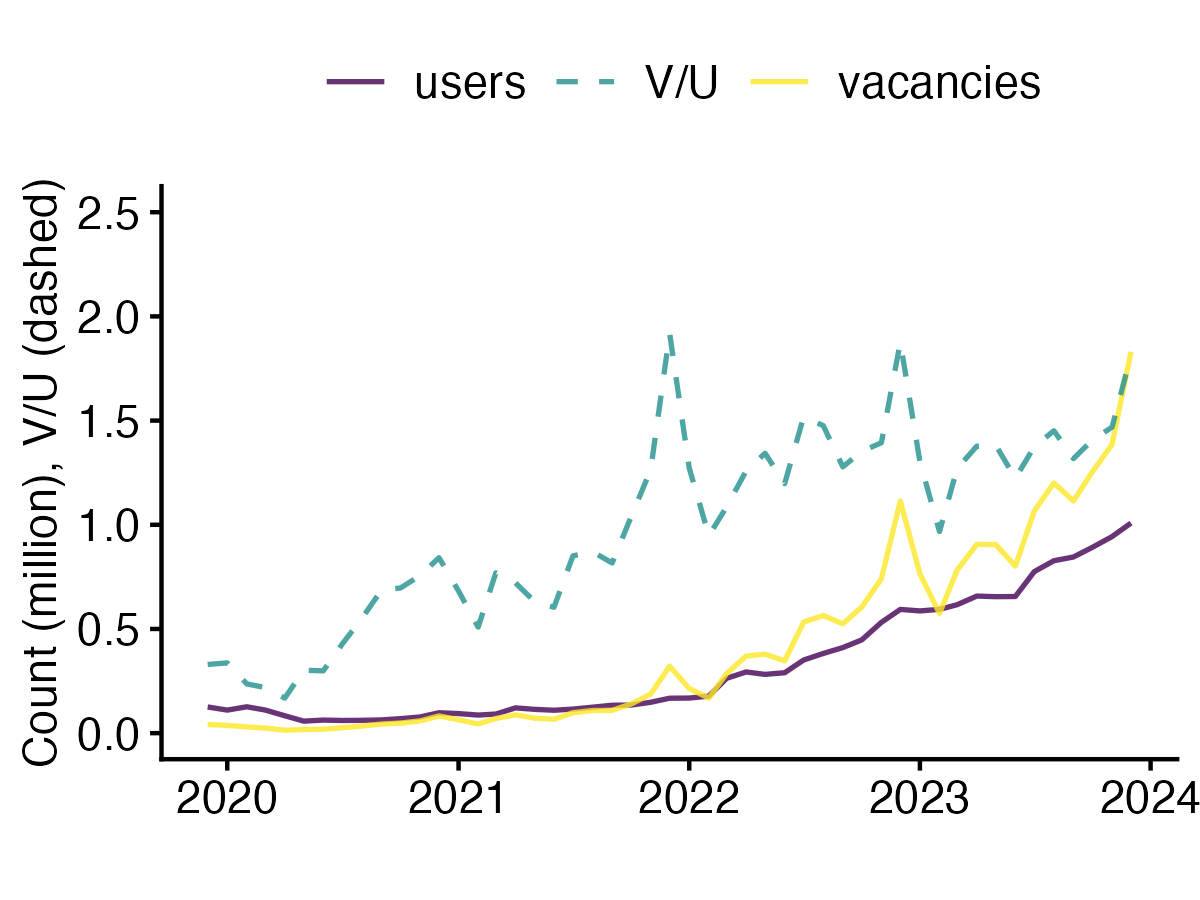}}\\
  \subfloat[Hire $H$]{\includegraphics[width = 0.33\textwidth]
  {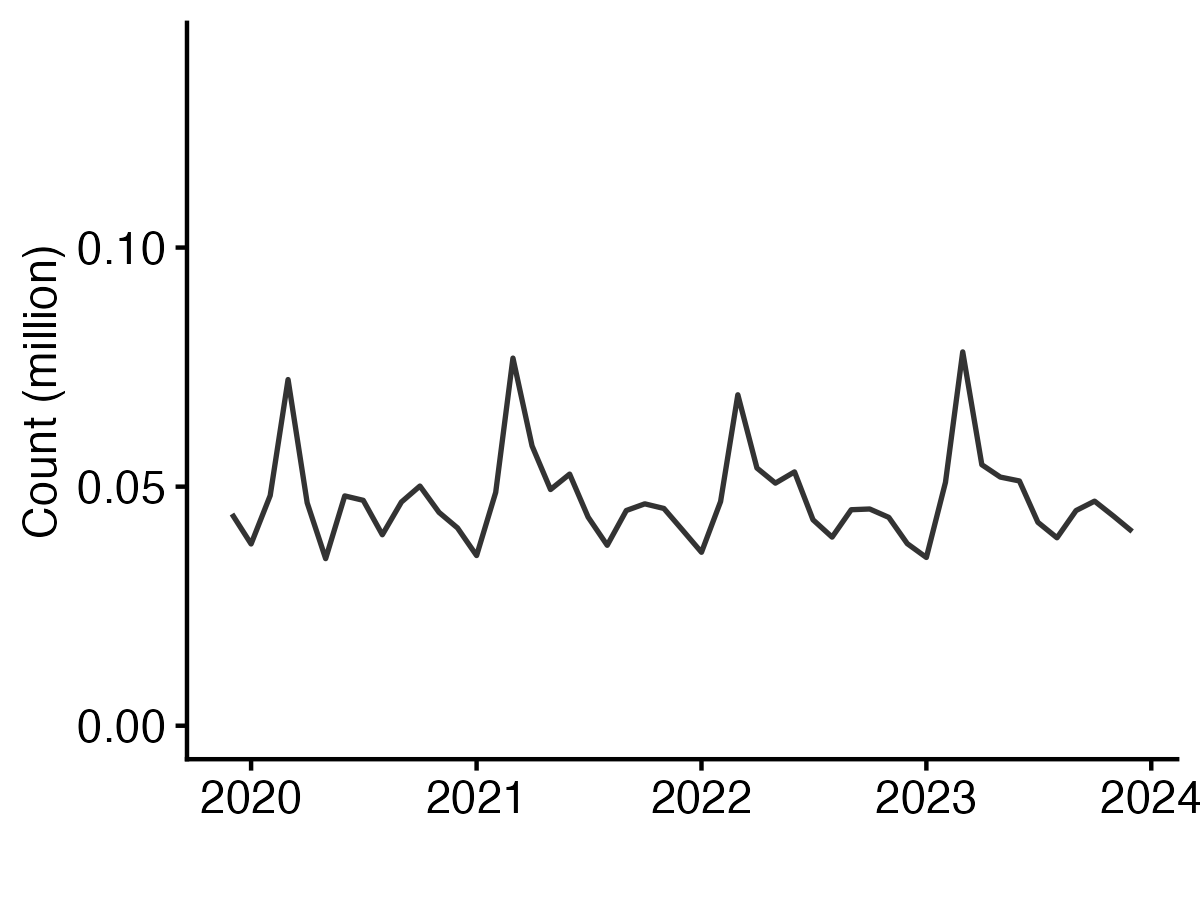}\includegraphics[width = 0.33\textwidth]
  {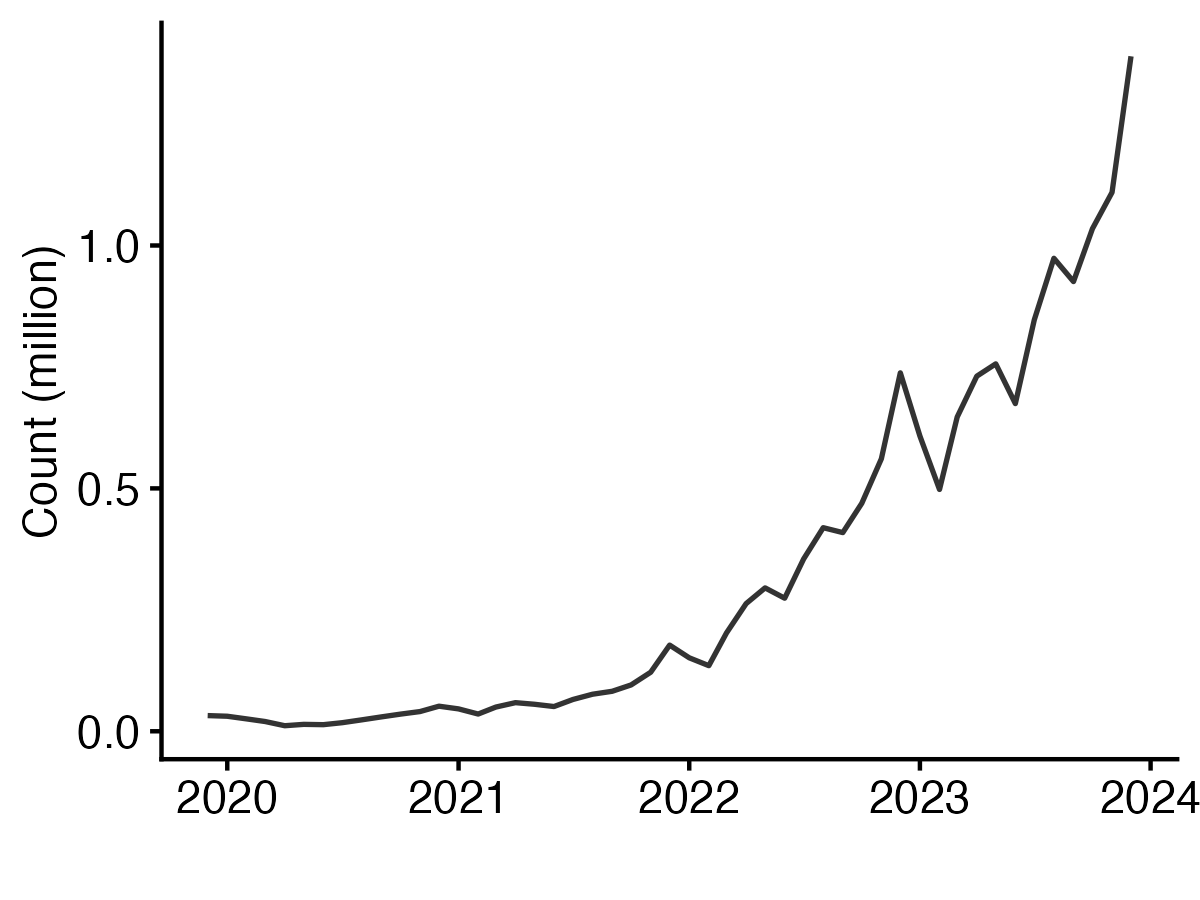}}
  \\
  \subfloat[Job and Worker Finding Rate ($\frac{H}{U}$,$\frac{H}{V}$)]{\includegraphics[width = 0.33\textwidth]
  {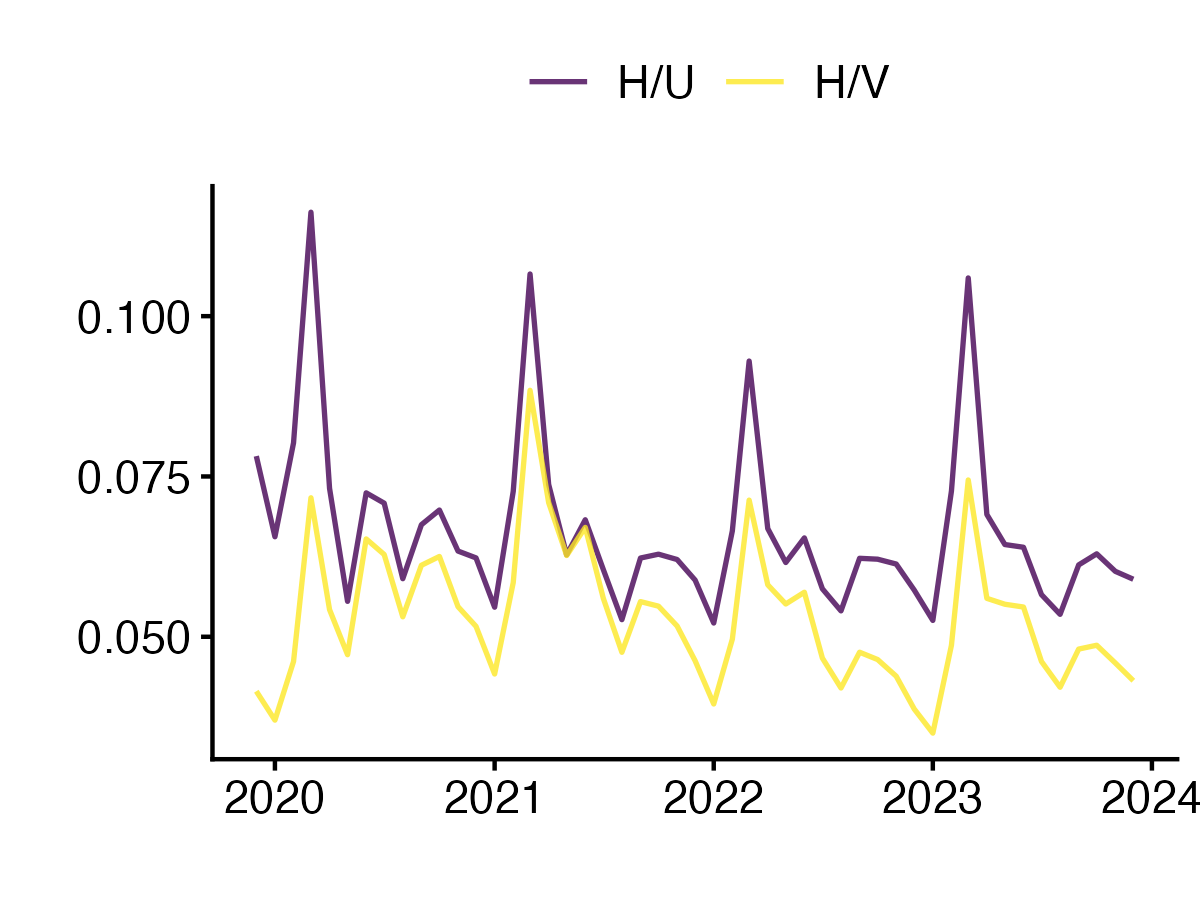}\includegraphics[width = 0.33\textwidth]
  {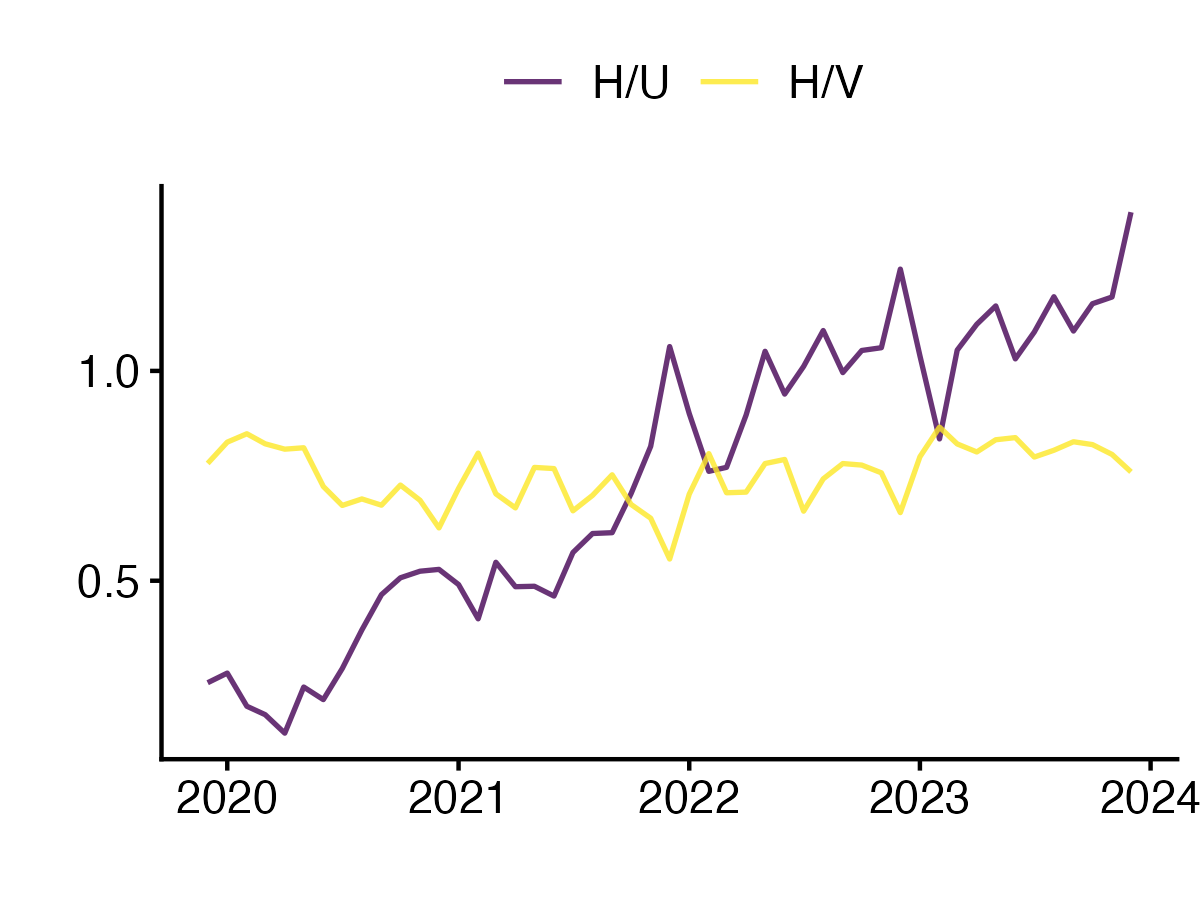}}
  \caption{Trends of Macro Variables: Hello Work Part-time (left) vs. Platform (right), 2019--2023}
  \label{fg:unemployed_vacancy_month_aggregate} 
  \end{center}
  \footnotesize
  Note: Details are in \cite{kanayama2026nonparametric}. The right panels are computed from the Timee contract-level data; ``users'' in the right panel of (a) denotes active users on the platform rather than registered workers. For the left panels, we use the Report on Employment Service (\textit{Shokugyo Antei Gyomu Tokei}) for the month-level aggregate data from December 2019 to December 2023 to capture trends in matching between unemployed workers and vacancies via a conventional platform. These datasets include the number of job openings, job seekers, and successful job placements, primarily sourced from the Ministry of Health, Labour and Welfare (MHLW) of Japan, which publishes monthly reports and statistical data on the Public Employment Security Office, commonly known as Hello Work. Hello Work is a government-operated institution in Japan and a conventional platform that provides job seekers with employment counseling, job placement services, and vocational training, playing a critical role in Japan's labor market. 
\end{figure}

Throughout the paper, ``spot work platform'' refers to Timee, a matching platform for spot work.\footnote{According to the Japan Spot Work Association (JASWA), ``spot work'' refers to employment contracts for short-hour, one-off work engagements that are facilitated through digital platforms. These platforms serve as intermediaries that match job offers with available workers---often on a same-day or near-term basis---and rely on digital technologies to streamline matching, contracting, and compensation. JASWA plays a central role in promoting sound practices in the sector by issuing guidelines for legal compliance, offering interpretations of labor laws, and supporting proper labor management. In particular, it emphasizes the use of digital infrastructure to reduce the administrative burden traditionally associated with employment, including time tracking, wage calculation, and performance evaluation. This description draws on information provided by the association's official overview and published guidelines (\url{https://www.jaswa.or.jp/about/}; \url{https://jaswa.or.jp/pdf/spotwork_gudeline_v2.pdf}).} Timee is the leading online platform in Japan's spot-work market, given its scale relative to comparable platforms.\footnote{As of 2024, major spot-work platforms exhibit the following scale: Timee (approximately 230{,}000 registered establishments; PR TIMES \url{https://prtimes.jp/main/html/rd/p/000000222.000036375.html}
), Sharefull (approximately 55{,}000 corporate registrations; Sharefull Info \url{https://sharefull.com/information/6602/}
), and Mercari Hallo (approximately 50{,}000 registered stores; HRog \url{https://hrog.net/knowledge/122159/}).}
As of 2024, more than 7 million workers were registered on the platform---more than 10\% of Japan's labor force of roughly 70 million---illustrating its significant reach in the country's labor market. Of the more than 7 million registered workers, roughly 1 million were active, that is, opened the Timee app in the month, at the end of 2023, as the right panel of Figure \ref{fg:unemployed_vacancy_month_aggregate}(a) shows. Because the job-finding rate in Panel (c) exceeds one, an active worker typically matches more than once a month, so the 8.7 million contracts in our sample reflect repeated matches by a much smaller set of workers. The platform connects businesses with temporary workers for short-term jobs through a mobile app.\footnote{For the platform's user interface and job recommendation system, see \citet{sekiya2026designing}.}

\paragraph{Labor Demand}

\textcolor{black}{On the labor-demand side, businesses must first undergo a screening and verification process before being allowed to post jobs. Once registered, they can easily create postings through a user-friendly interface by specifying job details such as tasks, wages, and shift times. The platform makes these listings immediately visible to workers and facilitates the matching and contracting process. }

\paragraph{Labor Supply}
\textcolor{black}{On the labor-supply side}, the primary users of the platform are individuals seeking flexible, task-based employment rather than long-term positions, while establishments use the platform to seek short-term workers.
Workers can register on the platform for free and immediately gain access to job offers from various companies, with the platform streamlining the process of matching workers with available shifts across industries such as food, retail, and logistics, \textcolor{black}{without the need for a traditional hiring process such as formal applications, interviews, or long-term employment contracts.}

Certain qualifications (nurses, caregivers, childcare workers, etc.) are required for some jobs, but these jobs are a minority of the total, and most are jobs that any worker can do.
Workers in this market can apply for jobs and perform their duties without having any special human capital.
Therefore, we treat this labor market as one requiring minimal skills.

\paragraph{Platform Objective}
\textcolor{black}{The platform's objective is to maximize revenue from successful matches between workers and businesses. Unlike digital platforms that rely on advertising, registration fees, or subscriptions, Timee does not charge either side for access. Instead, it collects a 30\% commission on each contracted payment between a business and a worker. Payments are remitted only when the worker actually completes the shift, so platform revenue is tied to realized matches rather than postings alone. This design aligns the platform's incentives with the volume of completed matches. Because the commission is proportional to contracted payments, it also scales establishments' per-shift labor cost proportionally, so a statutory wage-floor increase raises the effective cost of a spot shift by 1.3 times the statutory increase in yen terms, while the proportional increase is unchanged. Appendix \ref{sec:platform_revenue} shows that total posted pay on the platform did not decline after the October 2023 revision. This arrangement benefits both companies and workers by offering flexible, short-term employment opportunities while avoiding the formalities and commitments of long-term contracts.} 

\textcolor{black}{At the same time, the platform allows workers to take repeated jobs from the same establishment, and businesses may extend longer-term employment offers if they wish. In this sense, the platform does not preclude transitions into regular part-time employment and can serve as a trial channel through which businesses and workers explore longer-term arrangements. Its simplicity and immediacy are especially attractive to workers seeking to fit employment around their schedules.}

\paragraph{Transfers from Companies to Workers}
On this platform, workers may receive not only wage income based on hourly pay and hours worked but also transportation reimbursement. This reimbursement is intended to cover travel to and from the workplace, but firms are not legally required to provide it. Because the amount must be set when a job is posted, firms often lack precise information about workers' realized commuting costs, and workers are not required to reconcile differences ex post. Transportation reimbursement is therefore a salient non-wage margin along which firms could, in principle, adjust to higher labor costs following a minimum-wage increase.

\subsection{Institutional vs. Platform-Based Matching: Hello Work and Spot Work}

Figure \ref{fg:unemployed_vacancy_month_aggregate} provides a comparative analysis of labor market dynamics between the Hello Work public employment platform (left panel) and a private spot work platform (right panel) from December 2019 to December 2023. Hello Work is a government-operated institution in Japan and a conventional platform that provides job seekers with part-time and full-time employment counseling, job placement services, and vocational training, playing a critical role in Japan's labor market. 

First, in Figure \ref{fg:unemployed_vacancy_month_aggregate}, we discuss the left panels for Hello Work, which cover part-time workers and jobs.\footnote{For full-time jobs on the public and private job matching platforms, \cite{otani2025onthejob} estimates trends of matching efficiency and elasticity between 2014 and 2024. \cite{otani2024nonparametric} also provides the long-term trends of full-time and part-time job matching in Hello Work between 1972 and 2024.} In Panel (a), the number of unemployed individuals remains stable over the observed period, hovering around 0.8 million. Vacancies and tightness ($V/U$) show stable trends. Panel (b) shows that the hiring count through Hello Work remains low, with a small oscillating pattern throughout the period. This stable but limited hiring activity suggests that the platform may face constraints or inefficiencies in increasing the job match rate for part-time workers. In Panel (c), the job and worker finding rates ($H/U$ and $H/V$) exhibit a gradual decline over time, suggesting reduced effectiveness in matching job seekers with available vacancies via Hello Work. Here, $H/V$ is interpreted as hires per vacancy posting in aggregate data. This decline might reflect a growing preference for alternative job search methods such as the spot-work platform.

In contrast, the right panels in Figure \ref{fg:unemployed_vacancy_month_aggregate} reflect the dynamics of the private platform, where a distinctive pattern emerges. Unlike Hello Work's part-time flow from unemployed to employed, workers on this platform do not automatically leave after securing a match, which retains a larger pool of active users. The number of active users increases gradually, while the number of vacancies rises sharply, especially post-2022, leading to a notable rise in labor market tightness ($V/U$). This reflects an expanding demand for spot employment opportunities. The hiring count in Panel (b) shows a rapid upward trajectory starting around 2022, underscoring a marked increase in successful matches.

In Figure \ref{fg:unemployed_vacancy_month_aggregate}, Panel (c) illustrates the job and worker finding rates ($H/U$ and $H/V$), highlighting significant differences compared to Hello Work. The job finding rate ($H/U$) continues to rise, reaching values close to 1.3, indicating that each worker matches with multiple vacancies per month. On the other hand, the worker finding ratio ($H/V$) remains stable around 0.8. Because $V$ is the number of postings, this reflects hires per posting rather than a position-level fill probability. This pattern demonstrates the platform's increasing efficiency in matching workers to vacancies. It indicates a maturing spot labor market that still exhibits strong growth.

\subsection{Minimum Wages in Japan}

\begin{figure}
    \begin{center}
        \includegraphics[width=0.75\linewidth]{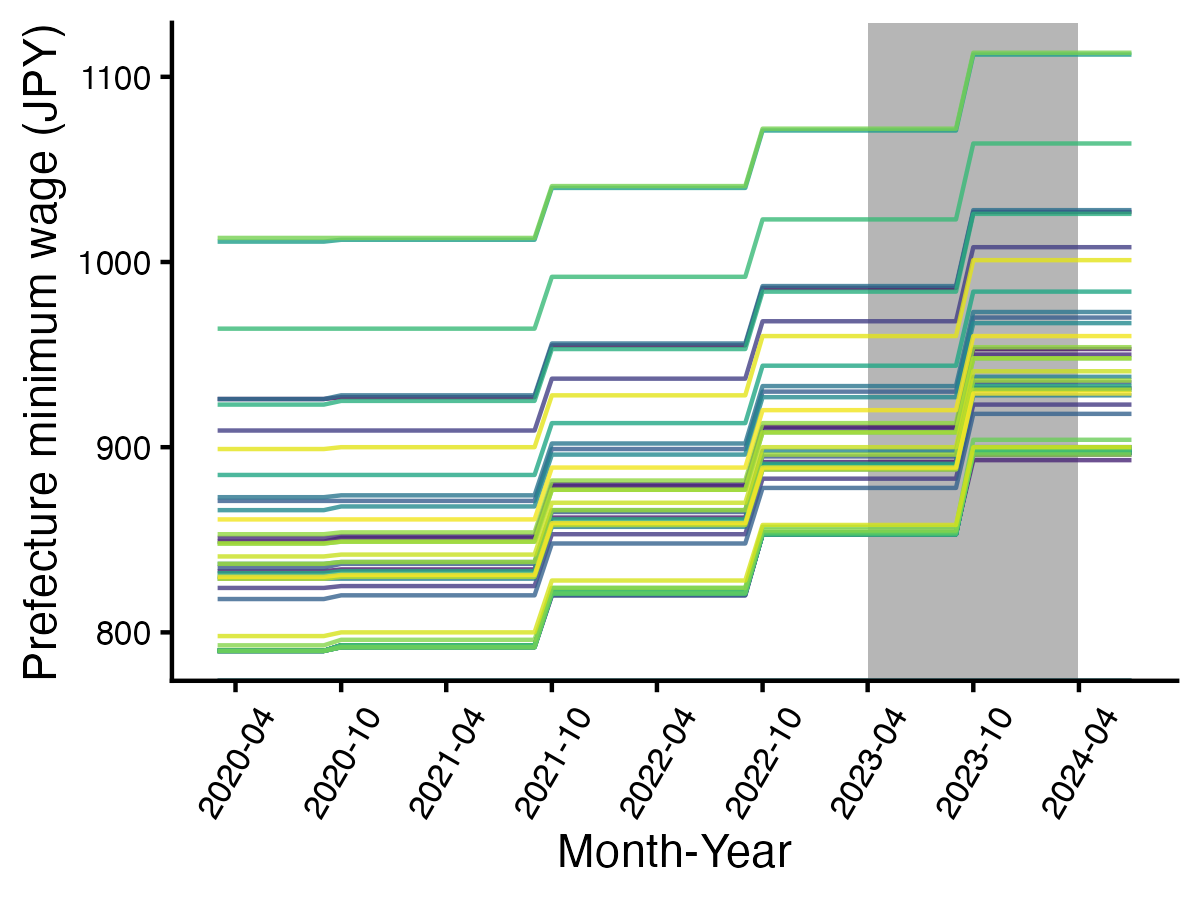}
        \caption{Dynamics of Regional Minimum Wage Changes}
        \label{fig:minimum_wage_year_month}
    \end{center}
    \footnotesize
    Note: The figure plots monthly regional minimum wages across prefectures. Until 2022, each prefecture was assigned to one of four categories based on workers' average income, and each category was given a guideline minimum wage. Many prefectures set their regional minimum wage close to that guideline. The shaded period is our focus.
\end{figure}

This section introduces the regional minimum wages set in each of the 47 prefectures in Japan, which we define as ``regional minimum wages.'' The introduction of regional minimum wages follows the framework described by \cite{Tamada_2009_en}, \cite{tamada2011analysis}, and \cite{Kawaguchi_Mori2021}. The Ministry of Health, Labour and Welfare established the Central Minimum Wage Council, which operates under the tripartite principle and consists of representatives from labor, management, and public interest groups.

The Council divides the 47 prefectures into \textcolor{black}{three} regional blocks based on regional welfare levels. It sets guidelines (``Meyasu'') for the recommended increase in the hourly minimum wage for each group.\footnote{\textcolor{black}{The Council had divided the 47 prefectures into four regional blocks before 2022.}} These recommendations primarily reference the average annual wage increase rate for workers at workplaces with fewer than 30 regular employees, as reported in the Minimum Wage Survey. The Council typically holds its first meeting in June and finalizes the guidelines by the end of July. Based on these guidelines, the Local Minimum Wage Councils in each prefecture determine the specific minimum wage increases in August, with the revised wages coming into effect during the first week of October. This two-stage process has been implemented annually since 1978, with a few exceptions, ensuring regular increases in the local minimum wage.

The Regional Minimum Wage Act applies to nearly all workers in Japan, regardless of industry, age, or gender. However, exceptions exist where the director of the labor bureau in each prefecture deems it appropriate to pay less than the regional minimum wage.\footnote{Examples where it is considered appropriate to pay less than the minimum wage include: 1) individuals with significant mental or physical disabilities, 2) employees in trial periods, 3) participants in certified basic vocational training programs, 4) workers engaged in light work, and 5) those performing intermittent work.}

Figure \ref{fig:minimum_wage_year_month} illustrates the changes in regional minimum wages over time, showing minimum wage variations across prefectures from 2020 to 2023. In October 2020, due to the negative employment shock caused by COVID-19, the minimum wage increased only slightly, by an average of 0.2\%. However, from 2021 onwards, the minimum wage increased consistently each year, with an average increase of 4.7\% in 2023. Notably, except for the period of economic stagnation during the COVID-19 pandemic, substantial minimum wage increases have occurred in October across all prefectures.

While minimum wage hikes in Japan are effective at raising wages \citep{aoyagi2016minimum}, they are also associated with negative employment effects.
Numerous studies have examined the impact of minimum wage policies on employment in Japan \citep{yugami, tachibanaki2006, Kawaguchi_Mori2009, HiguchiYoshio2013TDoP, kambayashi2013minimum, akesaka2017impact, okudaira2019minimum, izumicprc2020, Kawaguchi_Mori2021, Izumi_kawaguchi_okudaira2022}.
Many of these studies report negative employment effects for certain demographic groups \citep{Kawaguchi_Yamada2007, Kawaguchi_Mori2009, kambayashi2013minimum, akesaka2017impact, okudaira2019minimum, Kawaguchi_Mori2021, Izumi_kawaguchi_okudaira2022, mori2025higher}.
For example, \cite{Kawaguchi_Mori2021} examines the employment effects of minimum wages by leveraging the 2007 reform of Japan’s Minimum Wage Act.\footnote{This reform mandated that monthly minimum wage earnings must meet or exceed monthly public assistance levels. Prefectures where minimum wages fell below public assistance thresholds experienced substantial increases. Using the difference between public assistance levels and monthly minimum wage earnings in 2006 as an instrumental variable, they estimate the causal effect of the minimum wage on employment during the years following the reform. This approach has since been applied in several studies \citep{okudaira2019minimum, mizushima2021spillover, yamagishi2021minimum, Izumi_kawaguchi_okudaira2022, mori2025higher, yamanouchi2025minimum_wage}.}
They find that the employment elasticity of the minimum wage for less-educated young men is -1.2. This implies that although the minimum wage in Japan affects only certain demographic groups, the magnitude of its impact can be larger than what is observed in other countries.

\section{Data}\label{sec:data}

\subsection{Data Source and Construction}

\textcolor{black}{We base our analysis on contract- and posting-level records from Timee, a privately operated Japanese platform that matches spot workers and employers. As the oldest and largest player in Japan's spot labor market, Timee provides highly representative data for this setting. We focus on April 2023 through March 2024 to study the October 2023 minimum-wage revision in a period of broader platform activity.} Two further considerations motivate this choice. First, the 2023 revision is the largest increase among the 2020--2023 revisions shown in Figure \ref{fig:minimum_wage_year_month}, providing substantial wage-floor variation. Second, the platform had expanded substantially by 2023, providing broader coverage than in 2021 and 2022. The earlier years nonetheless serve as benchmarks: Appendix \ref{app:placebo_effect_2020} uses the October 2020 revision, when regional minimum wages rose by only 0--3 yen, as a placebo, and Appendix \ref{app:difference_in_difference_in_differences} uses 2020 as the reference year to assess sensitivity to wage-bin-specific seasonality. These checks do not rule out shocks specific to 2023 that affected low- and high-wage bins differently.

We aggregate the posting-level data to calculate key metrics at the prefecture-month-wage-bin and prefecture-month levels, including the number of posted positions (vacancies) and matches (employment). 
Unlike \cite{dustmann2022reallocation}, there is no difference between actual and contractual working hours, which is an advantage of our data. 
The dataset also includes not only detailed employer characteristics (industry, occupation, and location—prefecture, municipality, longitude, latitude) but also precise timestamps for vacancy posting and filling (and shift start), as well as the full job-description text. 
Measuring vacancy duration as the elapsed time a posting remains advertised—from appearance on the platform until it is filled, rather than until work begins—has the advantage of focusing on the time to find candidates, as in \cite{bassier2025vacancy}. 
Job descriptions set by employers provide rich textual signals—implicitly conveying skill requirements (e.g., “experience required,” heavy-lifting, certification cues), task intensity and shift constraints, and preferred worker attributes. These signals influence applicant self-selection, as shown in \cite{kuhn2020gender}, even when they are not binding screening criteria.
This rich dataset enables comprehensive heterogeneity analysis of how minimum wage changes impact the short-term dynamics of spot labor markets.

\begin{table}[!htbp]
  \begin{center}
      \caption{Summary Statistics}
      \label{tb:summary_statistics_wage_bin_prefecture_year_month}
      \subfloat[Posting Level]{
\begin{tabular}[t]{lrrrrrr}
\toprule
  & N & mean & median & sd & min & max\\
\midrule
Hourly wage & 8673942 & 1086.01 & 1100.00 & 123.12 & 853.00 & 2000.00\\
Working hour & 8673942 & 4.79 & 4.50 & 1.84 & 1.00 & 16.00\\
Num of positions & 8673942 & 1.67 & 1.00 & 4.60 & 1.00 & 8961.00\\
Transportation reimbursement (JPY) & 8673942 & 404.65 & 500.00 & 255.19 & 0.00 & 1500.00\\
Transportation reimbursement dummy & 8673942 & 0.81 & 1.00 & 0.39 & 0.00 & 1.00\\
Vacancy duration (Day) & 6707145 & 3.25 & 0.25 & 6.73 & 0.00 & 99.34\\
Posting-to-start lag (Day) & 6707145 & 9.04 & 5.85 & 9.69 & 0.00 & 99.99\\
Beginner-targeted job & 8673942 & 0.01 & 0.00 & 0.10 & 0.00 & 1.00\\
Experienced-targeted job & 8673942 & 0.30 & 0.00 & 0.46 & 0.00 & 1.00\\
Female-targeted job & 8673942 & 0.03 & 0.00 & 0.17 & 0.00 & 1.00\\
Male-targeted job & 8673942 & 0.03 & 0.00 & 0.17 & 0.00 & 1.00\\
\bottomrule
\end{tabular}
}\\
      \subfloat[Prefecture-month-wage-bin Level]{
\begin{tabular}[t]{lrrrrrr}
\toprule
  & N & mean & median & sd & min & max\\
\midrule
Employment & 61584 & 191.31 & 0.00 & 1787.87 & 0.00 & 83107.00\\
Vacancy & 61584 & 234.98 & 0.00 & 2171.39 & 0.00 & 111887.00\\
Mean transportation reimbursement (JPY) & 61584 & 111.73 & 0.00 & 222.09 & 0.00 & 1500.00\\
Mean transportation reimbursement dummy & 61584 & 0.21 & 0.00 & 0.38 & 0.00 & 1.00\\
\bottomrule
\end{tabular}
}\\
      \subfloat[Prefecture-month Level]{
\begin{tabular}[t]{lrrrrrr}
\toprule
  & N & mean & median & sd & min & max\\
\midrule
Employment & 564 & 20889.66 & 7150.00 & 34791.94 & 505.00 & 262867.00\\
Vacancy & 564 & 25657.84 & 9240.50 & 42700.28 & 633.00 & 350673.00\\
\bottomrule
\end{tabular}
}     
  \end{center}\footnotesize
  \textit{Note}: This table reports the mean, median, standard deviation, and minimum and maximum values of each outcome variable during the period of analysis, April 2023 through March 2024. 150 JPY is about 1 USD at the exchange rate in 2025. Vacancy duration and posting-to-start lag are defined only for vacancies that are filled.
\end{table} 

Table \ref{tb:summary_statistics_wage_bin_prefecture_year_month} reports summary statistics at the posting, prefecture–month–wage-bin, and prefecture–month levels.
At the posting level, hourly pay averages 1{,}086 JPY (median 1{,}100), with typical shifts of 4.79 hours. Employers commonly offer transportation reimbursement (mean 405 JPY; median 500), but the wide range underscores substantial heterogeneity by location and policy. About 81\% of postings include any reimbursement. For timing, vacancy duration is measured as the elapsed time from posting availability to fill; the posting-to-start lag (start lead time) is the elapsed time from posting availability to the shift start. Among fully filled vacancies, the mean duration is 3.25 days and the mean posting-to-start lag is 9.04 days. Job-description targeting flags indicate that beginner-targeted, experienced-worker-targeted, female-targeted, and male-targeted postings account for roughly 1\%, 30\%, 3\%, and 3\% of contracts, respectively.

At the prefecture–month–wage-bin level, employment and vacancies average 191.31 and 234.98 per bin, respectively; the mean transportation reimbursement is 111.73 JPY per contract in the cell, with a 21\% reimbursement incidence. At the prefecture–month level, spot employment averages 20{,}890 and vacancies 25{,}658, revealing substantial time-series and cross-prefecture variation and a persistent gap consistent with unfilled positions.

\subsection{Preliminary Evidence Before and After the Minimum Wage Change}

Leveraging the strengths of granular spot work platform data, we focus on Tokyo to illustrate how the minimum wage adjustment reshaped employment patterns, allowing for a detailed examination of wage shifts while revealing the stability of working hours and transportation reimbursement.

\begin{figure}[!ht]
  \begin{center}
  
  \subfloat[Daily]{\includegraphics[width = 0.48\textwidth]{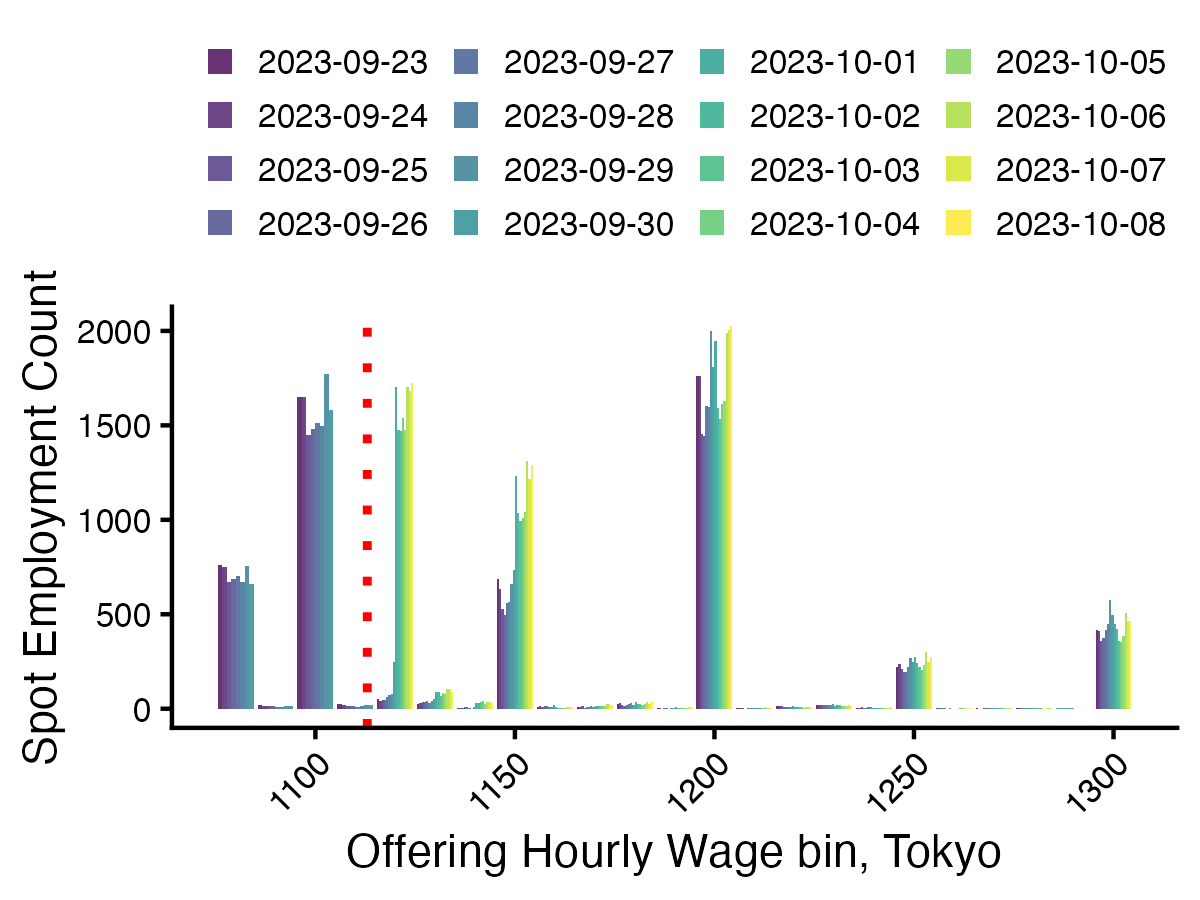}}
  \subfloat[Weekly]{\includegraphics[width = 0.48\textwidth]{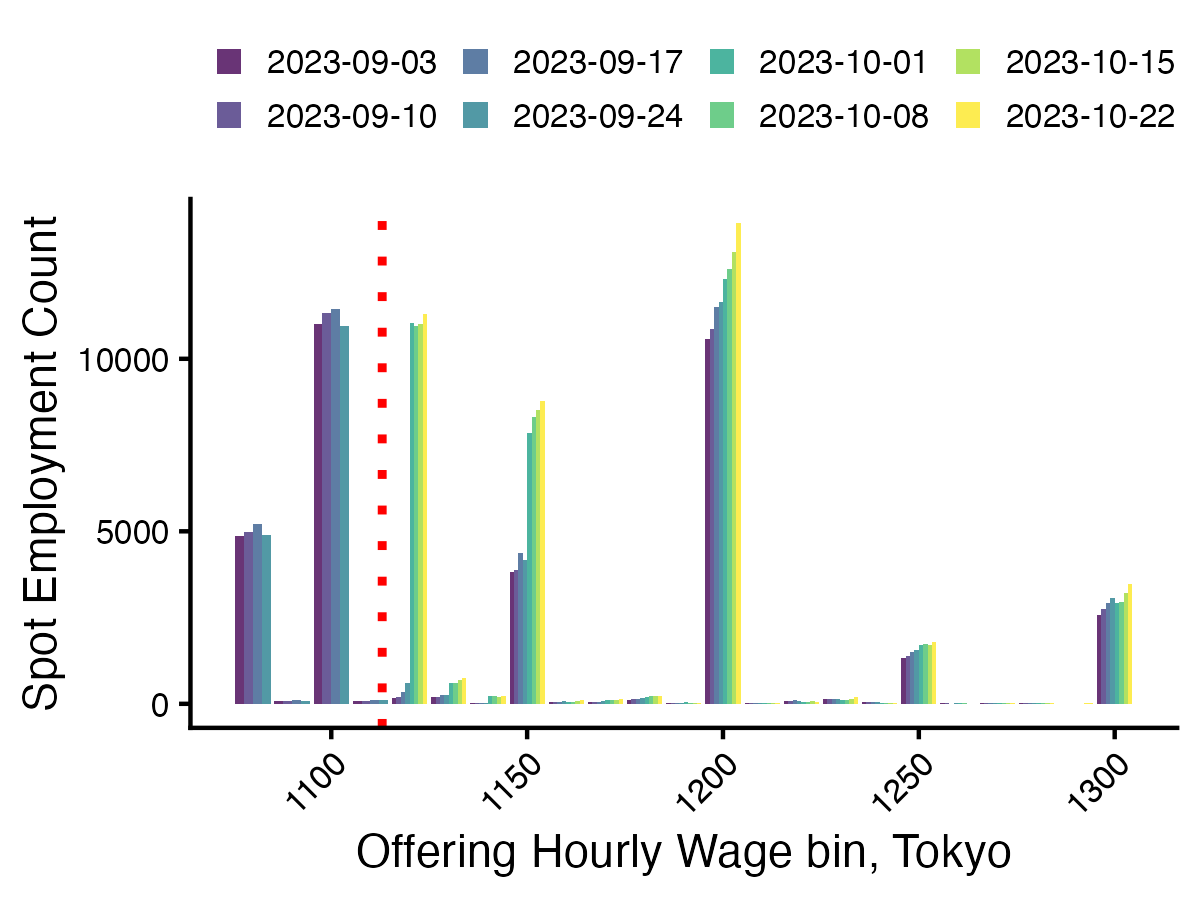}}\\
  \subfloat[Monthly]{\includegraphics[width = 0.48\textwidth]
  {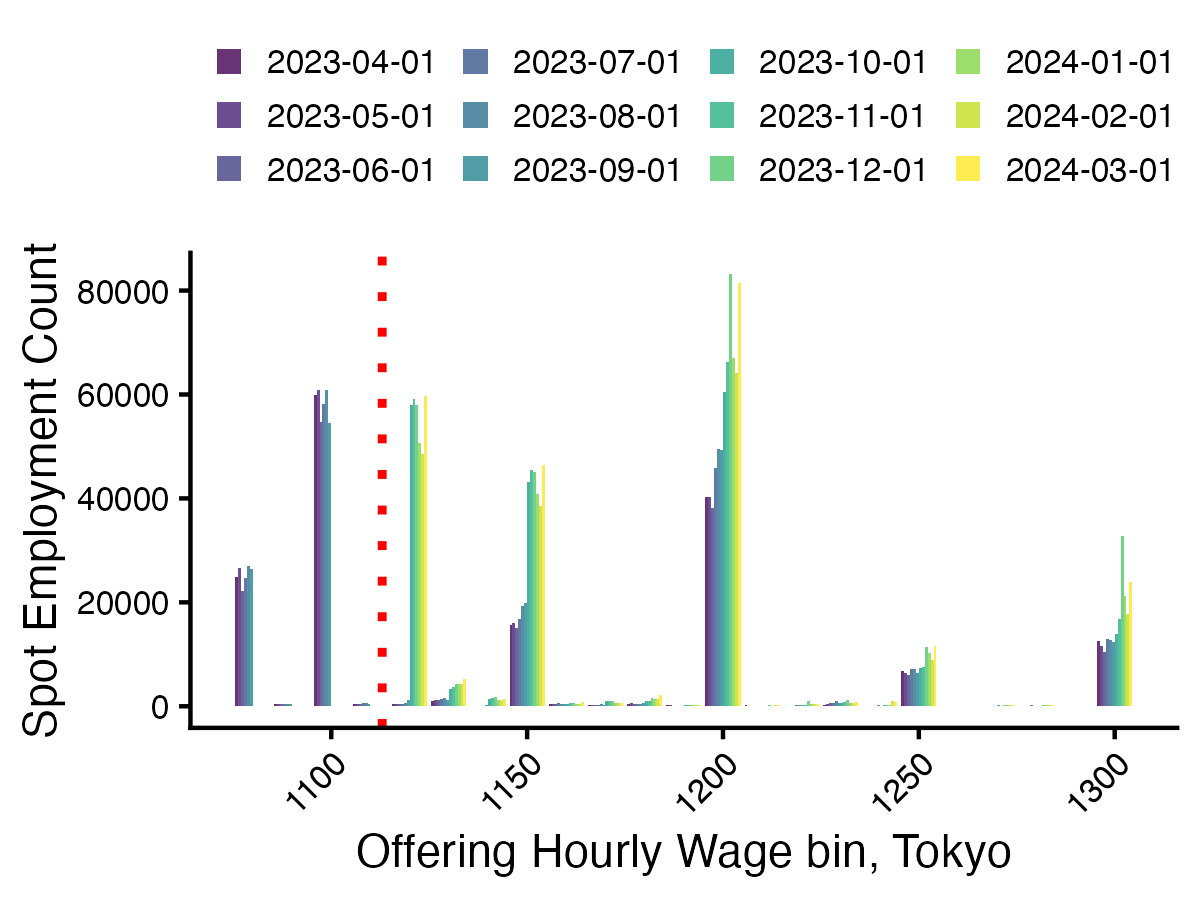}}
  \caption{Transition of Employment Distribution by Day, Week, and Month Before and After the Minimum Wage Change}
  \label{fg:wage_bin_year_month_week_offering_count} 
  \end{center}
  \footnotesize
  Note: The x-axis represents the wage bins, while the y-axis indicates the total count of spot employment. The colored bars represent different months, allowing for a clear comparison of employment pattern shifts over time.
\end{figure}

\paragraph{The Hourly Wage Distribution}
Figure \ref{fg:wage_bin_year_month_week_offering_count} illustrates the transition in employment distribution across hourly wage bins at both the daily and weekly levels before and after the minimum wage adjustment. The ability to observe such granular, high-frequency adjustments is a unique strength of our spot labor platform data, enabling us to capture firms' real-time responses to policy changes with day-to-day resolution. Panel (a) shows the daily evolution of the employment distribution in Tokyo, revealing a sharp and immediate decline in employment below the new minimum wage threshold (1,113 JPY) just one day after the policy took effect. Jobs previously offered at lower wages either disappeared or were adjusted upward, with a clear clustering of employment around the new minimum wage floor. Complementarily, Panel (b) aggregates these patterns at the weekly level, confirming the persistence of this shift over the course of a month. Panel (c) shows that the effect persists over a six-month period.

\begin{figure}[!ht]
  \begin{center}
  \subfloat[Posted Working Hour]{\includegraphics[width = 0.46\textwidth]
  {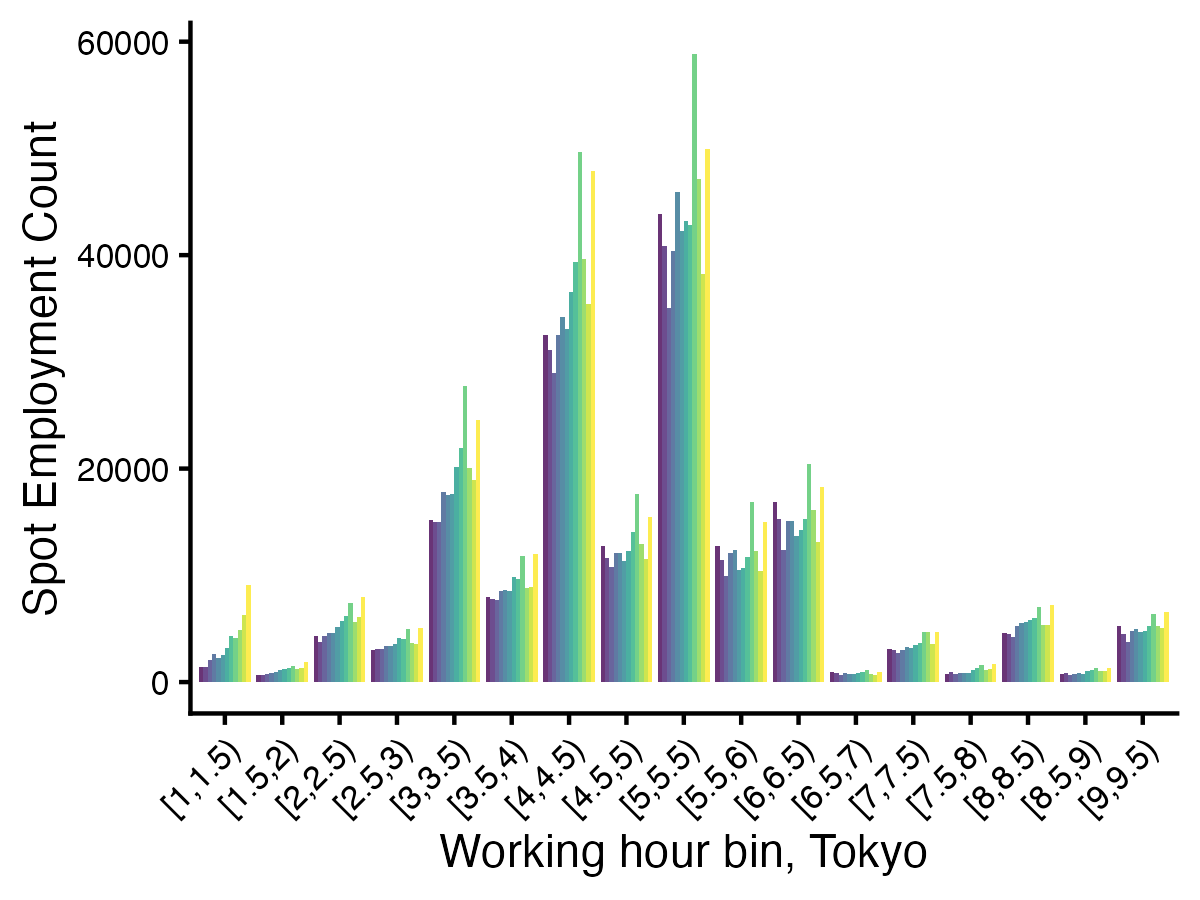}}
  \subfloat[Transportation Reimbursement]{\includegraphics[width = 0.46\textwidth]
  {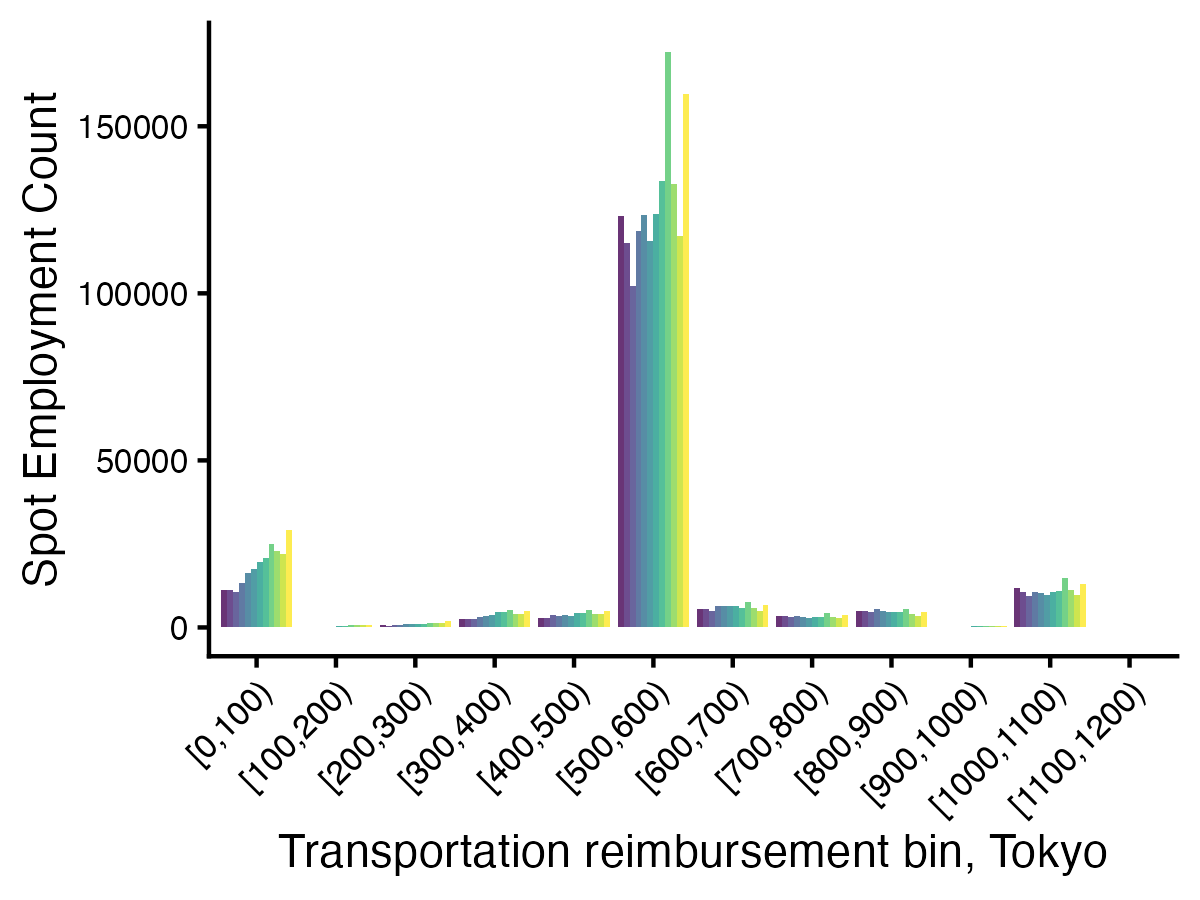}}
  \caption{Transition of Employment Distribution Before and After the Minimum Wage Change}
  \label{fg:wage_bin_year_month_offering_count}
  \end{center}
  \footnotesize
  Note: The x-axis represents working-hour bins (Panel (a)) and transportation reimbursement bins (Panel (b)), while the y-axis indicates the total count of spot employment. The colored bars represent different months, allowing for a clear comparison of employment pattern shifts over time.
\end{figure}

\paragraph{Adjustments in Working Hours and Transportation Reimbursement}
Panel (a) in Figure \ref{fg:wage_bin_year_month_offering_count} shows that the employment distribution across working-hour bins remains broadly stable. Common shift lengths, such as four and five hours, remain prominent, with at most modest changes in shorter shifts. Panel (b) shows a similarly stable distribution of transportation reimbursement, concentrated in the 500 JPY bin. Overall, the descriptive evidence suggests that firms mainly adjusted posted wages rather than other job attributes.

\begin{figure}[!ht]
  \begin{center}
  
  \subfloat[Working Hour and Wage]{\includegraphics[width = 0.33\textwidth]
  {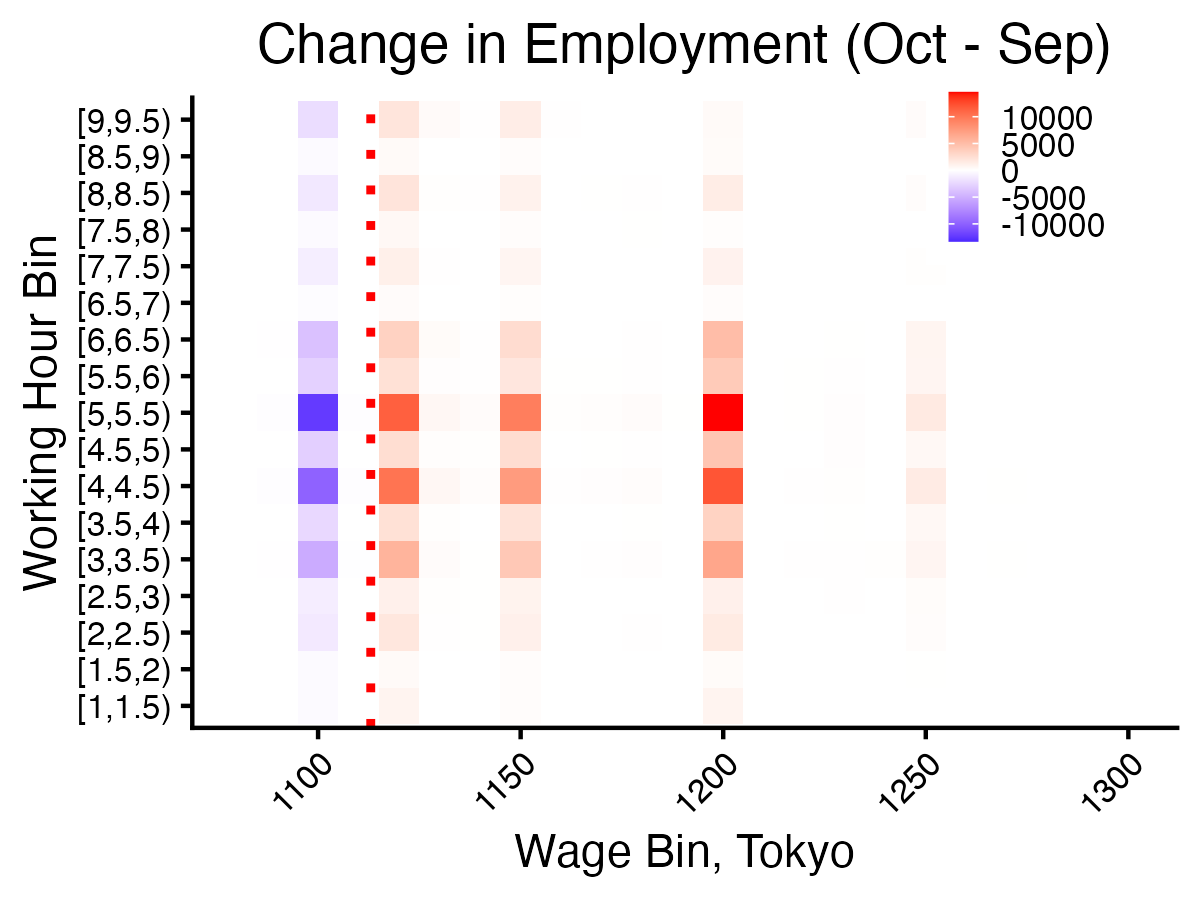}}
  \subfloat[Transportation Reimbursement and Wage]{\includegraphics[width = 0.33\textwidth]
  {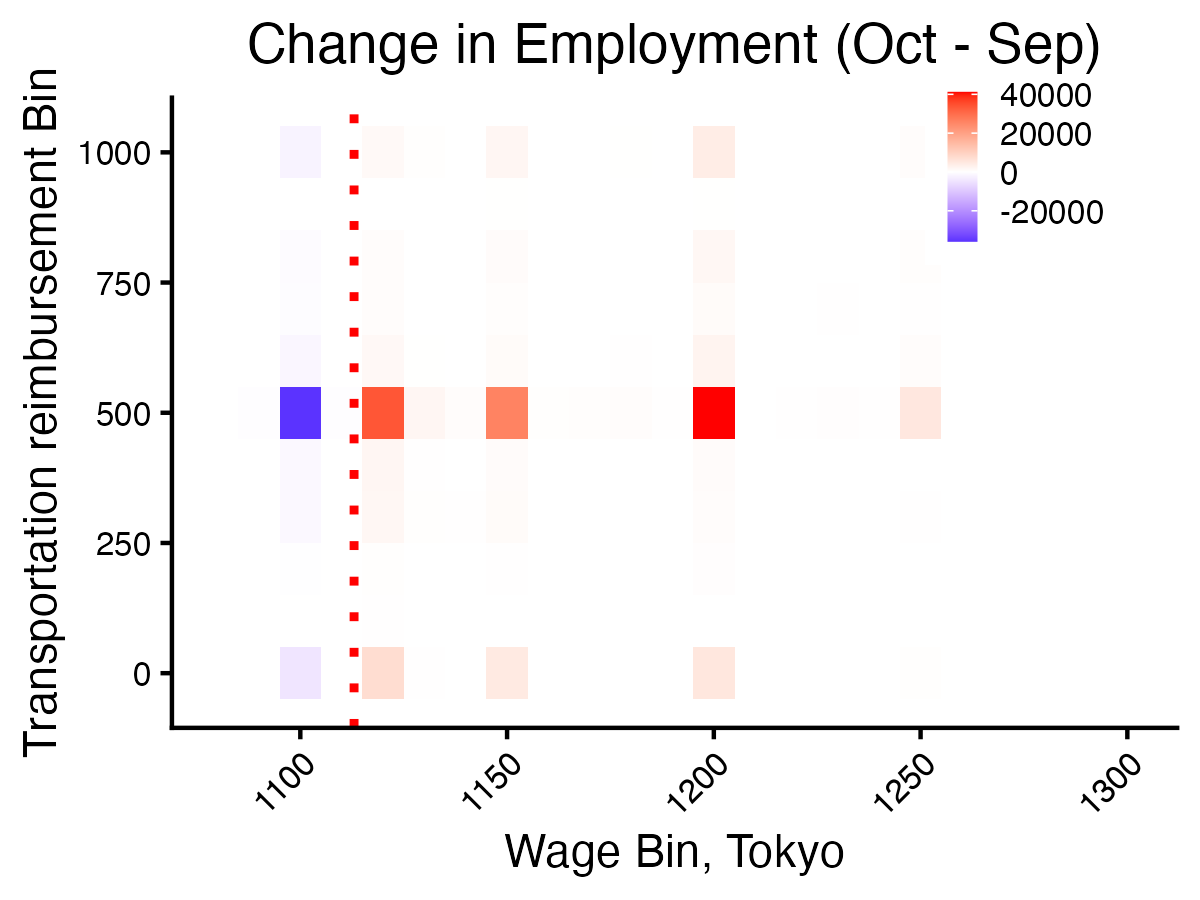}}
  \subfloat[Transportation Reimbursement and Working Hour]{\includegraphics[width = 0.33\textwidth]
  {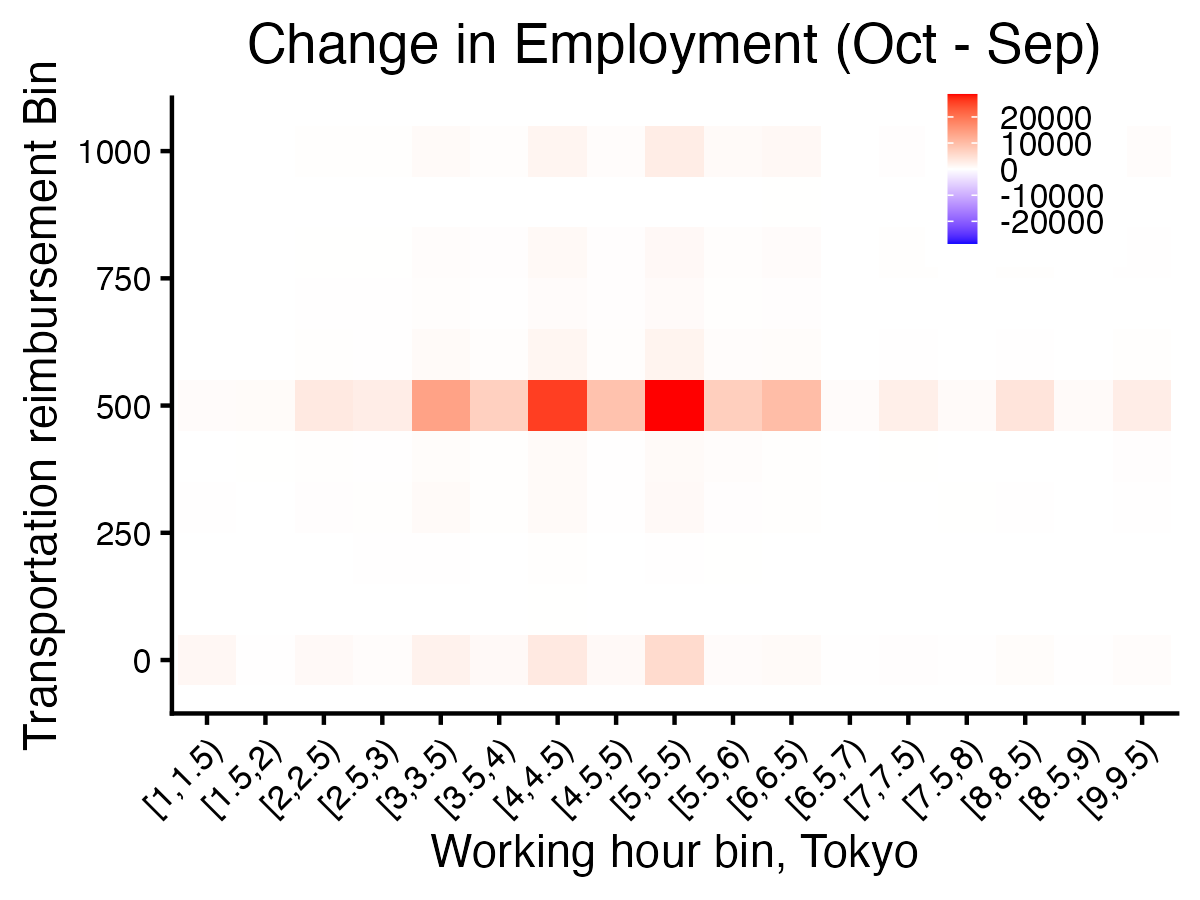}}
  \caption{Change in Employment Distribution Before and After the Minimum Wage Change}
  \label{fg:transportation_expense_wage_bin_offering_count_202310_minus_202309} 
  \end{center}
  \footnotesize
  Note: The color intensity reflects the change in spot employment between September and October, with red indicating an increase and blue indicating a decrease. The red dashed line denotes Tokyo's 2023 October minimum wage, i.e., 1,113 JPY.
\end{figure}

A natural question is whether the wage increase resulting from the minimum wage adjustment led to adjustments in working hours and transportation reimbursement, or whether firms primarily confined their response to wage adjustments. Figure \ref{fg:transportation_expense_wage_bin_offering_count_202310_minus_202309} demonstrates that the minimum wage adjustment primarily impacted the distribution of employment across wage bins, while working hours and transportation reimbursement remained largely unaffected. In Panel (a), employment previously concentrated in wage bins below the new minimum wage threshold shifted upward, clustering around the new minimum wage level, indicating compliance-driven wage adjustments. However, changes in the distribution of working hours are minimal, suggesting that firms maintained their established job structures despite increased labor costs. Similarly, transportation reimbursement patterns in Panel (b) exhibit stability, with no significant redistribution, reinforcing the notion that the overall distribution of this compensation component changed little.
\textcolor{black}{This pattern is further supported by Panel (c), which shows that even among jobs with short working hours—where transportation costs make up a larger share of compensation—there was little adjustment in reimbursement levels.}
These descriptive patterns show that the most visible response to the minimum wage hike was wage compliance; Appendix~\ref{app:other_measured_outcome} estimates the reimbursement response within the affected wage bins.

\paragraph{Summary}
These illustrative data patterns motivate us to move beyond descriptive analysis and investigate the causal impact of the minimum wage adjustment on employment outcomes. While the observed shifts in wage distribution suggest strong compliance effects and limited adjustment on other job attributes, such as working hours and transportation reimbursement, these patterns alone do not establish causality. In particular, they cannot account for concurrent platform growth or heterogeneous trends across prefectures, which may confound observed changes. To rigorously evaluate whether and to what extent the policy change affected employment dynamics—particularly in terms of spot hiring behavior—we turn to a difference-in-differences (DID) framework. This approach allows us to isolate the effect of the minimum wage hike by comparing changes in outcomes between treated and control groups over time, accounting for underlying trends and potential confounders.

\section{Empirical Strategy}\label{sec:empirical_strategy}
\subsection{Prefecture-month--wage-bin DID Estimates}
We compare employment shares near the minimum wage with those in the upper wage distribution, following the comparison-group logic of \citet{dustmann2022reallocation}. Shares remove proportional platform growth common to all wage bins within a prefecture, which can generate unequal absolute changes across bins. Our estimates describe relative changes in the wage distribution.

Specifically, we first decompose the wage distribution in each prefecture into wage bins, where each bin $j$ has a width of $10$ yen. We then partition each wage bin $j$ in prefecture $p$ into mutually exclusive groups $E_j(p)=e$ for $e \in \{-1,\dots,3,\infty\}$, based on its relative distance from the increased minimum wage $MW_{p}$\footnote{\textcolor{black}{In all prefectures, the increase in the minimum wage from 2022 to 2023 is less than 100 JPY. Therefore, any nominal wage bin that is 200 JPY or more below the new 2023 minimum wage would necessarily lie at least 100 JPY below the pre-revision (2022) minimum wage. In our setting, such bins are essentially never observed in posted wages and, in any case, lie outside our analysis window. Accordingly, we define ``missing jobs'' using the 100 JPY window immediately below the new minimum wage (the $e=-1$ group) and do not include bins further below (which would correspond to $e<-1$).}}.
Specifically, if wage bin $j$ in prefecture $p$ falls within the interval $[MW_{p}+e \times 100,MW_{p}+e\times100+99]$, it is assigned to group $e$. We define the threshold $\bar{W}_{p}=MW_{p}+400$, and assign $E_{j(p)}=\infty$ if wage bin $j$ falls within $[\bar{W}_{p},\infty]$. We assume that wage bins in group $\infty$ are unaffected by the minimum wage increase, an assumption we refer to as the limited spillover effect assumption. Under this assumption, we refer to group $\infty$ as the control group. Finally, for each group $e$, we compare the evolution of the outcome between group $e$ and group $\infty$ before and after the minimum wage increase.

We implement this DID identification strategy by running the following regression:
\begin{align}
\label{sec:empirical_strategy_DID_regression}
\frac{Y_{j(p),t}}{N_{p,t}}=\sum_{l \neq -1}\sum_{e \neq \infty}\mu_{l,e}\mathbf{1}\{t-7=l\}\mathbf{1}\{E_{j(p)}=e\}+\alpha_{e}+\lambda_{t}+\epsilon_{j,t},
\end{align}
where $t \in \{1,\dots,12\}$ denotes the calendar month, and $l \in \{-6,\dots,5\}$ denotes the relative time period before or after the minimum wage increase at time $t=7$.\footnote{$t=1$ corresponds to April $2023$, and $t=12$ corresponds to March $2024$.} $Y_{j(p),t}$ is employment for each bin $j$ in prefecture $p$ at time $t$, $N_{p,t}$ is employment in prefecture $p$ at time $t$, $\alpha_{e}$ is the group fixed effect, and $\lambda_{t}$ is the time fixed effect.\footnote{The notation $j(p)$ indicates that the wage bin $j$ belongs to prefecture $p$. We adopt this notation because the regression in \eqref{sec:empirical_strategy_DID_regression} exploits variation only at the group $e$ and time $t$ levels.} In additional analyses, we replace the left-hand-side outcome with corresponding measures such as vacancies or total hours. Note that this regression is saturated and each coefficient $\mu_{l,e}$ is the difference-in-differences estimand:
\begin{align*}
\mu_{l,e}=E\biggr[\frac{Y_{j(p),7+l}}{N_{p,7+l}}-\frac{Y_{j(p),6}}{N_{p,6}}\biggr|E_{j(p)}=e\biggr]-E\biggr[\frac{Y_{j(p),7+l}}{N_{p,7+l}}-\frac{Y_{j(p),6}}{N_{p,6}}\biggr|E_{j(p)}=\infty \biggr],
\end{align*}
where the pre-treatment period is $t=6$ and the control group is group $\infty$.\par 
Under parallel trends, each $\mu_{l,e}$ measures the dynamic share response of group $e$ relative to group $\infty$. Formally, we assume the three identifying assumptions below:

\subparagraph{Limited Spillover Effect Assumption}
This assumption requires that the effect of the minimum wage increase on the outcome vanishes at a certain point $\bar{W}_{p}$ in the wage distribution in each prefecture $p$. It ensures that group $\infty$ (the upper part of the wage distribution) serves as a valid control group. We will assess the validity of this assumption by examining the impact of the minimum wage increase on groups near $\bar{W}_{p}$ such as $e=2,3$.\footnote{As an additional robustness check, we report in Appendix \ref{app:control_definition} the results obtained when redefining the control group as wage bins at least $300$ yen above the new minimum wage ($MW_p + 300$) and, alternatively, at least $500$ yen above the new minimum wage ($MW_p + 500$).}

\subparagraph{Parallel Trends Assumption}
This assumption requires that for each group $e$, the outcome evolves similarly over time between group $e$ and the control group $\infty$.
This assumption is also not directly testable, but we assess the plausibility of this assumption by checking the pre-trend differences for the outcome. To assess the sensitivity of our estimates to potential heterogeneity in underlying trends between treated and control wage bins, we additionally estimate (i) an augmented version of our main specification that includes wage-bin fixed effects and wage-bin-by-time fixed effects, and (ii) a difference-in-difference-in-differences specification that removes wage-bin-specific seasonal trends, and report the corresponding results in Appendix \ref{app:additional_potential_issues}. To assess whether a similar October pattern appears in a year with little minimum-wage change, we also report in Appendix \ref{app:placebo_effect_2020} (iii) a placebo exercise around the October 2020 revision, when minimum wages were nearly unchanged.

\subparagraph{No Anticipation Assumption}
This assumption rules out the anticipation effect of the minimum wage increase on the outcome at a given period $l=-1$, i.e., before exposure to the policy shock.

\subparagraph{Additional Robustness Checks}
Beyond these identifying assumptions, we check the robustness of the main estimates to the composition of the comparison. To address the concern that treated and control wage bins may differ in job characteristics, we estimate a specification that adds occupation as an additional dimension, with treatment-group-by-occupation and month-by-occupation fixed effects, so that treated and control bins are compared within occupation and employment is normalized by prefecture--occupation--month totals; Appendix \ref{app:specification_with_occupation_controls} reports the results.

\subsection{Excess Jobs--Missing Jobs Approach}\label{sec:excess_job_missing_job_approach}
We aggregate the share effects using the missing--excess decomposition of \citet{Cengiz2019}. Throughout, employment effects and elasticities refer to this normalized share measure. \textcolor{black}{Let $K$ denote the prefecture average of the share of employment in the treated wage-bin groups ($e=-1,\dots,3$) in September 2023. We denote $\Delta a_{l}=\frac{10\sum_{e=0}^{3}\mu_{l,e}}{K}$ and $\Delta b_{l}=\frac{n_{-1}\,\mu_{l,-1}}{K}$ as the excess jobs and the missing jobs at a given relative period $l$, respectively, where 10 is the number of 10 JPY wage bins in each 100 JPY treated group and $n_{-1}$ is the prefecture average of the number of 10 JPY wage bins in the missing-job region; both quantities are normalized by the September employment share.\footnote{By comparison, \citet{Cengiz2019} use the sample-average employment-to-population ratio in treated states during the year (four quarters) prior to treatment, denoted by $EPOP_{-1}$, as a normalization benchmark.}}
Let $\Delta a=\frac{1}{6}\sum_{l=0}^{5}\Delta a_{l}$ and $\Delta b=\frac{1}{6}\sum_{l=0}^{5}\Delta b_{l}$ denote the average normalized excess and missing shares, respectively.
Also, let $\Delta a_e = \frac{10}{6K}\sum_{l = 0}^{5} \mu_{l, e}$ and $\Delta b_{-1} = \frac{n_{-1}}{6K}\sum_{l = 0}^{5} \mu_{l, -1}$ define the excess jobs and the missing jobs at a given relative group $e$, respectively.
We compute the net share effect at relative period $l$, denoted as $\Delta e_{l}$, by summing the excess and missing jobs at that period: $\Delta e_{l}=\Delta a_{l}+ \Delta b_{l}$.\footnote{\interlinepenalty=10000\relax For the reported Excess estimates, we multiply the estimated coefficient for each treated wage-bin group $e=0,\dots,3$ by 10, the number of 10 JPY wage bins in each 100 JPY group, and sum these contributions. For Missing, we multiply the coefficient for $e=-1$ by the prefecture average of the number of 10 JPY wage bins belonging to the missing-job region. Both quantities are then divided by $K$. Thus, Missing, Excess, and Net share the same denominator, and reported post-treatment statistics average the six monthly effects rather than cumulating them. The wage-distribution bars use the same bin-count weights, normalization, and time averaging. Vacancy, total-hours, and subgroup results use their corresponding reference stocks; employer-history vacancy components share the full-sample vacancy benchmark so that they add to the baseline. Weekly and daily results use analogous reference stocks at their own frequencies. Alternative control thresholds use their own lower-tier groups and reference stocks, while the DDD uses the September 2023 employment-share stock, not the 2023-minus-2020 reference difference. Standard errors use the full prefecture-clustered coefficient covariance, retaining covariance between Missing and Excess when calculating Net.} 
Furthermore, we calculate the average net share effect after the increase, denoted as $\Delta e$, by averaging $\Delta e_{l}$ over all relevant periods: $\Delta e= \frac{1}{6}\sum_{l=0}^{5}\Delta e_{l}$.

\section{Results}\label{sec:results} 

\subsection{Prefecture-month--wage-bin DID Estimates on Employment}\label{sec:prefecture_month_wage_bin_results} 

\begin{figure}[!ht]
  \begin{center}
  \subfloat[Missing and Excess Jobs over Time]{\includegraphics[width = 0.46\textwidth]
  {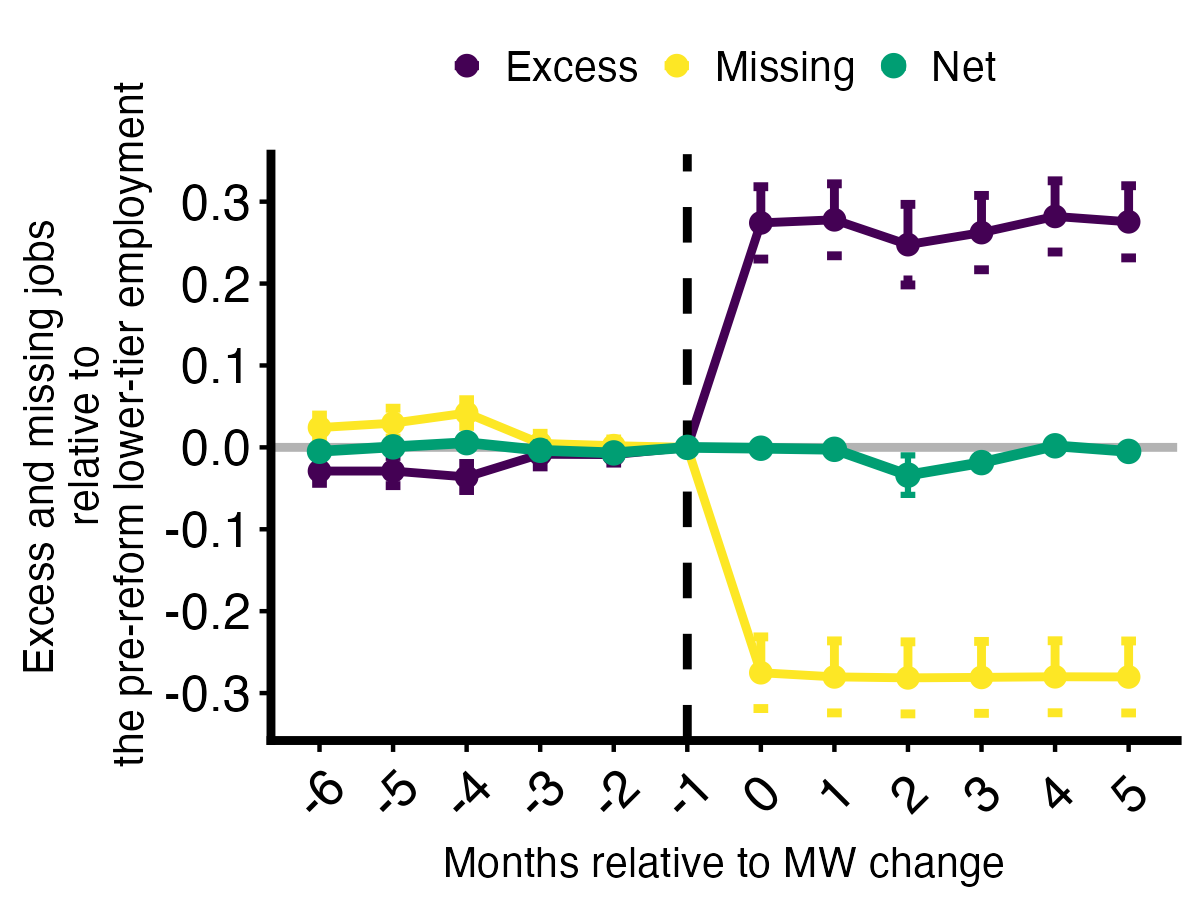}}
  \subfloat[Wage Distribution]{\includegraphics[width = 0.46\textwidth]
  {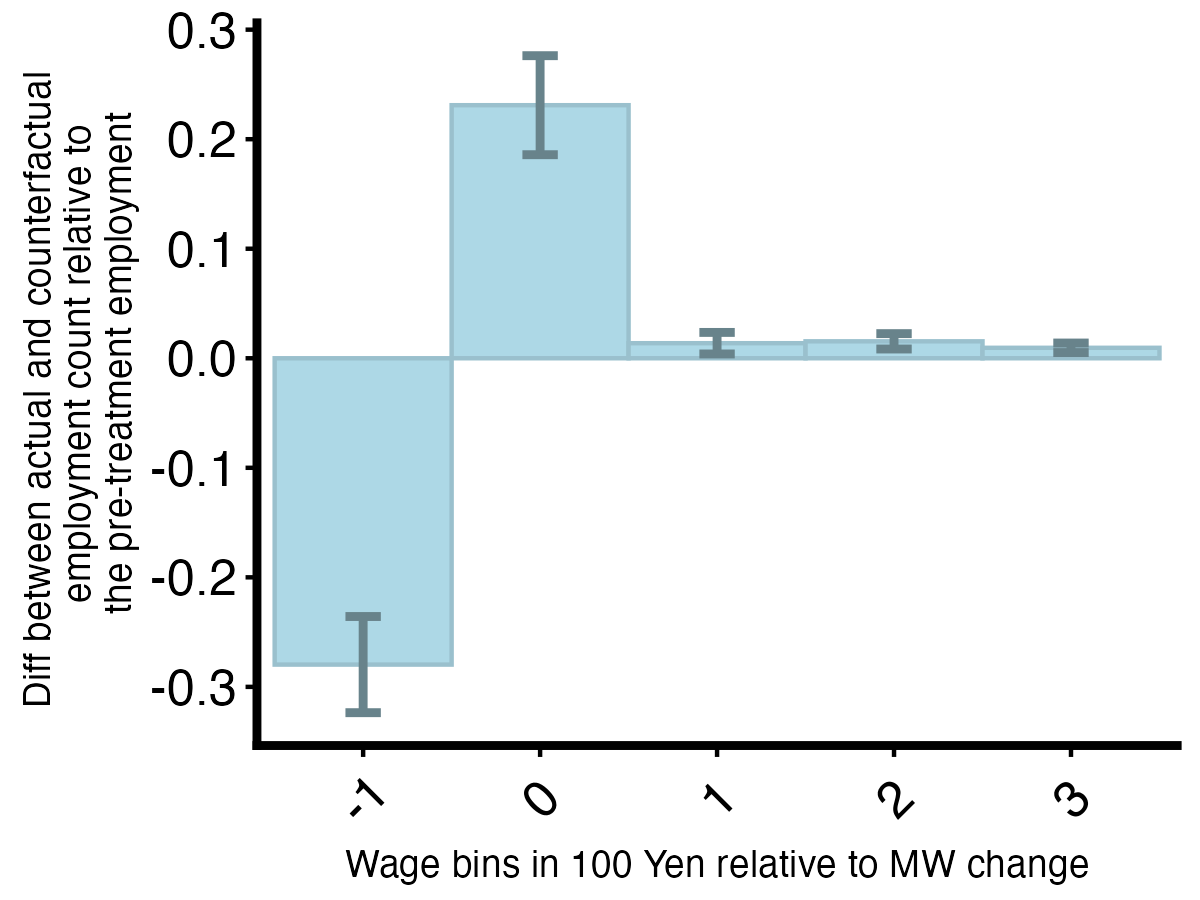}}
  \caption{Impact of Minimum Wages on Employment}
  \label{fg:wage_distribution_moving_employment_overall_cengiz_original_form} 
  \end{center}
  \footnotesize
  Note: Panel (a) shows the estimated excess jobs and missing jobs for each month before and after the minimum wage increase. Panel (b) shows the estimated impact of the minimum wage increase on employment for each group $e$ in the wage distribution. 
\end{figure}

We begin by presenting the impact of minimum wage increases on wage bins below and above the minimum wage in the wage distribution. 
Panel (a) in Figure \ref{fg:wage_distribution_moving_employment_overall_cengiz_original_form} presents the estimated excess jobs and missing jobs for each month before and after the minimum wage increase. 
Panel (b) in Figure \ref{fg:wage_distribution_moving_employment_overall_cengiz_original_form} presents the estimated impact of the minimum wage increase on employment for each group $e$ in the wage distribution, reflecting the average effect across all months following the increase.

Three key findings emerge from Figure \ref{fg:wage_distribution_moving_employment_overall_cengiz_original_form}. 
First, employment shares fall below the new minimum wage and rise above it immediately after the increase. 
Second, these missing and excess jobs remain stable across the post-treatment months in our sample.\footnote{The post-treatment missing-job series is also markedly less volatile than the pre-treatment series and than the excess-job series. This reflects compliance: once the new floor takes effect, employment in wage bins below it essentially disappears, as Figure \ref{fg:wage_bin_year_month_week_offering_count} shows, so the missing-job coefficient in each post-treatment month is approximately the negative of the pre-reform employment share in those bins, a fixed quantity, net of the small change in the control group. Before the revision, by contrast, these bins were active and their shares fluctuated, and the excess-job series continues to reflect ongoing posting decisions at and above the new floor.} 
Consistent with the longer-run persistence documented by \citet{Cengiz2019}, this pattern suggests that distributional effects need not dissipate quickly. 
Finally, Figure \ref{fg:wage_distribution_moving_employment_overall_cengiz_original_form} shows that the magnitude of missing shares ($-28.0\%$) exceeds excess shares ($27.0\%$) on the common September reference scale.\footnote{The corresponding share-based minimum-wage elasticity is $-0.211$; Appendix \ref{sec:elasticity_results} gives details.}

Panels (a) and (b) in Figure \ref{fg:wage_distribution_moving_employment_overall_cengiz_original_form} also provide evidence relevant to the identifying assumptions. 
From Panel (a), we observe that the estimated excess jobs and missing jobs are close to zero in the period before the minimum wage increase. 
This supports the plausibility of the parallel trends assumption. 
From Panel (b), we observe that excess jobs are hardly generated in the upper distribution, far from the minimum wage. 
This supports the plausibility of the limited spillover effect assumption.\footnote{
Similar to our findings, \cite{autor2016contribution} and \cite{Cengiz2019} also show that the spillover effects of the minimum wage on higher wages are limited.
}

\paragraph{Weekly and Daily Estimates}
We exploit the granularity of our platform data to probe the time dimension more directly than in the baseline monthly analysis. Specifically, we re-estimate the wage-bin DID at weekly and daily frequencies for three purposes: (i) to assess robustness to the temporal aggregation of the data, (ii) to provide a stricter pre-trend check by examining higher-frequency dynamics immediately before the minimum-wage implementation, and (iii) to measure the speed of adjustment by tracing how quickly missing and excess jobs emerge and stabilize after the policy takes effect. Because spot work can adjust rapidly, these higher-frequency estimates may reveal short-run dynamics that are smoothed out at the monthly level.

\begin{figure}[!ht]
  \begin{center}
  \subfloat[Missing and Excess Jobs over Time per Week]{\includegraphics[width = 0.46\textwidth]{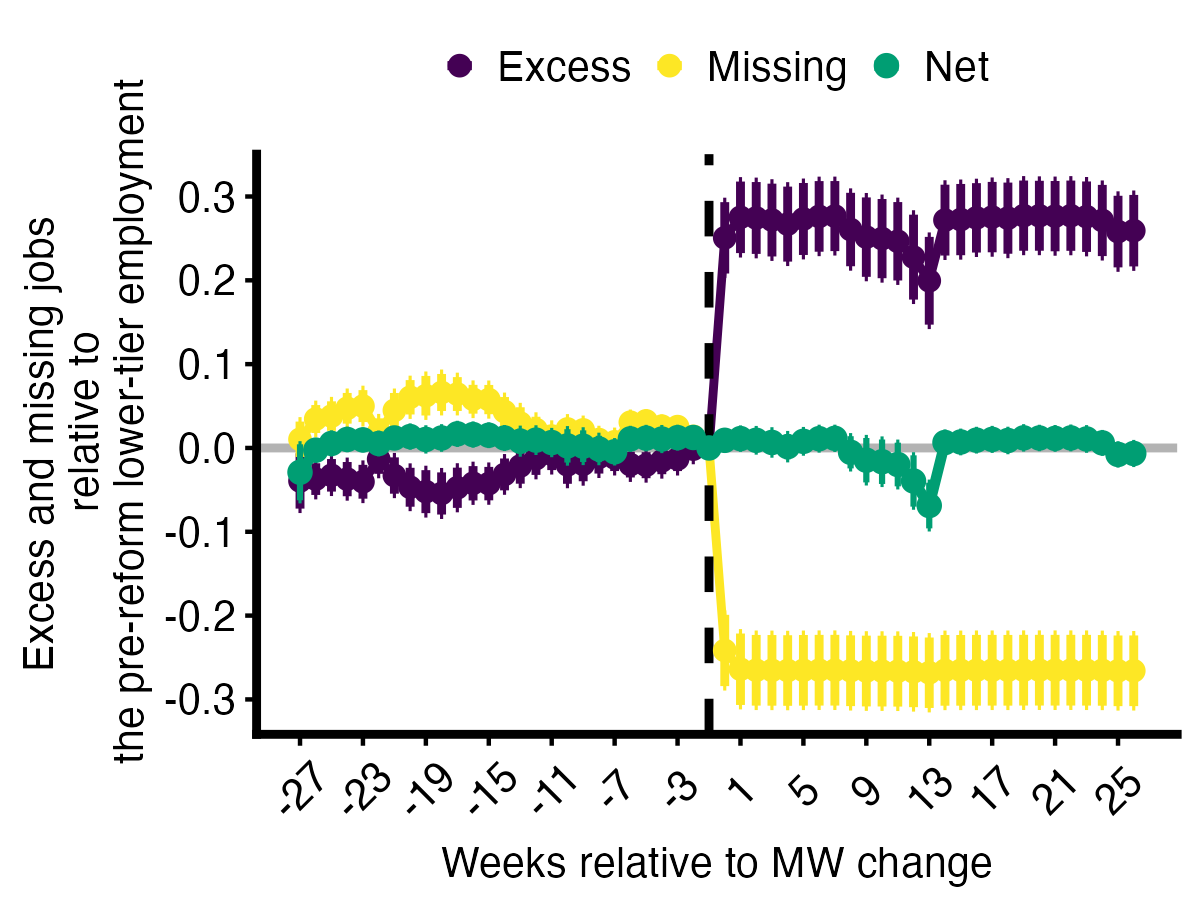}}
  \subfloat[Missing and Excess Jobs over Time per Day]{\includegraphics[width = 0.46\textwidth]{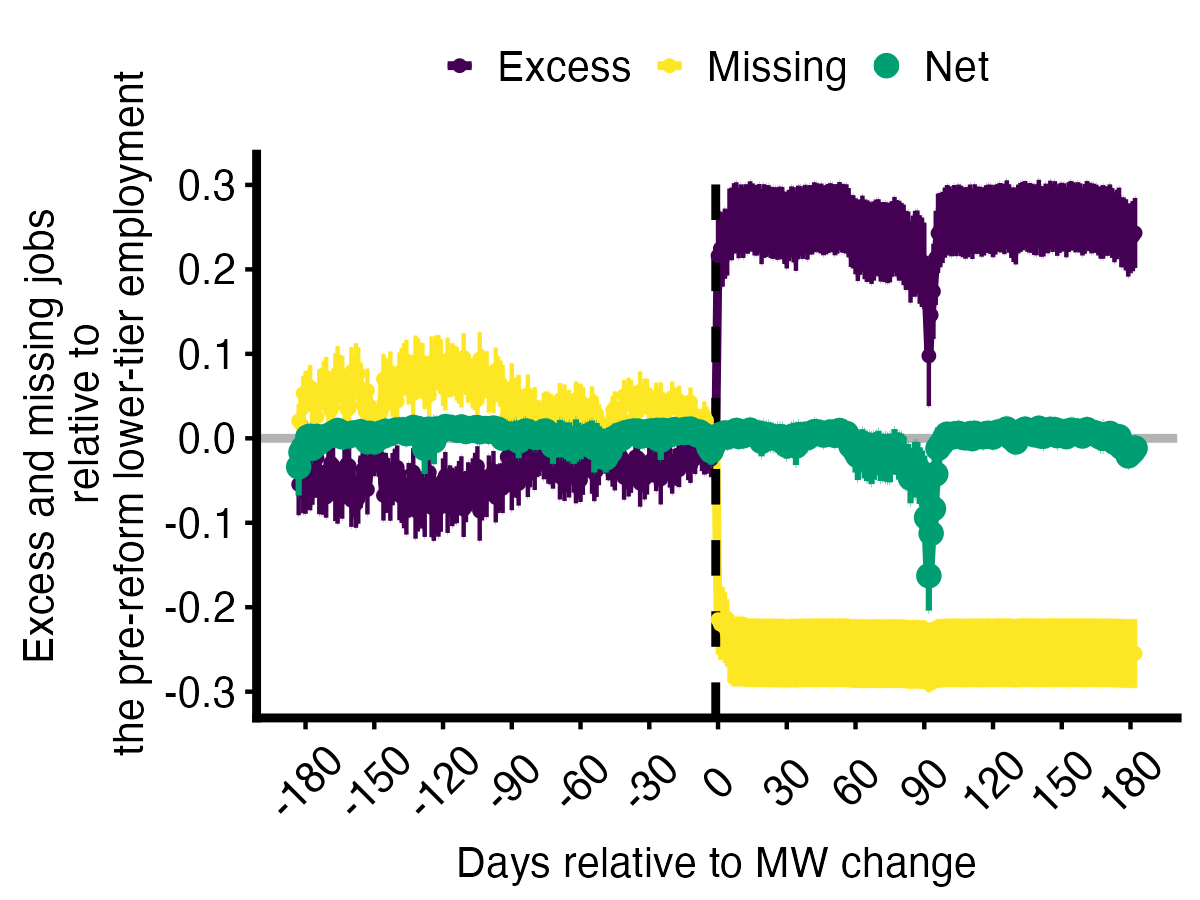}}
  \caption{Impact of Minimum Wages on Employment}
  \label{fg:wage_distribution_moving_employment_overall_cengiz_original_form_week_day} 
  \end{center}
  \footnotesize
  Note: Panel (a) shows the estimated excess jobs and missing jobs for each week before and after the minimum wage increase. Panel (b) shows the estimated excess jobs and missing jobs for each day before and after the minimum wage increase. The daily estimates in Panel (b) use data from 183 days before to 182 days after the minimum-wage increase (relative days $-183$ to $182$).
\end{figure}

Panel (a) in Figure~\ref{fg:wage_distribution_moving_employment_overall_cengiz_original_form_week_day} reports weekly estimates of missing and excess jobs. 
The pre-treatment coefficients fluctuate but remain close to zero, reinforcing the plausibility of parallel trends. 
Immediately after implementation, missing and excess jobs appear and then remain stable, with magnitudes comparable to the baseline monthly estimates in Figure~\ref{fg:wage_distribution_moving_employment_overall_cengiz_original_form}. 
Panel (b) reports daily estimates using data from relative day $-183$ through relative day $182$ around the minimum-wage increase. 
As in the weekly analysis, the pre-treatment series stays near zero and the post-treatment effects materialize quickly and persist, indicating rapid adjustment in this market. Overall, the weekly and daily results corroborate the baseline findings, strengthen the pre-trend evidence, and show that most of the adjustment occurs immediately after the minimum-wage change rather than gradually over subsequent months.

\subsection{Impact on Spot Labor Demand: Vacancy Counts and Total Working Hours}

The employment effects documented above naturally raise the question of whether adjustments to minimum wage increases operate primarily through labor demand or labor supply in spot labor markets. Unlike standard employment settings, spot work allows firms to adjust hiring margins flexibly along multiple dimensions, including the number of vacancies posted and the total hours offered. To examine these margins, we separately estimate the responses of vacancy counts and total working hours using the same DID framework. Table~\ref{tb:missing_excess_net_by_category} summarizes the six-month averaged missing, excess, and net effects for these outcomes. It also reports the employer-history decomposition discussed in Section~\ref{sec:firm_entry_exit_dynamics} and the category-specific employment effects analyzed in Section~\ref{sec:heterogeneity_analysis_across_prefecture_occupation}; the underlying outcome and job-category figures are collected in Appendix~\ref{app:event_study_figures_by_category}.

\begin{table}[!ht]
  \begin{center}
      \caption{Missing, Excess, and Net Effects of the Minimum Wage}
      \label{tb:missing_excess_net_by_category}
      \footnotesize
      
\begin{tabular}{lccc}
\toprule
 & Missing & Excess & Net\\
\midrule

\multicolumn{4}{l}{\textbf{Outcome}}\\
\hspace{1em}Employment & -0.280 (0.022) & 0.270 (0.023) & -0.010 (0.004)\\
\hspace{1em}Vacancies & -0.280 (0.022) & 0.271 (0.022) & -0.009 (0.004)\\
\hspace{1em}Total working hours & -0.273 (0.024) & 0.264 (0.024) & -0.009 (0.005)\\

\multicolumn{4}{l}{\textbf{Employer posting history (vacancies)}}\\
\hspace{1em}Posted before and after & -0.272 (0.022) & 0.053 (0.026) & -0.219 (0.010)\\
\hspace{1em}Posted only before & -0.008 (0.001) & -0.016 (0.001) & -0.023 (0.002)\\
\hspace{1em}Posted only after & 0.000 (0.000) & 0.233 (0.009) & 0.233 (0.009)\\

\multicolumn{4}{l}{\textbf{Occupation (employment)}}\\
\hspace{1em}Restaurant & -0.204 (0.021) & 0.190 (0.022) & -0.014 (0.004)\\
\hspace{1em}Light Work & -0.255 (0.025) & 0.244 (0.026) & -0.011 (0.006)\\
\hspace{1em}Retail & -0.401 (0.030) & 0.398 (0.031) & -0.003 (0.005)\\
\hspace{1em}Customer Service & -0.216 (0.029) & 0.200 (0.034) & -0.017 (0.018)\\
\hspace{1em}Professional & -0.187 (0.054) & 0.217 (0.057) & 0.030 (0.031)\\
\hspace{1em}Logistics & -0.320 (0.043) & 0.318 (0.044) & -0.002 (0.024)\\
\hspace{1em}Entertainment & -0.595 (0.076) & 0.568 (0.095) & -0.027 (0.067)\\
\hspace{1em}Office Work & -0.343 (0.067) & 0.399 (0.085) & 0.057 (0.059)\\
\hspace{1em}Event Staff & -0.197 (0.040) & 0.236 (0.050) & 0.039 (0.037)\\

\multicolumn{4}{l}{\textbf{Time of day (employment)}}\\
\hspace{1em}Morning (6:00--12:00) & -0.301 (0.024) & 0.295 (0.025) & -0.006 (0.004)\\
\hspace{1em}Evening (12:00--18:00) & -0.286 (0.022) & 0.277 (0.022) & -0.009 (0.004)\\
\hspace{1em}Night (18:00--24:00) & -0.222 (0.025) & 0.214 (0.025) & -0.008 (0.007)\\
\hspace{1em}Midnight (24:00--6:00) & -0.209 (0.037) & 0.165 (0.040) & -0.044 (0.020)\\

\multicolumn{4}{l}{\textbf{Transportation reimbursement (employment)}}\\
\hspace{1em}Provided & -0.274 (0.022) & 0.262 (0.022) & -0.012 (0.005)\\
\hspace{1em}Not provided & -0.312 (0.031) & 0.310 (0.031) & -0.002 (0.004)\\

\multicolumn{4}{l}{\textbf{Gender targeting (employment)}}\\
\hspace{1em}Female-targeted & -0.318 (0.039) & 0.310 (0.043) & -0.009 (0.017)\\
\hspace{1em}Male-targeted & -0.164 (0.033) & 0.231 (0.056) & 0.067 (0.048)\\

\multicolumn{4}{l}{\textbf{Skill targeting (employment)}}\\
\hspace{1em}Beginner-targeted & -0.323 (0.049) & 0.303 (0.061) & -0.020 (0.034)\\
\hspace{1em}Experienced-targeted & -0.318 (0.027) & 0.304 (0.027) & -0.014 (0.006)\\
\bottomrule
\end{tabular}

  \end{center}\footnotesize
  \textit{Note}: Estimates average the October 2023--March 2024 effects from Equation \eqref{sec:empirical_strategy_DID_regression}. Missing refers to $e=-1$, Excess sums $e=0,\dots,3$, and Net is their sum, normalized as in Section \ref{sec:excess_job_missing_job_approach}. The Outcome panel uses employment, vacancy, and total-hours shares of prefecture-month totals. Employer posting history partitions vacancies by posting in April--September 2023 and October 2023--March 2024; its rows use the Vacancies row's denominator and normalization and add to that row before rounding. Job-category panels report category-specific share effects. Standard errors clustered by prefecture are in parentheses. Underlying outcome and job-category figures appear in Appendix \ref{app:event_study_figures_by_category}.
\end{table}

\paragraph{Vacancy Counts} The Vacancies row of Table~\ref{tb:missing_excess_net_by_category} shows a pattern consistent with adjustment in spot labor demand. Missing posted vacancies below the minimum wage and excess vacancies just above it appear immediately and persist over time, and these changes are concentrated around the minimum wage threshold, with little evidence of spillovers to higher wage bins. These patterns are consistent with changes in vacancy creation around the minimum wage threshold. Vacancy creation is therefore a plausible demand-side margin of adjustment. This is the margin predicted by the newsvendor framework in Appendix \ref{sec:newsvendor}, in which a binding wage floor lowers the number of vacancies an establishment commits to before demand is realized. Section \ref{sec:firm_entry_exit_dynamics} examines whether this vacancy response reflects employer exit or reduced posting by continuing employers.

\paragraph{Total Working Hours} The Total working hours row of Table~\ref{tb:missing_excess_net_by_category} examines adjustments in total working hours following the minimum wage increase. The normalized missing, excess, and net estimates for total working hours are similar to those for employment and vacancies. Total hours are calculated from shift lengths and realized matches and enter as the cell's share of total hours in the prefecture-month, in parallel with the employment and vacancy shares.\footnote{In Appendix \ref{sec:platform_revenue}, we examine total posted pay.}

\subsection{Vacancy Adjustment within and across Employers}\label{sec:firm_entry_exit_dynamics}

We next decompose the negative net vacancy-share estimate by employer posting history. We examine employers' observed posting histories, which do not distinguish permanent exit from temporary inactivity.

\begin{figure}[H]
  \begin{center}
  \includegraphics[width = 0.85\textwidth]{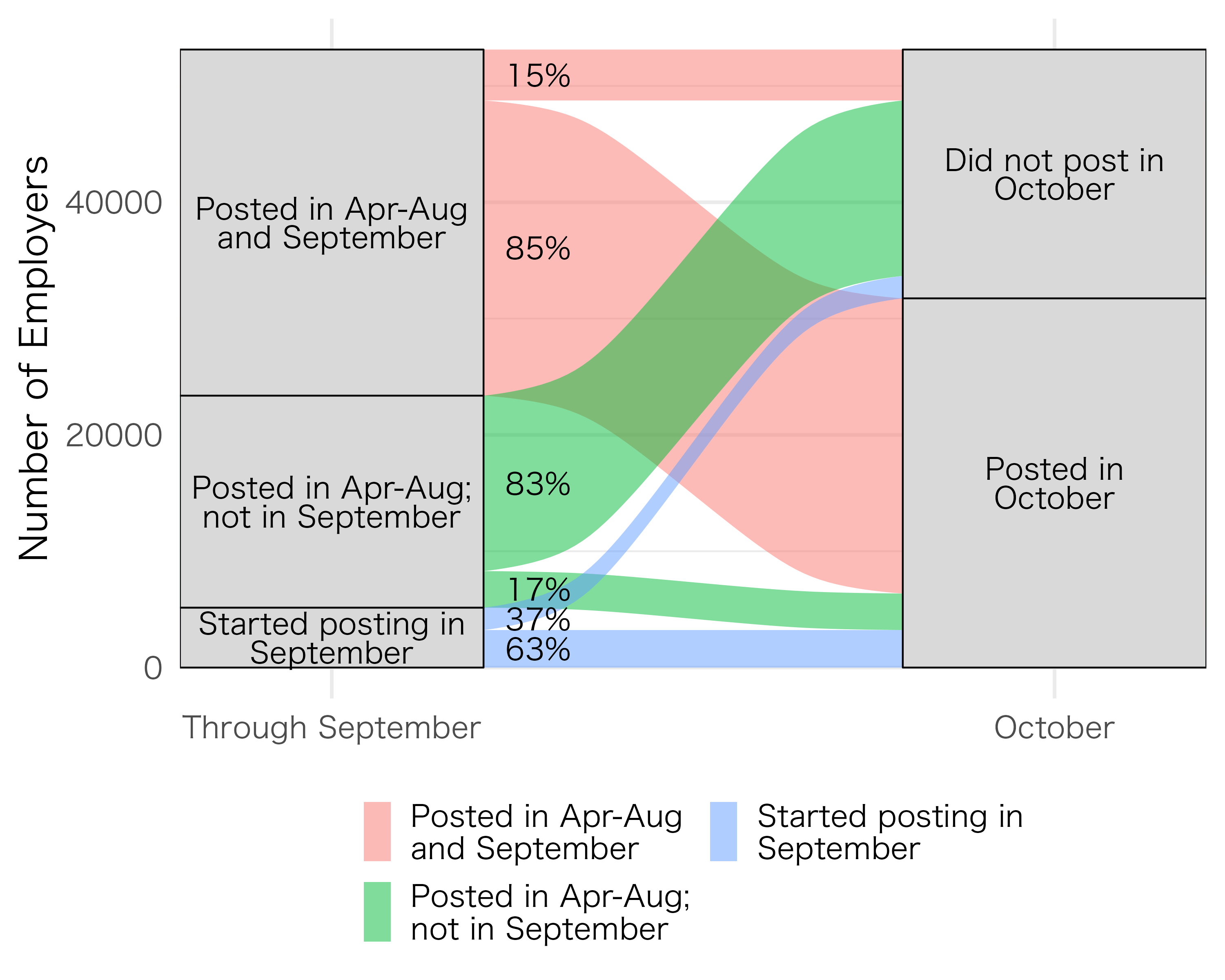}
  \caption{Employer Posting Transitions around the October 2023 Minimum-Wage Revision}
  \label{fg:client_sankey_dynamic_before_after_mw_change}
  \end{center}
  \footnotesize
  Note: The left-hand side classifies employers that posted at least once from April through September 2023 by their posting histories. The right-hand side records whether each employer posted a vacancy in October 2023. The width of each flow represents the number of employers making the corresponding transition, and the label on each flow is the share of the left-hand group that makes that transition.
\end{figure}

Figure \ref{fg:client_sankey_dynamic_before_after_mw_change} counts employers by their posting transitions and shows that most employers posting immediately before the revision continued to post in October. Among employers that posted both in April--August and in September, 85 percent also posted in October; among those that started posting in September, 63 percent did so.

The employer posting history panel of Table~\ref{tb:missing_excess_net_by_category} uses a coarser classification than Figure~\ref{fg:client_sankey_dynamic_before_after_mw_change}: employers are grouped by whether they posted at least once in April--September 2023 and at least once in October 2023--March 2024, so posting in both periods does not require posting every month. This panel decomposes the baseline missing, excess, and net vacancy effects by these histories. The negative net contribution is concentrated among employers posting in both periods. Employers posting only after the revision contribute positively and partly offset it; employers posting only before make a smaller negative contribution. The three contributions sum to the baseline estimate.

The missing vacancies below the new minimum wage come almost entirely from employers posting in both periods, whereas these employers account for only a small part of the excess vacancies just above it; most excess vacancies are posted by employers that did not post in the six months before the revision. The vacancy response therefore reflects reduced posting by incumbent employers and entry of other employers at the new floor; relabeling of postings by the same employers accounts for only a small part of the excess vacancies.

\subsection{Worker Employment Patterns and Year-End Adjustment}\label{sec:labor_supply_discussion}

A natural concern is whether minimum wage increases affect spot labor supply \citep{godoy2024parental}, particularly in the Japanese context where labor supply often exhibits seasonal adjustment toward year-end \citep{Akabayashi2006, Takahashi2009, Bessho_Hayashi2014, Yokoyama_Kodama2015, Yokoyama_2018}. In principle, workers may respond to higher minimum wages by reducing participation or reallocating effort across jobs \citep{KodamaMomota2024, mori2025higher}. In standard employment relationships, however, tracking short-run labor supply adjustments at daily or weekly frequencies is challenging, as hours and participation are typically observed only at coarse intervals. In contrast, while spot work platforms capture worker behavior only within the platform, they offer a unique advantage: high-frequency observation of worker-side actions throughout the year. We use employment patterns by worker group to describe year-end adjustment.

\begin{figure}
    \begin{center}
        \includegraphics[width=1\linewidth]{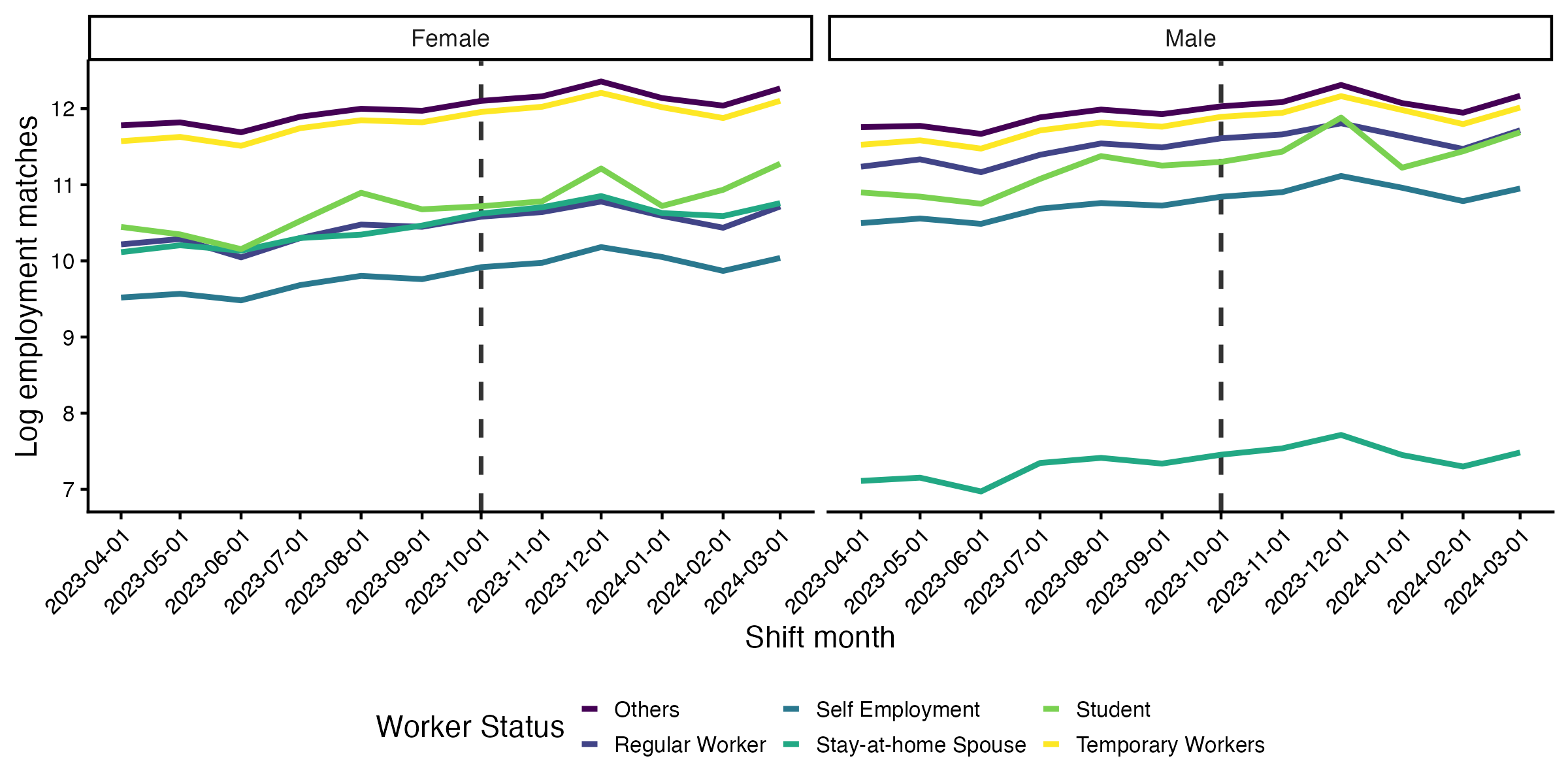}
        \caption{Employment Dynamics by Worker Status}
        \label{fig:employment_dynamic_by_user_occupation}
    \end{center}
    \footnotesize
    Note: The figure plots the log number of realized employment matches by worker status and gender, aggregated by the month of shift start (Japan Standard Time). Repeated matches by the same worker are counted separately.
\end{figure}

The left panel of Figure~\ref{fig:employment_dynamic_by_user_occupation} reports the log number of realized employment matches involving female workers by worker status, while the right panel reports the corresponding series for males. Across all gender–status groups, matches display a broadly similar upward trend through December 2023, with no obvious year-end decline specific to groups that are more likely to face tax-related work disincentives. Although temporary female workers and stay-at-home spouses are more likely to qualify for spousal tax deductions in Japan, employment matches continue to rise through November and December 2023 for both women and men. These realized-match patterns are consistent with the labor-demand interpretation of the vacancy results. They do not separately identify worker participation or labor supply.

\section{Heterogeneous Job-Posting Responses to the Minimum Wage}\label{sec:heterogeneity_analysis_across_prefecture_occupation}

This section investigates heterogeneity in firms' responses to minimum wage increases, motivated by the idea that adjustment costs and bindingness vary systematically across locations, occupations, and job attributes, leading employers to reallocate employment and modify postings along margins that are cheapest to adjust in the short run.\footnote{Additional heterogeneity analysis is provided in the Appendix.}

\subsection{Prefecture-month and Occupation-month Level DID Estimates}
In Section \ref{sec:prefecture_month_wage_bin_results}, we estimate the effect of the minimum wage increase on the outcome by aggregating policy events across prefectures. However, the minimum-wage bite varies across prefectures, and the relationship between the employment effect and the Kaitz index (the ratio of the minimum wage to the median wage) is of independent interest. We therefore implement the DID strategy separately for each prefecture $p$ and show that estimated employment effects are more negative in prefectures with a higher Kaitz index.

Specifically, we first construct a separate dataset for each prefecture $p$ and wage bin $j$ and run the following regression within each dataset:
\begin{align}
\label{section3.3_DID_regression}
\frac{Y^{p}_{j,t}}{N^{p}_{t}}=\sum_{l \neq -1}\sum_{e \neq \infty}\mu^{p}_{l,e}\mathbf{1}\{t-7=l\}\mathbf{1}\{E^{p}_{j}=e\}+\alpha^{p}_{e}+\lambda^{p}_{t}+\epsilon^{p}_{j,t}.
\end{align}
Here, the superscript $p$ indicates that this regression is estimated separately for each prefecture $p$, and $N^{p}_{t}$ is employment in prefecture $p$ at time $t$, as in Equation \eqref{sec:empirical_strategy_DID_regression}. As in Equation \eqref{sec:empirical_strategy_DID_regression}, each coefficient $\mu^{p}_{l,e}$ is the DID estimand, capturing the dynamic effect of the minimum wage increase on the outcome at a given period $l$ in prefecture $p$ under the DID identifying assumptions:
\begin{align*}
\mu^{p}_{l,e}=E\biggr[\frac{Y^{p}_{j,7+l}}{N^{p}_{7+l}}-\frac{Y^{p}_{j,6}}{N^{p}_{6}}\biggr|E^{p}_j=e\biggr]-E\biggr[\frac{Y^{p}_{j,7+l}}{N^{p}_{7+l}}-\frac{Y^{p}_{j,6}}{N^{p}_{6}}\biggr|E^{p}_j=\infty\biggr],
\end{align*}
where the pre-treatment period is $l=-1$ and the control group is group $\infty$ in prefecture $p$.

Next, as in Section \ref{sec:excess_job_missing_job_approach}, we compute the missing jobs, excess jobs, and the overall employment effect for each prefecture $p$ by aggregating $\mu^{p}_{l,e}$ in the first step:
\begin{align*}
\Delta a^{p}=\frac{1}{6}\sum_{l=0}^{5}\Delta a^{p}_{l},\hspace{2mm}\Delta b^{p}=\frac{1}{6}\sum_{l=0}^{5}\Delta b^{p}_{l},\hspace{2mm}\Delta e^{p}=\frac{1}{6}\sum_{l=0}^{5}\Delta e^{p}_{l},
\end{align*}
where the superscript $p$ indicates that each quantity is defined within prefecture $p$.

\subsection{Prefecture-month Level Results}
\begin{figure}[!ht]
  \begin{center}
  \subfloat[Missing and Excess Jobs]{\includegraphics[width = 0.46\textwidth]{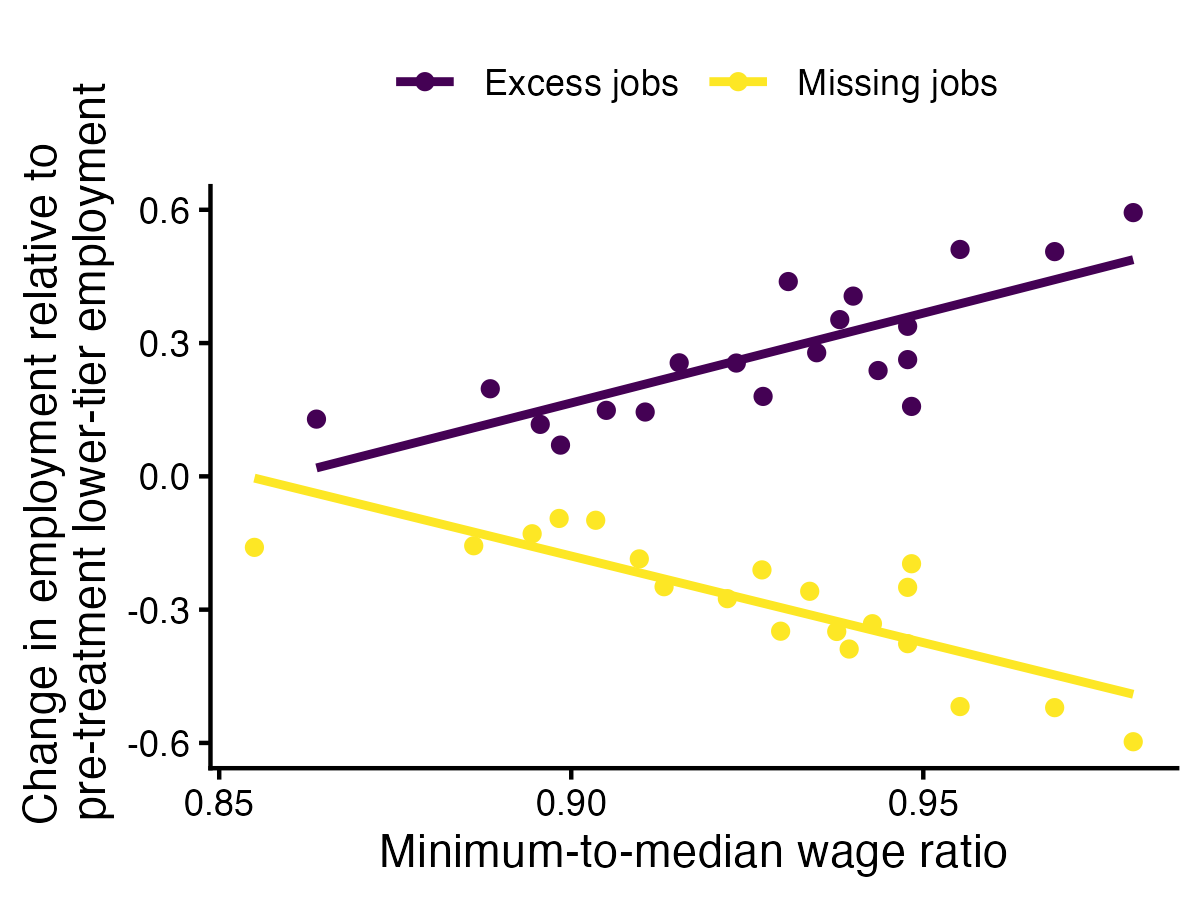}}
  \subfloat[Employment Change]{\includegraphics[width = 0.46\textwidth]{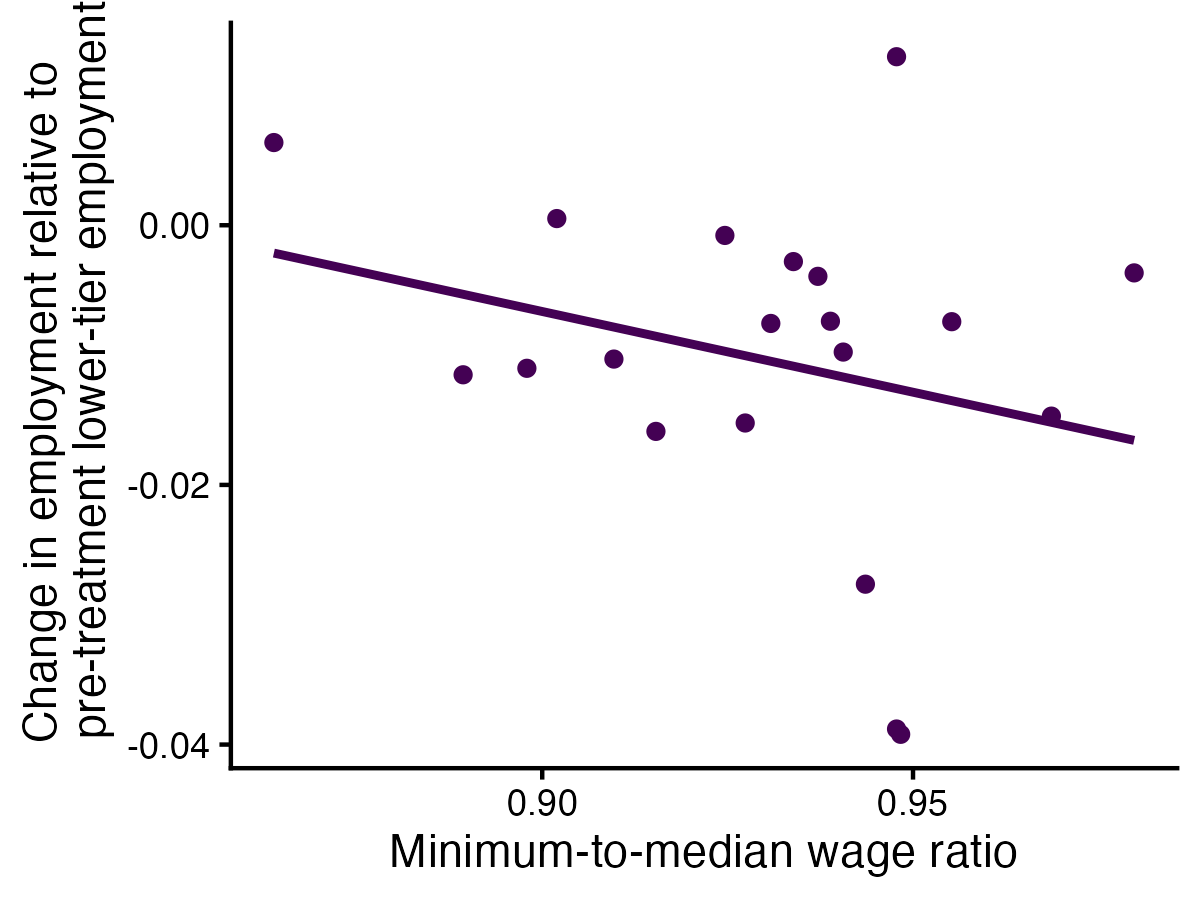}}
  \caption{Relationship between Excess Jobs, Missing Jobs, Employment Change, and the Kaitz Index (Minimum-to-Median Wage Ratio) across Prefectures}
  \label{fg:prefecture_employment_change_kaitz_index} 
  \end{center}
  \footnotesize
  Note: Figure \ref{fg:prefecture_employment_change_kaitz_index} shows the binned scatter plots for missing and excess jobs in Panel (a), and net employment changes in Panel (b) by the value of the minimum-to-median wage ratio (Kaitz index) for the 47 prefecture-specific estimates. The Kaitz index is calculated for each prefecture as the ratio of the September 2023 minimum wage to the median wage. 
\end{figure}

Figure \ref{fg:prefecture_employment_change_kaitz_index} illustrates the relationship between the Kaitz index and employment outcomes across prefectures. 
The Kaitz index is calculated for each prefecture as the ratio of the September 2023 minimum wage to the median wage. 
Panel (a) plots normalized missing and excess shares against the Kaitz index. 
It shows a positive association between the Kaitz index and excess jobs, and a negative association with missing jobs. 
When the minimum wage is high relative to the median wage, minimum wage hikes tend to have a stronger impact. 
Panel (b) presents the relationship between the Kaitz index and the overall employment effect, revealing a negative correlation. 
Higher minimum-wage bite is associated with more negative net share estimates.

\subsection{Occupation-month Level Results}

While prior studies have often concentrated on sectors like restaurants—where the minimum wage is most binding and effects are more easily detected—we adopt a distributional approach that captures wage and employment responses even in industries with limited direct exposure. 
This enables a more comprehensive evaluation of the policy’s impact across a broader range of occupations.\footnote{\cite{lordan2018people} and \cite{aaronson2019wage} find that the minimum wage reduces the employment of the lowest-wage workers in routine automatable occupations. \cite{dustmann2022reallocation} find that minimum wages could also affect productivity by improving the composition of jobs in the economy; in response to minimum wage increases, small, inefficient firms exited the market, and workers at these firms found employment at more productive firms.
Also, \cite{forsythe2023effect} finds that there is no heterogeneity across occupations.}

The Occupation panel of Table~\ref{tb:missing_excess_net_by_category} shows large missing and excess shares in every occupation. The normalized net estimate is significantly negative only for Restaurant, at $-1.4\%$ of the category's pre-reform lower-tier employment share; the other eight occupations show no significant net change.

\subsection{Heterogeneity Across Job Attributes}

Beyond aggregate employment and vacancy responses, minimum wage increases may alter the composition of jobs posted on the platform. We therefore examine heterogeneity across time slot and job attributes inferred from posting descriptions, which capture salient characteristics such as amenity provision and gender or skill targeting. These attributes allow us to classify postings into economically meaningful groups and estimate responses separately for each group. The newsvendor framework generates compositional selection when posting productivity is correlated with these attributes; the attribute-level estimates describe responses within job types, without measuring posting productivity. By estimating employment responses within these description-based categories, we describe group-specific short-run patterns, complementing evidence on longer-run reallocation documented in \cite{dustmann2022reallocation}.

\paragraph{Time of Day}

We examine how employment responses to minimum-wage increases differ by shift timing, a salient non-wage attribute of spot jobs. Specifically, we classify postings into four time-of-day categories---morning (6:00--12:00), evening (12:00--18:00), night (18:00--24:00), and midnight (24:00--6:00)---and estimate the distributional DID effects separately for each category.  The Time of day panel of Table~\ref{tb:missing_excess_net_by_category} shows normalized net share declines of $4.4\%$ for Midnight shifts and $0.9\%$ for Evening shifts, whereas Morning and Night shifts show no significant net change. Missing and excess jobs are large in all four categories, so the reallocation across wage bins occurs at every time of day, and the net estimate is most negative for midnight shifts.

\paragraph{Transportation Reimbursement Provision}

The Transportation reimbursement panel of Table~\ref{tb:missing_excess_net_by_category} examines how minimum wage increases affect employment dynamics by whether a job posting offers transportation reimbursement. Both groups show share declines below the new minimum wage and increases just above it. Postings offering transportation reimbursement show a significant normalized net share decline of $1.2\%$, while postings without reimbursement show no significant net change. Although the missing-job effect is smaller in magnitude for postings with reimbursement, excess-job gains offset a smaller proportion of those losses.

These subgroup estimates describe wage-distribution responses separately for postings with and without transportation reimbursement.\footnote{Figure \ref{fg:wage_distribution_moving_transporation_expenses_reimbursement_overall_cengiz_original_form} in Appendix \ref{app:other_measured_outcome} separately examines changes in reimbursement amounts and provision probabilities in wage groups at and above the new minimum wage.} The newsvendor framework explains the contraction in vacancy creation, and the relative response of postings with and without reimbursement depends on how amenity provision is associated with posting productivity and costs.

\paragraph{Job Description and Composition: Gender and Skill Targeting}

We next estimate employment responses separately for groups defined by worker-targeting language in job descriptions. Although firms cannot screen or reject applicants on this platform, descriptions convey information about desired worker attributes and task requirements, shaping which workers are more likely to apply \citep{kuhn2020gender}. We report estimates separately for each targeting group.\footnote{\textcolor{black}{Our targeting measures are based on keyword mentions in the job description. Specifically, we set a female-targeted (male-targeted) dummy equal to one if the description contains an explicit reference to women (men)—for example, language indicating that women are active in the workplace (e.g., “many women are actively employed”)—and equal to zero otherwise. These indicators are not mutually exclusive: postings that mention both women and men are coded as one for both dummies. Importantly, a “female-targeted” posting in our definition does \emph{not} imply that the job is restricted to women or that male applicants are excluded; rather, it captures descriptive language that may influence worker self-selection.}}\footnote{For text variables, we observe the recorded description for each posting instance but do not observe complete within-ID version histories. Hence, we cannot separately identify within-posting text edits from compositional entry/exit of postings.}

The Gender targeting panel of Table~\ref{tb:missing_excess_net_by_category} shows that female-targeted postings lose nearly twice as many jobs below the new minimum wage as male-targeted postings, but the excess jobs just above it offset these losses in both groups, so neither group shows a significant net change. These estimates describe responses within each targeting group and carry no information about workers' underlying productivity.

The Skill targeting panel of Table~\ref{tb:missing_excess_net_by_category} documents the same pattern along the skill-targeting dimension. The volumes of missing and excess jobs are nearly identical across beginner- and experienced-worker-targeted postings; the normalized net share estimate is $-1.4\%$ for experienced-targeted postings, which account for about 30 percent of contracts, whereas beginner-targeted postings show no significant net change.\footnote{Female-targeted, male-targeted, and beginner-targeted postings are small shares of contracts (about 3, 3, and 1 percent, respectively; Table \ref{tb:summary_statistics_wage_bin_prefecture_year_month}), whereas experienced-worker-targeted postings account for about 30 percent. The gender- and beginner-targeting results therefore characterize small job types rather than a quantitatively important margin of the aggregate employment response.}

\subsection{Limitations}

While our contract-level data allow us to examine the immediate responses of employers to minimum wage hikes in spot labor markets, several limitations remain. 

First, our data are limited to labor demand and supply within a single job-matching platform, which restricts the external validity of our findings. This limitation is shared by many studies relying on platform data \citep{horton2025price,melo2025minimum,stanton2021benefits}. Spot work platforms are rapidly expanding worldwide, so future research using alternative data from such platforms will be valuable for assessing external validity and mapping how minimum-wage adjustments generalize across settings.

Second, firm-level exposure designs are informative in other settings \citep{card1994minimum,machin2003minimum,draca2011minimum,harasztosi2019pays} and are useful for understanding firms' margins of adjustment, but they are difficult to implement here because the set of affected firms cannot be tracked consistently over time. Platform usage is highly seasonal, many firms do not post continuously, and it is hard to distinguish permanent exit from temporary inactivity. As a result, firm-month panels constructed from continuous posters are likely to be non-representative, making firm-level analysis difficult to interpret. As indirect evidence on this participation margin, Section \ref{sec:firm_entry_exit_dynamics} documents employers' posting transitions around the revision: most employers that posted immediately before the revision continued to post in October 2023, and most employers that did not post in October had already been inactive in September. These transitions count employers rather than vacancies. Section \ref{sec:firm_entry_exit_dynamics} also decomposes the baseline vacancy effects by observed posting history, with the components adding to the baseline estimate. Estimating causal effects on employer entry and exit requires a firm-level panel that distinguishes permanent exit from temporary inactivity. Relatedly, Appendix \ref{app:specification_with_occupation_controls} reports qualitatively similar results when treated and control bins are compared within occupation.

Third, our data are well suited to measuring short-run adjustments \emph{within} the spot market—such as vacancy creation, wages, and selected posting attributes—but they do not allow us to trace substitution across broader production margins or to quantify the \emph{timing} and \emph{sequence} of such substitution. In particular, we cannot observe whether establishments substitute away from spot hiring toward capital investment or toward long-term contract workers, nor can we determine how rapidly these substitutions occur. Prior work suggests that capital--labor substitution may arise following minimum wage hikes \citep{harasztosi2019pays}, but such investment responses typically unfold over longer horizons. Labor--labor substitution may operate more quickly—for example, establishments may reduce spot postings and reallocate tasks to incumbent workers, potentially alongside shifts in skill requirements \citep{clemens2021dropouts}.\footnote{In Appendix \ref{sec:labor_substitution_note}, we examine substitution across hiring channels (Timee vs. Hello Work) using staggered platform entry. We do not find prefecture-month evidence that Timee expansion reduces public-office vacancies. Testing substitution toward incumbent workers within firms requires establishment-level staffing data.} More broadly, reallocating long-term workers can act as a form of wage retrenchment \citep{dube2024minimum}. Identifying \emph{which} margin adjusts first and \emph{how fast} substitution propagates across inputs requires linked data that connect spot hiring to establishments' broader staffing and investment decisions. Nor do we observe worker- or match-level productivity. The newsvendor framework treats posting productivity as latent, and identifying productivity or selective exit from the compositional results requires match-level productivity data.

Taken together, our findings provide valuable evidence on short-term spot-market adjustments to minimum wage hikes, but a fuller account of firm behavior—especially substitution across inputs and the dynamics of adjustment—remains an important direction for future research using richer linked employer--worker and investment data.

\section{Conclusion}\label{sec:conclusion}

This paper studies short-run minimum-wage effects in a spot labor market using high-frequency posting- and contract-level data from Timee. Using a wage-bin DID design, we find that missing jobs below the new minimum wage exceed excess jobs just above it, yielding a normalized net share effect of $-0.010$ and a share-based elasticity of $-0.211$. The negative net employment effects are concentrated in December and January. The similar patterns in vacancy and employment shares are consistent with a labor-demand response, and the normalized total-hours estimates are similar. Our newsvendor model, based on \citet{mas2026newsvendor}, rationalizes the vacancy response and the excess of missing over excess jobs in the short run. These elasticities summarize distributional responses within the platform.

We also document heterogeneity across prefectures, occupations, and posting types, and reallocation across employers. Net share estimates are more negative where the minimum-wage bite is higher. Across occupations, the net employment effect is negative for restaurant work, while the other occupations show no significant net change. The group-specific net estimates are significantly negative for postings offering transportation reimbursement, midnight and evening shifts, and experienced-worker-targeted postings, whereas female-, male-, and beginner-targeted postings show no significant net change. Decomposing vacancies by employers' posting histories, we find that incumbent employers cut postings below the new floor and recreate few of them just above it, while employers that did not post before the revision account for most of the excess jobs. Taken together, our results provide a high-frequency picture of how minimum-wage hikes reshape employment and vacancy distributions in the short run and offer a foundation for linking immediate adjustments to longer-run reallocation dynamics.

\bibliographystyle{ecca}
\bibliography{utmd_timee}

\begin{thebibliography}{103}
\providecommand{\natexlab}[1]{#1}

\bibitem[{Aaronson and French(2007)}]{aaronson2007product}
\textsc{Aaronson, D.} and \textsc{French, E.} (2007). Product market evidence
  on the employment effects of the minimum wage. \textit{Journal of Labor
  Economics}, \textbf{25}~(1), 167--200.

\bibitem[{Aaronson \textit{et~al.}(2018)Aaronson, French, Sorkin and
  To}]{aaronson2018industry}
\textsc{---}, \textsc{---}, \textsc{Sorkin, I.} and \textsc{To, T.} (2018).
  Industry dynamics and the minimum wage: a putty-clay approach.
  \textit{International Economic Review}, \textbf{59}~(1), 51--84.

\bibitem[{Aaronson and Phelan(2019)}]{aaronson2019wage}
\textsc{---} and \textsc{Phelan, B.~J.} (2019). Wage shocks and the
  technological substitution of low-wage jobs. \textit{The Economic Journal},
  \textbf{129}~(617), 1--34.

\bibitem[{Adams-Prassl \textit{et~al.}(2020)Adams-Prassl, Balgova and
  Qian}]{adams2020flexible}
\textsc{Adams-Prassl, A.}, \textsc{Balgova, M.} and \textsc{Qian, M.} (2020).
  \textit{Flexible work arrangements in low wage jobs: Evidence from job
  vacancy data}. Tech. rep., IZA Discussion Papers.

\bibitem[{Addison \textit{et~al.}(2009)Addison, Blackburn and
  Cotti}]{addison2009minimum}
\textsc{Addison, J.~T.}, \textsc{Blackburn, M.~L.} and \textsc{Cotti, C.~D.}
  (2009). Do minimum wages raise employment? evidence from the {US}
  retail-trade sector. \textit{Labour Economics}, \textbf{16}~(4), 397--408.

\bibitem[{Akabayashi(2006)}]{Akabayashi2006}
\textsc{Akabayashi, H.} (2006). The labor supply of married women and spousal
  tax deductions in {Japan}—a structural estimation. \textit{Review of
  Economics of the Household}, \textbf{4}, 349--378.

\bibitem[{Akesaka \textit{et~al.}(2017)Akesaka, Ito and
  Ohtake}]{akesaka2017impact}
\textsc{Akesaka, M.}, \textsc{Ito, Y.} and \textsc{Ohtake, F.} (2017).
  \textit{Impact of Change in Minimum Wage on Employment and Poverty in {Japan}
  (in {Japanese})}. Tech. rep., Institute of Social and Economic Research,
  Osaka University.

\bibitem[{Allegretto \textit{et~al.}(2017)Allegretto, Dube, Reich and
  Zipperer}]{allegretto2017credible}
\textsc{Allegretto, S.}, \textsc{Dube, A.}, \textsc{Reich, M.} and
  \textsc{Zipperer, B.} (2017). Credible research designs for minimum wage
  studies: A response to {Neumark}, {Salas}, and {Wascher}. \textit{ILR
  Review}, \textbf{70}~(3), 559--592.

\bibitem[{Allegretto \textit{et~al.}(2011)Allegretto, Dube and
  Reich}]{allegretto2011minimum}
\textsc{Allegretto, S.~A.}, \textsc{Dube, A.} and \textsc{Reich, M.} (2011). Do
  minimum wages really reduce teen employment? accounting for heterogeneity and
  selectivity in state panel data. \textit{Industrial Relations: A Journal of
  Economy and Society}, \textbf{50}~(2), 205--240.

\bibitem[{Angrist \textit{et~al.}(2021)Angrist, Caldwell and
  Hall}]{angrist2021uber}
\textsc{Angrist, J.~D.}, \textsc{Caldwell, S.} and \textsc{Hall, J.~V.} (2021).
  {Uber} versus taxi: A driver’s eye view. \textit{American Economic Journal:
  Applied Economics}, \textbf{13}~(3), 272--308.

\bibitem[{Aoyagi \textit{et~al.}(2016)Aoyagi, Ganelli and
  Tawk}]{aoyagi2016minimum}
\textsc{Aoyagi, C.}, \textsc{Ganelli, G.} and \textsc{Tawk, N.} (2016). Minimum
  wage as a wage policy tool in {Japan}. \textit{The Japanese Political
  Economy}, \textbf{42}~(1-4), 72--88.

\bibitem[{Autor \textit{et~al.}(2016)Autor, Manning and
  Smith}]{autor2016contribution}
\textsc{Autor, D.~H.}, \textsc{Manning, A.} and \textsc{Smith, C.~L.} (2016).
  The contribution of the minimum wage to {US} wage inequality over three
  decades: A reassessment. \textit{American Economic Journal: Applied
  Economics}, \textbf{8}~(1), 58--99.

\bibitem[{Balgova(2024)}]{maria2024death}
\textsc{Balgova, M.} (2024). The death of distance in hiring. \textit{Job
  Market Paper, IZA}.

\bibitem[{Bassier \textit{et~al.}(2025)Bassier, Manning and
  Petrongolo}]{bassier2025vacancy}
\textsc{Bassier, I.}, \textsc{Manning, A.} and \textsc{Petrongolo, B.} (2025).
  Vacancy duration and wages. \textit{Review of Economics and Statistics}, pp.
  1--28.

\bibitem[{Bell and Machin(2018)}]{bell2018minimum}
\textsc{Bell, B.} and \textsc{Machin, S.} (2018). Minimum wages and firm value.
  \textit{Journal of Labor Economics}, \textbf{36}~(1), 159--195.

\bibitem[{Bessho and Hayashi(2014)}]{Bessho_Hayashi2014}
\textsc{Bessho, S.-i.} and \textsc{Hayashi, M.} (2014). Intensive margins,
  extensive margins, and spousal allowances in the {Japanese} system of
  personal income taxes: A discrete choice analysis. \textit{Journal of the
  Japanese and International Economies}, \textbf{34}, 162--178.

\bibitem[{Boeri \textit{et~al.}(2020)Boeri, Giupponi, Krueger and
  Machin}]{boeri2020solo}
\textsc{Boeri, T.}, \textsc{Giupponi, G.}, \textsc{Krueger, A.~B.} and
  \textsc{Machin, S.} (2020). Solo self-employment and alternative work
  arrangements: A cross-country perspective on the changing composition of
  jobs. \textit{Journal of Economic Perspectives}, \textbf{34}~(1), 170--195.

\bibitem[{Bren{\v{c}}i{\v{c}}(2010)}]{brenvcivc2010employers}
\textsc{Bren{\v{c}}i{\v{c}}, V.} (2010). Do employers respond to the costs of
  continued search? \textit{Oxford Bulletin of Economics and Statistics},
  \textbf{72}~(2), 221--245.

\bibitem[{Bren{\v{c}}i{\v{c}}(2016)}]{brenvcivc2016impact}
\textsc{Bren{\v{c}}i{\v{c}}, V.} (2016). The impact of {Craigslist’s} entry
  on competing employment websites. \textit{IZA Journal of Labor Economics},
  \textbf{5}~(1), 7.

\bibitem[{Brochu \textit{et~al.}(2025)Brochu, Green, Lemieux and
  Townsend}]{brochu2025minimum}
\textsc{Brochu, P.~R.}, \textsc{Green, D.~A.}, \textsc{Lemieux, T.} and
  \textsc{Townsend, J.~H.} (2025). \textit{The minimum wage, turnover, and the
  shape of the wage distribution}. Tech. rep., National Bureau of Economic
  Research.

\bibitem[{Buchholz \textit{et~al.}(2023)Buchholz, Shum and
  Xu}]{buchholz2023rethinking}
\textsc{Buchholz, N.}, \textsc{Shum, M.} and \textsc{Xu, H.} (2023).
  \textit{Rethinking reference dependence: Wage dynamics and optimal taxi labor
  supply}. Tech. rep., working paper, Princeton University Economics Dept.

\bibitem[{Callaway and Sant'Anna(2021)}]{callaway2021difference}
\textsc{Callaway, B.} and \textsc{Sant'Anna, P.~H.} (2021).
  Difference-in-differences with multiple time periods. \textit{Journal of
  econometrics}, \textbf{225}~(2), 200--230.

\bibitem[{Card(1992)}]{card1992minimum}
\textsc{Card, D.} (1992). Do minimum wages reduce employment? a case study of
  {California}, 1987--89. \textit{ILR Review}, \textbf{46}~(1), 38--54.

\bibitem[{Card and Krueger(1994)}]{card1994minimum}
\textsc{---} and \textsc{Krueger, A.~B.} (1994). Minimum wages and employment:
  A case study of the fast-food industry in {New} {Jersey} and {Pennsylvania}.
  \textit{The American Economic Review}, \textbf{84}~(4), 772--793.

\bibitem[{Cengiz \textit{et~al.}(2019)Cengiz, Dube, Lindner and
  Zipperer}]{Cengiz2019}
\textsc{Cengiz, D.}, \textsc{Dube, A.}, \textsc{Lindner, A.} and
  \textsc{Zipperer, B.} (2019). The effect of minimum wages on low-wage jobs.
  \textit{The Quarterly Journal of Economics}, \textbf{134}~(3), 1405--1454.

\bibitem[{Chava \textit{et~al.}(2023)Chava, Oettl and Singh}]{chava2023does}
\textsc{Chava, S.}, \textsc{Oettl, A.} and \textsc{Singh, M.} (2023). Does a
  one-size-fits-all minimum wage cause financial stress for small businesses?
  \textit{Management Science}, \textbf{69}~(11), 7095--7117.

\bibitem[{Clemens(2021)}]{clemens2021firms}
\textsc{Clemens, J.} (2021). How do firms respond to minimum wage increases?
  understanding the relevance of non-employment margins. \textit{Journal of
  Economic Perspectives}, \textbf{35}~(1), 51--72.

\bibitem[{Clemens \textit{et~al.}(2018)Clemens, Kahn and
  Meer}]{clemens2018minimum}
\textsc{---}, \textsc{Kahn, L.~B.} and \textsc{Meer, J.} (2018). \textit{The
  minimum wage, fringe benefits, and worker welfare}. Tech. rep., National
  Bureau of Economic Research.

\bibitem[{Clemens \textit{et~al.}(2021)Clemens, Kahn and
  Meer}]{clemens2021dropouts}
\textsc{---}, \textsc{---} and \textsc{---} (2021). Dropouts need not apply?
  the minimum wage and skill upgrading. \textit{Journal of Labor Economics},
  \textbf{39}~(S1), S107--S149.

\bibitem[{Cullen and Farronato(2021)}]{cullen2021outsourcing}
\textsc{Cullen, Z.} and \textsc{Farronato, C.} (2021). Outsourcing tasks
  online: Matching supply and demand on peer-to-peer internet platforms.
  \textit{Management Science}, \textbf{67}~(7), 3985--4003.

\bibitem[{Cunningham \textit{et~al.}(2024)Cunningham, DeAngelo and
  Tripp}]{cunningham2024did}
\textsc{Cunningham, S.}, \textsc{DeAngelo, G.} and \textsc{Tripp, J.} (2024).
  Did {Craigslist’s} erotic services reduce female homicide and rape?
  \textit{Journal of Human Resources}, \textbf{59}~(1), 280--315.

\bibitem[{Djourelova \textit{et~al.}(2025)Djourelova, Durante and
  Martin}]{djourelova2025impact}
\textsc{Djourelova, M.}, \textsc{Durante, R.} and \textsc{Martin, G.~J.}
  (2025). The impact of online competition on local newspapers: Evidence from
  the introduction of {Craigslist}. \textit{Review of Economic Studies},
  \textbf{92}~(3), 1738--1772.

\bibitem[{Doh \textit{et~al.}(2024)Doh, Kim, Lee, Song
  \textit{et~al.}}]{doh2024economic}
\textsc{Doh, T.}, \textsc{Kim, S.}, \textsc{Lee, H.}, \textsc{Song, K.}
  \textit{et~al.} (2024). The economic effects of a rapid increase in the
  minimum wage: Evidence from {South} {Korea} experiments. \textit{Federal
  Reserve Bank of Kansas City Working Paper}, ~(22-13).

\bibitem[{Draca \textit{et~al.}(2011)Draca, Machin and
  Van~Reenen}]{draca2011minimum}
\textsc{Draca, M.}, \textsc{Machin, S.} and \textsc{Van~Reenen, J.} (2011).
  Minimum wages and firm profitability. \textit{American Economic Journal:
  Applied Economics}, \textbf{3}~(1), 129--151.

\bibitem[{Dube \textit{et~al.}(2010)Dube, Lester and Reich}]{Dube_etal2010}
\textsc{Dube, A.}, \textsc{Lester, T.~W.} and \textsc{Reich, M.} (2010).
  Minimum wage effects across state borders: Estimates using contiguous
  counties. \textit{Review of Economics and Statistics}, \textbf{92}~(4),
  945--964.

\bibitem[{Dube \textit{et~al.}(2016)Dube, Lester and Reich}]{dube2016minimum}
\textsc{---}, \textsc{---} and \textsc{---} (2016). Minimum wage shocks,
  employment flows, and labor market frictions. \textit{Journal of labor
  economics}, \textbf{34}~(3), 663--704.

\bibitem[{Dube and Lindner(2024)}]{dube2024minimum}
\textsc{---} and \textsc{Lindner, A.~S.} (2024). \textit{Minimum wages in the
  21st century}. Tech. rep., National Bureau of Economic Research.

\bibitem[{Dustmann \textit{et~al.}(2022)Dustmann, Lindner, Sch{\"o}nberg,
  Umkehrer and Vom~Berge}]{dustmann2022reallocation}
\textsc{Dustmann, C.}, \textsc{Lindner, A.}, \textsc{Sch{\"o}nberg, U.},
  \textsc{Umkehrer, M.} and \textsc{Vom~Berge, P.} (2022). Reallocation effects
  of the minimum wage. \textit{The Quarterly Journal of Economics},
  \textbf{137}~(1), 267--328.

\bibitem[{Forsythe(2023)}]{forsythe2023effect}
\textsc{Forsythe, E.} (2023). The effect of minimum wage policies on the wage
  and occupational structure of establishments. \textit{Journal of Labor
  Economics}, \textbf{41}~(S1), S291--S324.

\bibitem[{Fukai \textit{et~al.}(2021)Fukai, Ichimura and
  Kawata}]{fukai2021describing}
\textsc{Fukai, T.}, \textsc{Ichimura, H.} and \textsc{Kawata, K.} (2021).
  Describing the impacts of {COVID-19} on the labor market in {Japan} until
  june 2020. \textit{The Japanese Economic Review}, \textbf{72}~(3), 439--470.

\bibitem[{Fukui and Kikuchi(2020)}]{fukui2020job}
\textsc{Fukui, M.} and \textsc{Kikuchi, S.} (2020). Job creation during the
  {COVID-19} pandemic in {Japan}. Working paper.

\bibitem[{Glasner(2023)}]{glasner2023minimum}
\textsc{Glasner, B.} (2023). The minimum wage, self-employment, and the online
  gig economy. \textit{Journal of Labor Economics}, \textbf{41}~(1), 103--127.

\bibitem[{God{\o}y \textit{et~al.}(2024)God{\o}y, Reich, Wursten and
  Allegretto}]{godoy2024parental}
\textsc{God{\o}y, A.}, \textsc{Reich, M.}, \textsc{Wursten, J.} and
  \textsc{Allegretto, S.} (2024). Parental labor supply: Evidence from minimum
  wage changes. \textit{Journal of Human Resources}, \textbf{59}~(2), 416--442.

\bibitem[{Goldfarb and Tucker(2019)}]{goldfarb2019digital}
\textsc{Goldfarb, A.} and \textsc{Tucker, C.} (2019). Digital economics.
  \textit{Journal of economic literature}, \textbf{57}~(1), 3--43.

\bibitem[{Gopalan \textit{et~al.}(2021)Gopalan, Hamilton, Kalda and
  Sovich}]{gopalan2021state}
\textsc{Gopalan, R.}, \textsc{Hamilton, B.~H.}, \textsc{Kalda, A.} and
  \textsc{Sovich, D.} (2021). State minimum wages, employment, and wage
  spillovers: Evidence from administrative payroll data. \textit{Journal of
  Labor Economics}, \textbf{39}~(3), 673--707.

\bibitem[{Harasztosi and Lindner(2019)}]{harasztosi2019pays}
\textsc{Harasztosi, P.} and \textsc{Lindner, A.} (2019). Who pays for the
  minimum wage? \textit{American Economic Review}, \textbf{109}~(8),
  2693--2727.

\bibitem[{Helleseter \textit{et~al.}(2020)Helleseter, Kuhn and
  Shen}]{helleseter2020age}
\textsc{Helleseter, M.~D.}, \textsc{Kuhn, P.} and \textsc{Shen, K.} (2020). The
  age twist in employers’ gender requests: Evidence from four job boards.
  \textit{Journal of Human Resources}, \textbf{55}~(2), 428--469.

\bibitem[{Higuchi(2013)}]{HiguchiYoshio2013TDoP}
\textsc{Higuchi, Y.} (2013). The dynamics of poverty and the promotion of
  transition from non-regular to regular employment in {Japan}: Economic
  effects of minimum wage revision and job training support. \textit{Japanese
  Economic Review}, \textbf{64}~(2), 147--200.

\bibitem[{Horton(2025)}]{horton2025price}
\textsc{Horton, J.~J.} (2025). Price floors and employer preferences: Evidence
  from a minimum wage experiment. \textit{American Economic Review},
  \textbf{115}~(1), 117–46.

\bibitem[{Izumi \textit{et~al.}(2020)Izumi, Kodama and Kwon}]{izumicprc2020}
\textsc{Izumi, A.}, \textsc{Kodama, N.} and \textsc{Kwon, H.~U.} (2020).
  \textit{Labor Market Concentration on Wage, Employment, and Exit of Plants:
  Empirical Evidence with Minimum Wage Hike}. Tech. rep., Japan Fair Trade
  Commission.

\bibitem[{Izumi \textit{et~al.}(2022)Izumi, Okudaira and
  Kawaguchi}]{Izumi_kawaguchi_okudaira2022}
\textsc{---}, \textsc{Okudaira, H.} and \textsc{Kawaguchi, D.} (2022).
  \textit{Minimum Wage and Firms' Hiring Behavior}. CREPE Discussion Paper 129,
  Center for Research and Education for Policy Evaluation (CREPE), The
  University of Tokyo, in Japanese.

\bibitem[{Kambayashi \textit{et~al.}(2013)Kambayashi, Kawaguchi and
  Yamada}]{kambayashi2013minimum}
\textsc{Kambayashi, R.}, \textsc{Kawaguchi, D.} and \textsc{Yamada, K.} (2013).
  Minimum wage in a deflationary economy: The {Japanese} experience,
  1994--2003. \textit{Labour Economics}, \textbf{24}, 264--276.

\bibitem[{Kanayama and Otani(2026)}]{kanayama2026nonparametric}
\textsc{Kanayama, H.} and \textsc{Otani, S.} (2026). Nonparametric estimation
  of matching efficiency and elasticity in a spot gig work platform:
  2019--2023. \textit{Journal of the Japanese and International Economies},
  \textbf{81}, 101439.

\bibitem[{K{\"a}ssi and Lehdonvirta(2018)}]{kassi2018online}
\textsc{K{\"a}ssi, O.} and \textsc{Lehdonvirta, V.} (2018). Online labour
  index: Measuring the online gig economy for policy and research.
  \textit{Technological Forecasting and Social Change}, \textbf{137}, 241--248.

\bibitem[{Katz and Krueger(1992)}]{katz1992effect}
\textsc{Katz, L.~F.} and \textsc{Krueger, A.~B.} (1992). The effect of the
  minimum wage on the fast-food industry. \textit{ILR Review}, \textbf{46}~(1),
  6--21.

\bibitem[{Katz and Krueger(2019)}]{katz2019rise}
\textsc{---} and \textsc{---} (2019). The rise and nature of alternative work
  arrangements in the {United States}, 1995--2015. \textit{ILR Review},
  \textbf{72}~(2), 382--416.

\bibitem[{Kawaguchi and Mori(2009)}]{Kawaguchi_Mori2009}
\textsc{Kawaguchi, D.} and \textsc{Mori, Y.} (2009). Is minimum wage an
  effective anti-poverty policy in {Japan}? \textit{Pacific Economic Review},
  \textbf{14}~(4), 532--554.

\bibitem[{Kawaguchi and Mori(2021)}]{Kawaguchi_Mori2021}
\textsc{---} and \textsc{---} (2021). Estimating the effects of the minimum
  wage using the introduction of indexation. \textit{Journal of Economic
  Behavior \& Organization}, \textbf{184}, 388--408.

\bibitem[{Kawaguchi and Yamada(2007)}]{Kawaguchi_Yamada2007}
\textsc{---} and \textsc{Yamada, K.} (2007). The impact of the minimum wage on
  female employment in {Japan}. \textit{Contemporary Economic Policy},
  \textbf{25}~(1), 107--118.

\bibitem[{Kikuchi \textit{et~al.}(2021)Kikuchi, Kitao and
  Mikoshiba}]{kikuchi2021suffers}
\textsc{Kikuchi, S.}, \textsc{Kitao, S.} and \textsc{Mikoshiba, M.} (2021). Who
  suffers from the {COVID-19} shocks? labor market heterogeneity and welfare
  consequences in {Japan}. \textit{Journal of the Japanese and International
  Economies}, \textbf{59}, 101117.

\bibitem[{Kodama and Momota(2024)}]{KodamaMomota2024}
\textsc{Kodama, N.} and \textsc{Momota, S.} (2024). Do minimum wage increases
  reduce working hours? \textit{Japanese Journal of Labour Studies},
  \textbf{66}~(10), 84--100, in Japanese.

\bibitem[{Kroft and Pope(2014)}]{kroft2014does}
\textsc{Kroft, K.} and \textsc{Pope, D.~G.} (2014). Does online search crowd
  out traditional search and improve matching efficiency? evidence from
  {Craigslist}. \textit{Journal of Labor Economics}, \textbf{32}~(2), 259--303.

\bibitem[{Kuhn and Shen(2013)}]{kuhn2013gender}
\textsc{Kuhn, P.} and \textsc{Shen, K.} (2013). Gender discrimination in job
  ads: Evidence from {China}. \textit{The Quarterly Journal of Economics},
  \textbf{128}~(1), 287--336.

\bibitem[{Kuhn \textit{et~al.}(2020)Kuhn, Shen and Zhang}]{kuhn2020gender}
\textsc{---}, \textsc{---} and \textsc{Zhang, S.} (2020). Gender-targeted job
  ads in the recruitment process: Facts from a {Chinese} job board.
  \textit{Journal of Development Economics}, \textbf{147}, 102531.

\bibitem[{Kuhn and Skuterud(2004)}]{kuhn2004internet}
\textsc{---} and \textsc{Skuterud, M.} (2004). Internet job search and
  unemployment durations. \textit{American Economic Review}, \textbf{94}~(1),
  218--232.

\bibitem[{Kuroda and Onishi(2023)}]{kuroda2023exploring}
\textsc{Kuroda, S.} and \textsc{Onishi, K.} (2023). Exploring the gig economy
  in {Japan}: a bank data-driven analysis of food delivery gig workers.
  \textit{RIETI Discussion Paper Series 23-E-025}.

\bibitem[{Lee(1999)}]{lee1999wage}
\textsc{Lee, D.~S.} (1999). Wage inequality in the {UNITED STATES}. \textit{The
  Quarterly Journal of Economics}, \textbf{114}~(3), 977--1023.

\bibitem[{Leung(2021)}]{leung2021minimum}
\textsc{Leung, J.~H.} (2021). Minimum wage and real wage inequality: Evidence
  from pass-through to retail prices. \textit{Review of Economics and
  Statistics}, \textbf{103}~(4), 754--769.

\bibitem[{Lordan and Neumark(2018)}]{lordan2018people}
\textsc{Lordan, G.} and \textsc{Neumark, D.} (2018). People versus machines:
  The impact of minimum wages on automatable jobs. \textit{Labour Economics},
  \textbf{52}, 40--53.

\bibitem[{Luca and Luca(2019)}]{luca2019survival}
\textsc{Luca, D.~L.} and \textsc{Luca, M.} (2019). \textit{Survival of the
  fittest: The impact of the minimum wage on firm exit}. Tech. rep., National
  Bureau of Economic Research.

\bibitem[{Machin \textit{et~al.}(2003)Machin, Manning and
  Rahman}]{machin2003minimum}
\textsc{Machin, S.}, \textsc{Manning, A.} and \textsc{Rahman, L.} (2003). Where
  the minimum wage bites hard: Introduction of minimum wages to a low wage
  sector. \textit{Journal of the European Economic Association},
  \textbf{1}~(1), 154--180.

\bibitem[{Manning(2021)}]{Manning_2021}
\textsc{Manning, A.} (2021). The elusive employment effect of the minimum wage.
  \textit{Journal of Economic Perspectives}, \textbf{35}~(1), 3--26.

\bibitem[{Marinescu and Wolthoff(2020)}]{marinescu2020opening}
\textsc{Marinescu, I.} and \textsc{Wolthoff, R.} (2020). Opening the black box
  of the matching function: The power of words. \textit{Journal of Labor
  Economics}, \textbf{38}~(2), 535--568.

\bibitem[{Mas(2025)}]{mas2025non}
\textsc{Mas, A.} (2025). \textit{Non-Wage Amenities}. Tech. Rep. 33643,
  National Bureau of Economic Research.

\bibitem[{Mas(2026)}]{mas2026newsvendor}
\textsc{---} (2026). \textit{The Labor Market as an Equilibrium Newsvendor
  Problem}. NBER Working Paper 35148, National Bureau of Economic Research.

\bibitem[{Mas and Pallais(2020)}]{mas2020alternative}
\textsc{---} and \textsc{Pallais, A.} (2020). Alternative work arrangements.
  \textit{Annual Review of Economics}, \textbf{12}~(1), 631--658.

\bibitem[{Mayneris \textit{et~al.}(2018)Mayneris, Poncet and
  Zhang}]{mayneris2018improving}
\textsc{Mayneris, F.}, \textsc{Poncet, S.} and \textsc{Zhang, T.} (2018).
  Improving or disappearing: Firm-level adjustments to minimum wages in
  {China}. \textit{Journal of Development Economics}, \textbf{135}, 20--42.

\bibitem[{Melo \textit{et~al.}(2025)Melo, Kaiser, Neumark, Palagashvili and
  Farren}]{melo2025minimum}
\textsc{Melo, V.~C.}, \textsc{Kaiser, C.}, \textsc{Neumark, D.},
  \textsc{Palagashvili, L.} and \textsc{Farren, M.~D.} (2025). \textit{Minimum
  Wage Laws and Job Search}. Tech. rep., National Bureau of Economic Research.

\bibitem[{Miyamoto(2025)}]{miyamoto2025macroeconomic}
\textsc{Miyamoto, H.} (2025). Macroeconomic facts in the {Japanese} labor
  market: survey. \textit{The Japanese Economic Review}, pp. 1--46.

\bibitem[{Mizushima and Noguchi(2021)}]{mizushima2021spillover}
\textsc{Mizushima, Y.} and \textsc{Noguchi, H.} (2021). \textit{Spillover
  effects of minimum wages on suicide mortality: Evidence from {Japan}}. Waseda
  INstitute of Political EConomy, Waseda University.

\bibitem[{Mori and Okudaira(2025)}]{mori2025higher}
\textsc{Mori, Y.} and \textsc{Okudaira, H.} (2025). Higher minimum wage,
  stagnant income? the case of women's work hours in {Japan}. \textit{Research
  Institute of Economy, Trade and Industry (RIETI)}.

\bibitem[{Neumark \textit{et~al.}(2014)Neumark, Salas and
  Wascher}]{Neumark2014}
\textsc{Neumark, D.}, \textsc{Salas, J.~I.} and \textsc{Wascher, W.} (2014).
  Revisiting the minimum wage—employment debate: Throwing out the baby with
  the bathwater? \textit{ILR Review}, \textbf{67}~(3\_suppl), 608--648.

\bibitem[{Neumark and Shirley(2022)}]{neumark2022myth}
\textsc{---} and \textsc{Shirley, P.} (2022). Myth or measurement: What does
  the new minimum wage research say about minimum wages and job loss in the
  {United States}? \textit{Industrial Relations: A Journal of Economy and
  Society}, \textbf{61}~(4), 384--417.

\bibitem[{Neumark and Wascher(1992)}]{neumark1992employment}
\textsc{---} and \textsc{Wascher, W.} (1992). Employment effects of minimum and
  subminimum wages: panel data on state minimum wage laws. \textit{ILR Review},
  \textbf{46}~(1), 55--81.

\bibitem[{Neumark and Wascher(2008)}]{Neumark_Wascher2008}
\textsc{---} and \textsc{Wascher, W.~L.} (2008). \textit{Minimum wages}.
  Cambridge, Mass: MIT Press.

\bibitem[{Okudaira \textit{et~al.}(2019)Okudaira, Takizawa and
  Yamanouchi}]{okudaira2019minimum}
\textsc{Okudaira, H.}, \textsc{Takizawa, M.} and \textsc{Yamanouchi, K.}
  (2019). Minimum wage effects across heterogeneous markets. \textit{Labour
  Economics}, \textbf{59}, 110--122.

\bibitem[{Otani(2024)}]{otani2024nonparametric}
\textsc{Otani, S.} (2024). Nonparametric estimation of matching efficiency and
  mismatch in labor markets via public employment security offices in {Japan},
  1972-2024. \textit{arXiv preprint arXiv:2407.20931}.

\bibitem[{Otani(2025)}]{otani2025onthejob}
\textsc{---} (2025). Nonparametric estimation of matching efficiency and
  elasticity on a private on-the-job search platform: Evidence from {Japan},
  2014-2024. \textit{Journal of the Japanese and International Economies}, p.
  101394.

\bibitem[{Portugal and Cardoso(2006)}]{portugal2006disentangling}
\textsc{Portugal, P.} and \textsc{Cardoso, A.~R.} (2006). Disentangling the
  minimum wage puzzle: An analysis of worker accessions and separations.
  \textit{Journal of the European Economic Association}, \textbf{4}~(5),
  988--1013.

\bibitem[{Sabia(2009{\natexlab{a}})}]{sabia2009effects}
\textsc{Sabia, J.~J.} (2009{\natexlab{a}}). The effects of minimum wage
  increases on retail employment and hours: New evidence from monthly {CPS}
  data. \textit{Journal of Labor Research}, \textbf{30}, 75--97.

\bibitem[{Sabia(2009{\natexlab{b}})}]{sabia2009identifying}
\textsc{---} (2009{\natexlab{b}}). Identifying minimum wage effects: New
  evidence from monthly {CPS} data. \textit{Industrial Relations: A Journal of
  Economy and Society}, \textbf{48}~(2), 311--328.

\bibitem[{Sekiya \textit{et~al.}(2026)Sekiya, Otani, Komatsu, Fujii, Ozeki and
  Noda}]{sekiya2026designing}
\textsc{Sekiya, K.}, \textsc{Otani, S.}, \textsc{Komatsu, Y.}, \textsc{Fujii,
  Y.}, \textsc{Ozeki, S.} and \textsc{Noda, S.} (2026). Designing
  recommendation exposure and favorite lists: A field experiment in a spot-work
  platform. \textit{arXiv preprint arXiv:2606.17397}.

\bibitem[{Simon and Kaestner(2004)}]{simon2004minimum}
\textsc{Simon, K.~I.} and \textsc{Kaestner, R.} (2004). Do minimum wages affect
  non-wage job attributes? evidence on fringe benefits. \textit{ILR Review},
  \textbf{58}~(1), 52--70.

\bibitem[{Stanton and Thomas(2021)}]{stanton2021benefits}
\textsc{Stanton, C.~T.} and \textsc{Thomas, C.} (2021). \textit{Who Benefits
  from Online Gig Economy Platforms?} Tech. rep., National Bureau of Economic
  Research.

\bibitem[{Tachibanaki and Urakawa(2006)}]{tachibanaki2006}
\textsc{Tachibanaki, T.} and \textsc{Urakawa, K.} (2006). \textit{Poverty Study
  in {Japan}(in {Japanese})}. University of Tokyo Press.

\bibitem[{Takahashi \textit{et~al.}(2009)Takahashi, Kawade and
  Kato}]{Takahashi2009}
\textsc{Takahashi, S.}, \textsc{Kawade, M.} and \textsc{Kato, R.~R.} (2009).
  Labor supply of {Japanese} married women, sensitivity analysis and a new
  estimate. \textit{Economic analysis \& policy series}, \textbf{9}, 1--43.

\bibitem[{Tamada(2009)}]{Tamada_2009_en}
\textsc{Tamada, K.} (2009). How is minimum wage determined? \textit{Jpn. J.
  Labor.stud}, \textbf{593}, 16--28, in Japanese.

\bibitem[{Tamada(2011)}]{tamada2011analysis}
\textsc{---} (2011). Analysis of the determinants of minimum wages in {Japan}.

\bibitem[{Yamagishi(2021)}]{yamagishi2021minimum}
\textsc{Yamagishi, A.} (2021). Minimum wages and housing rents: Theory and
  evidence. \textit{Regional Science and Urban Economics}, \textbf{87}, 103649.

\bibitem[{Yamanouchi \textit{et~al.}(2025)Yamanouchi, Okudaira, Takizawa and
  Hosono}]{yamanouchi2025minimum_wage}
\textsc{Yamanouchi, K.}, \textsc{Okudaira, H.}, \textsc{Takizawa, M.} and
  \textsc{Hosono, K.} (2025). Who pays for the minimum wage in the {Japanese}
  manufacturing sector? \textit{Research Institute of Economy, Trade and
  Industry (RIETI)}.

\bibitem[{Yokoyama(2018)}]{Yokoyama_2018}
\textsc{Yokoyama, I.} (2018). How the tax reform on the special exemption for
  spouse affected the work-hour distribution. \textit{Journal of the Japanese
  and International Economies}, \textbf{49}, 69--84.

\bibitem[{Yokoyama and Kodama(2016)}]{Yokoyama_Kodama2015}
\textsc{---} and \textsc{Kodama, N.} (2016). Women's labor and taxes: A
  data-based analysis of the current situation. \textit{Financial
  Review/Institute for Financial Policy Studies, Ministry of Finance, ed.},
  \textbf{2016}~(2), 49--76, japanese.

\bibitem[{Yugami(2005)}]{yugami}
\textsc{Yugami, K.} (2005). \textit{Analysis of local labor markets using
  prefecture-level data - on the regional differences of unemployment and
  non-employment (in {Japanese})}. Tech. rep., The Japan Institute of Labour
  Policy and Training.

\end{thebibliography}

\newpage
\appendix

\section{Additional Analysis}

\subsection{Elasticities Related to Minimum Wages}\label{sec:elasticity_results}

Adapting \citet{Cengiz2019}, we scale the normalized net employment-share effect by minimum-wage growth and lower-tier average-wage growth.\footnote{The reported elasticities use the full lower-tier region, $e=-1,\dots,3$, and the common September employment-share stock described in the implementation note in Section \ref{sec:excess_job_missing_job_approach}. The employment effect is not divided again by the pre-reform share below the new minimum wage. For the wage calculation, the same event-study regression is estimated for actual hourly-wage bills: sums of hourly wages across realized matches divided by prefecture-month employment, without multiplying by shift lengths. The average-wage change is obtained from the ratio of the updated wage-bill and employment stocks. The minimum-wage elasticity divides the net employment effect by the equally weighted mean prefecture-level minimum-wage growth rate; the own-wage elasticity divides it by the average-wage change. Nonlinear standard errors use the delta method with the joint prefecture-clustered covariance of employment and wage-bill estimates, including their cross-outcome covariance. Reference stocks and minimum-wage growth rates are treated as fixed, and confidence intervals use the normal approximation.}

\begin{table}[!htbp]
  \begin{center}
      \caption{Employment-Share and Wage Responses}
      \label{tb:employment_elasticity_table}
      
\begin{tabular}{lc}
\toprule
 & Estimate (SE)\\
\midrule
Missing jobs below new MW & -0.280\\
 & (0.022)\\
Excess jobs above new MW & 0.270\\
 & (0.023)\\
Affected wages & 0.018\\
 & (0.002)\\
Affected employment & -0.010\\
 & (0.004)\\
Emp. elasticity w.r.t MW & -0.211\\
 & (0.096)\\
Emp. elasticity w.r.t affected wage & -0.537\\
 & (0.257)\\
Job below new MW & 0.278\\
\% MW changes & 0.047\\
\bottomrule
\end{tabular}
  
  \end{center}\footnotesize
  \textit{Note}: The table reports the effects of a minimum wage increase based on the event study analysis exploiting 47 prefecture-level minimum wage changes in 2023 (i.e., about a 4.7\% minimum wage increase). Employment effects are normalized share effects; elasticities divide Net by minimum-wage or average-wage growth. Estimates average October 2023--March 2024. Robust standard errors in parentheses are clustered by prefecture.
\end{table} 

Table \ref{tb:employment_elasticity_table} reports six-month–averaged post-treatment effects for the baseline specification depicted in Figure \ref{fg:wage_distribution_moving_employment_overall_cengiz_original_form}. 
The magnitude of missing jobs below new minimum wages exceeds the magnitude of excess jobs above new minimum wages in the six-month window. 
Table \ref{tb:employment_elasticity_table} reports share-based elasticities of $-0.211$ with respect to the minimum wage and $-0.537$ with respect to the lower-tier average wage.

\subsection{Event-Study and Wage-Distribution Figures by Outcome and Job Category}\label{app:event_study_figures_by_category}

\begin{figure}[!ht]
  \begin{center}
  \subfloat[Missing and Excess Posted Vacancies over Time]{\includegraphics[width = 0.46\textwidth]{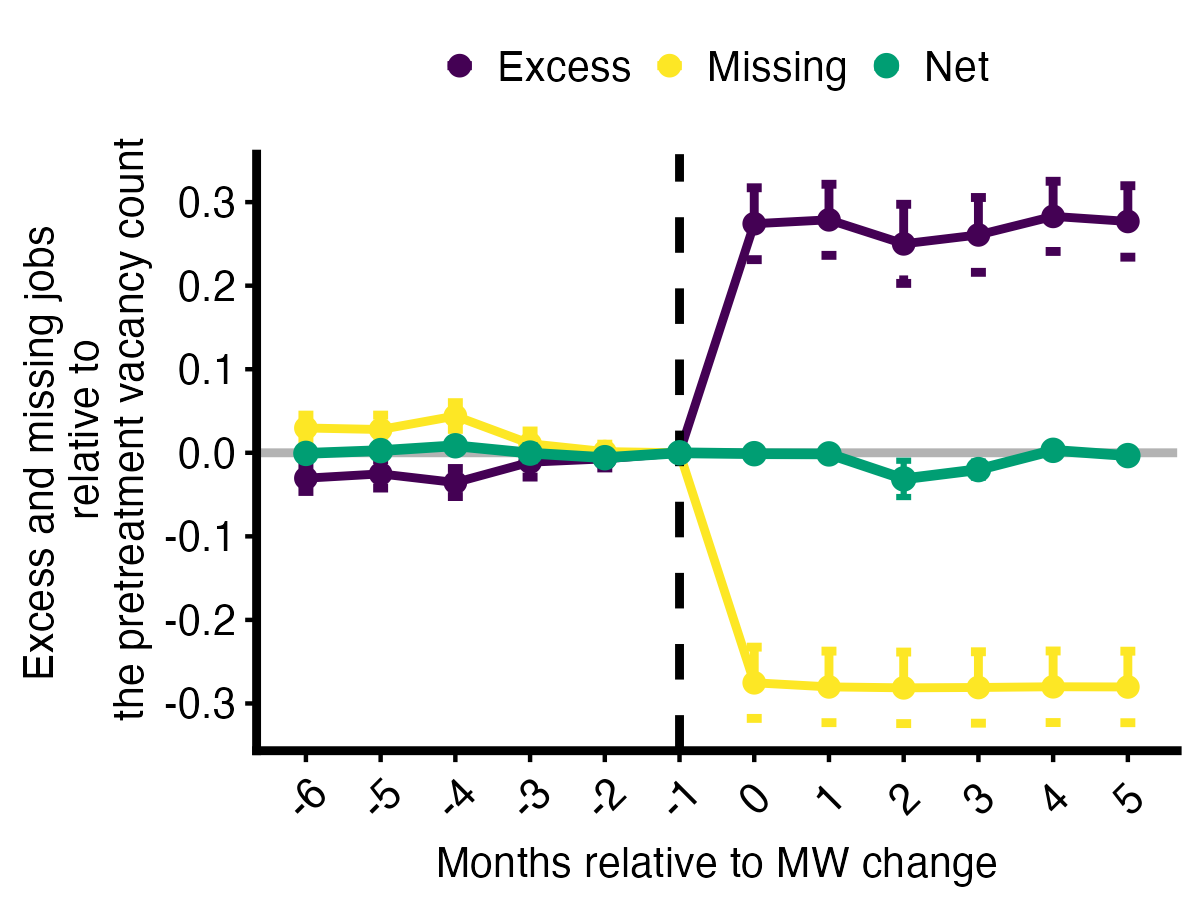}}
  \subfloat[Wage Distribution]{\includegraphics[width = 0.46\textwidth]{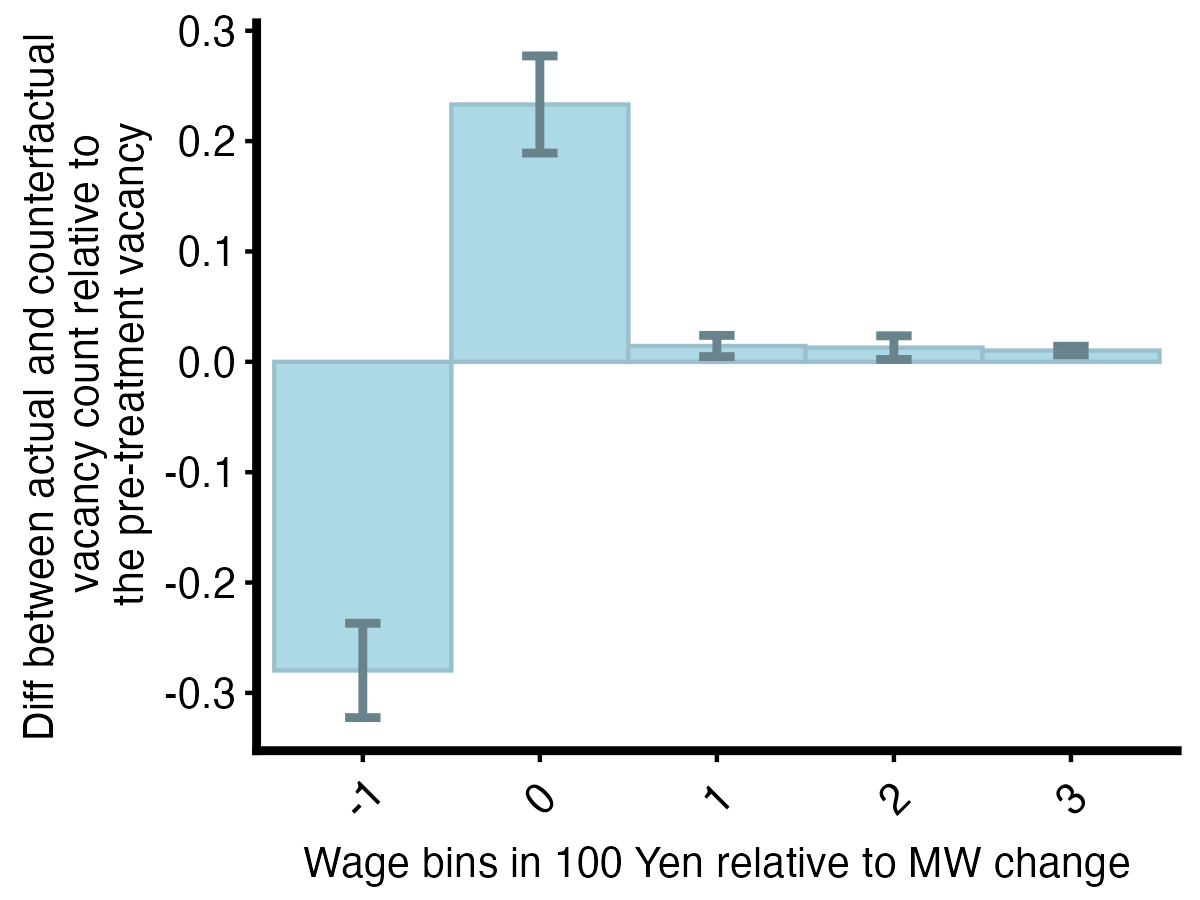}}
  \caption{Impact of Minimum Wages on Vacancies}
  \label{fg:wage_distribution_moving_vacancy_overall_cengiz_original_form}
  \end{center}
  \footnotesize
  Note: Panel (a) shows the estimated excess and missing posted vacancies for each month before and after the minimum wage increase. Panel (b) shows the estimated impact of the minimum wage increase on vacancies for each group $e$ in the wage distribution.
\end{figure}

\begin{figure}[!ht]
  \begin{center}
  \subfloat[Missing and Excess Working Hours over Time]{\includegraphics[width = 0.46\textwidth]{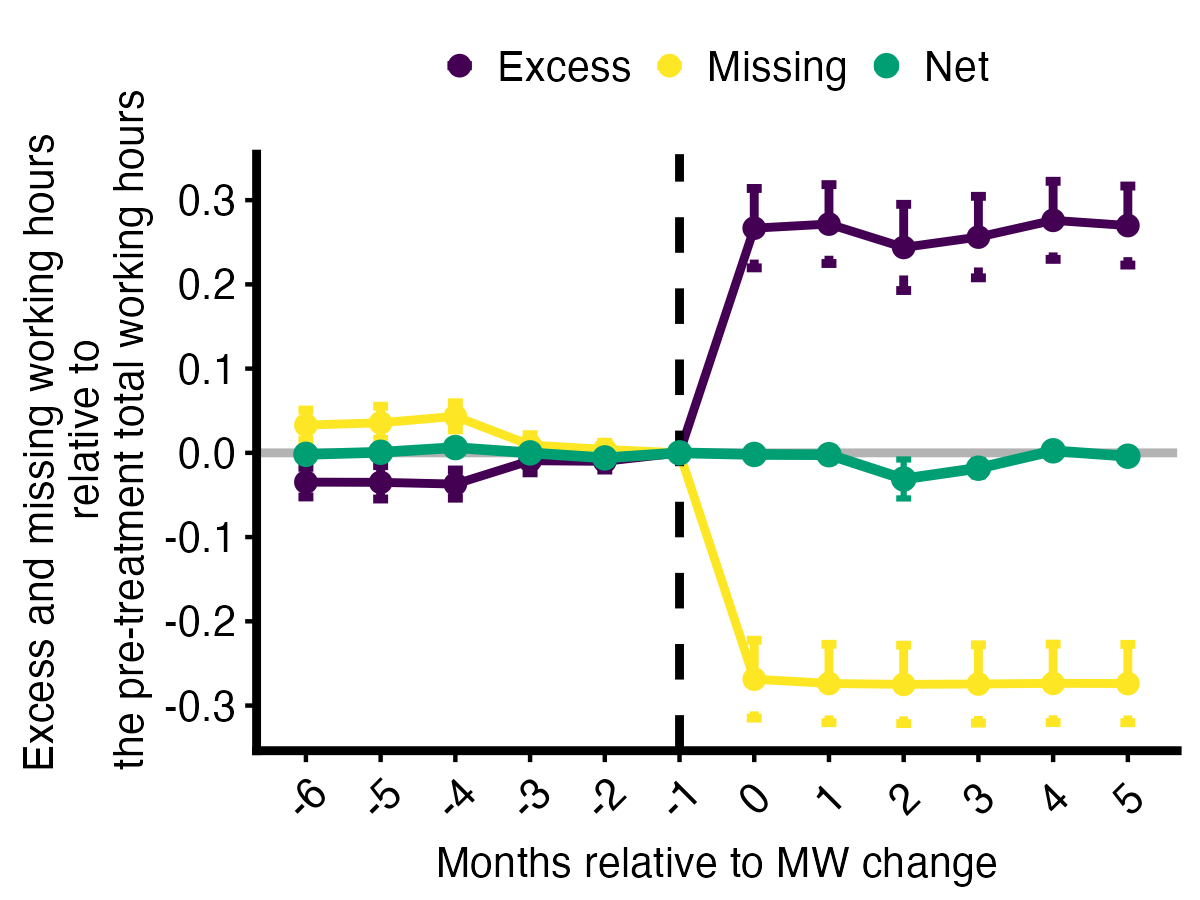}}
  \subfloat[Wage Distribution]{\includegraphics[width = 0.46\textwidth]{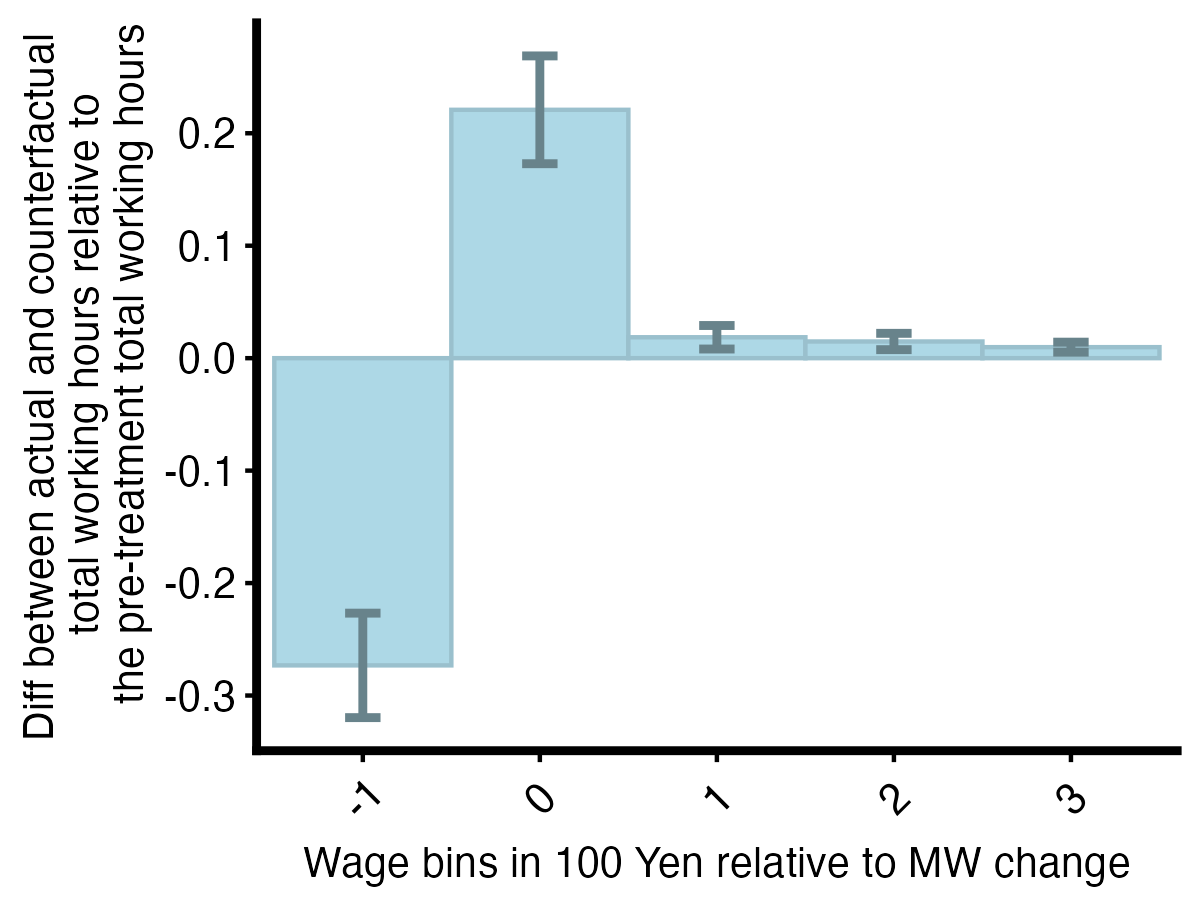}}
  \caption{Impact of Minimum Wages on Total Working Hours}
  \label{fg:wage_distribution_moving_total_working_hours_overall_cengiz_original_form}
  \end{center}
  \footnotesize
  Note: Panel (a) shows the estimated excess and missing working hours for each month before and after the minimum wage increase. Panel (b) shows the estimated impact of the minimum wage increase on total working hours for each group $e$ in the wage distribution. The outcome is the cell's share of total working hours in the prefecture-month, in parallel with the employment and vacancy shares.
\end{figure}

\begin{figure}[!ht]
  \begin{center}
  \subfloat[Restaurant]{\includegraphics[width = 0.33\textwidth]{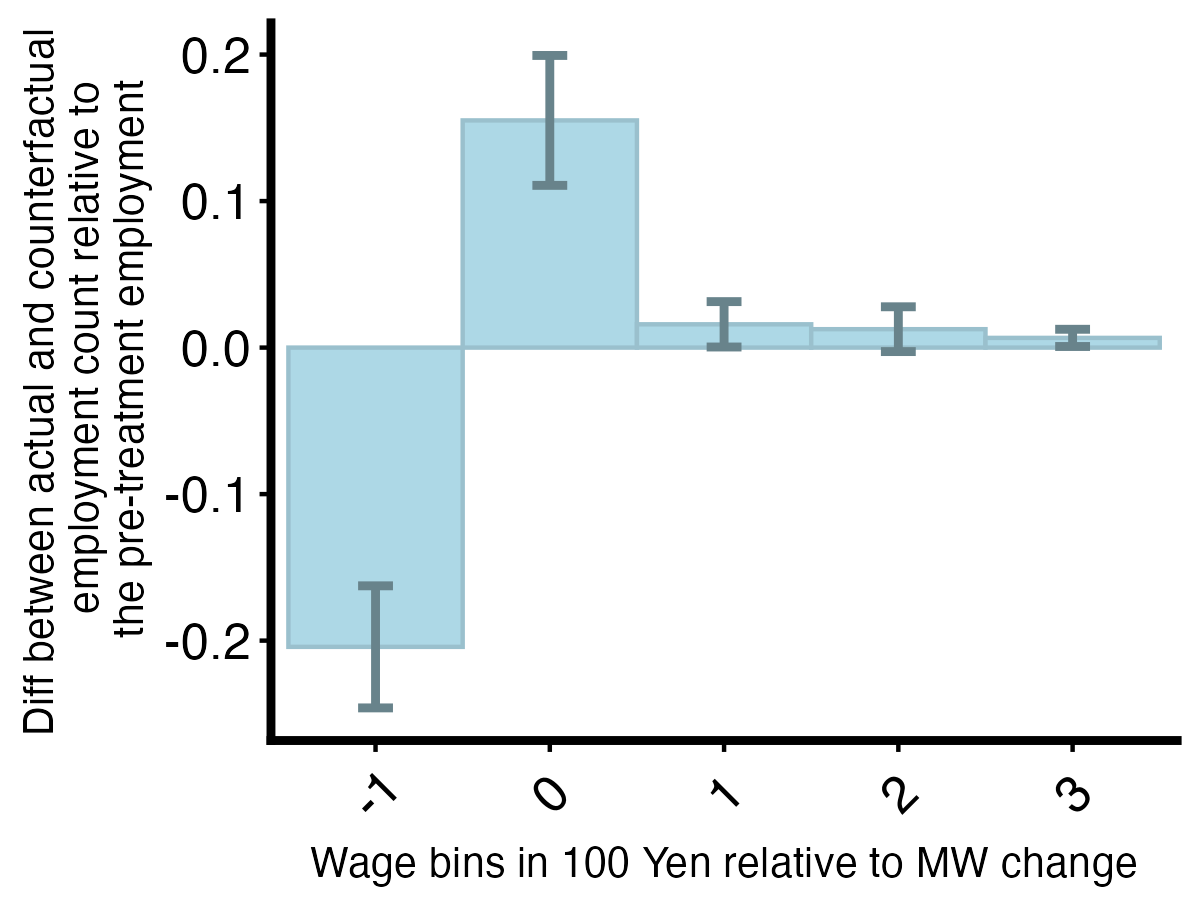}}
  \subfloat[Light Work]{\includegraphics[width = 0.33\textwidth]{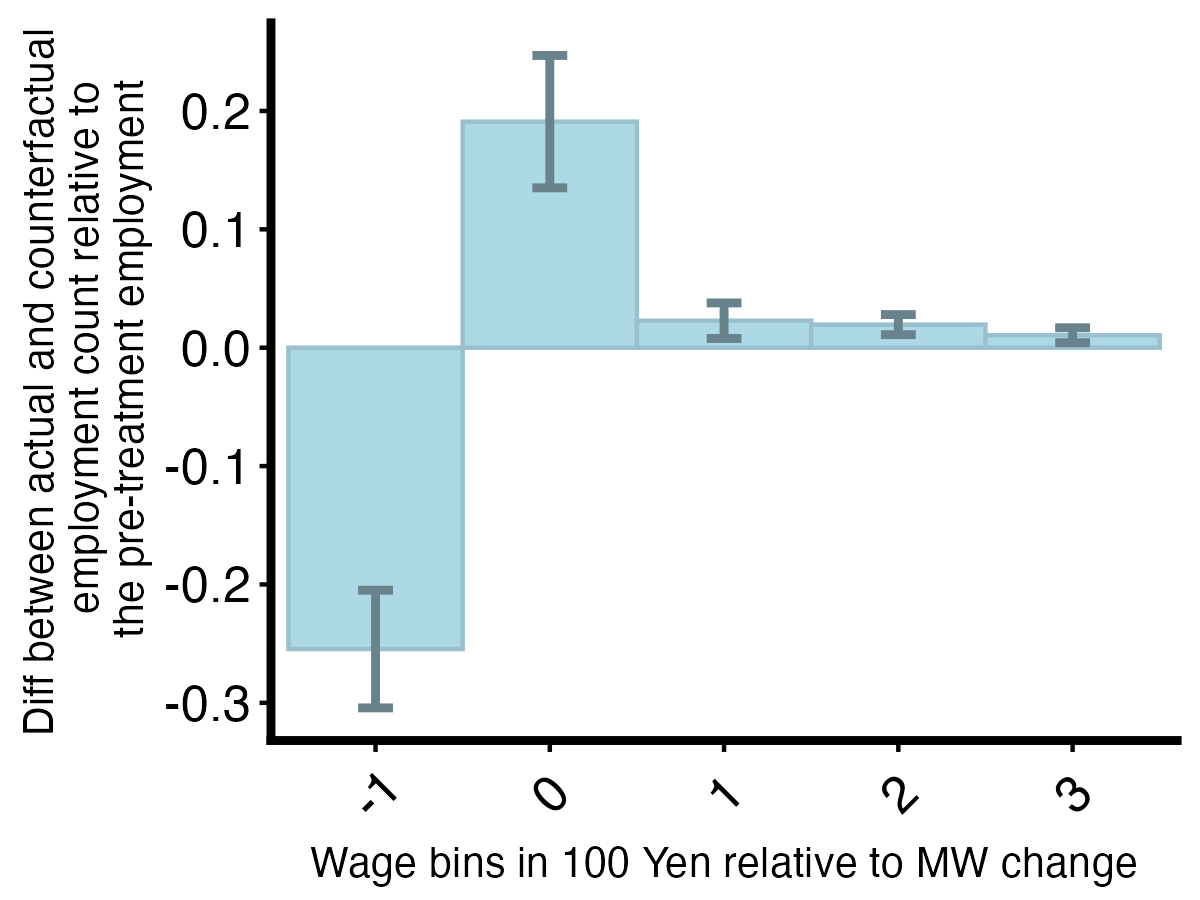}}
  \subfloat[Retail]{\includegraphics[width = 0.33\textwidth]{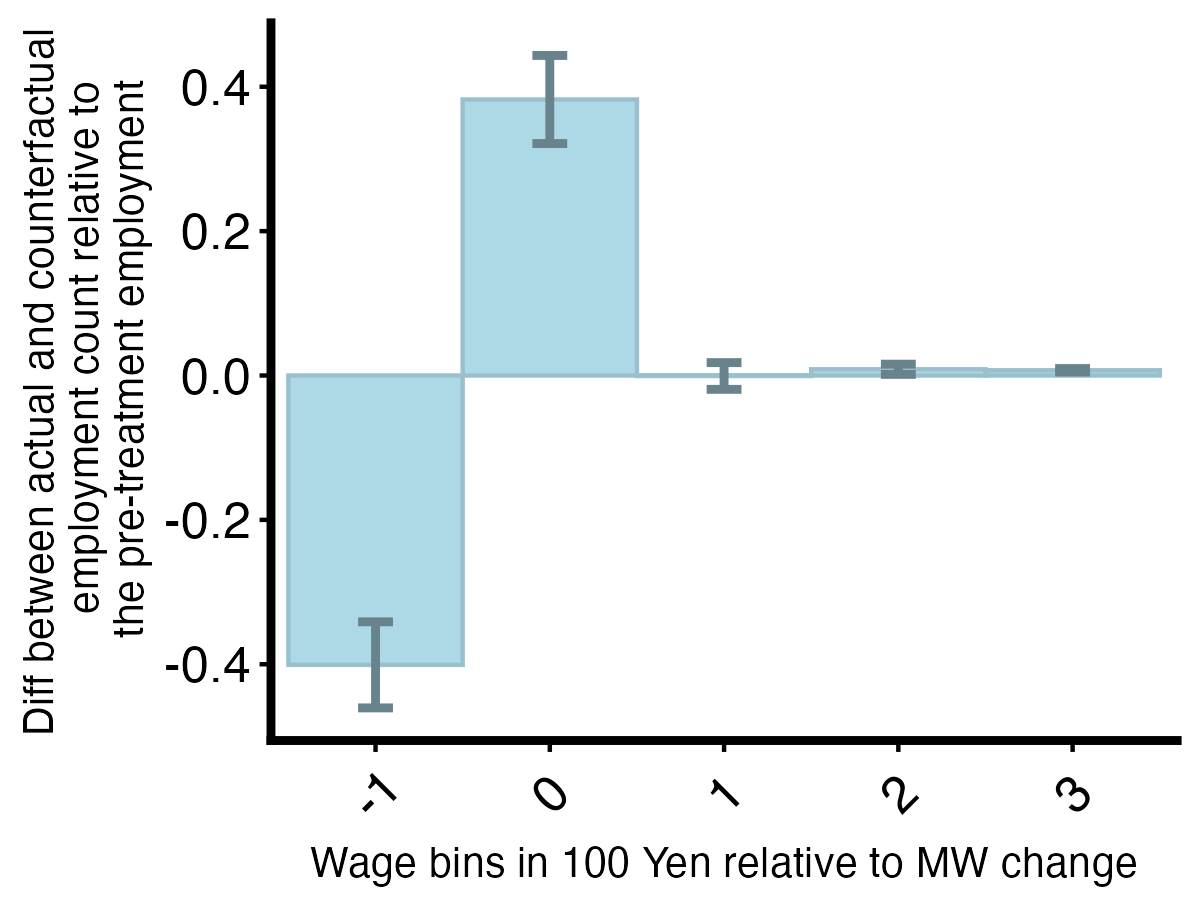}}\\
  \subfloat[Customer Service]{\includegraphics[width = 0.33\textwidth]{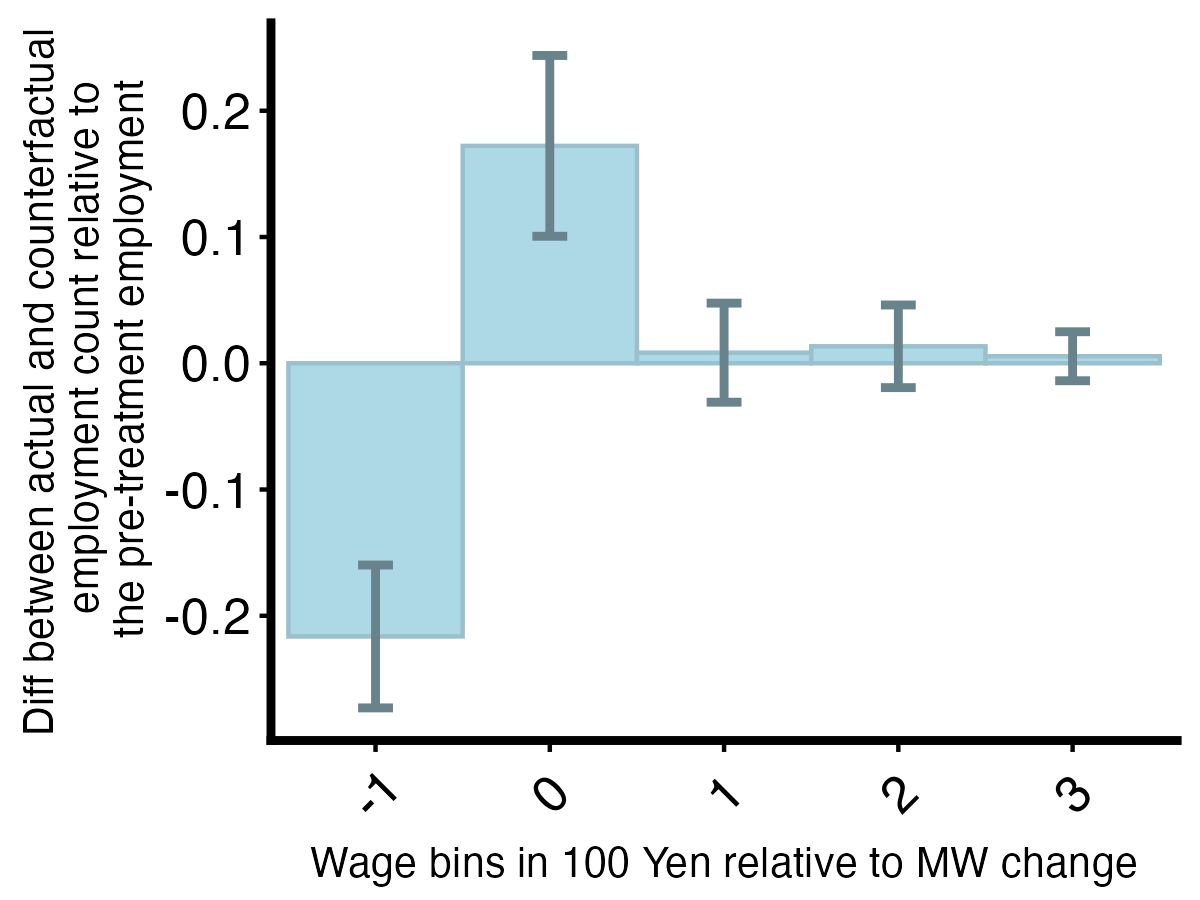}}
  \subfloat[Professional]{\includegraphics[width = 0.33\textwidth]{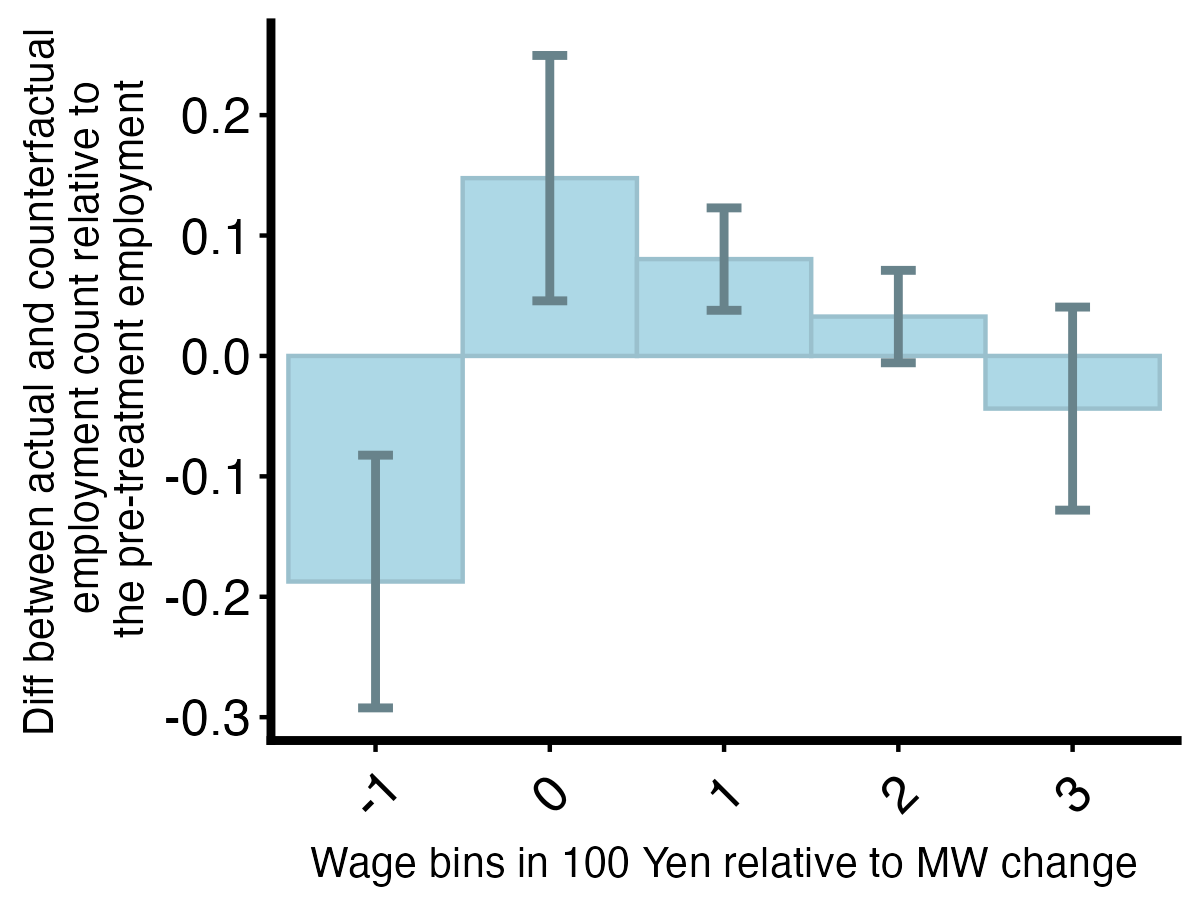}}
  \subfloat[Logistics]{\includegraphics[width = 0.33\textwidth]{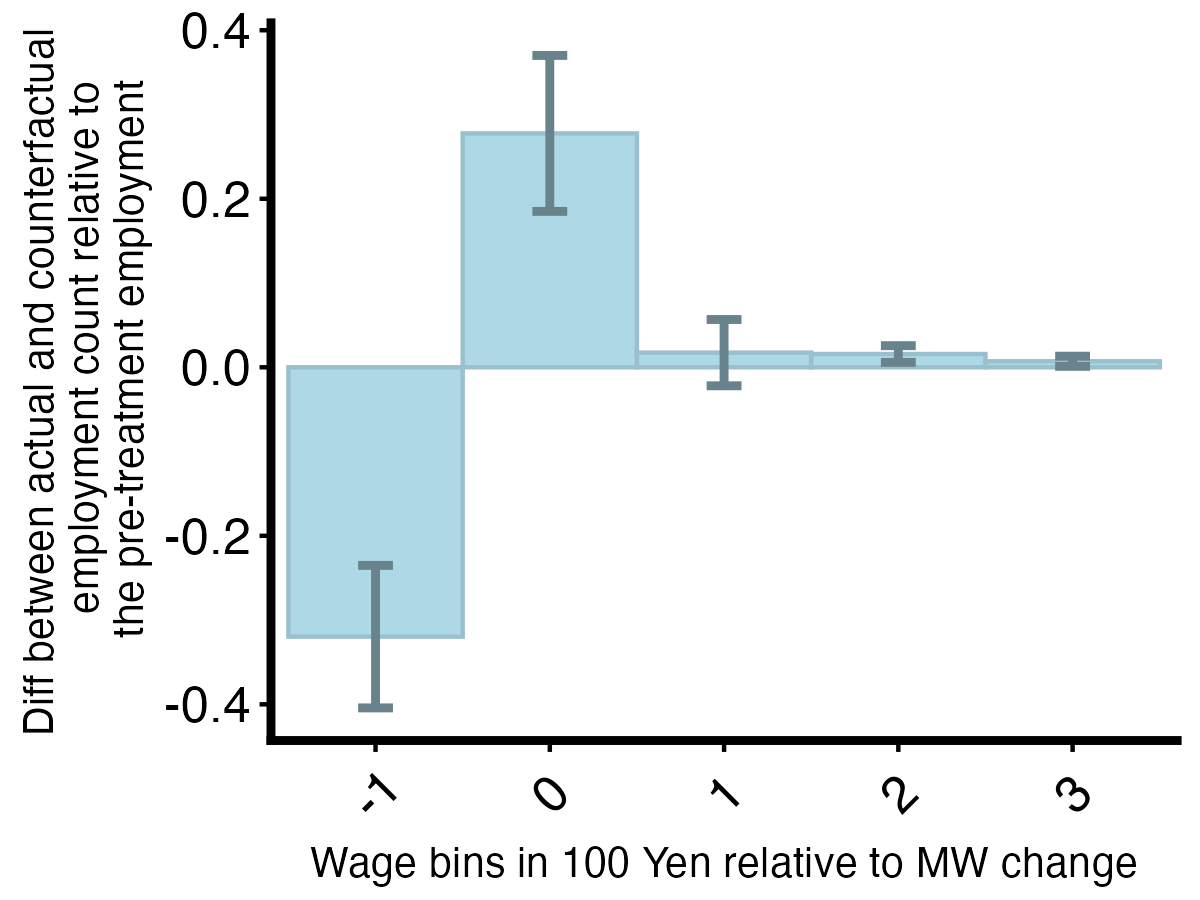}}\\
  \subfloat[Entertainment]{\includegraphics[width = 0.33\textwidth]{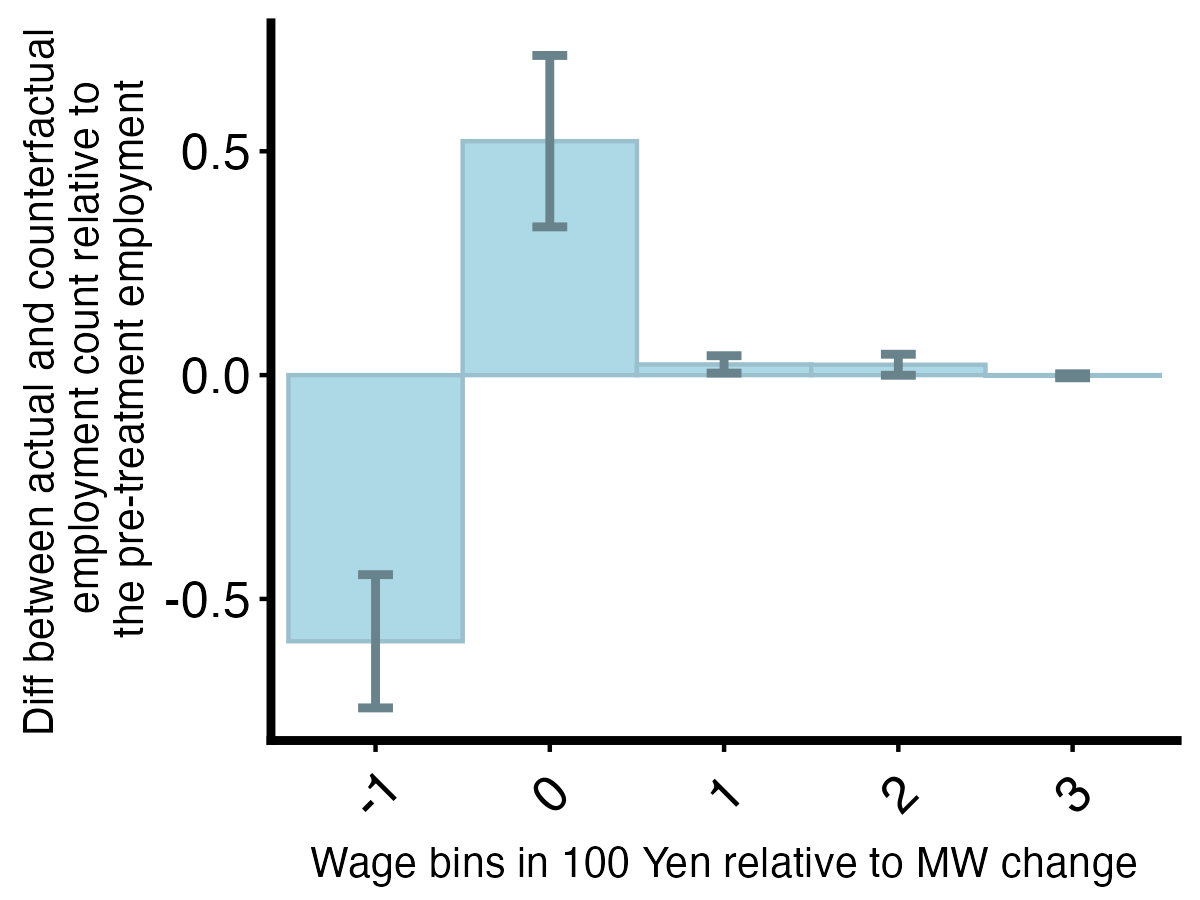}}
  \subfloat[Office Work]{\includegraphics[width = 0.33\textwidth]{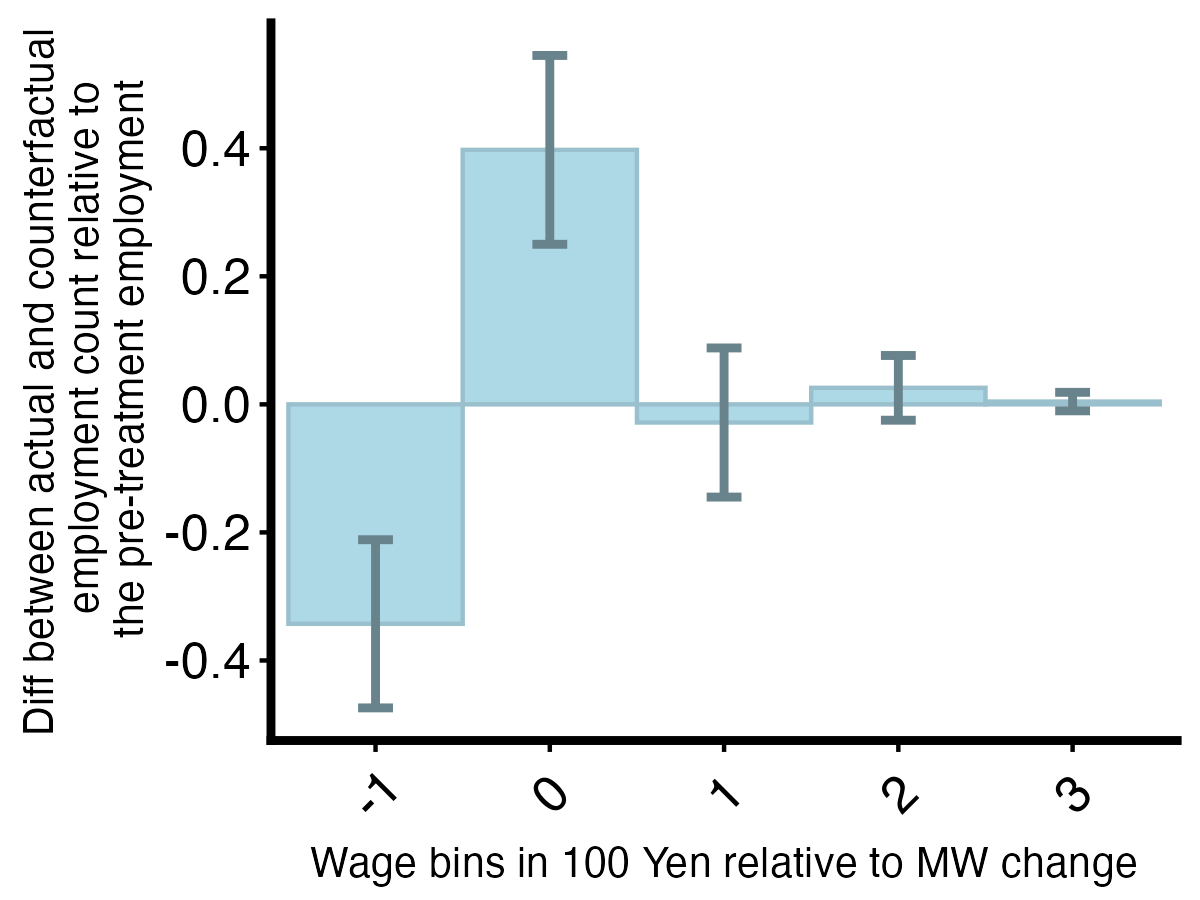}}
  \subfloat[Event Staff]{\includegraphics[width = 0.33\textwidth]{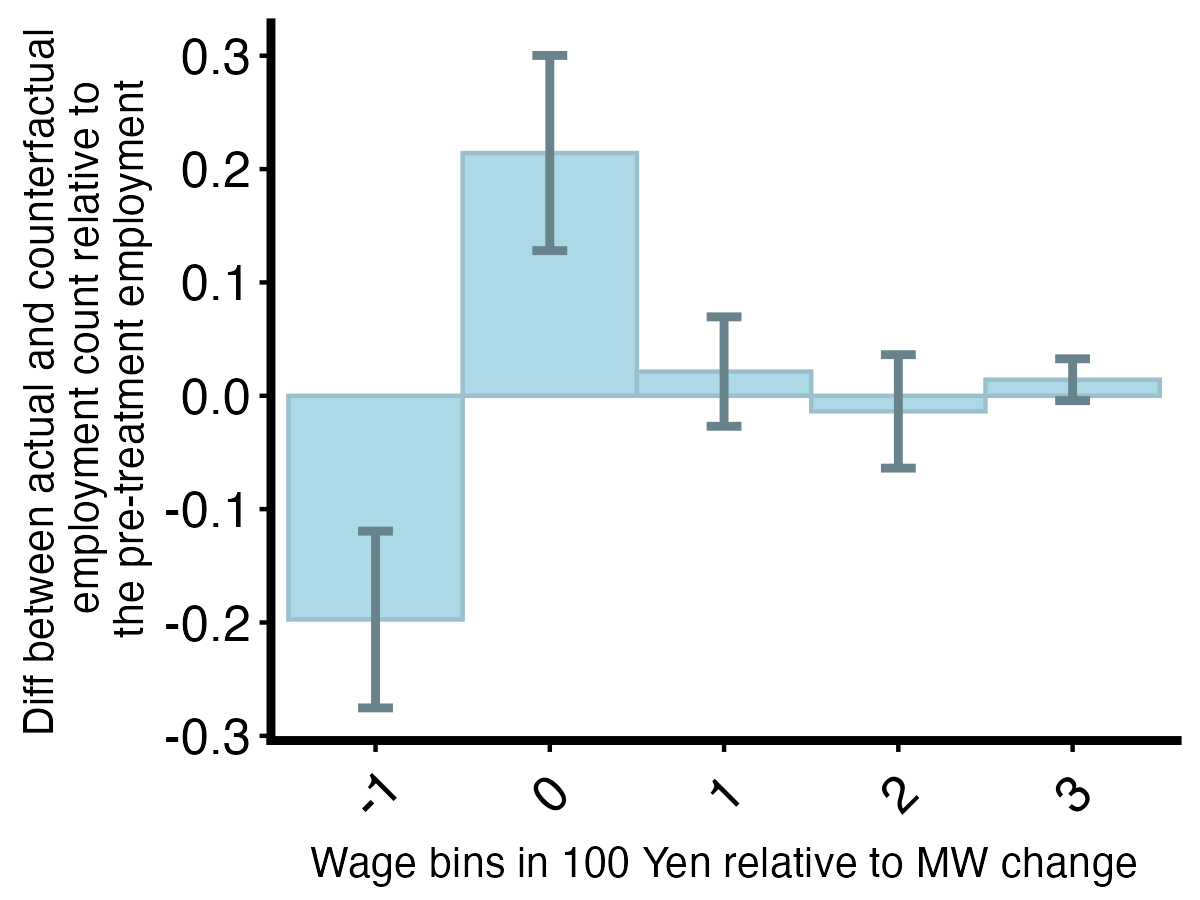}}
  \caption{Impact of Minimum Wages on Employment by Occupation}
  \label{fg:wage_distribution_moving_employment_event_staff_cengiz_original_form}
  \end{center}
  \footnotesize
  Note: Figure \ref{fg:wage_distribution_moving_employment_event_staff_cengiz_original_form} shows the estimated impact of the minimum wage increase on employment for each group $e$ in the wage distribution for each occupation in the platform.
\end{figure}

\begin{figure}[!ht]
  \begin{center}
  \subfloat[Morning]{\includegraphics[width = 0.46\textwidth]  {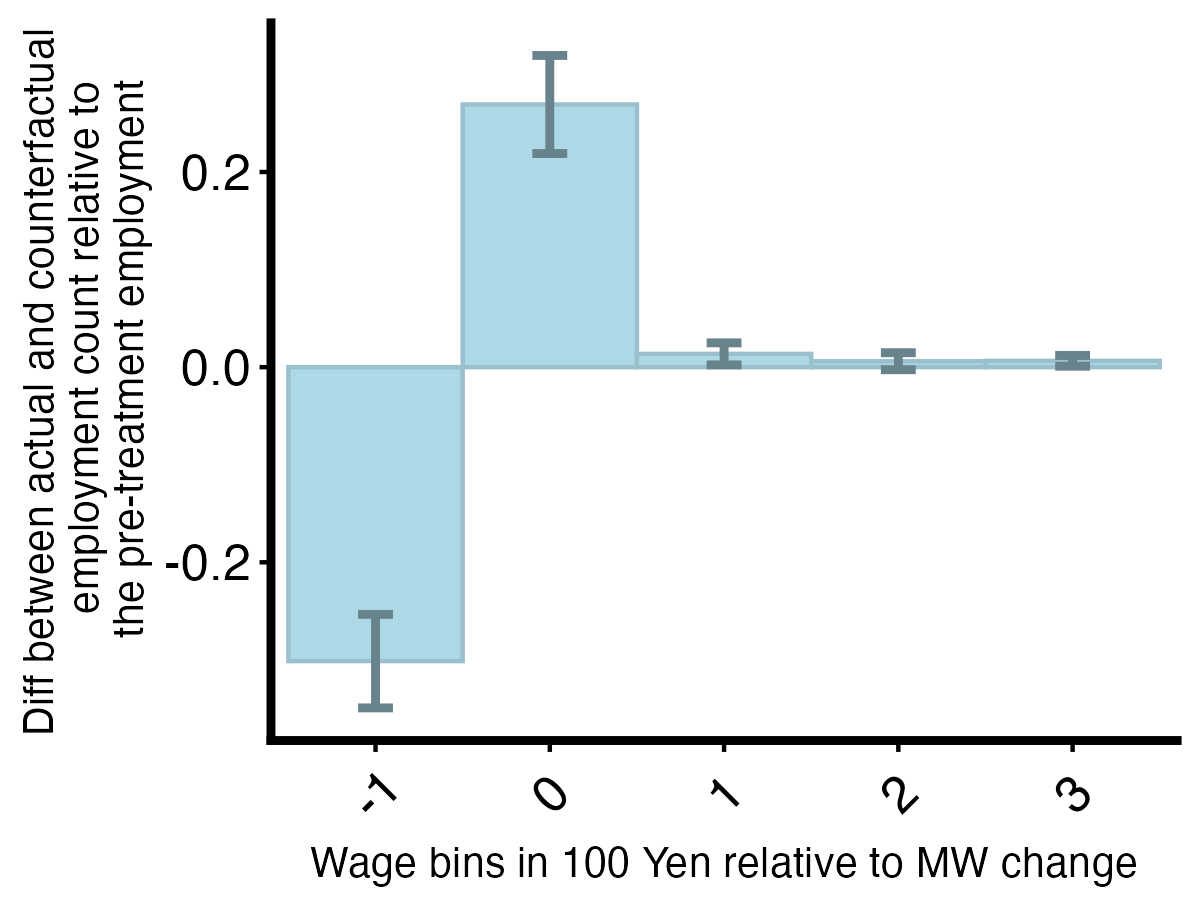}}
  \subfloat[Evening]{\includegraphics[width = 0.46\textwidth]  {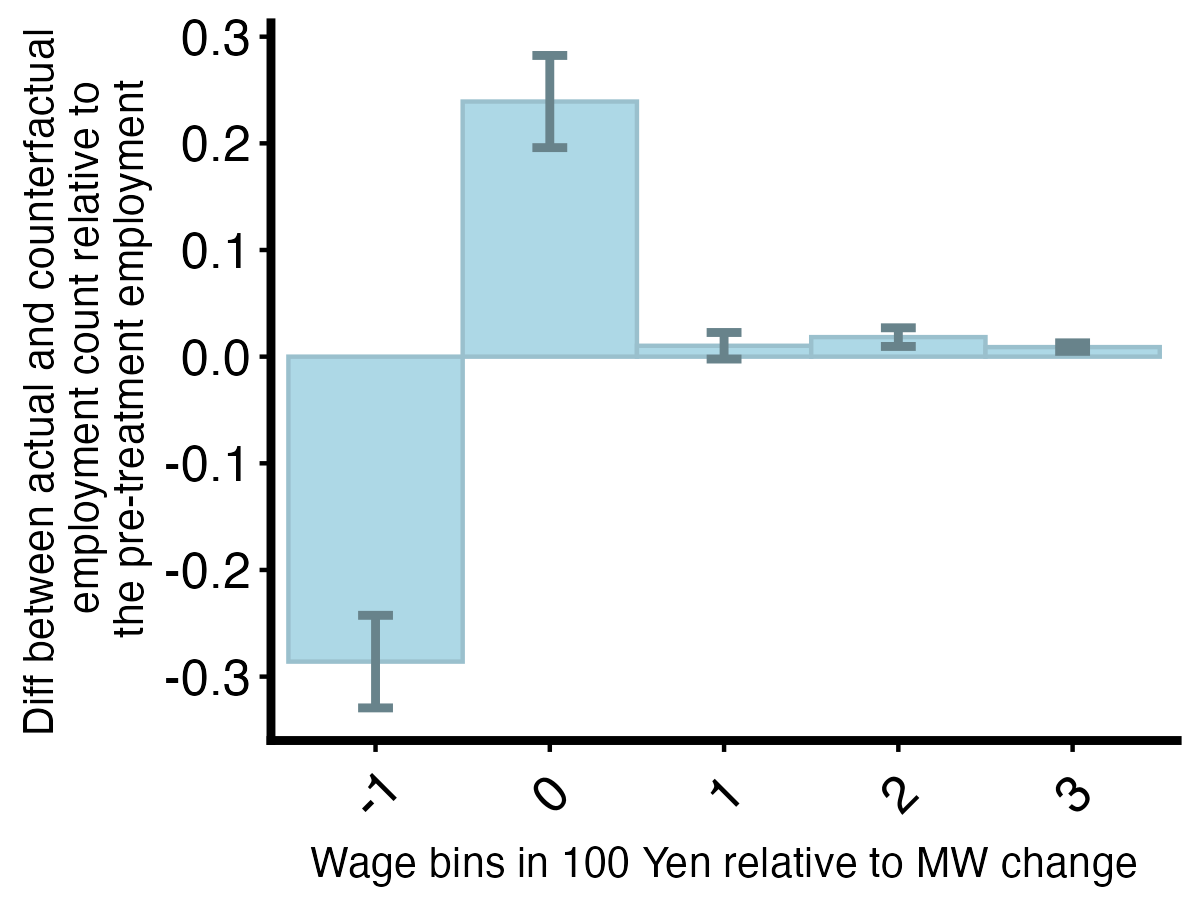}}\\
  \subfloat[Night]{\includegraphics[width = 0.46\textwidth]  {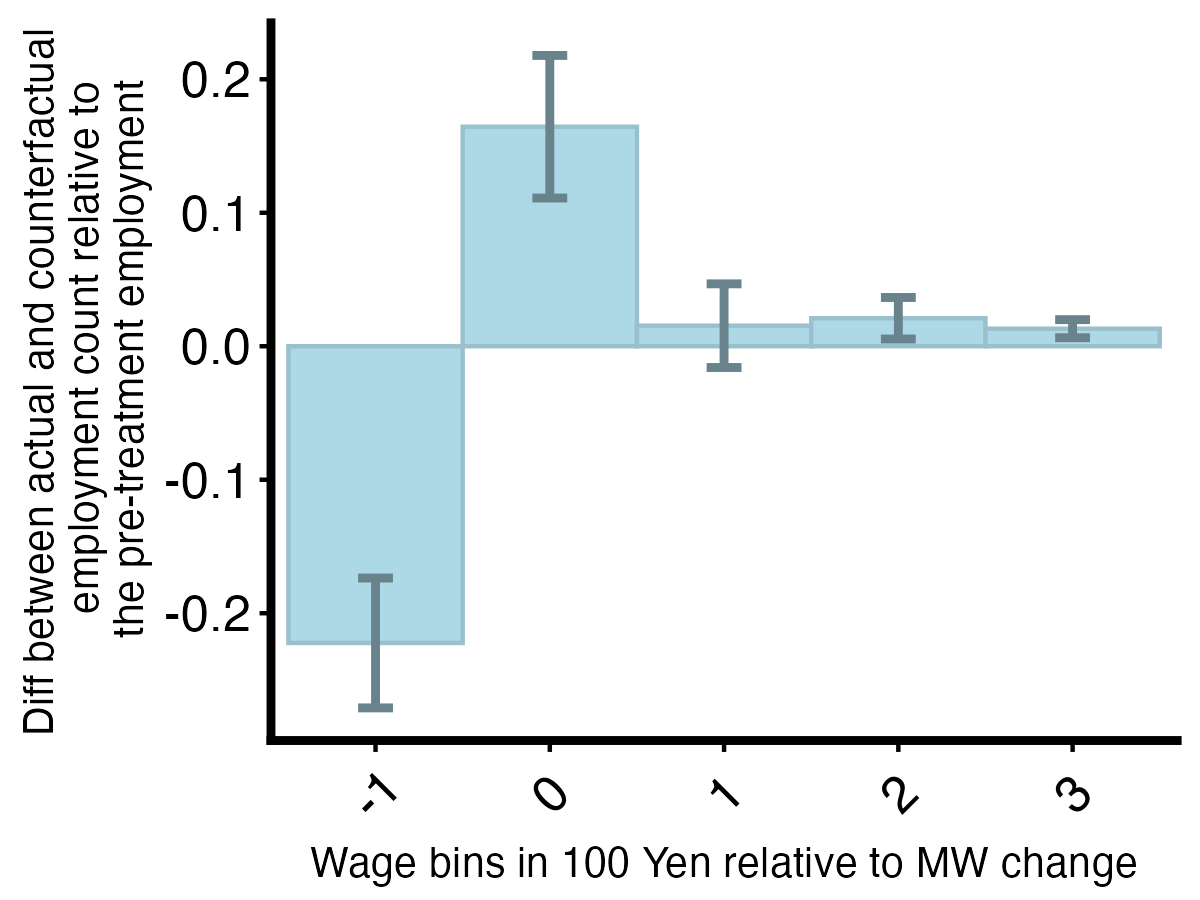}}
  \subfloat[Midnight]{\includegraphics[width = 0.46\textwidth]  {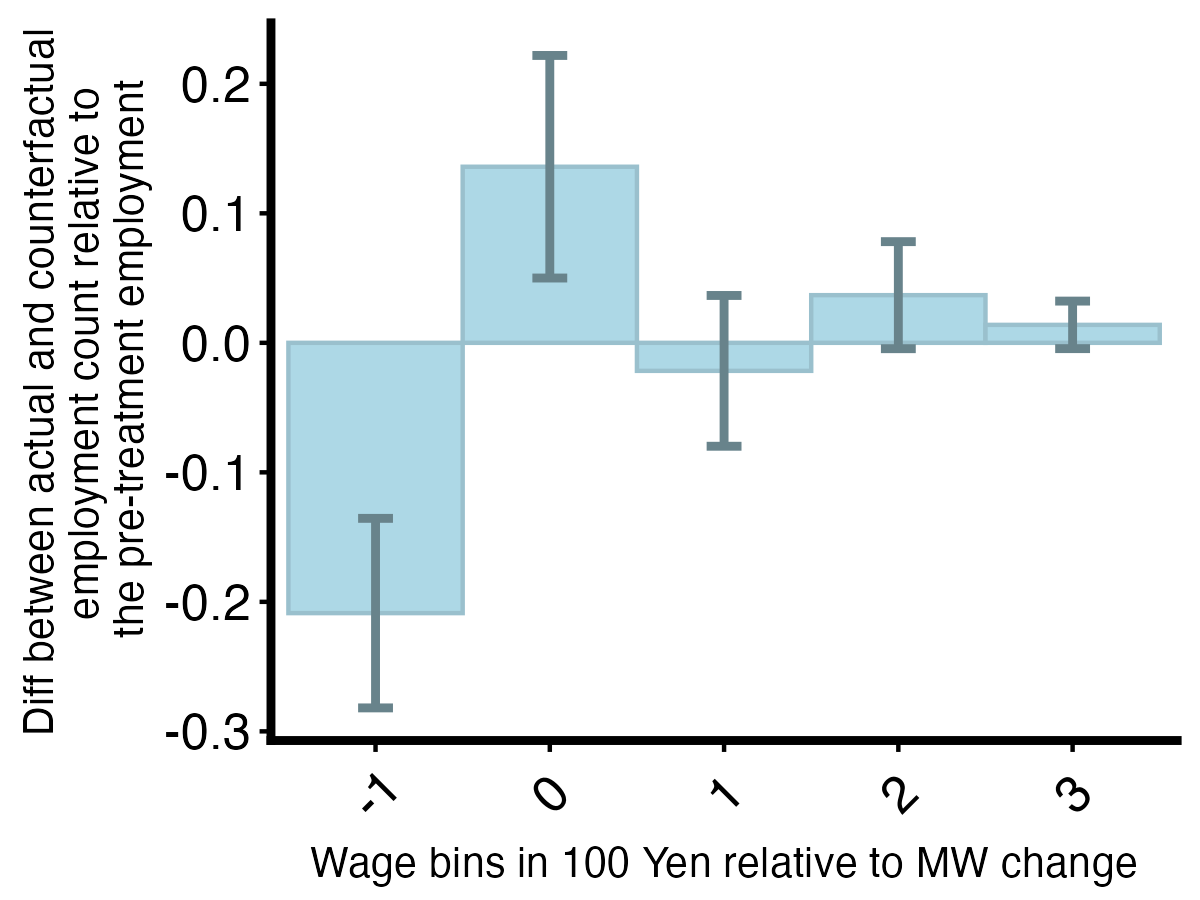}}
  \caption{Impact of Minimum Wages on Employment by Time-of-Day Category}
  \label{fg:wage_distribution_moving_employment_midnight_overall_cengiz_original_form}
  \end{center}
  \footnotesize
  Note: We divide the 24-hour day into four periods—morning (6:00–12:00), evening (12:00–18:00), night (18:00–24:00), and midnight (24:00–6:00).
\end{figure}

\begin{figure}[!ht]
  \begin{center}
  \subfloat[Amenity Provision: Missing and Excess Jobs over Time]{\includegraphics[width = 0.46\textwidth]{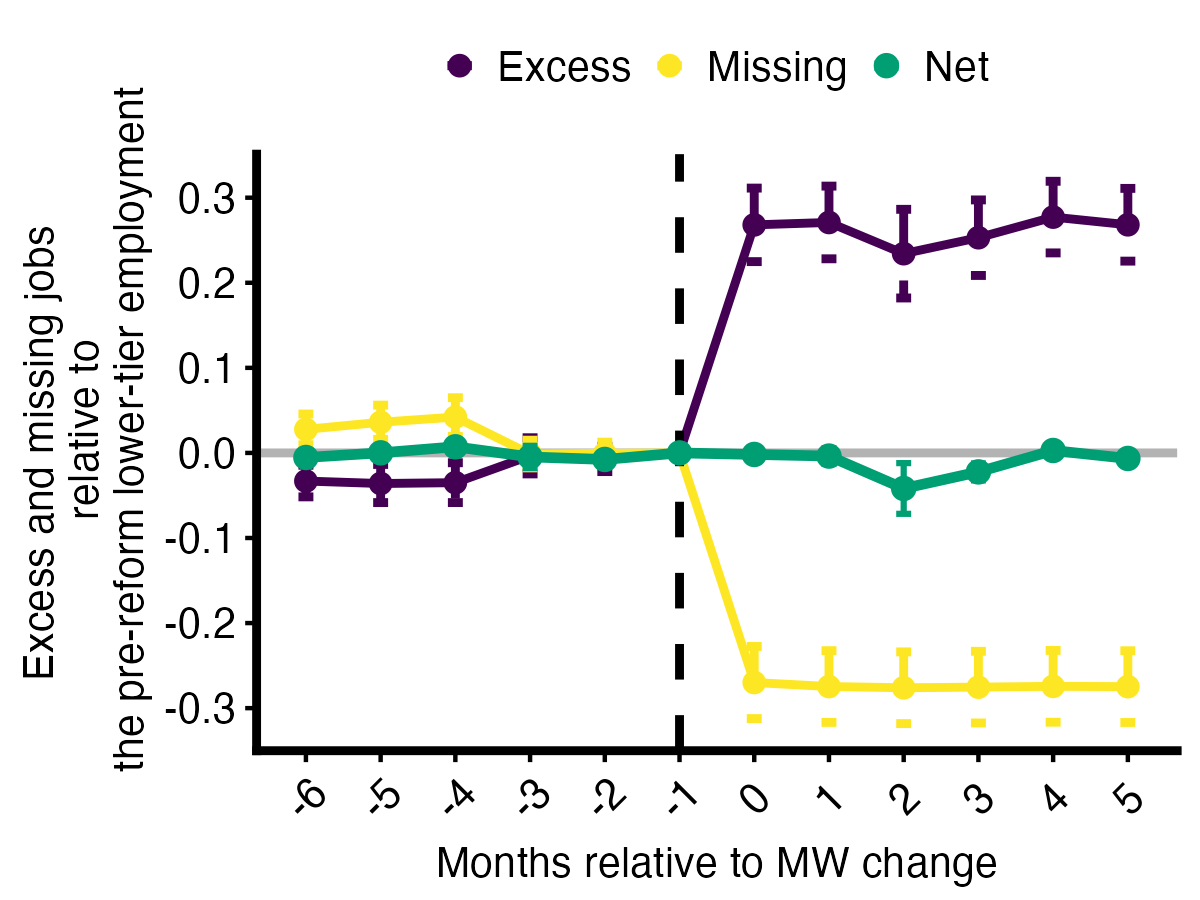}}
  \subfloat[Amenity Provision: Wage Distribution]{\includegraphics[width = 0.46\textwidth]{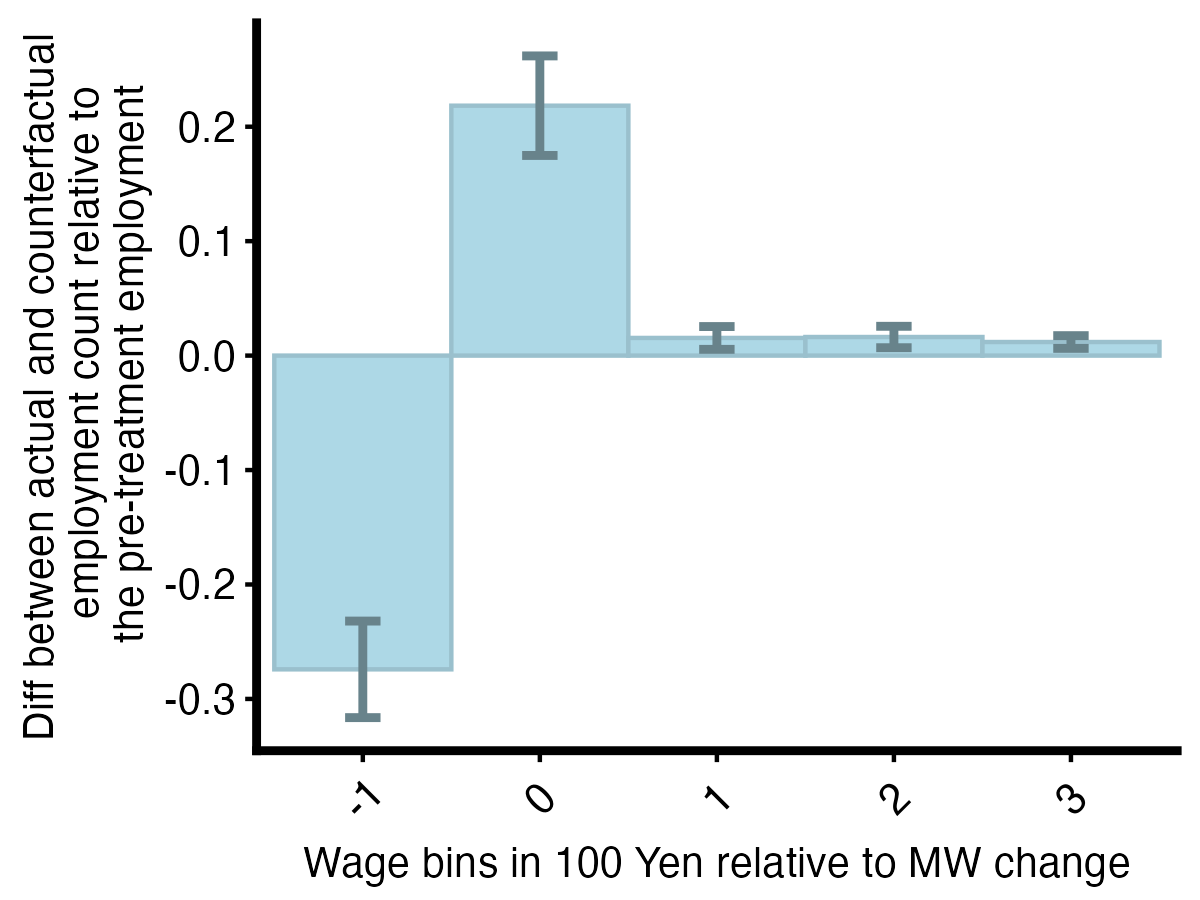}}\\
  \subfloat[No Amenity Provision: Missing and Excess Jobs over Time]{\includegraphics[width = 0.46\textwidth]{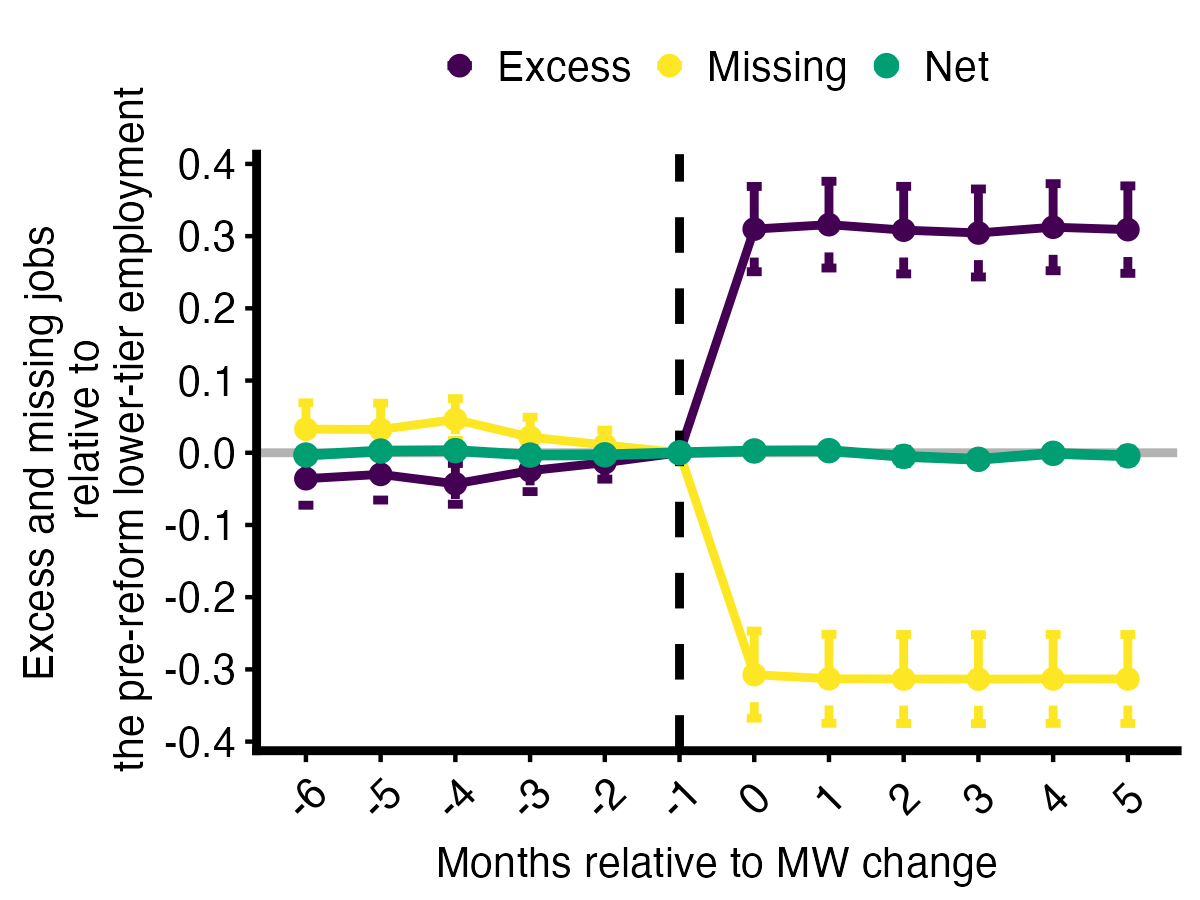}}
  \subfloat[No Amenity Provision: Wage Distribution]{\includegraphics[width = 0.46\textwidth]{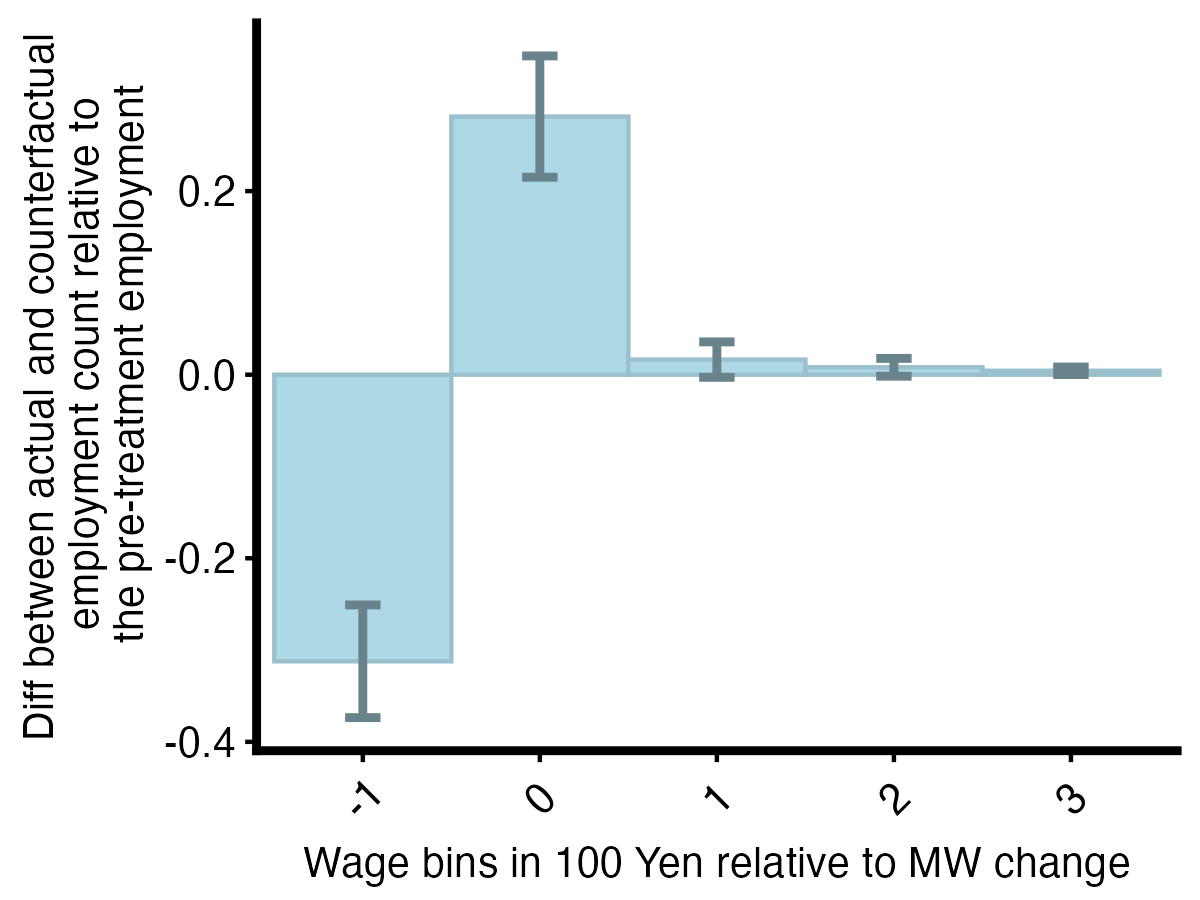}}
  \caption{Impact of Minimum Wages on Employment Separated by Transportation Reimbursement Provision}
  \label{fg:wage_distribution_moving_employment_provision_cengiz_original_form}
  \end{center}
  \footnotesize
  Note: Panels (a) and (c) show the estimated excess and missing jobs before and after the minimum wage increase for postings with and without transportation reimbursement, respectively. Panels (b) and (d) show the corresponding wage-distribution effects for each group $e$.
\end{figure}

\begin{figure}[!ht]
  \begin{center}
  \subfloat[Female-Targeted: Missing and Excess Jobs over Time]{\includegraphics[width = 0.46\textwidth]{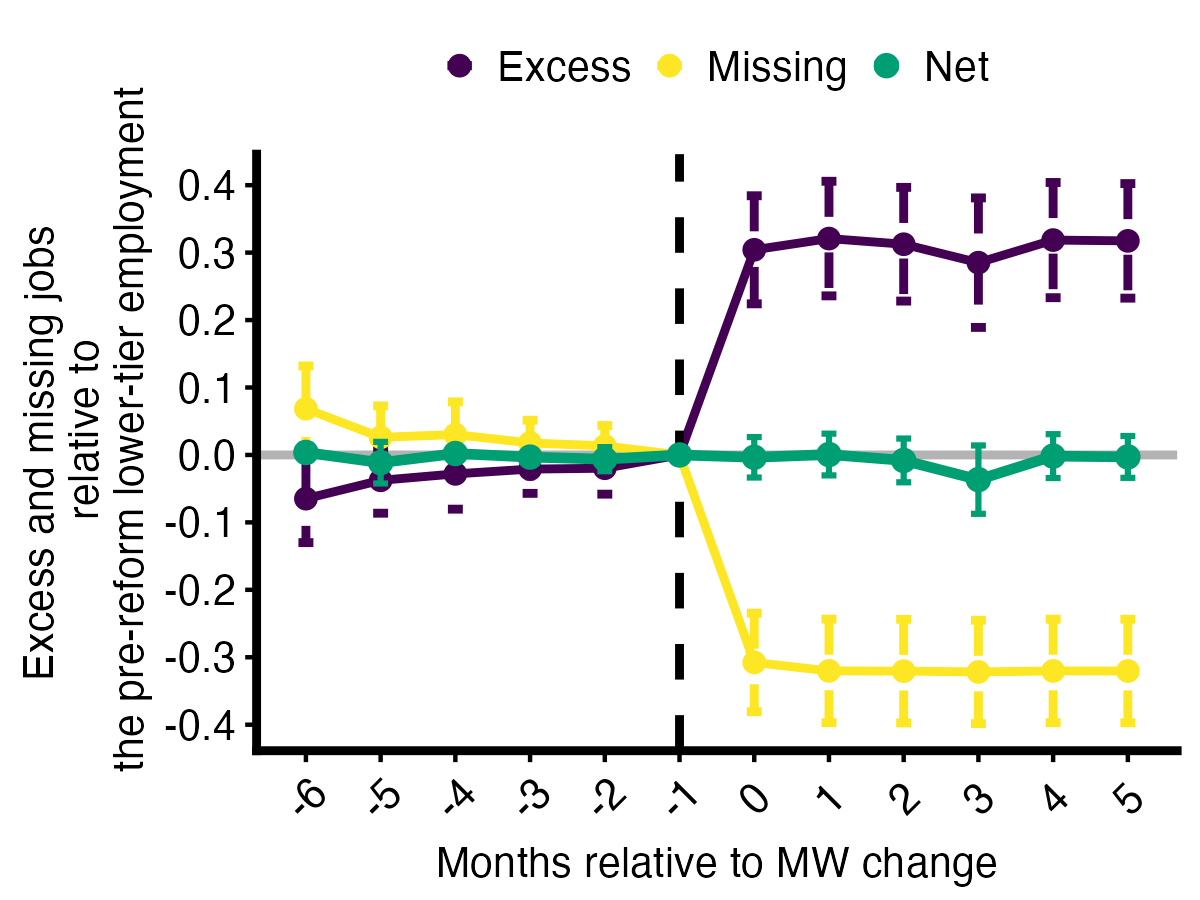}}
  \subfloat[Female-Targeted: Wage Distribution]{\includegraphics[width = 0.46\textwidth]{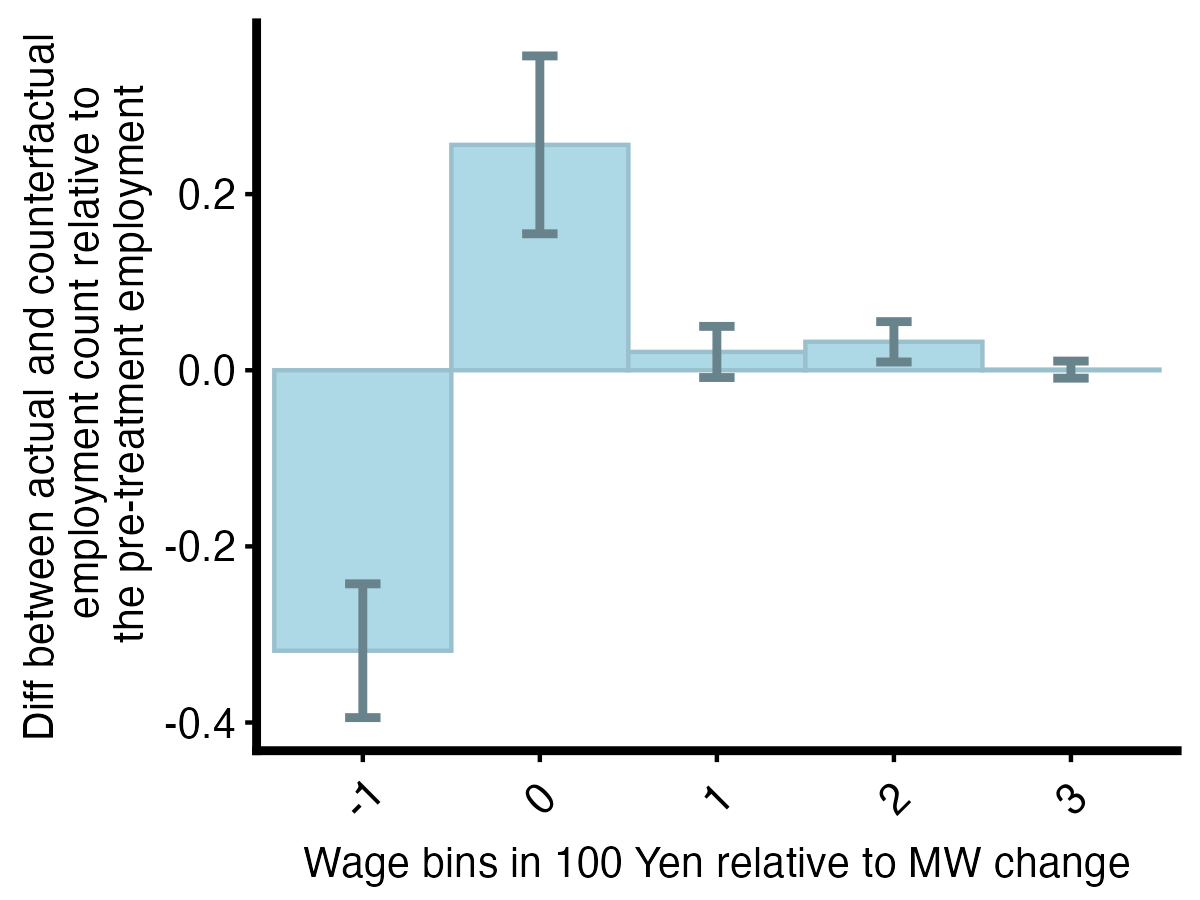}}\\
  \subfloat[Male-Targeted: Missing and Excess Jobs over Time]{\includegraphics[width = 0.46\textwidth]{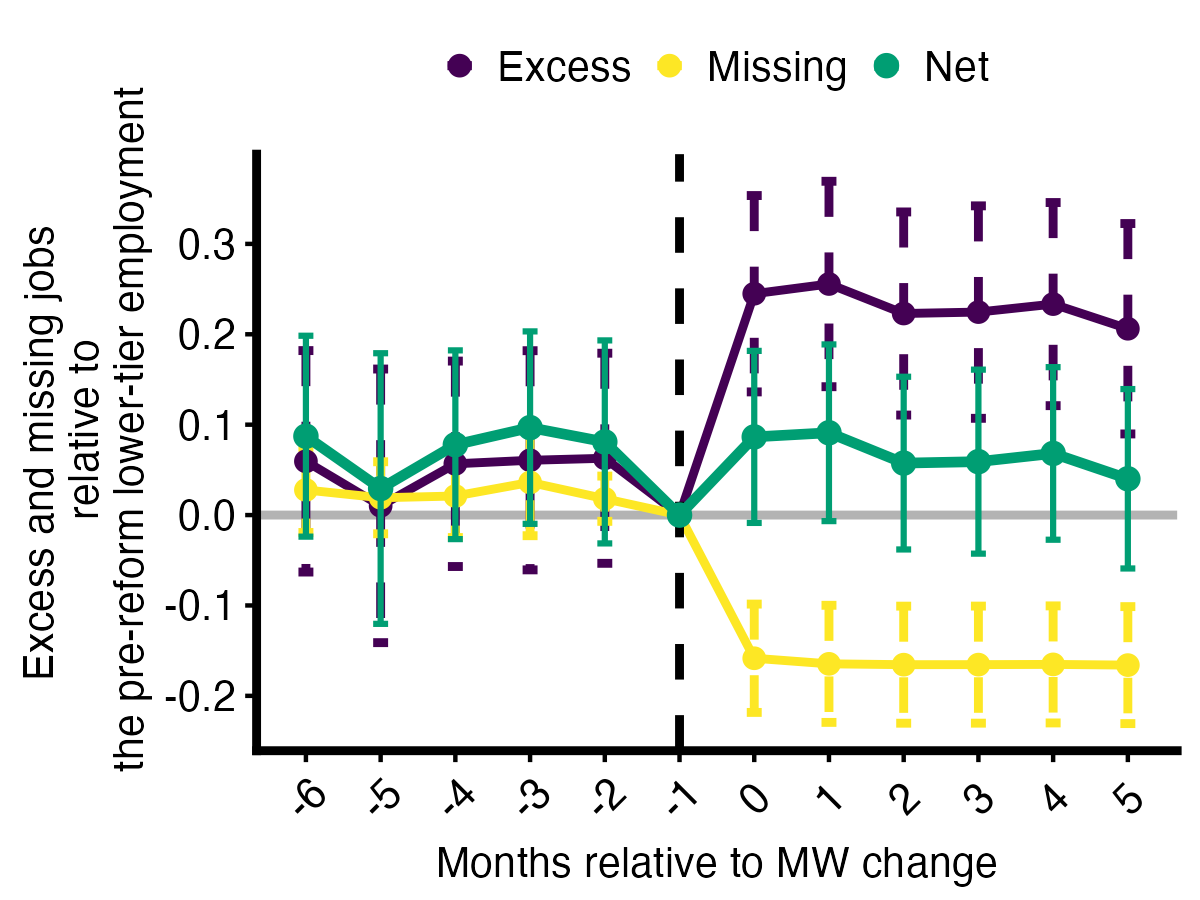}}
  \subfloat[Male-Targeted: Wage Distribution]{\includegraphics[width = 0.46\textwidth]{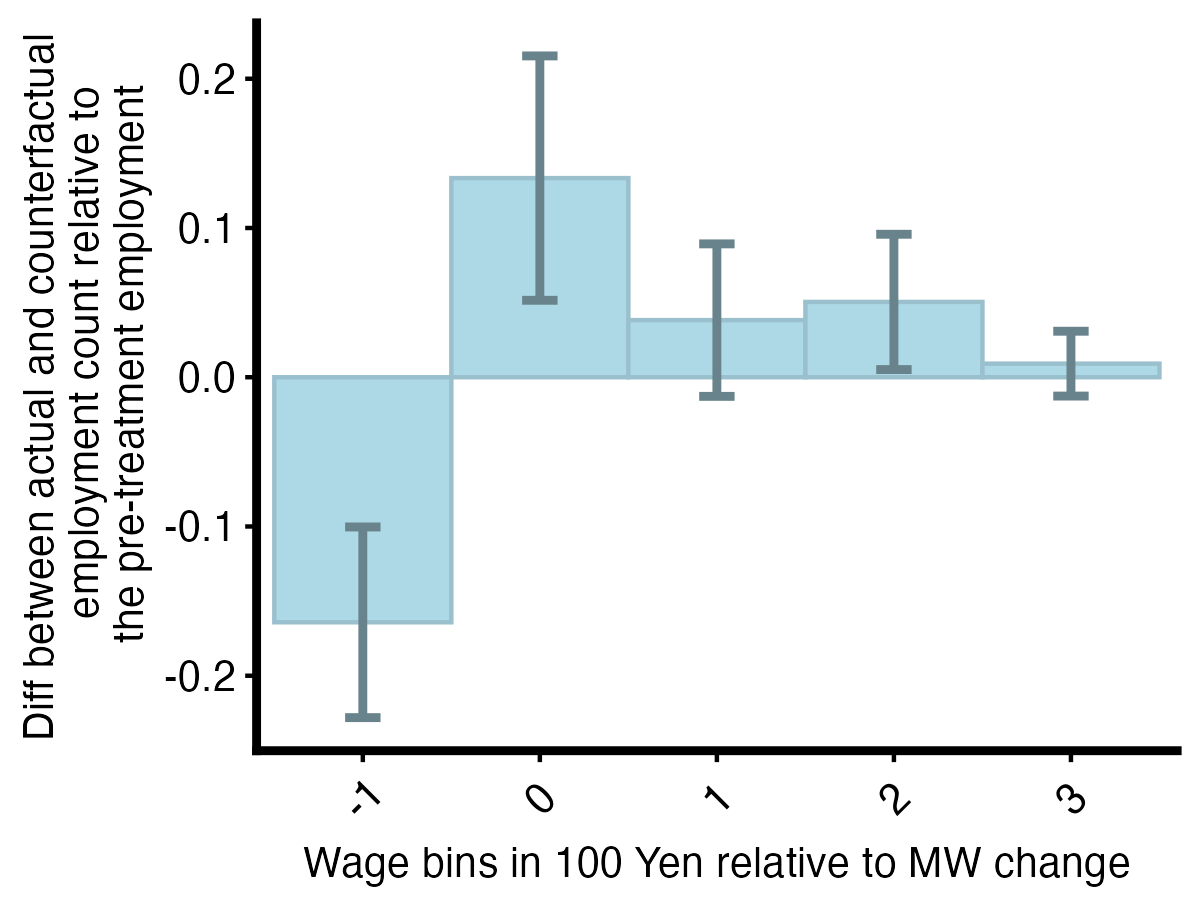}}
  \caption{Impact of Minimum Wages on Employment Separated by Gender}
  \label{fg:wage_distribution_moving_employment_documented_female_cengiz_original_form}
  \end{center}
  \footnotesize
  Note: Panel (a) plots the changes in female-targeted jobs before and after the minimum wage increase. Panel (b) plots the estimated effects of minimum wage increases on female-targeted jobs for each group $e \in \{-1,\dots,3\}$. Panels (c) and (d) provide the corresponding results for male-targeted jobs.
\end{figure}

\begin{figure}[!ht]
  \begin{center}
  \subfloat[Beginner-Targeted: Missing and Excess Jobs over Time]{\includegraphics[width = 0.46\textwidth]{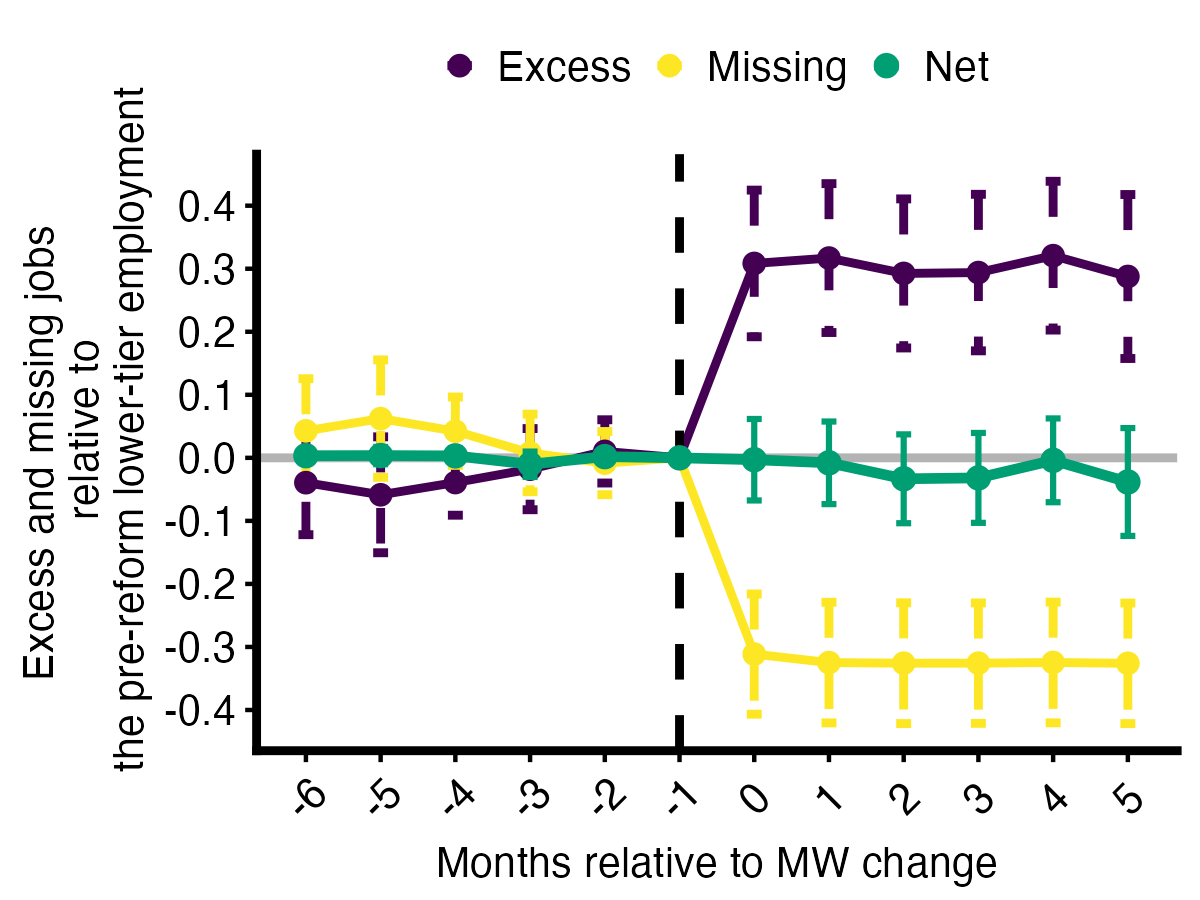}}
  \subfloat[Beginner-Targeted: Wage Distribution]{\includegraphics[width = 0.46\textwidth]{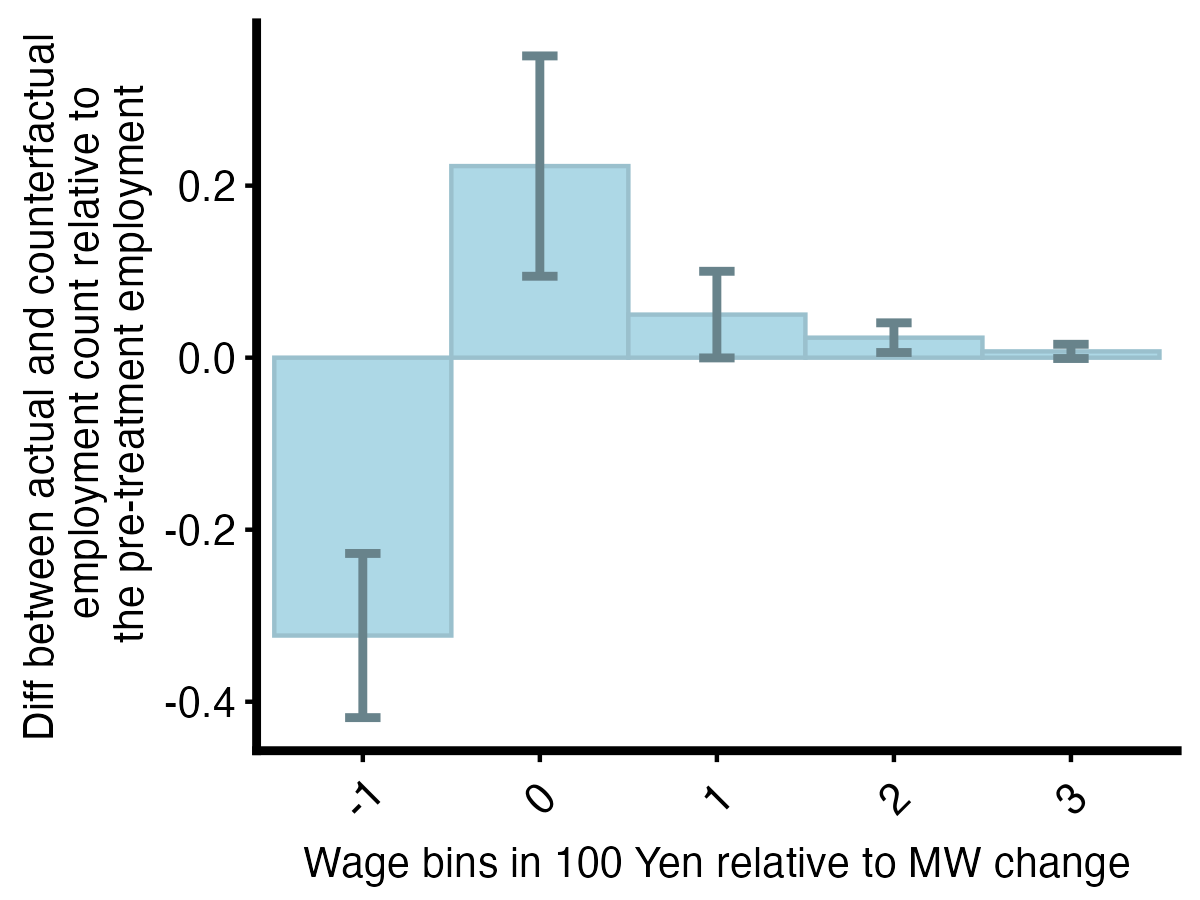}}\\
  \subfloat[Experienced-Worker-Targeted: Missing and Excess Jobs over Time]{\includegraphics[width = 0.46\textwidth]{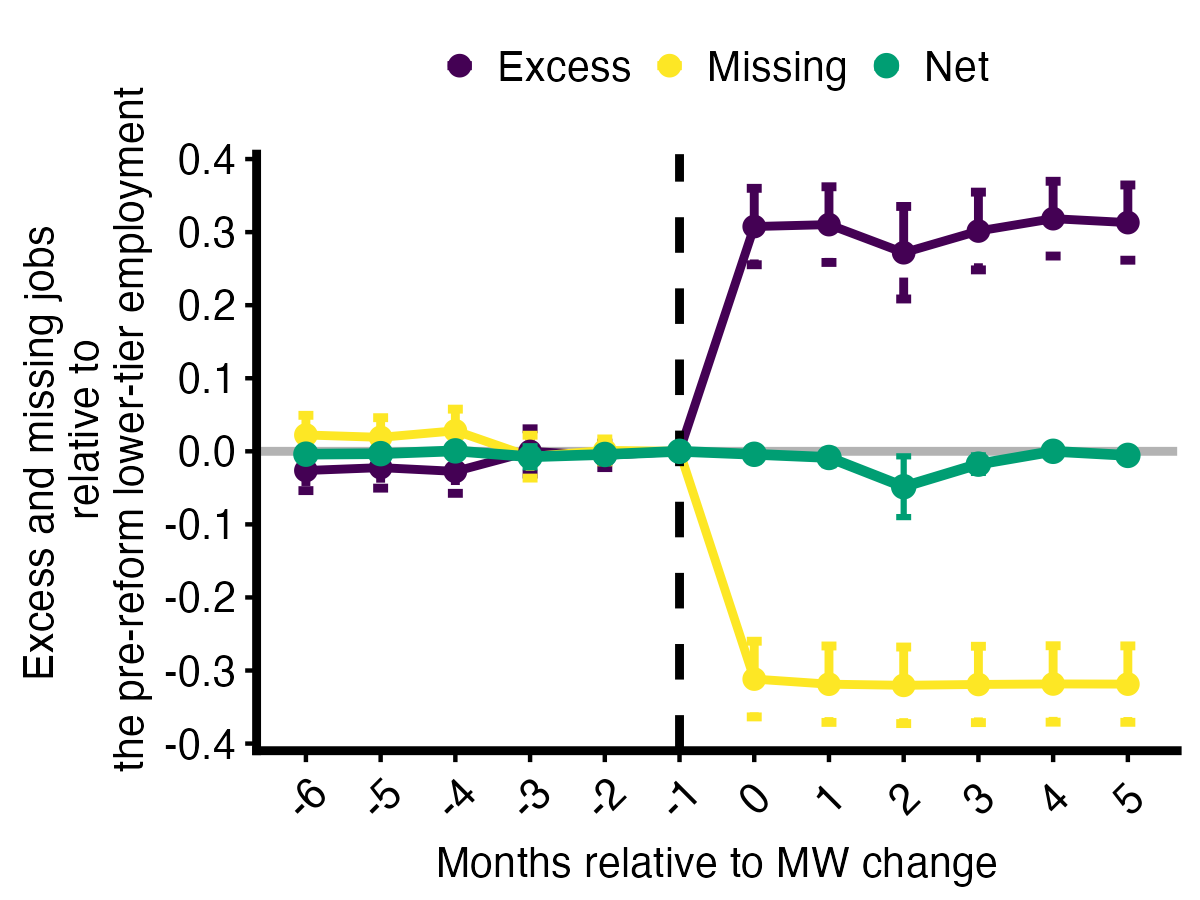}}
  \subfloat[Experienced-Worker-Targeted: Wage Distribution]{\includegraphics[width = 0.46\textwidth]{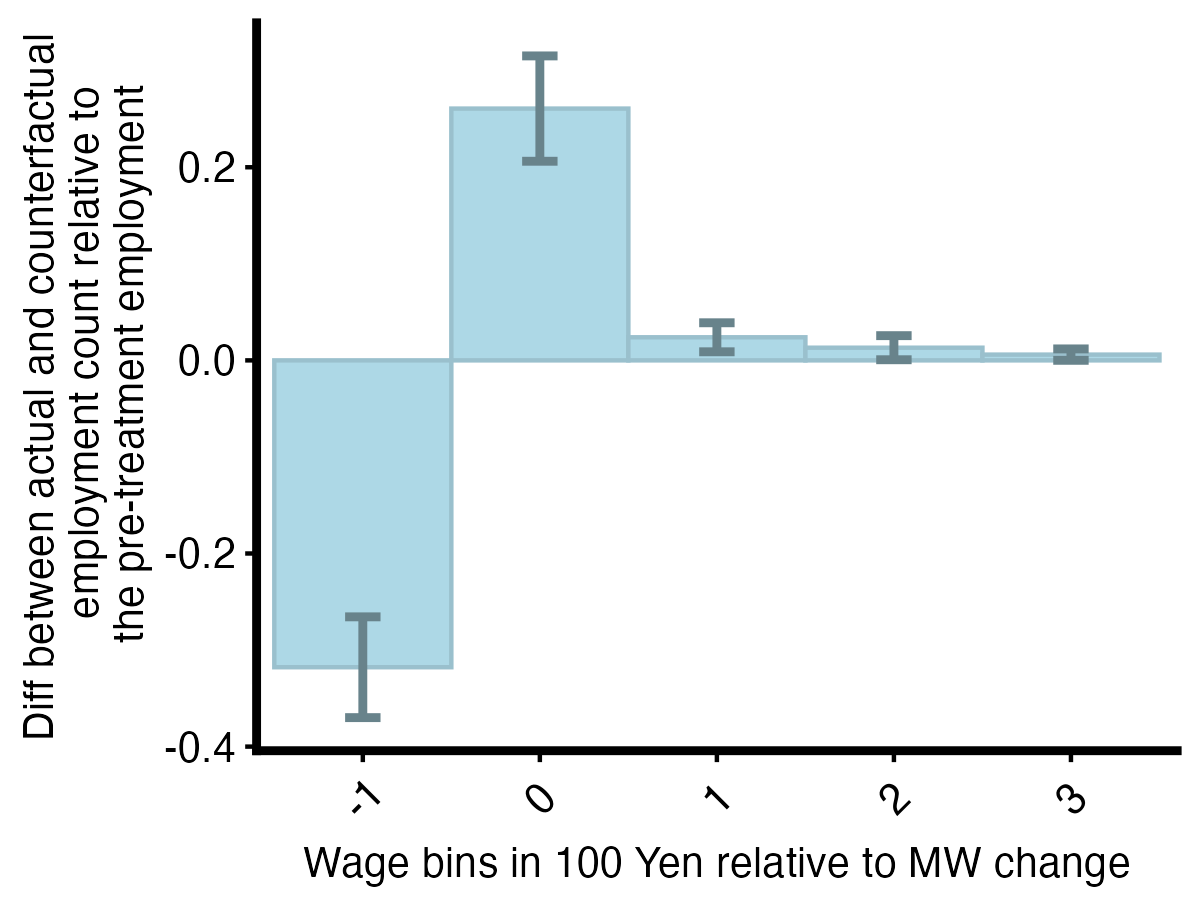}}
  \caption{Impact of Minimum Wages on Employment Separated by Skill Targeting}
  \label{fg:wage_distribution_moving_employment_documented_beginer_cengiz_original_form}
  \end{center}
  \footnotesize
  Note: Panel (a) plots the changes in beginner-targeted jobs before and after the minimum wage increase. Panel (b) plots the estimated effects of minimum wage increases on beginner-targeted jobs for each group $e \in \{-1,\dots,3\}$. Panels (c) and (d) provide the corresponding results for experienced-worker-targeted jobs.
\end{figure}

This appendix collects the event-study and wage-distribution figures that underlie Table~\ref{tb:missing_excess_net_by_category}. Figures~\ref{fg:wage_distribution_moving_vacancy_overall_cengiz_original_form} and \ref{fg:wage_distribution_moving_total_working_hours_overall_cengiz_original_form} report posted vacancies and total working hours, Figure~\ref{fg:wage_distribution_moving_employment_event_staff_cengiz_original_form} reports occupations, Figure~\ref{fg:wage_distribution_moving_employment_midnight_overall_cengiz_original_form} reports time-of-day categories, Figure~\ref{fg:wage_distribution_moving_employment_provision_cengiz_original_form} reports transportation reimbursement provision, and Figures~\ref{fg:wage_distribution_moving_employment_documented_female_cengiz_original_form} and \ref{fg:wage_distribution_moving_employment_documented_beginer_cengiz_original_form} report gender and skill targeting. Each figure is estimated with the specification in Equation \eqref{sec:empirical_strategy_DID_regression} on the indicated outcome or subsample. In the axis labels of these figures and of the figures in the main text, ``pre-treatment employment,'' ``pre-treatment vacancy count,'' ``pre-treatment total working hours,'' and ``pre-reform lower-tier employment'' all denote the September 2023 stock of the corresponding outcome in the treated wage-bin groups ($e=-1,\dots,3$), as defined in Section~\ref{sec:excess_job_missing_job_approach}.

\clearpage

\subsection{Other Measured Outcome} \label{app:other_measured_outcome}

\paragraph{Amenity as Outcome}
We estimate changes in transportation reimbursement among observed postings in excess wage bins. The estimation method follows that of Section \ref{sec:empirical_strategy}, with the outcome variable being transportation reimbursement linked to employment (set to zero if no transportation reimbursement is provided) and the probability of providing any transportation reimbursement. Since employment in group $e=-1$ disappears after the minimum wage revision, we focus on estimating and aggregating the effects on transportation reimbursement for groups with $e \geq 0$. We normalize the estimates by $K^{a}$, the September 2023 mean of the corresponding outcome in the excess group. We plot the average of $\mu_{l,e}/K^{a}$ across $e=0,\dots,3$, expressed as a percentage.

\begin{figure}[!ht]
  \begin{center}
  \subfloat[Amenity Amount over Time]{\includegraphics[width = 0.46\textwidth]{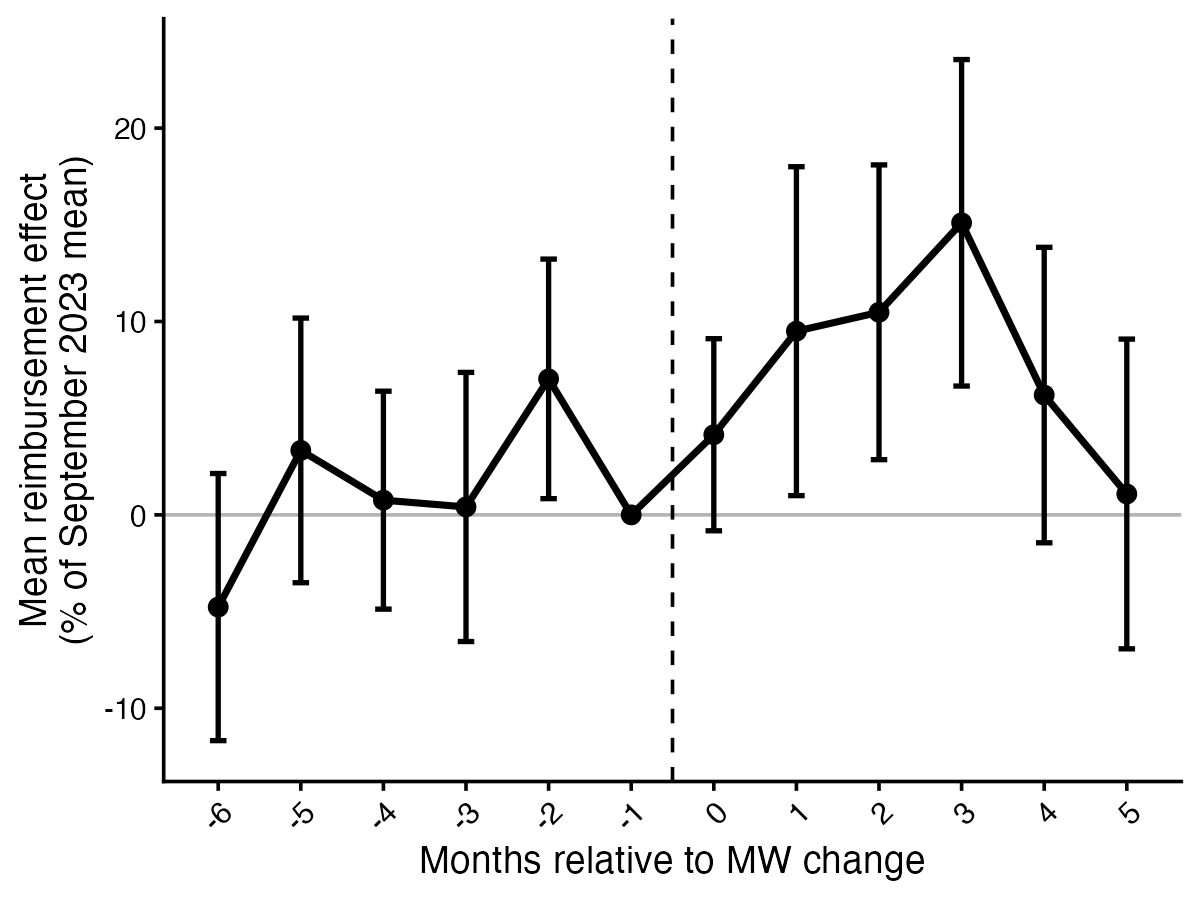}}
  \subfloat[Amenity Provision Probability over Time]{\includegraphics[width = 0.46\textwidth]{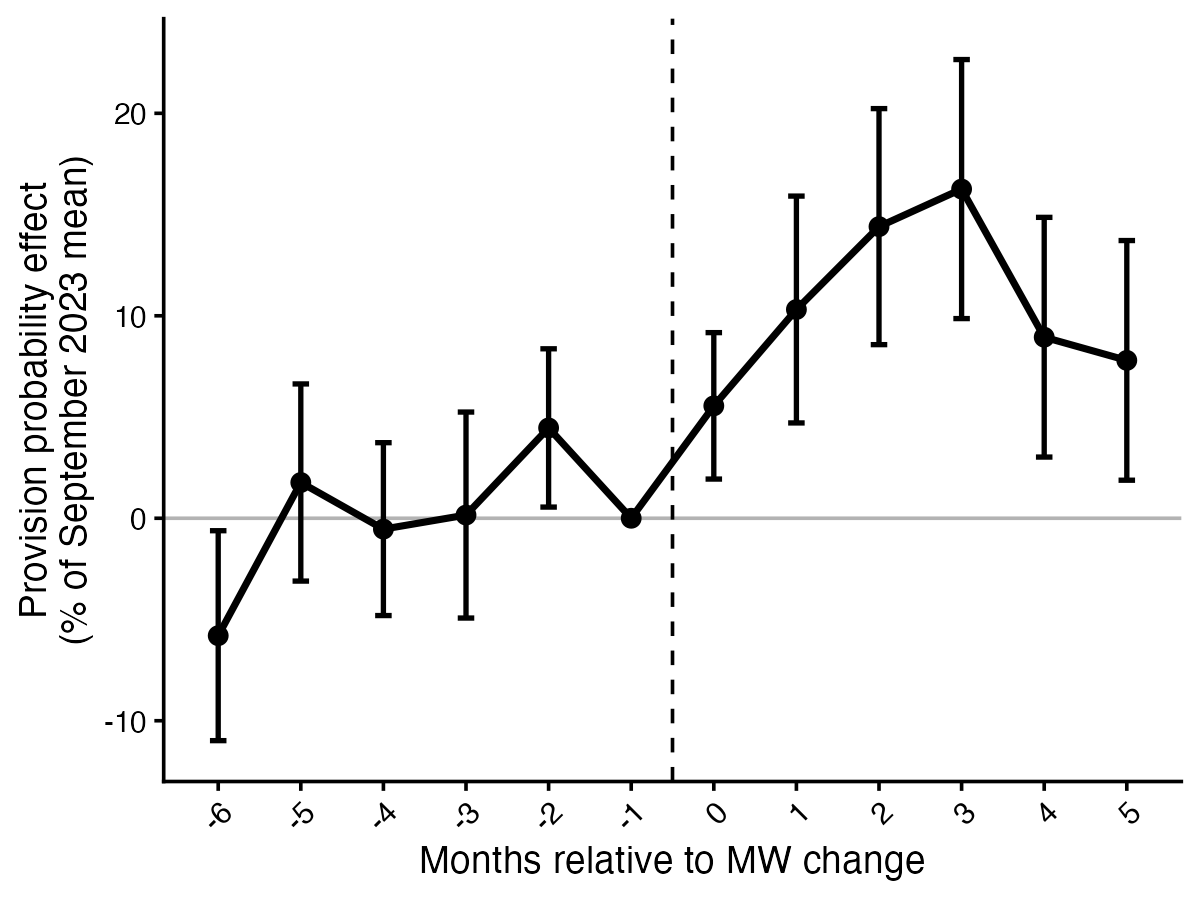}}
  \caption{Impact of Minimum Wages on Amount and Provision of Transportation Reimbursement as Amenities}
  \label{fg:wage_distribution_moving_transporation_expenses_reimbursement_overall_cengiz_original_form} 
  \end{center}
  \footnotesize
  Note: Panel (a) plots changes in transportation reimbursement across excess wage bins before and after the minimum wage increase. Panel (b) provides the corresponding probabilities of providing transportation reimbursement.
\end{figure}
Panel (a) in Figure \ref{fg:wage_distribution_moving_transporation_expenses_reimbursement_overall_cengiz_original_form} displays the estimated changes in transportation reimbursement, measured relative to the timing of the minimum wage increase. The estimates are close to zero in the pre-treatment period, apart from a positive estimate two months before the revision, and are generally positive following the policy change.
Some estimates are statistically significant at conventional levels---peaking at an increase of approximately 15.1 percent---and are economically meaningful in the post-treatment period.
These estimates describe changes in observed reimbursement amounts in excess wage bins.

Panel (b) in Figure \ref{fg:wage_distribution_moving_transporation_expenses_reimbursement_overall_cengiz_original_form} presents the dynamic effects of the minimum wage increase on the provision probability of transportation reimbursement. The estimates gradually increase in the post-treatment period, peaking around 16.3 percent, indicating a positive change in the observed provision measure in these wage bins. These repeated cross-sectional estimates do not distinguish changes within individual postings from changes in the mix of observed postings.

\subsection{Total Posted Pay}\label{sec:platform_revenue}

A key advantage of using platform data is the ability to observe not only posted hourly wages but also the total pay attached to each posting, hourly wage multiplied by working hours, on a daily basis. The negative net employment-share estimate can coexist with rising total pay.

\begin{figure}
    \begin{center}
        \includegraphics[width=0.6\linewidth]{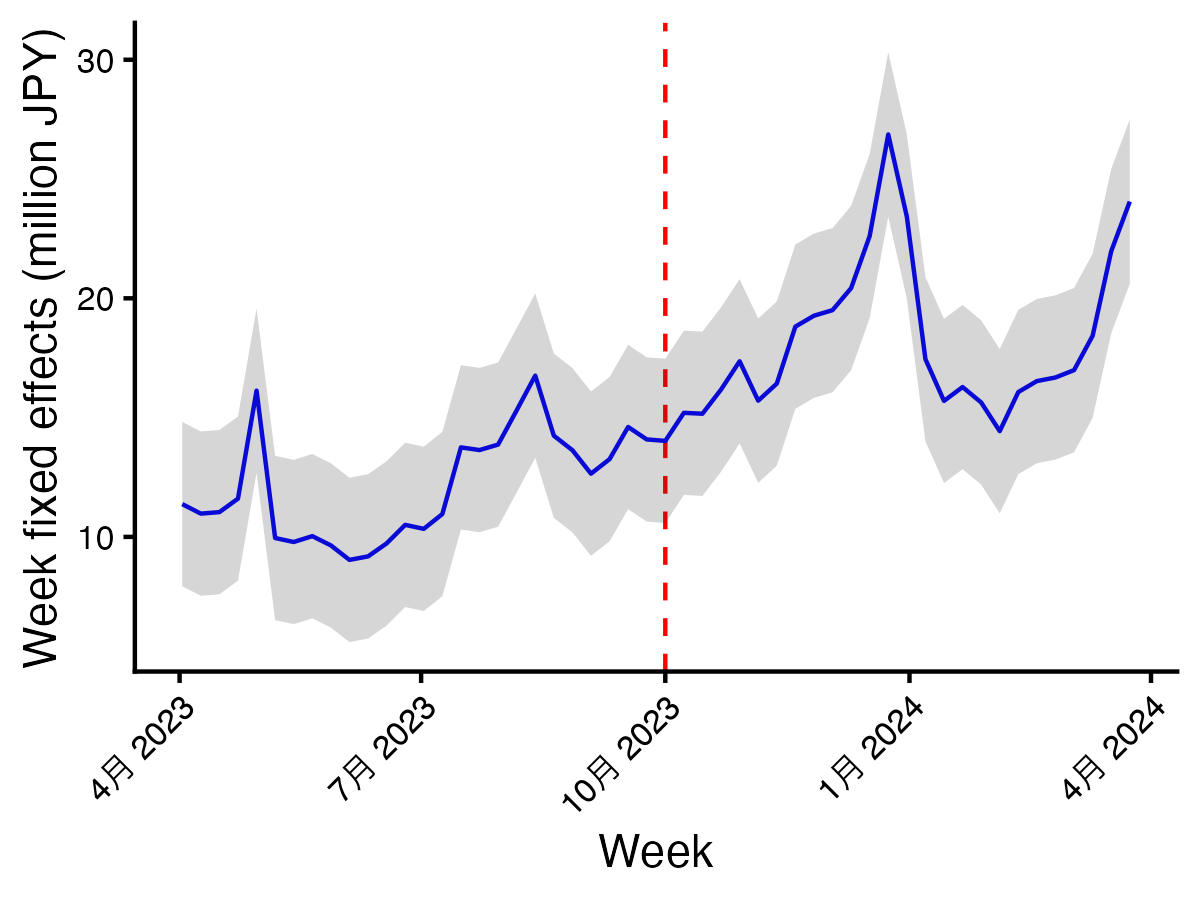}
        \caption{Estimated Week Fixed Effects in Total Posted Pay}
        \label{fig:year_month_week_prefecture_offering_salary_week_fixed_effect}
    \end{center}
    \footnotesize
    Note: The figure plots week fixed effects with confidence intervals from a regression of total posted pay at the prefecture-week level, in millions of JPY, on prefecture and week indicators. Total posted pay is calculated as hourly wages multiplied by working hours and summed over the postings starting in each prefecture-week.
\end{figure}

Figure \ref{fig:year_month_week_prefecture_offering_salary_week_fixed_effect} illustrates the temporal dynamics of weekly fixed effects from a regression of total posted pay at the prefecture-week level on prefecture and week dummy variables. Weekly total pay continues the upward trend of the preceding months through December and remains at or above the September level thereafter, so we observe no decline after the October revision.






\clearpage

\subsection{Addressing Potential Issues}\label{app:additional_potential_issues}

This appendix collects the robustness checks referenced in Section \ref{sec:empirical_strategy}. Each subsection addresses one concern about the baseline design: Appendix \ref{app:robust_control_group} examines the validity of the control group, Appendix \ref{app:robust_wage_bin_trends} heterogeneous trends across wage bins, and Appendix \ref{app:robust_composition} the composition of treated and control bins.

\subsubsection{Validity of the Control Group}\label{app:robust_control_group}

The limited spillover effect assumption requires that minimum-wage effects vanish above the threshold that defines the control group. This subsection varies that threshold.

\paragraph{The Definition of the Control Group}\label{app:control_definition}

\begin{figure}[!ht]
  \begin{center}
  \subfloat[Missing and Excess Jobs over Time]{\includegraphics[width = 0.46\textwidth]{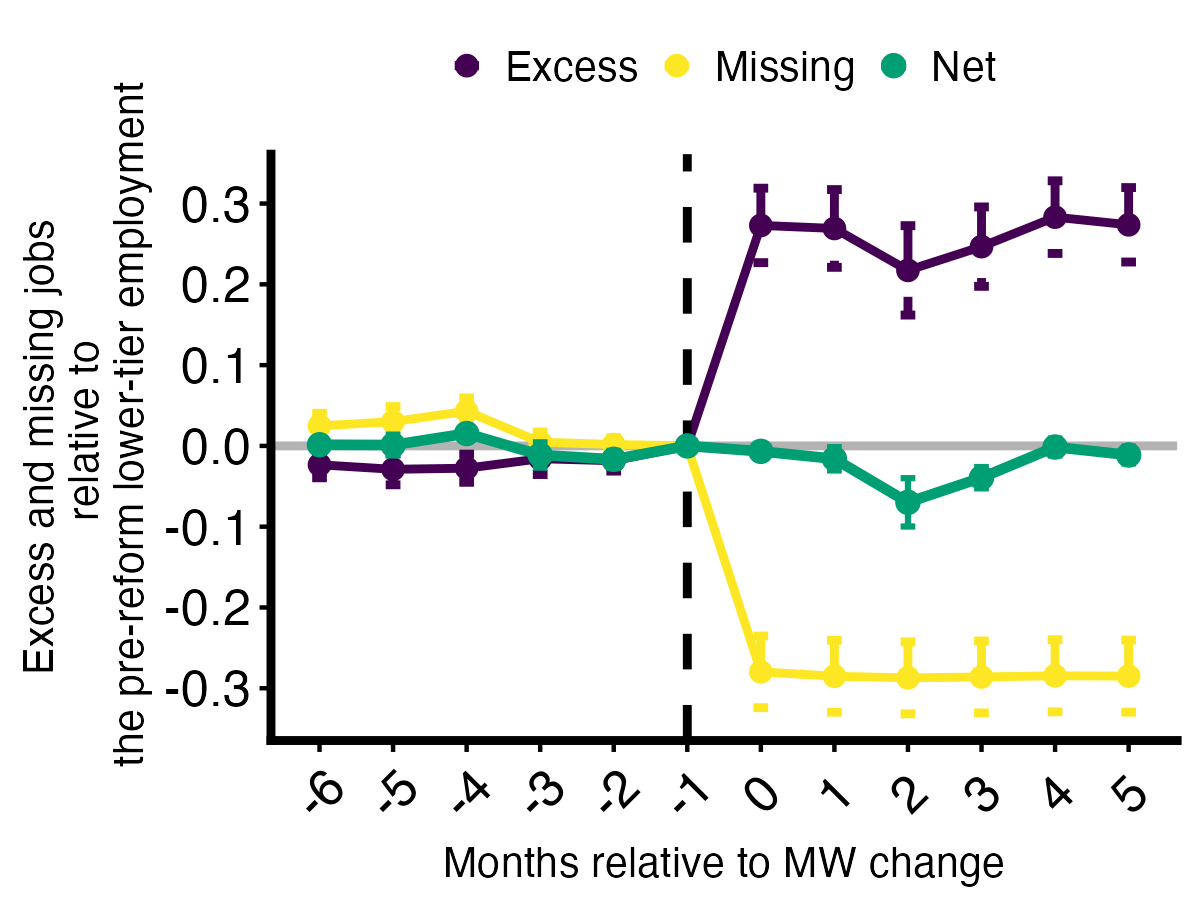}}
  \subfloat[Wage Distribution]{\includegraphics[width = 0.46\textwidth]{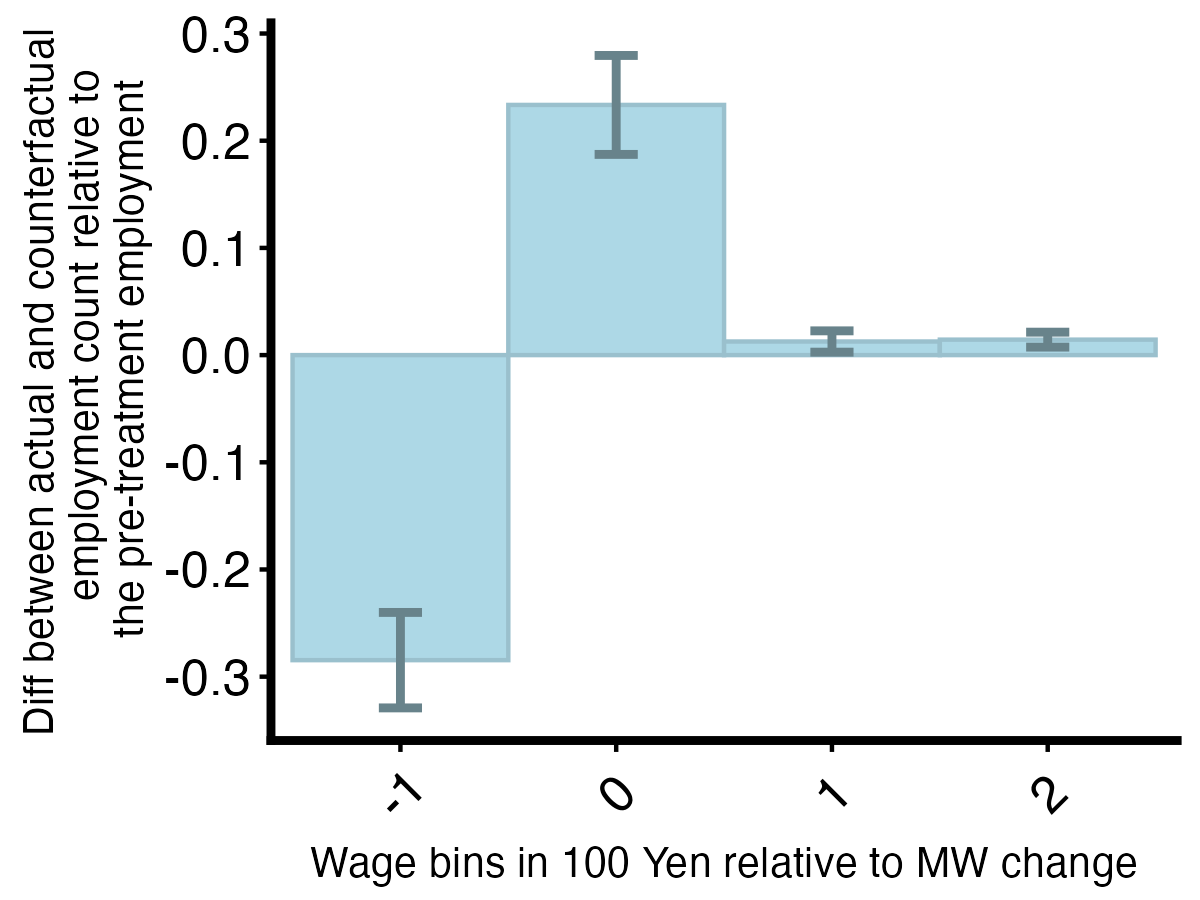}} \\
  \subfloat[Missing and Excess Jobs over Time]{\includegraphics[width = 0.46\textwidth]{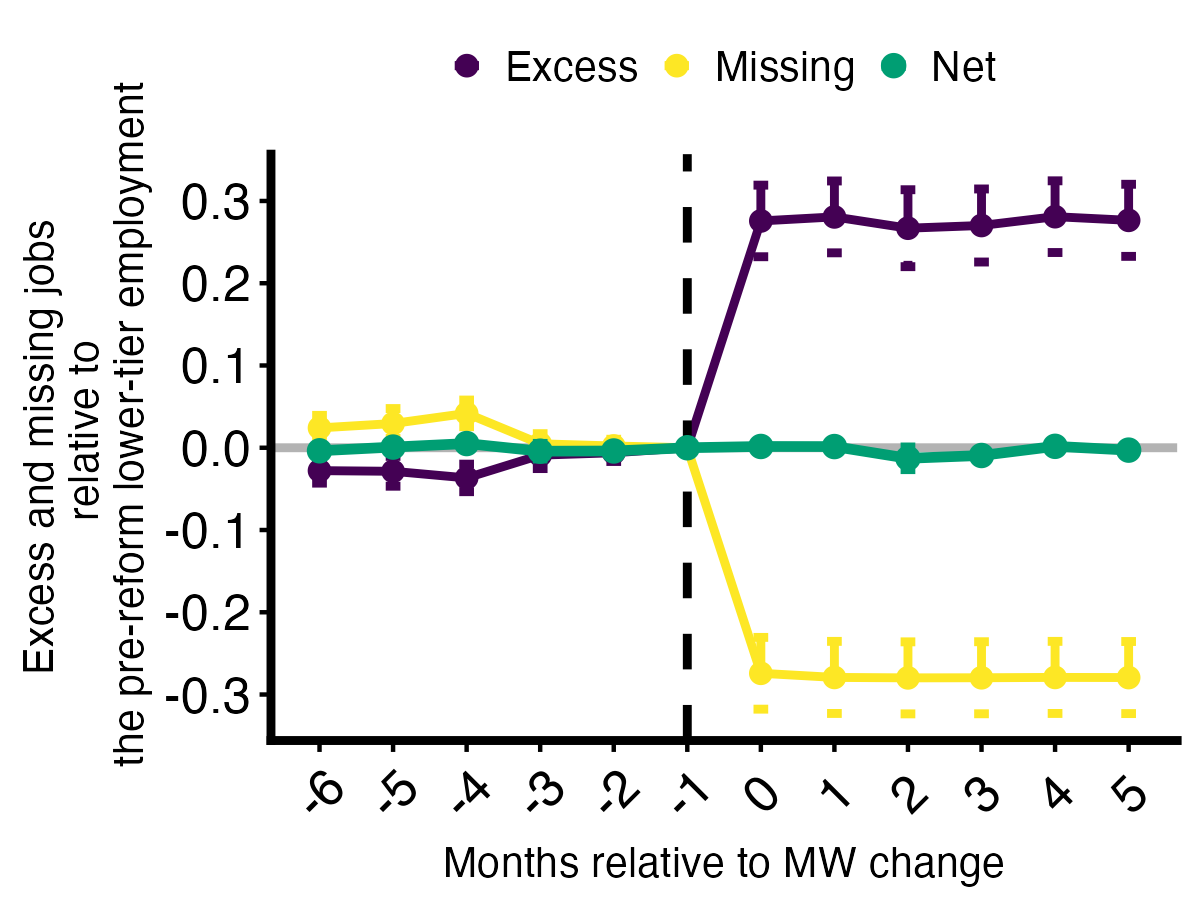}}
  \subfloat[Wage Distribution]{\includegraphics[width = 0.46\textwidth]{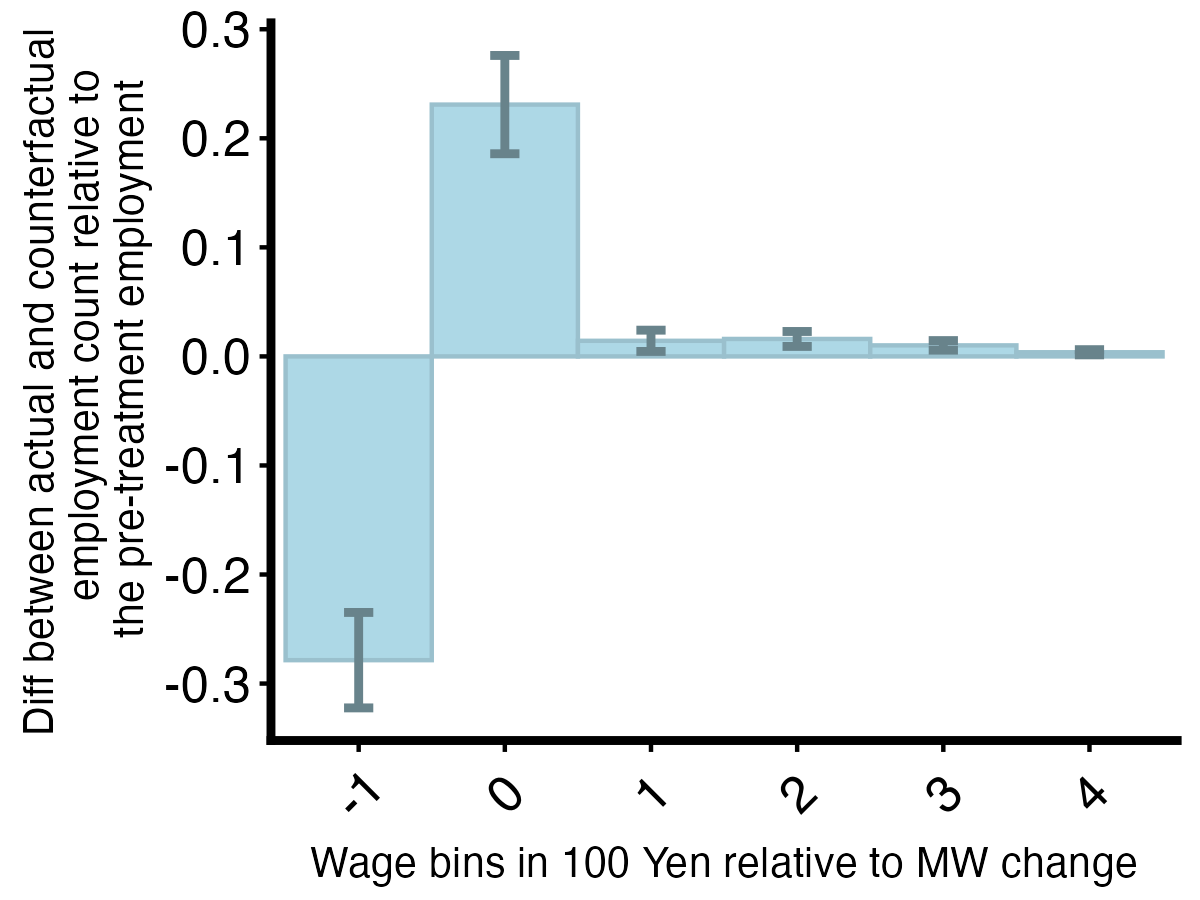}}
  \caption{Impact of Minimum Wages on Employment: Alternative Definition of the Control Group}
  \label{fg:wage_distribution_moving_employment_overall_cengiz_original_form_control_change} 
  \end{center}
  \footnotesize
  Note: 
  This figure presents robustness checks using alternative definitions of the control wage bin in the event-study DID design. 
  Panels (a) and (b) report the results when the control group is defined as wage bins with wages $\geq MW_p + 300$. Panels (c) and (d) report the results when the control group is defined as wage bins with wages $ \geq MW_p + 500$. Panels (a) and (c) plot the monthly estimates of missing and excess jobs, while Panels (b) and (d) plot the average post-treatment effect for each wage group. 

\end{figure} 

Our baseline event-study DID design defines treatment status based on each wage bin's distance to the new prefecture-level minimum wage and uses the upper part of the wage distribution as the control group. 
Concretely, we set a threshold $\overline{W}_p = MW_p + 400$ and assign all wage bins at or above $\overline{W}_p$ to the control group (group $\infty$). 
This choice is motivated by the ``limited spillover'' assumption: minimum-wage effects should vanish sufficiently far above the new wage floor so that higher-wage bins provide a valid counterfactual trend for bins near the threshold.
Under this assumption, the identifying variation comes from comparing changes in outcomes for bins near the minimum wage to changes for bins in group $\infty$ around the timing of the reform.

A natural concern is that the estimated effects could be sensitive to where the control region begins. 
If spillovers extend further up the wage distribution than assumed, defining the control group too close to the minimum wage (e.g., $MW_p + 300$) could contaminate the control bins and mechanically attenuate the estimated treatment effects. 
Conversely, defining the control group farther away (e.g., $MW_p + 500$) reduces the risk of contamination but may change precision by shifting the composition of the control region. 
To assess this sensitivity, we re-estimate our baseline specification under two alternative thresholds for $\overline{W}_p$: $MW_p + 300$ and $MW_p + 500$. 

Figure \ref{fg:wage_distribution_moving_employment_overall_cengiz_original_form_control_change} reports the corresponding event-study paths for missing and excess jobs and the associated wage-distribution patterns under alternative control-threshold definitions.
Across both alternative control definitions, the main qualitative patterns remain intact. 
First, pre-treatment coefficients stay close to zero, supporting the plausibility of the parallel-trends assumption under each control threshold. 
Second, the post-reform dynamics continue to show a sharp contraction of employment below the new minimum wage (missing jobs) of about 28 percent of pre-reform lower-tier employment and an expansion just above it (excess jobs) of 22 to 28 percent, so the net effect is small in every month: under the $MW_p + 500$ threshold it shows no significant change, while under the $MW_p + 300$ threshold it declines significantly in December and January, by 7 and 4 percent, as in the baseline.
Finally, the distributional profile of the effects remains concentrated near the wage floor, consistent with limited spillovers to higher wage bins—the key condition needed for group $\infty$ to serve as a valid control. 
Overall, the immediate job losses below the new minimum wage and the offsetting gains just above it are not driven by the baseline choice of $\overline{W}_p = MW_p + 400$.

\subsubsection{Heterogeneous Trends across Wage Bins}\label{app:robust_wage_bin_trends}

The parallel trends assumption requires that treated and control wage bins would have evolved similarly in the absence of the revision. This subsection allows for wage-bin-specific trends, runs a placebo exercise in a year without a meaningful revision, and nets out that year's pattern in a triple-difference design.

\paragraph{Main Specification with Wage-bin Specific Trends}\label{app:wage_bin_specific_trend}

A potential concern with our baseline event-study DID design is that wage bins far above the minimum wage (which form the control group) may follow different underlying time patterns than wage bins close to the minimum wage. In particular, if high-wage and low-wage bins exhibit systematically different trends even absent the policy change, the standard parallel-trends assumption may be violated, and our baseline estimates could partly reflect differential underlying dynamics rather than the minimum-wage revision.

To address this concern, we estimate an augmented specification that flexibly absorbs wage-bin-specific heterogeneity. 
Specifically, we re-estimate our main event-study regression while including wage-bin fixed effects, \(J_j\), and wage-bin-by-time fixed effects, \(\Omega_{j,t}\), in addition to the baseline set of controls:

\begin{align}
\label{eq:augmented_DID_regression}
\frac{Y_{j(p),t}}{N_{p,t}}
=
\sum_{l \neq -1}\sum_{e \neq \infty}
\mu_{l,e}\mathbf{1}\{t-7=l\}\mathbf{1}\{E_{j(p)}=e\}
+\alpha_{e}+\lambda_{t}+J_j+\Omega_{j,t}+\epsilon_{j,t}.
\end{align}

The inclusion of \(J_j\) controls for time-invariant differences across wage bins, while \(\Omega_{j,t}\) allows each wage bin to have its own time trends, thereby relaxing the requirement that high-wage and low-wage bins must share common trends. 

\textcolor{black}{
Because the mapping from a nominal wage bin $j$ to an exposure group $E_{j(p)}$ depends on prefecture-specific minimum wages, the same bin $j$ can belong to different exposure groups across prefectures. 
As a result, even after absorbing wage-bin-by-time shocks via $\Omega_{j,t}$, there remains within-$(j,t)$ cross-prefecture variation in $\mathbf{1}\{E_{j(p)}=e\}$. 
Hence the event-study interaction $\mathbf{1}\{t-7=l\}\mathbf{1}\{E_{j(p)}=e\}$ is not perfectly collinear with $\Omega_{j,t}$, and the coefficients $\mu_{l,e}$ are identified from this residual variation.
}

This specification compares changes around the minimum-wage revision within wage bins after netting out wage-bin-specific time variation, and thus provides a stringent robustness check against confounding differential trends across the wage distribution.

\begin{figure}[!ht]
  \begin{center}
  \subfloat[Missing and Excess Jobs over Time]{\includegraphics[width = 0.46\textwidth]{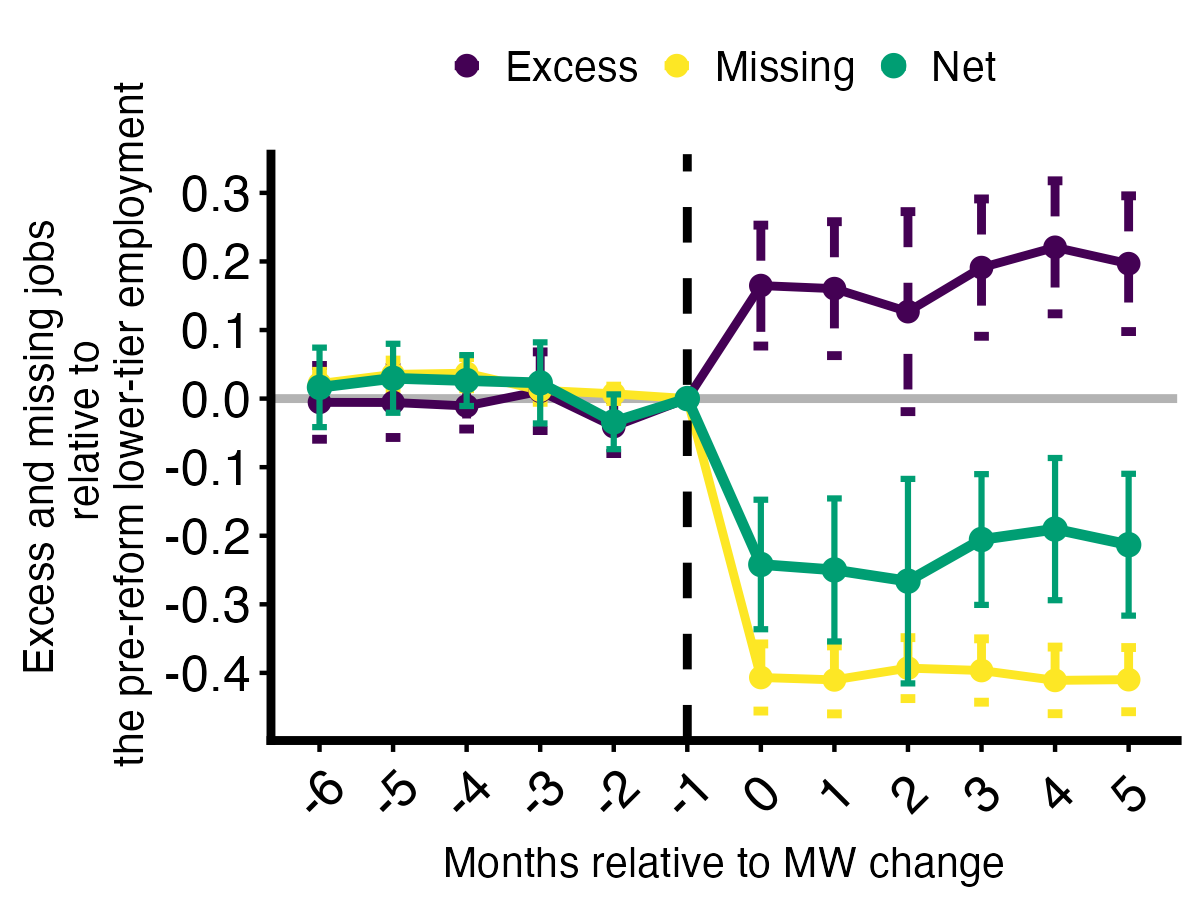}}
  \subfloat[Wage Distribution]{\includegraphics[width = 0.46\textwidth]{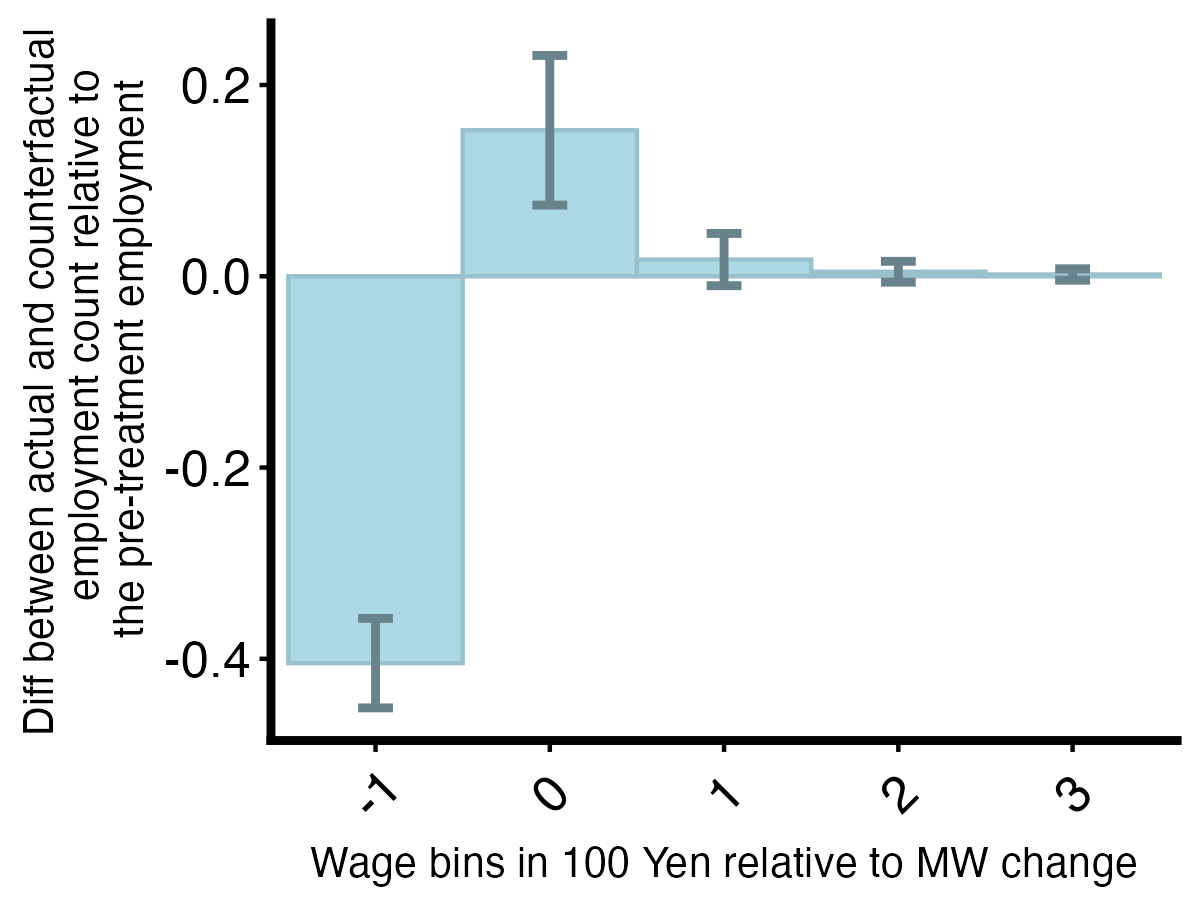}}
  \caption{Impact of Minimum Wages on Employment: Controlling Wage-bin Specific Trends}
  \label{fg:wage_distribution_moving_employment_overall_cengiz_original_form_with_wage_bin_trend} 
  \end{center}
  \footnotesize
  Note: Panel (a) shows the estimated excess jobs and missing jobs for each month before and after the minimum wage increase. Panel (b) shows the estimated impact of the minimum wage increase on employment for each group $e$ in the wage distribution. 
\end{figure} 

Figure \ref{fg:wage_distribution_moving_employment_overall_cengiz_original_form_with_wage_bin_trend} reports the resulting estimates. 
The estimated effects remain concentrated at the new minimum wage, with the largest adjustments in the wage bins just below and at the wage floor and small changes farther above it. Missing jobs are about 40 percent of pre-reform lower-tier employment, larger than the baseline 28 percent, and excess jobs are 13 to 22 percent, smaller than the baseline 27 percent, so the net effect is a decline of 19 to 27 percent, significantly negative in every post-treatment month, compared with the baseline net decline of 1 percent. The pre-revision estimates show no significant change. Allowing for wage-bin fixed effects \(J_j\) and wage-bin-by-time fixed effects \(\Omega_{j,t}\) therefore preserves the immediate redistribution of employment away from the lowest wage bins and enlarges the net decline.

\paragraph{Placebo Effect of Minimum Wage Hikes utilizing 2020's variation}\label{app:placebo_effect_2020}

We next conduct a placebo exercise to examine whether the employment changes observed around the 2023 minimum-wage increase could instead reflect seasonal movements in employment. Our baseline identification strategy assumes that employment proportions in low- and high-wage bins would have followed parallel trends in the absence of the increase. Although month fixed effects absorb common seasonal changes, they do not account for seasonal movements that differ across wage bins. Such differences around October could therefore be mistaken for effects of the minimum-wage increase. The year 2020 provides a useful setting for investigating this concern because prefecture-level minimum wages increased by only 0--3 yen, leaving little scope for substantial employment adjustments in response to the revision. By applying the same event-study design around October 2020, we examine whether a similar pattern of missing jobs below the minimum wage and excess jobs above it emerges even when the minimum wage remains nearly unchanged. This exercise provides a check on whether the baseline pattern reflects seasonal changes across wage bins rather than the substantial minimum-wage increase in 2023.

Using data from April 2020 through March 2021, we estimate the baseline specification in Equation \eqref{sec:empirical_strategy_DID_regression}, defining wage groups relative to each prefecture's October 2020 minimum wage. We set October 2020 as relative month zero and September 2020 as the reference month. We calculate excess jobs ($\Delta a_l$) and missing jobs ($\Delta b_l$) following the aggregation and normalization procedure in Section \ref{sec:excess_job_missing_job_approach}, applied to the 2020 sample.

\begin{figure}[H]
  \begin{center}
  \subfloat[2023 Minimum-Wage Revision]{\includegraphics[width = 0.46\textwidth]{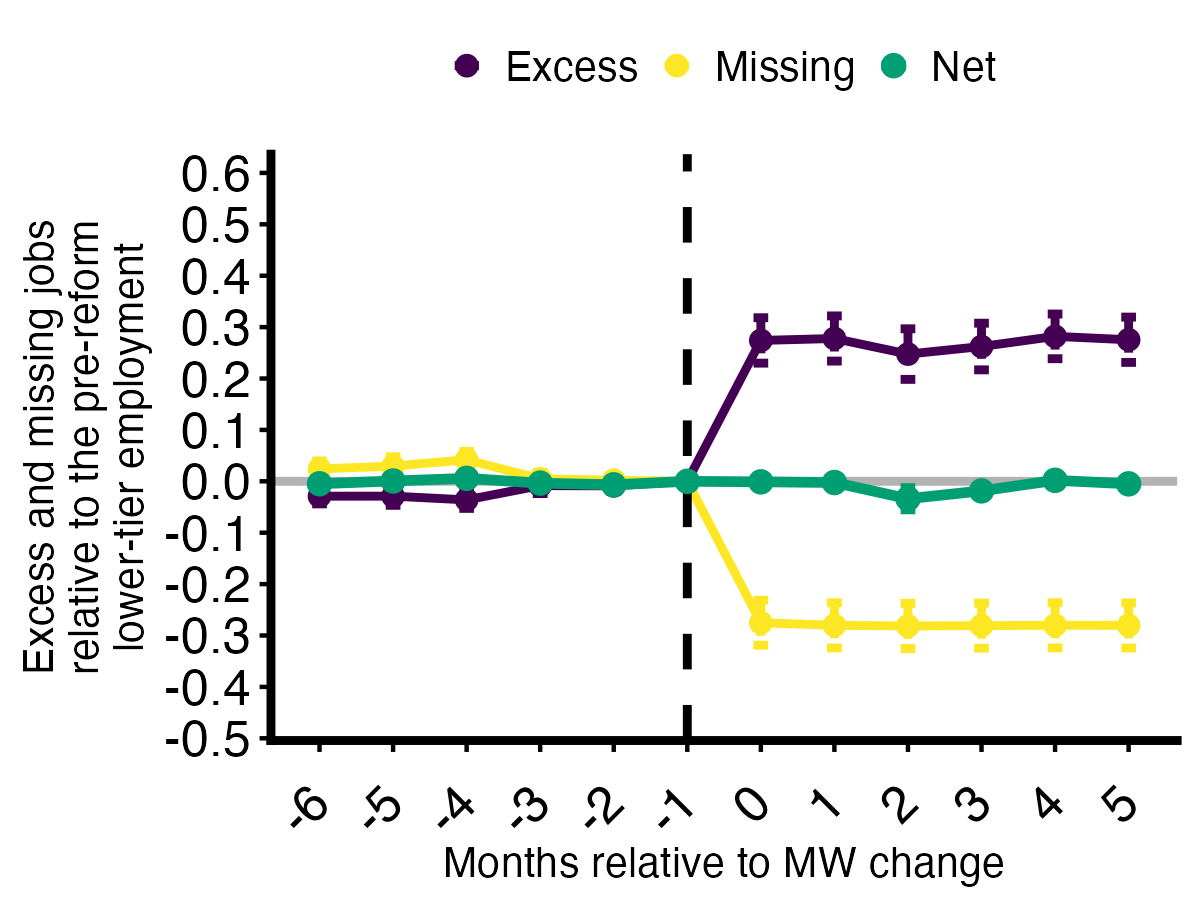}}
  \subfloat[2020 Placebo Exercise]{\includegraphics[width = 0.46\textwidth]{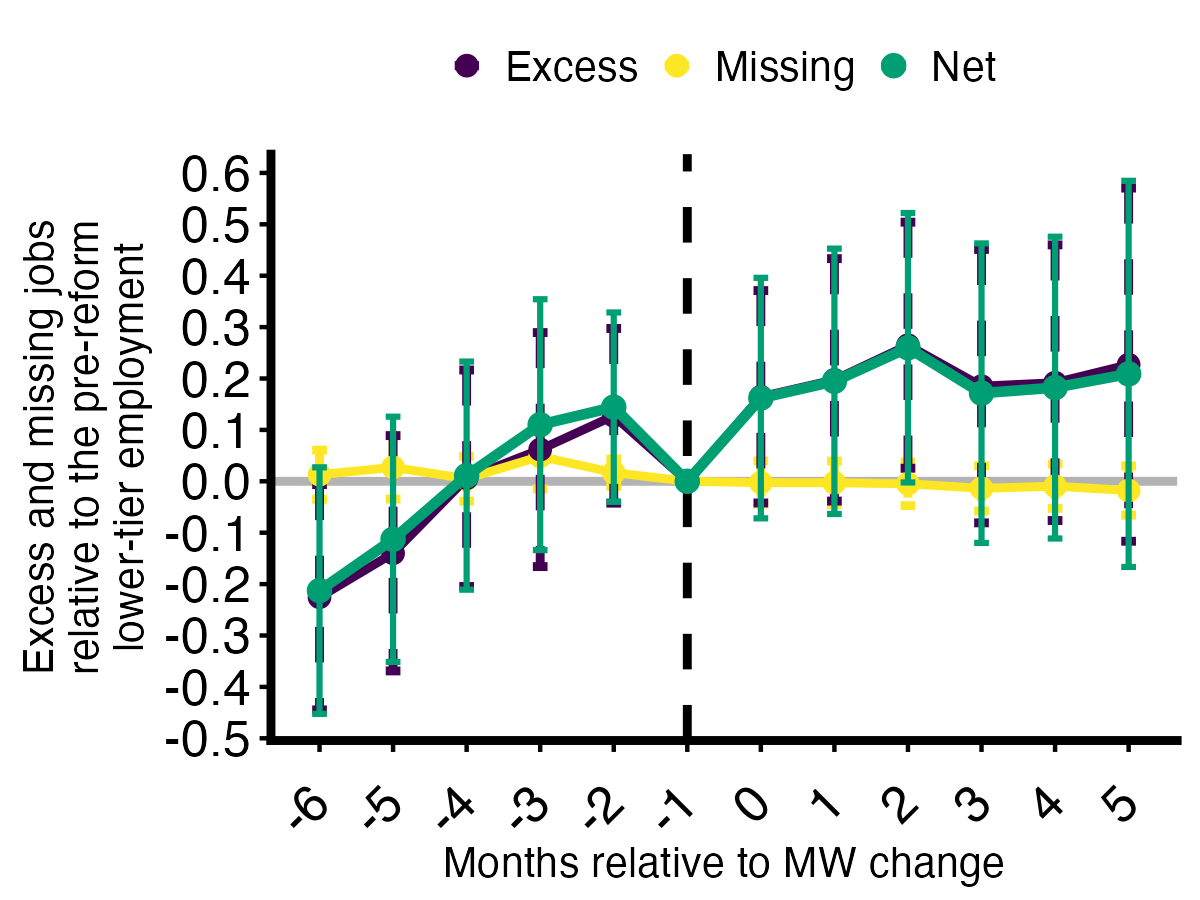}}
  \caption{Employment Effects of the 2023 Minimum-Wage Revision and the 2020 Placebo Exercise}
  \label{fg:employment_effects_2023_and_placebo_2020}
  \end{center}
  \footnotesize
  Note: Both panels use identical axis labels, scales, line widths, and confidence-interval formatting. Panel (a) shows the estimated excess and missing jobs around the October 2023 minimum-wage revision. Panel (b) repeats the exercise around October 2020 using wage groups defined relative to the 2020 minimum wage. The 2020 minimum-wage changes were substantially smaller than those in 2023.
\end{figure}

Panel (a) of Figure \ref{fg:employment_effects_2023_and_placebo_2020} presents the monthly estimates of excess and missing jobs around the October 2023 minimum-wage increase, while Panel (b) reports the corresponding estimates from the 2020 placebo exercise. In both panels, the horizontal axis indicates months relative to October of the respective year, and the vertical axis shows excess jobs ($\Delta a_l$) and missing jobs ($\Delta b_l$), normalized by pre-treatment employment as in the main analysis.

The two panels reveal a clear difference between the years. In Panel (a), the missing-job estimates fall sharply and the excess-job estimates rise in October 2023, with both changes persisting throughout the post-treatment period. In Panel (b), the missing-job estimates remain close to zero throughout the event window and display no comparable October break. The excess-job estimates drift upward over the whole window, from about 20 percent of pre-reform lower-tier employment below the reference month six months before October to about 20 percent above it after October, with wide confidence intervals. The placebo exercise thus does not reproduce the abrupt, opposite-signed changes in missing and excess jobs observed in 2023, so the missing-job response in 2023 does not reflect seasonal movements across wage bins around October. The upward drift in excess jobs in 2020 shows that the excess margin moves for reasons unrelated to a revision, which the triple-difference design below nets out.

\paragraph{Difference-in-Difference-in-Differences Approach}\label{app:difference_in_difference_in_differences}

We introduce a difference-in-difference-in-differences (DDD) approach as a sensitivity check for our identification strategy.
Our baseline difference-in-differences (DID) specification relies on the assumption that, absent the policy change, the treated and control wage bins would have followed parallel trends.
However, in practice, it is plausible that employment proportions in wage bins close to the minimum wage threshold and those in higher-wage bins exhibit different underlying trends even before the policy change.
Such wage-bin-specific trends may arise from differential demand conditions, compositional changes, or other factors that are correlated with wage levels.

To mitigate concerns about differential wage-bin trends, we propose a DDD specification that uses three-year differences in outcomes---i.e., the difference between outcomes in 2023 and those in the corresponding month in 2020.
We use 2020 as the reference year because prefecture-level minimum wages increased by only 0--3 yen that year and, as the placebo exercise above shows, missing jobs display no October break while excess jobs drift upward over the year, so 2020 provides a benchmark for wage-bin-specific movements that are unrelated to a revision.
The key idea is to net out wage-bin-specific seasonal and persistent components by comparing each month's outcome to the same calendar month three years earlier.
The nominal 10-yen wage bin $j$ is held fixed across the two years.
We then estimate an event-study regression identical to our main specification, except that the outcome is the three-year change in the employment rate in each wage bin.
The DDD regression is given by:

\begin{align}
\label{sec:empirical_strategy_DID_regression_difference}
\frac{Y_{j(p),t}}{N_{p,t}} -\frac{Y_{j(p),t-36}}{N_{p,t-36}} = \sum_{l \neq -1}\sum_{e \neq \infty}\widetilde{\mu}_{l,e}\mathbf{1}\{t-7=l\}\mathbf{1}\{E_{j(p)}=e\}+\alpha_{e}+\lambda_{t}+\epsilon_{j,t},
\end{align}

This specification is identical to our main event-study DID regression, except that the outcome is defined as a three-year difference.
Specifically, the outcome is the change in the employment rate in wage bin $j$ of prefecture $p$ between month $t$ and the corresponding month three years earlier ($t-36$).

The parameters of interest are the set of $\widetilde{\mu}_{l,e}$:
\begin{align*}
\widetilde{\mu}_{l,e}
&= E\biggr[\biggr(\frac{Y_{j(p),7+l}}{N_{p,7+l}} - \frac{Y_{j(p),7+l-36}}{N_{p,7+l-36}}\biggr)-\biggr(\frac{Y_{j(p),6}}{N_{p,6}} - \frac{Y_{j(p),6-36}}{N_{p,6-36}} \biggr)\biggr|E_{j(p)}=e\biggr] \\
& \;\;\;\; - E\biggr[\biggr(\frac{Y_{j(p),7+l}}{N_{p,7+l}} - \frac{Y_{j(p),7+l-36}}{N_{p,7+l-36}} \biggr) - \biggr(\frac{Y_{j(p),6}}{N_{p,6}} - \frac{Y_{j(p),6-36}}{N_{p,6-36}} \biggr)\bigg|E_{j(p)}=\infty \biggr] \\
&= \; \mu_{l, e} - \mu_{l, e}^{2020},
\end{align*}

where $\mu_{l ,e}^{2020}$ denotes $E\biggr[\frac{Y_{j(p),7+l-36}}{N_{p,7+l-36}}-\frac{Y_{j(p),6-36}}{N_{p,6-36}}\bigg|E_{j(p)}=e\biggr]-E\biggr[\frac{Y_{j(p),7+l-36}}{N_{p,7+l-36}}-\frac{Y_{j(p),6-36}}{N_{p,6-36}}\bigg|E_{j(p)}=\infty \biggr]$.
The term $\mu_{l,e}^{2020}$ can be interpreted as a placebo DID parameter evaluated three years earlier: it captures the treated--control differential change between months $6 - 36$ and $7 + l - 36$, i.e., in the placebo year 2020 when the policy change of interest occurs in 2023.
In an ideal setting with no anticipation and no spurious differential dynamics, this placebo effect should be zero.
More importantly, if treated and control groups are subject to different seasonal patterns, $\mu_{l,e}^{2020}$ may be non-zero even in the absence of the policy.
In that case, subtracting this placebo component from the baseline DID estimand yields $\widetilde{\mu}_{l, e} = \mu_{l, e} - \mu_{l,e}^{2020}$, which removes differential seasonality and provides a more robust estimate of the event-time treatment effect.

Note that this analysis requires employment shares in each wage bin to follow the same underlying calendar-month trends in 2020 and 2023, absent the 2023 minimum-wage increase and after accounting for common time effects.
In particular, the changes relative to the high-wage control group that would have occurred without the reform must be stable across the two years.
This assumption may not fully hold because of the disruptions to labor demand caused by COVID-19 in 2020.
The COVID-19 pandemic disrupted labor demand in Japan as economic activity contracted, particularly in face-to-face service industries that rely heavily on part-time workers \citep{fukui2020job, kikuchi2021suffers, fukai2021describing}.
To the extent that these disruptions affected sectors and occupations differently, differences in their composition across wage bins could generate employment-share trends in 2020 that differ from those in 2023.
Subtracting the 2020 changes could therefore remove pandemic-specific demand movements as well as recurring seasonality, so the DDD estimates should be interpreted as a sensitivity check conditional on the stability of wage-bin trends across years.

\begin{figure}[H]
  \begin{center}
  \subfloat[Missing and Excess Jobs over Time]{\includegraphics[width = 0.46\textwidth]{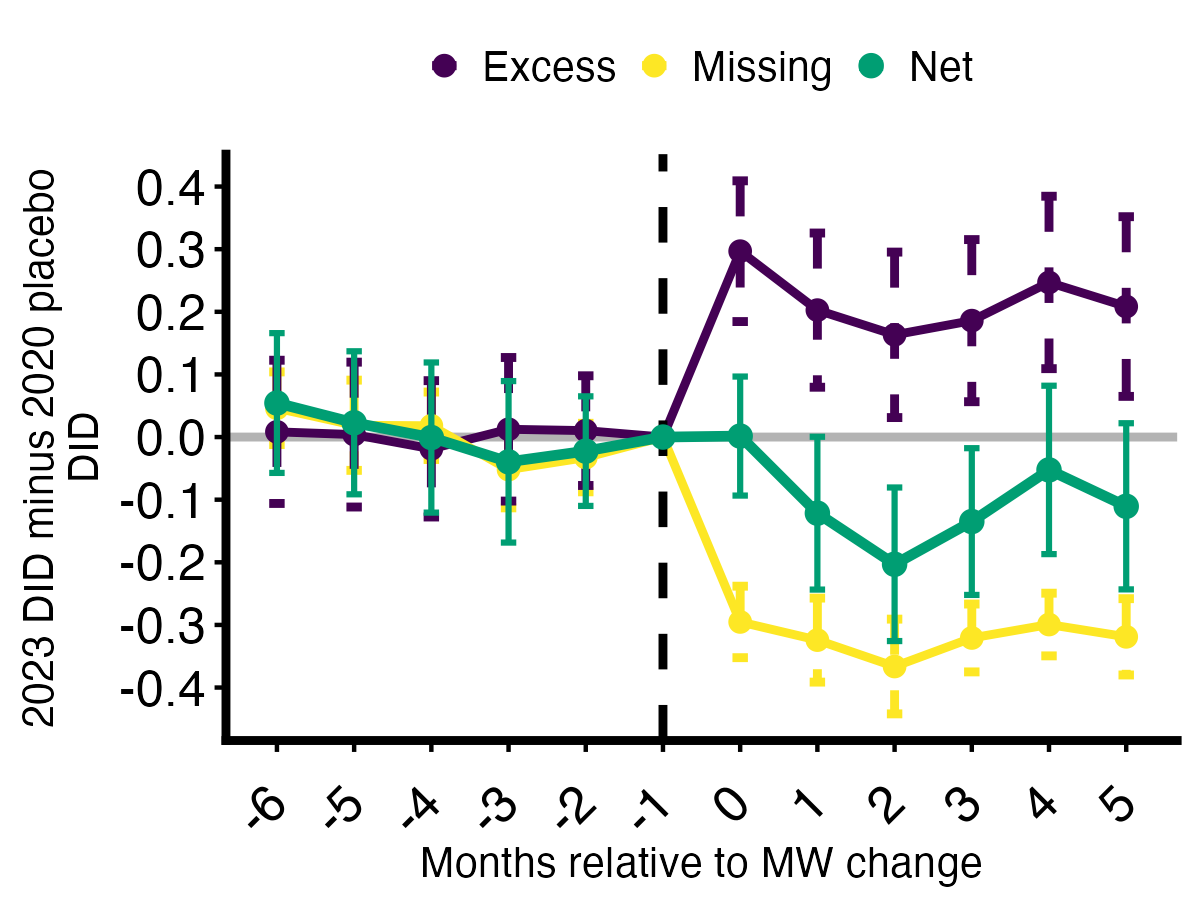}}
  \subfloat[Wage Distribution]{\includegraphics[width = 0.46\textwidth]{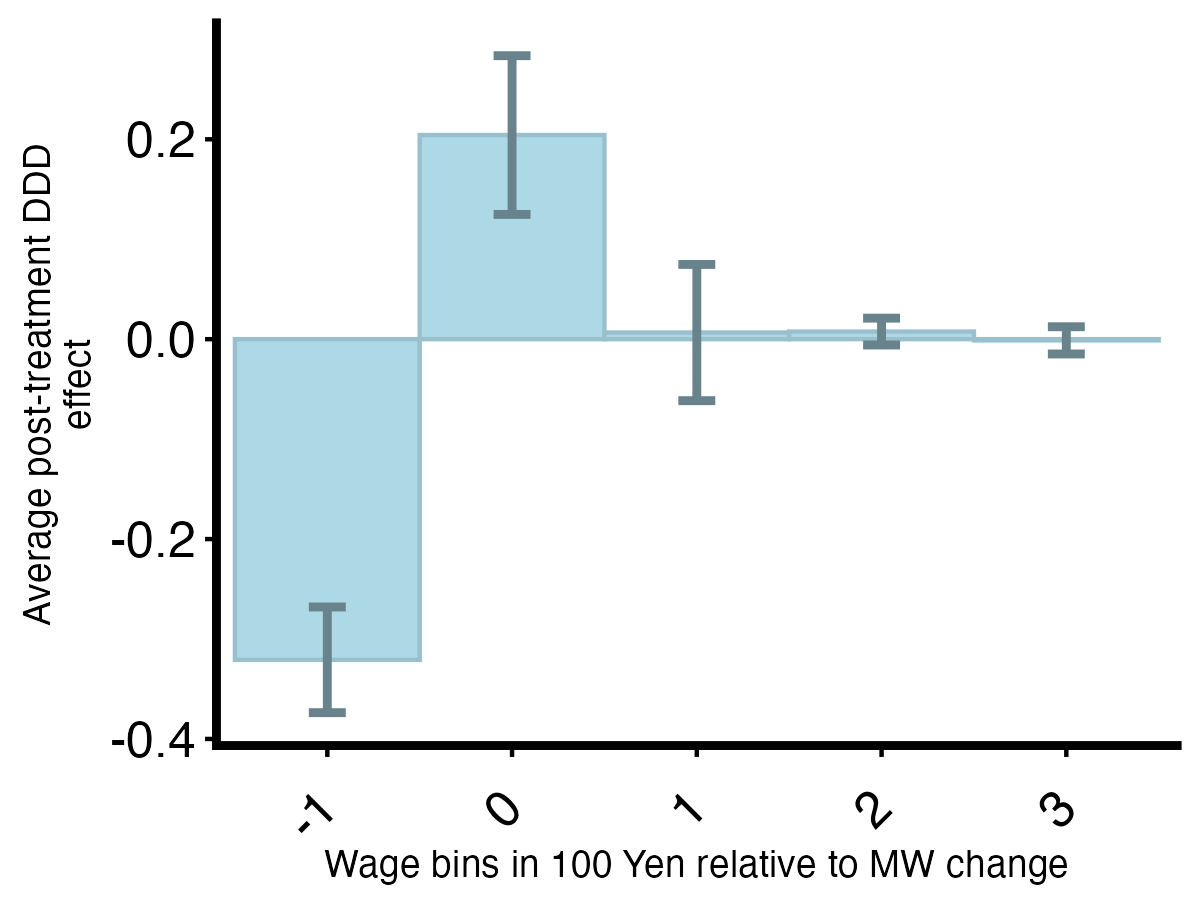}}
  \caption{Impact of Minimum Wages on Employment: DDD Specification}
  \label{fg:wage_distribution_moving_difference_employment_overall_cengiz_original_form}
  \end{center}
  \footnotesize
  Note: Panel (a) shows the estimated excess jobs and missing jobs for each month before and after the minimum wage increase. Panel (b) shows the estimated impact of the minimum wage increase on employment for each group $e$ in the wage distribution. The DDD holds nominal wage bins fixed and subtracts the corresponding 2020 outcome from the 2023 outcome. Both panels are normalized by the September 2023 employment share in the treated wage-bin groups.
\end{figure}

Panel (a) in Figure \ref{fg:wage_distribution_moving_difference_employment_overall_cengiz_original_form} shows the estimated excess jobs and missing jobs for each month before and after the minimum wage increase in our DDD specification.
Panel (b) presents the estimated impact of the minimum wage increase on employment for each group $e$ in the wage distribution.

Three findings emerge from Panels (a) and (b) of Figure \ref{fg:wage_distribution_moving_difference_employment_overall_cengiz_original_form}.
First, the estimated excess and missing jobs show no significant change in the pre-treatment period.
Second, missing and excess jobs are of similar size to the baseline DID estimates of Equation \eqref{sec:empirical_strategy_DID_regression}: missing jobs are about 32 percent of pre-reform lower-tier employment and excess jobs about 22 percent, against 28 and 27 percent in the baseline, and both emerge immediately after the minimum-wage increase and persist throughout the post-treatment period. The net effect is zero in October and negative thereafter, and it is significantly negative in December and January, when it reaches 20 and 13 percent of pre-reform lower-tier employment, compared with the baseline net decline of 1 percent.
Finally, the effects for $e \geq 1$ are small, so spillovers to higher wage tiers are limited.
Netting out the 2020 pattern therefore preserves the missing and excess jobs of the baseline and yields a larger net employment decline.

\subsubsection{Composition of Treated and Control Bins}\label{app:robust_composition}

Treated and control wage bins may differ in job characteristics. This subsection compares them within occupation.

\paragraph{Specification with Occupation Controls}\label{app:specification_with_occupation_controls}

We next estimate the effects of the minimum-wage increase on employment proportions using a specification that controls for occupation. Because wages differ across occupations, low- and high-wage bins contain jobs with different occupational compositions, so the control bins may differ from the treated bins in job characteristics as well as in wage levels. To address this concern, we reconstruct the estimation sample at the wage-bin--prefecture--occupation--month level and estimate our main specification with occupation-related fixed effects, so that treated and control bins are compared within occupation.

Let $o$ index occupation categories and let $N_{p,o,t}$ denote total employment in prefecture $p$, occupation $o$, and month $t$. The estimating equation is

\begin{align}
\label{eq:occupation_control_specification}
\frac{Y_{j(p),o,t}}{N_{p,o,t}}
=
\sum_{l \neq -1}\sum_{e \neq \infty}
\mu^{\mathrm{occ}}_{l,e}
\mathbf{1}\{t-7=l\}
\mathbf{1}\{E_{j(p)}=e\}
+\alpha_{e,o}+\lambda_{t,o}+\gamma_o+\epsilon_{j,p,o,t}.
\end{align}

This is the baseline event-study equation augmented with occupation as an additional dimension. The specification includes treatment-group-by-occupation fixed effects $\alpha_{e,o}$, month-by-occupation fixed effects $\lambda_{t,o}$, and occupation fixed effects $\gamma_o$. Hence, identification compares treated and control wage bins within occupation while allowing occupation-specific levels and occupation-specific monthly shocks.\footnote{In this implementation, the number of 10 JPY wage bins in each treated group and the September 2023 lower-tier employment share are averaged over the same set of prefecture--occupation pairs, and the Missing, Excess, and Net estimates divide by that common average share and average the six post-treatment months, as in the baseline. Standard errors are clustered by prefecture, allowing dependence across occupations, wage bins, and months within a prefecture.}

\begin{figure}[H]
  \begin{center}
  \subfloat[Missing and Excess Jobs over Time]{\includegraphics[width = 0.46\textwidth]{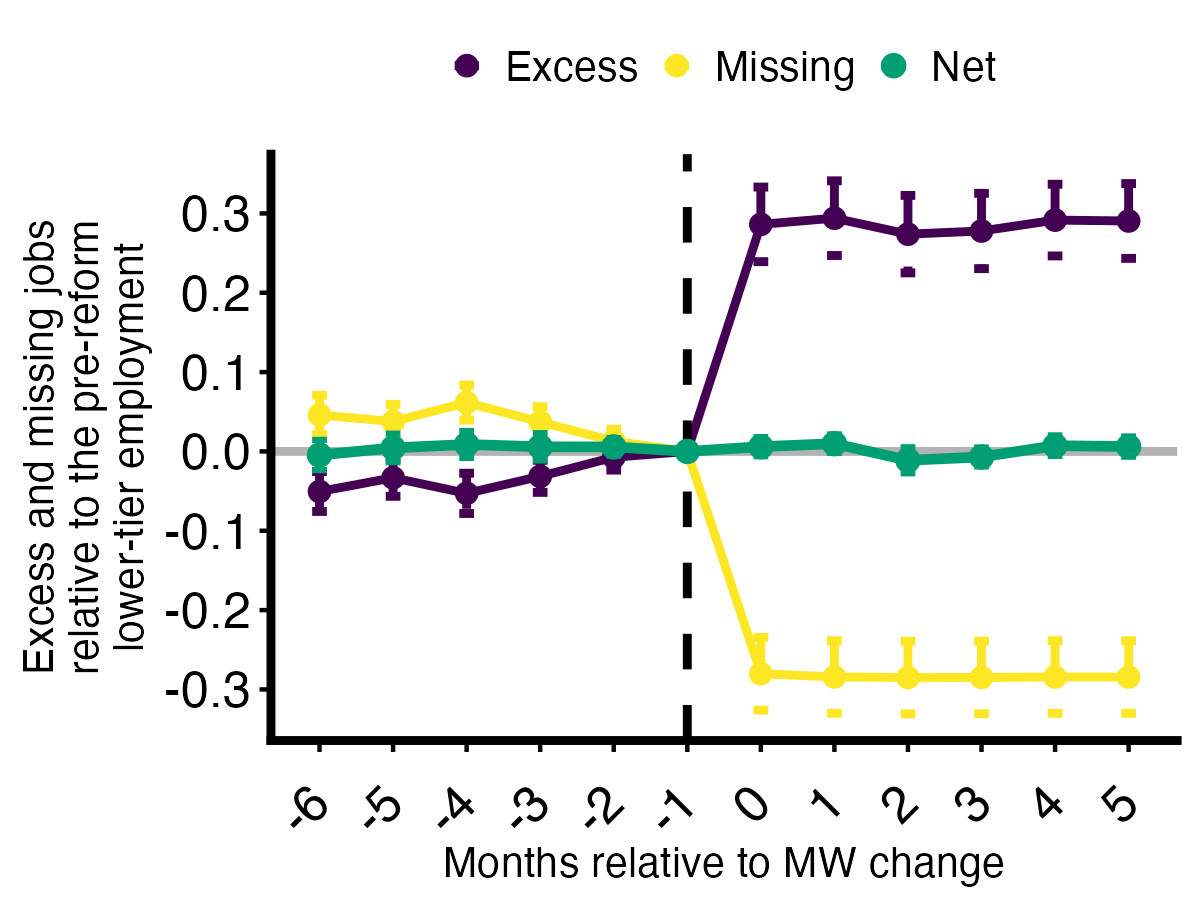}}
  \subfloat[Wage Distribution]{\includegraphics[width = 0.46\textwidth]{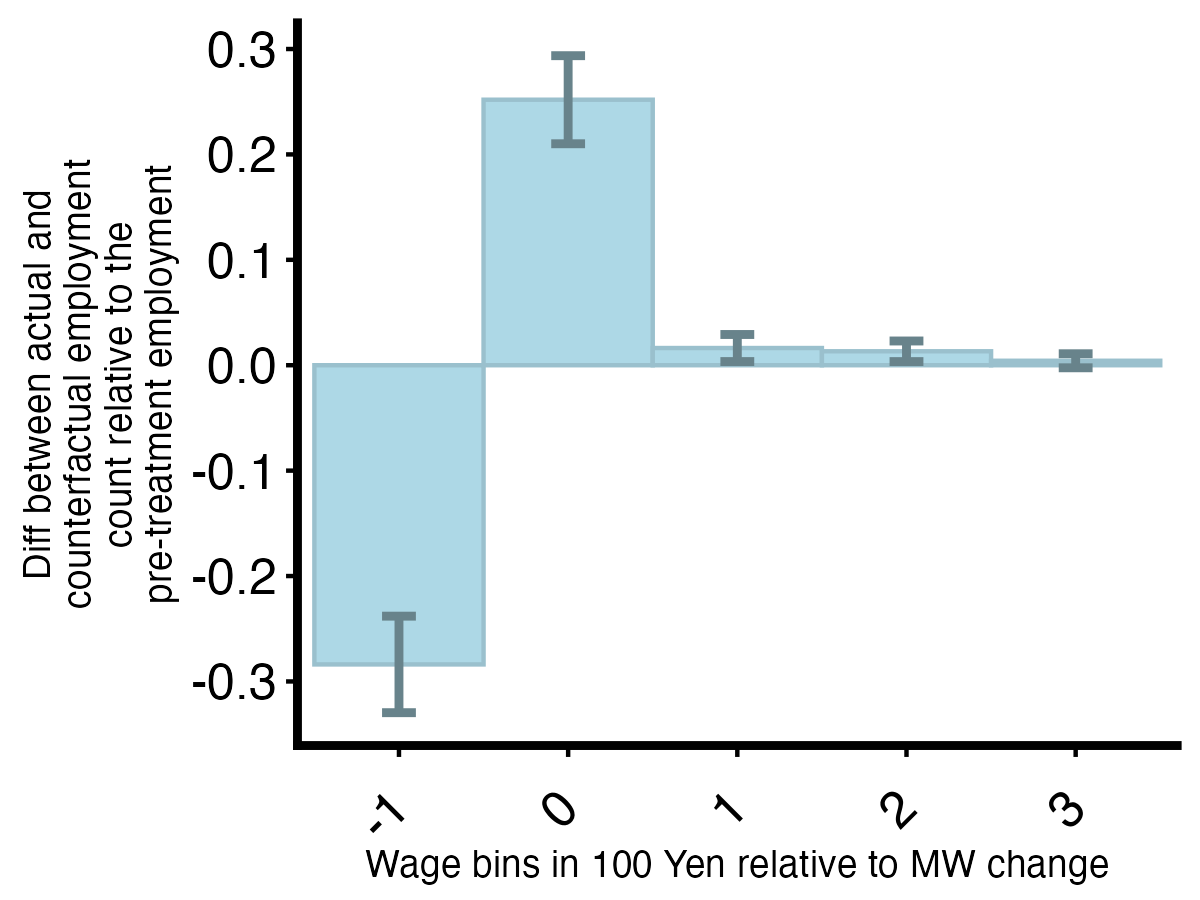}}
  \caption{Impact of Minimum Wages on Employment: Specification with Occupation Controls}
  \label{fg:employment_effects_with_occupation_controls}
  \end{center}
  \footnotesize
  Note: The estimation sample is organized at the wage-bin--prefecture--occupation--month level. The outcome is employment in each cell divided by total employment in the corresponding prefecture--occupation--month. Panel (a) shows the estimated excess and missing jobs before and after the minimum-wage increase. Panel (b) shows the average post-treatment effect for each wage group relative to the new minimum wage.
\end{figure}

Panel (a) of Figure \ref{fg:employment_effects_with_occupation_controls} plots the monthly estimates of excess and missing jobs before and after the minimum-wage increase from the specification with occupation controls. Panel (b) reports the estimated employment effect for each wage group $e$, averaged over the post-treatment period.

Three findings emerge from these panels. First, the pre-treatment estimates of excess and missing jobs are small relative to the post-treatment changes. Second, the estimates with occupation controls are similar in magnitude to those from the baseline DID specification (Equation \eqref{sec:empirical_strategy_DID_regression}). In both specifications, missing jobs below the new minimum wage and excess jobs above it appear immediately after the increase and persist throughout the post-treatment period. Third, most of the change in excess jobs is accounted for by group $e=0$, while the effects on exclusive groups with $e \geq 1$ are limited. Overall, these findings suggest that the baseline results are robust to accounting for differences across occupations in employment-share levels and monthly shocks.

\section{Hiring Channel Substitution and Timee Rollout}\label{sec:labor_substitution_note}

In this section, we examine whether app-based spot work platforms substitute for or complement traditional hiring channels such as public employment offices. Using confidential contract-level data from Timee and public vacancy records, we exploit the platform's staggered regional rollout from 2018 to 2020. We conduct difference-in-differences analysis to assess whether Timee's expansion affected monthly active vacancies in the public system. The estimates do not detect a clear reduction in public-office vacancies, but they do not establish the absence of substitution between hiring channels.

Understanding how employers and job seekers choose between standard hiring channels (e.g., public employment offices) and digital labor platforms is central to modern labor-market design \citep{kroft2014does,kuhn2004internet,goldfarb2019digital}. As platform adoption accelerates while conventional systems persist, questions about hiring--channel substitution---between app-based and public-office vacancies---have become increasingly salient. Existing studies show that digital postings on platforms like Craigslist, CareerBuilder, and Indeed can shift local outcomes: unemployment, vacancy rates, and rental markets \citep{kroft2014does}, crime \citep{cunningham2024did}, media and political behavior \citep{djourelova2025impact}, substitution across job boards \citep{brenvcivc2016impact}, and migration \citep{maria2024death}. In contrast, little is known about substitution between spot gig work and standard employment. At its core, this raises a fundamental choice for employers: whether to fill labor needs via standard hiring (stable contracts) or via app-based spot work (short-term, demand-responsive).

We fill the gap with a 2018--2024 prefecture--month panel covering Japan, combining (i) confidential \emph{Timee} data for short, on-site spot shifts and (ii) public \emph{Hello Work} records on conventional vacancies, registered unemployed, and successful matches. Timee rolled out heterogeneously across prefectures and expanded rapidly. We exploit this staggered timing and intensity variation to test whether platform entry and growth reduced public-office vacancies.

Two features of the Timee data are central. First, we observe confidential contract-level matches linking each vacancy to a realized hire with full job detail. This labor-demand granularity allows us to study a substitution margin distinct from prior Craigslist-type studies \citep{kroft2014does,brenvcivc2016impact}: whether firms meet needs via fast, flexible spot contracts or slower standard hires, with an explicit cost--flexibility tradeoff (e.g., Timee charges about 30\% commission per contract). These logs also allow us to construct a prefecture-specific Platform Popularity Index (posts per 1,000 residents) a la \cite{kroft2014does}, and track substitution on both the posting (vacancies) and matching (employment) margins over time. Second, unlike household-service apps (e.g., TaskRabbit in \cite{cullen2021outsourcing}), Timee spans a wide range of on-site, shift-based jobs---retail, food service, logistics, events, and light manufacturing---providing broad labor-market coverage.

Using the prefecture--month panel and staggered rollout, we estimate dynamic difference-in-differences models testing whether Timee entry and growth affect Hello Work vacancies. We use monthly active vacancies, which include vacancies carried over from the preceding month and newly registered vacancies, with new vacancies counted once. The dynamic estimates do not detect a clear reduction after entry. A two-way fixed effects model using Platform Popularity as a continuous regressor also shows no detectable association with this vacancy measure. These findings concern vacancies recorded by public offices; they do not identify substitution toward incumbent workers or the role of spot work in absorbing temporary workload shocks.

\subsection{Prefecture-month-level Timee Introduction and Expansion}\label{sec:labor_sub_timee_introduction}
To exploit prefecture-month variation in Timee's expansion, we focus on the calendar month in which job postings first appear in a given prefecture (the rollout month). This is defined as the first month that Timee postings are recorded in platform logs for that prefecture and serves as the binary treatment in our difference-in-differences design.

\begin{figure}[!ht]
  \begin{center}
  \subfloat[Geographical Variation of Timee Rollout]{\includegraphics[width=0.73\linewidth]{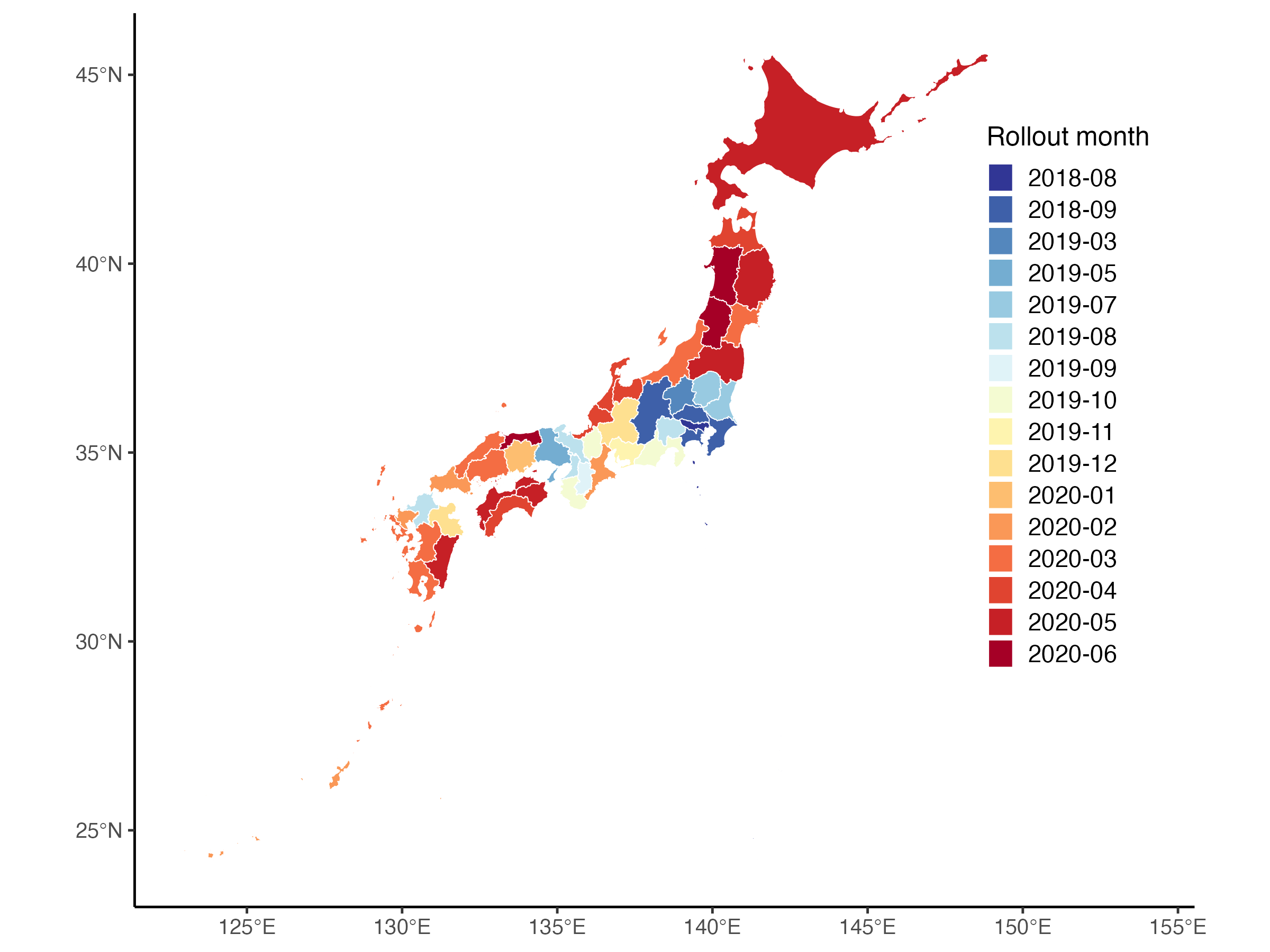}}\\
  \subfloat[Timee Rollout Timing (ordered by Resident Counts (2018))]{\includegraphics[width=0.73\linewidth]{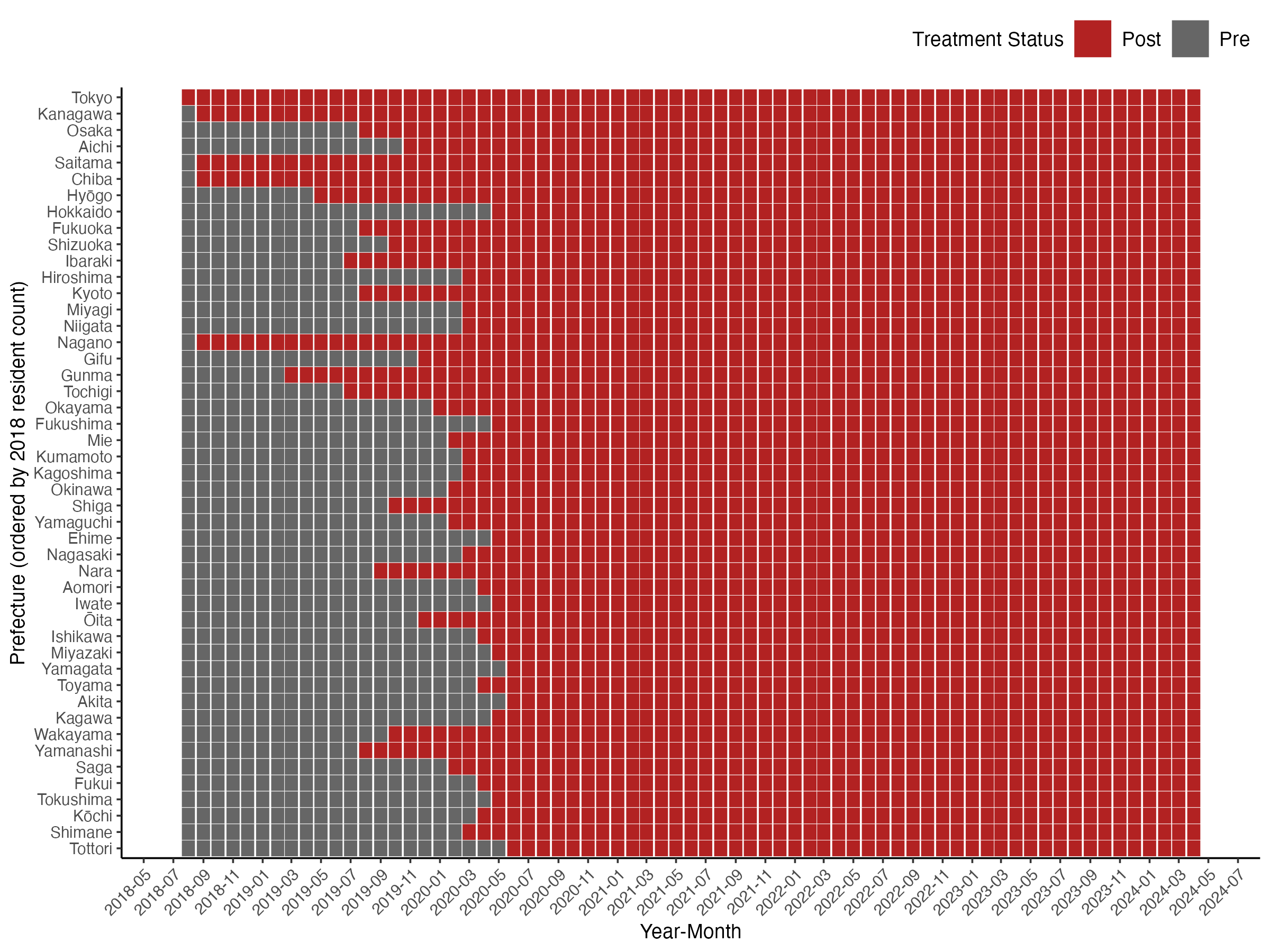}}
  \caption{Timee Rollout Across Prefectures}
  \label{fig:labor_sub_prefecture_month_timee_rollout_map}
  \end{center}
  \footnotesize
\end{figure}

Figure \ref{fig:labor_sub_prefecture_month_timee_rollout_map} summarizes Timee's spatial rollout and intensity patterns.
Panel (a) shows that the platform first launched in Tokyo in August 2018, followed in September 2018 by the neighboring prefectures of Chiba, Saitama, and Kanagawa and by Nagano, and then expanded to the remaining prefectures by June 2020---reflecting a core-to-periphery diffusion strategy typical of digital platforms.
Panel (b) orders prefectures by 2018 population and shows that higher-population areas generally adopted earlier. While this pattern holds overall, some lower-ranked prefectures, such as Nagano, received the rollout earlier than more populous regions, suggesting that other factors also shaped the rollout sequence.

\subsection{Empirical Exercise: Staggered Rollout of Timee}\label{sec:labor_sub_empirical_strategy}

\subsubsection{Binary Treatment Staggered DID}

Our main identification strategy for estimating the impact of Timee's rollout on Hello Work vacancy counts exploits the expansion of Timee across Japanese prefectures during our sample period. As discussed in Section \ref{sec:labor_sub_timee_introduction}, the timing of Timee's entry varies across prefectures and is not systematically related to stronger past or predicted economic conditions at the prefecture level. Accordingly, we identify the treatment effect on traditional vacancies using difference-in-differences by comparing prefecture--month outcomes between treated and not-yet-treated prefectures while accounting for the staggered timing of Timee's entry.

To align notation with the staggered-adoption framework, we write the event-time parameterization below, but estimation is conducted using the Callaway and Sant'Anna \citeyearpar{callaway2021difference} group-time ATT procedure rather than OLS on this equation:
\begin{equation}
\label{eq:labor_sub_timee_eventstudy}
Y_{p(g),t}
= \alpha_{p(g)} + \beta_t
  + \sum_{g \in G}\sum_{j \neq -1} \delta_{g, j} D_{p(g)}^{g, j}
  + \varepsilon_{p(g),t},
\end{equation}

where $Y_{p(g),t}$ is the log of monthly active Hello Work vacancies for prefecture $p(g)$ in month $t$; $\alpha_{p(g)}$ and $\beta_t$ denote prefecture and month fixed effects; and $D_{p(g)}^{g, j}$ is an event--time indicator equal to one for prefecture $p(g)$ when $t$ is $j$ months relative to its first Timee entry month $g$.

The parameters of interest are event-time average treatment effects across adoption cohorts. We estimate group-time ATT$(g,t)$ using not-yet-treated prefectures as comparisons and then aggregate to event time following \cite{callaway2021difference}. Because every prefecture had entered by June 2020, the estimation sample covers January 2018 through May 2020, and the three prefectures that entered in June 2020 remain in the comparison group throughout.

\begin{figure}[!ht]
  \begin{center}
  \includegraphics[width=0.75\linewidth]{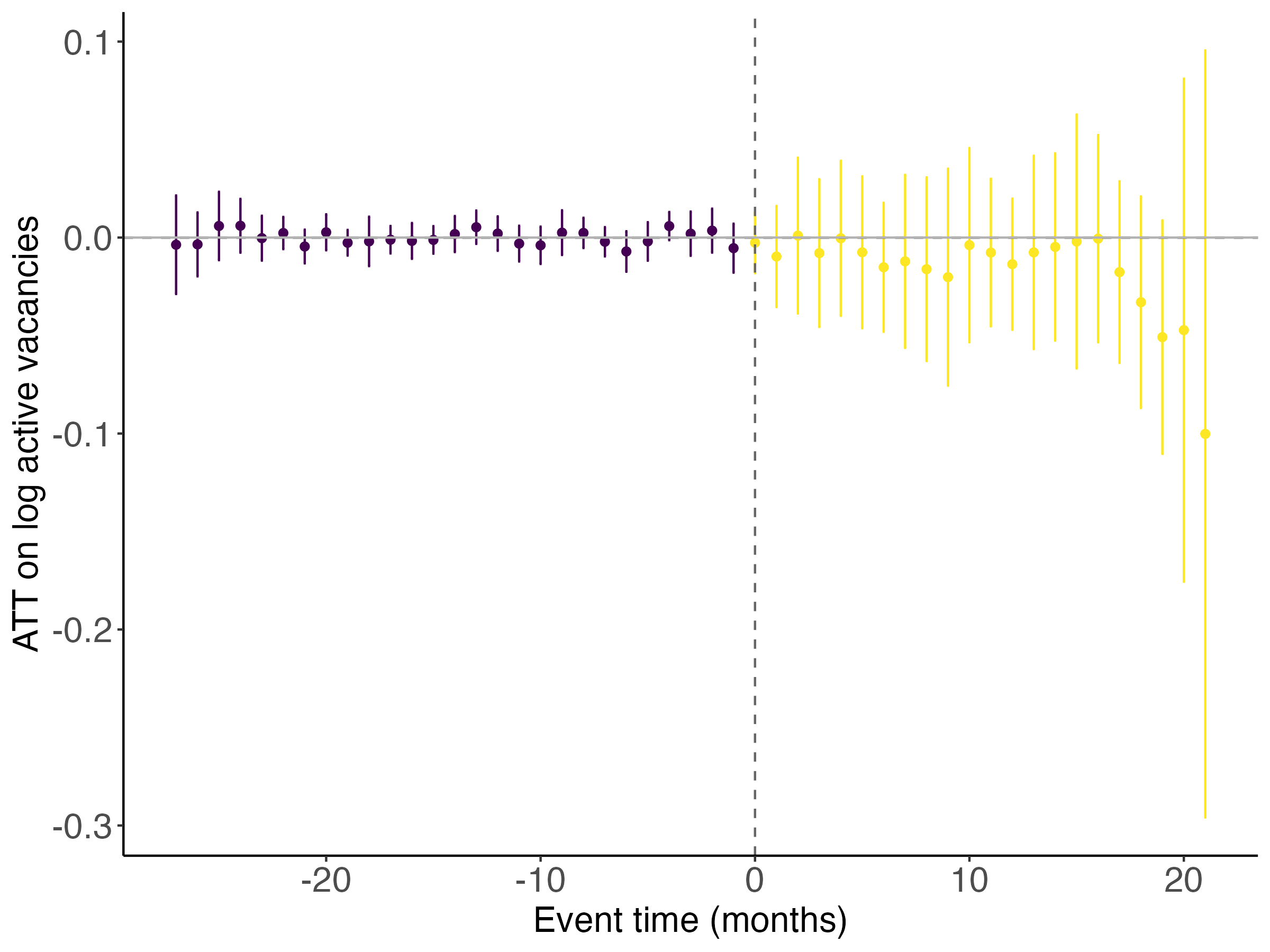}
  \caption{Staggered Timee Entry and Log Monthly Active Hello Work Vacancies}
  \label{fig:labor_sub_estimate_log_prefecture_year_month_vacancy_timee_entry_callaway_santanna}
  \end{center}
  \footnotesize \textit{Notes:} The figure reports event-time averages of ATT$(g,t)$ from a difference-in-differences design with staggered adoption. The outcome is the log of monthly active vacancies, including carried-over and newly registered vacancies. Not-yet-treated prefectures form the comparison set. Pre-entry estimates compare adjacent months; post-entry estimates use the last untreated month as the baseline. Vertical lines denote simultaneous 95\% confidence bands based on 1,000 multiplier bootstrap repetitions following \cite{callaway2021difference}.
\end{figure}

Figure \ref{fig:labor_sub_estimate_log_prefecture_year_month_vacancy_timee_entry_callaway_santanna} plots event-time average treatment effects on the treated (ATT) for log monthly active Hello Work vacancies around Timee entry. Most pre-entry estimates are close to zero, although small adoption cohorts limit the precision of this diagnostic. Following entry, the simultaneous confidence bands include zero throughout the reported event window. The estimates at 18 and 19 months after entry are negative but are not statistically significant. These horizons rely on the five prefectures that entered in 2018 and fall in February--April 2020, the first months of the COVID-19 pandemic. The figure therefore provides no clear evidence of a reduction in public-office vacancies after rollout, while leaving economically meaningful negative effects possible.

\subsection{Timee Popularity Index and Continuous-Intensity Estimates}\label{app:labor_sub_popularity}
To complement the binary rollout treatment, we construct a time-varying prefecture-month Timee Popularity Index that captures relative platform penetration. Specifically, $\text{Pop}_{pt}$ is the number of posted spot jobs in prefecture $p$ and month $t$ per 1,000 residents, where the denominator is fixed at prefecture population from the 2018 Basic Resident Registration. This construction mirrors the Craigslist penetration measure used by \citet{kroft2014does}, who define platform popularity as the number of online job advertisements per 1,000 people in each Metropolitan Statistical Area (MSA). In their study, this index is used to exploit cross-sectional variation in online job search intensity and to study substitution between online and offline job ads. Similarly, our index captures cross-prefecture and over-time variation in Timee adoption. For robustness, we also construct an alternative version based on realized contracts (spot employment matches per 1,000 residents), using detailed contract-level data linked to administrative identifiers.

\begin{figure}[!ht]
  \begin{center}
  \subfloat[Timee Popularity]{\includegraphics[width=0.7\linewidth]{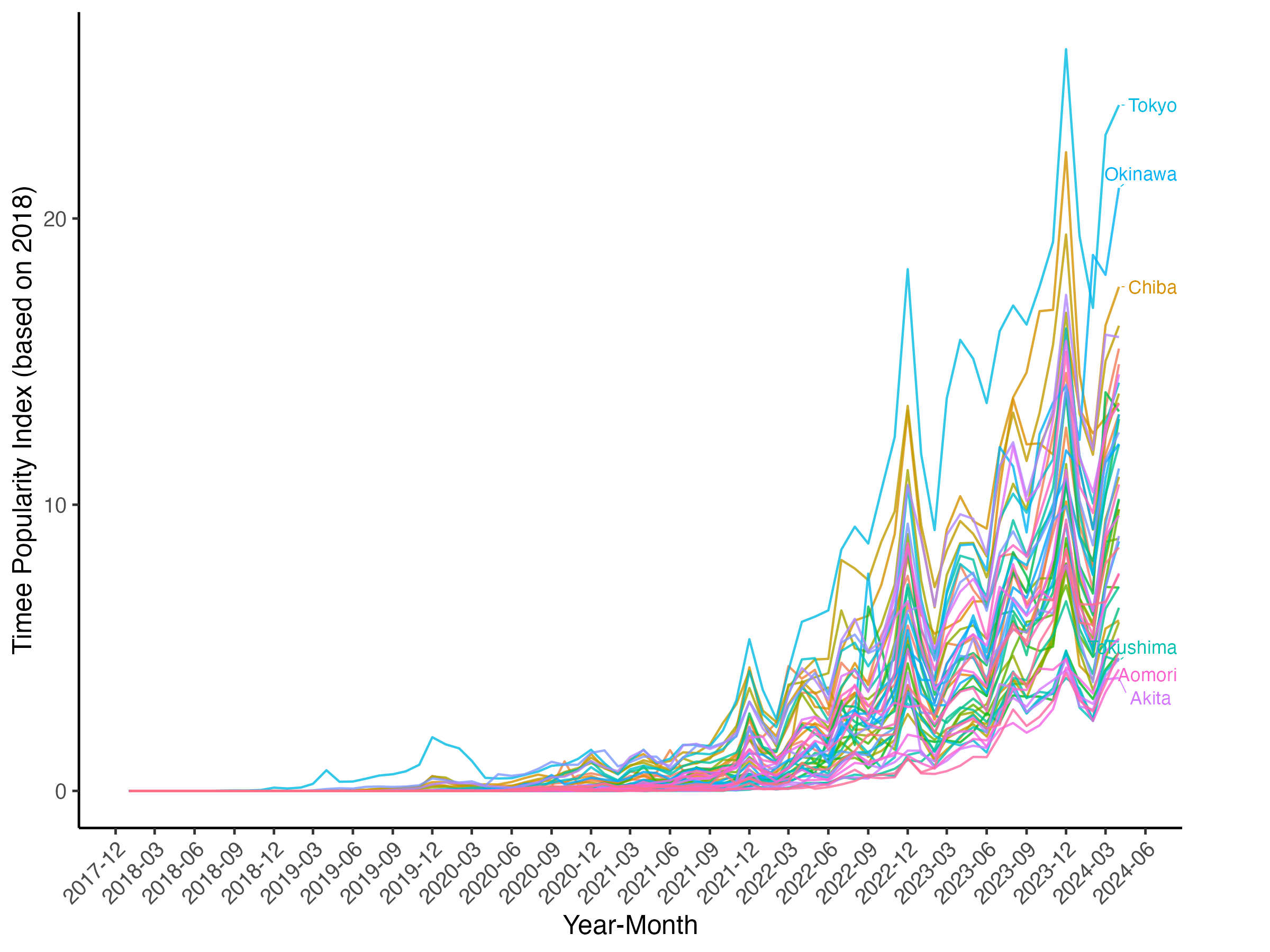}}\\
  \subfloat[Timee Popularity (April 2024) vs Resident Counts]{\includegraphics[width=0.7\linewidth]{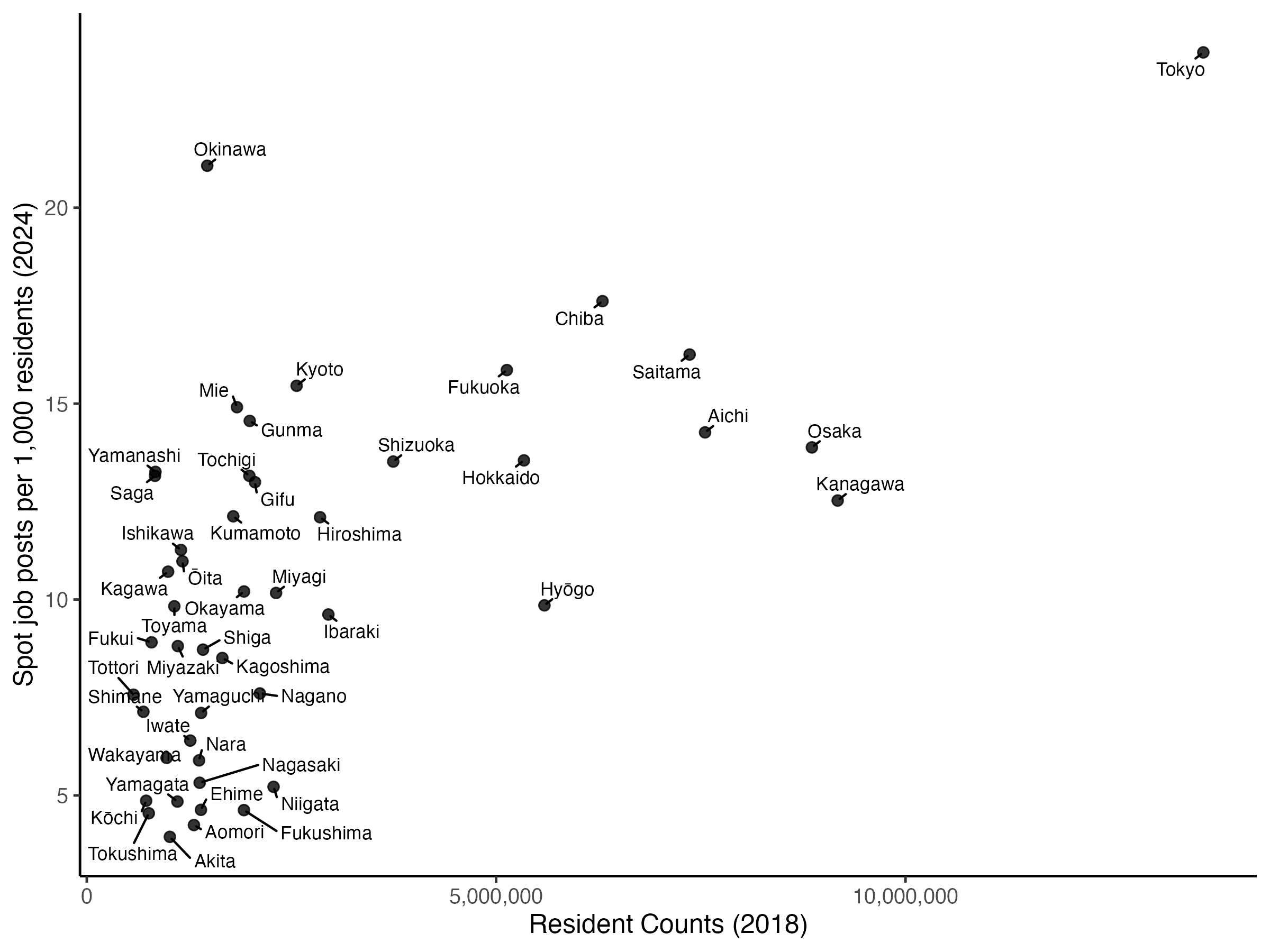}}
  \caption{Timee Popularity Trajectories and Scale}
  \label{fig:labor_sub_prefecture_month_timee_popularity_map}
  \end{center}
  \footnotesize
  Note: Timee's popularity index is defined monthly as the number of spot-job posts per 1,000 residents, with 2018 resident counts used as the fixed denominator.
\end{figure}

Figure \ref{fig:labor_sub_prefecture_month_timee_popularity_map} summarizes the spatial rollout and intensity patterns. Panel (a) illustrates heterogeneous popularity trajectories, with several prefectures experiencing pronounced surges after 2022, while Panel (b) plots a modest, noisy correlation between April 2024 popularity and population size.

As an intensity-based robustness check, we estimate a two--way fixed effects model that treats Timee popularity as a continuous regressor:
\begin{equation}
\label{eq:labor_sub_twfe_pop_same}
Y_{pt}
= \alpha_{p} + \beta_t
+ \gamma\,\text{Pop}_{pt}
+ \varepsilon_{pt},
\end{equation}
where $\text{Pop}_{pt}$ denotes the Timee Popularity Index in prefecture $p$ and month $t$, and $\gamma$ captures the within-prefecture association between popularity and Hello Work vacancies after accounting for prefecture and month shocks. Because popularity and vacancies are observed contemporaneously at the prefecture-month level, we interpret $\gamma$ as a descriptive association rather than a causal parameter. As a robustness check, column (2) of Table \ref{tb:labor_sub_result_estimate_log_aggregate_vacancy_2wfe} replaces the posting-based index with the alternative index based on realized matches.

Table \ref{tb:labor_sub_result_estimate_log_aggregate_vacancy_2wfe} shows that Timee popularity has no detectable association with monthly active Hello Work vacancies. This result is a descriptive association and does not establish the absence of hiring-channel substitution.

\begin{table}[!htbp]
  \begin{center}
      \caption{Timee Popularity Index and Monthly Active Hello Work Vacancies}
      \label{tb:labor_sub_result_estimate_log_aggregate_vacancy_2wfe}
      
\begin{tabular}[t]{lcc}
\toprule
  & (1) & (2)\\
\midrule
Timee vacancy counts per 1000 residents & 0.002 & \\
 & (0.003) & \\
Timee matching counts per 1000 residents &  & 0.002\\
 &  & (0.004)\\
\midrule
FE (Month) & Y & Y\\
FE (Prefecture) & Y & Y\\
Num.Obs. & 3572 & 3572\\
R2 & 0.996 & 0.996\\
R2 Adj. & 0.996 & 0.996\\
\bottomrule
\end{tabular}

  \end{center}\footnotesize
  \textit{Note}: The dependent variable is the log of monthly active Hello Work vacancies, which include vacancies carried over from the preceding month and newly registered vacancies. Standard errors are in parentheses, two-way clustered by prefecture and month.
\end{table}

This section finds no clear reduction in monthly active public-office vacancies associated with Timee's early rollout, and no detectable association with its subsequent popularity. These estimates do not rule out substitution across hiring channels and do not test whether firms use spot work to absorb temporary workload shocks. They also concern a different period and shock from the October 2023 minimum-wage revision.

\newpage

\section{A Newsvendor Interpretation of Spot Labor Demand under a Wage Floor}\label{sec:newsvendor}

This appendix develops a minimal model that organizes our empirical findings around a single idea: in a spot labor market, the establishment decides how much flexible labor to commit \emph{before} demand is realized. This is the same trade-off as in the classic newsvendor problem. The framework adapts \citet{mas2026newsvendor}, who studies a firm's choice between holding a stable buffer of permanent workers and hiring from a spot market whose frictions depend on market participation. We use the short-run, pure-spot region of that model as an interpretive benchmark, holding permanent staffing, capital, and task reorganization fixed. The model organizes the vacancy response and explains why missing jobs can exceed excess jobs. It also illustrates possible selection across posting attributes when those attributes are correlated with latent productivity. The amenity discussion draws on the wage--amenity logic in \citet{mas2025non}: non-wage amenities are part of the job bundle, so that under regulated wages and search frictions the direct amenity response is theoretically ambiguous, which is why we read the observed amenity shift as compositional.

\subsection{Setup}

Consider a single establishment posting a homogeneous spot job with minimal skill requirements. Each productively filled shift generates value \(r>0\), where \(r\) summarizes task-specific productivity. Across postings, \(r\) is distributed with continuous cumulative distribution function \(\Phi\) on \([\underline r,\bar r]\). The term \(r\) is the analogue of position productivity in \citet{mas2026newsvendor}.

The establishment's period labor need \(D\) is uncertain, with continuous cumulative distribution function \(G\) and density \(g\) on \([0,\infty)\). Demand is realized only after postings are made. Because work is spot and must be advertised in advance, the establishment commits to a number of vacancies \(V\ge 0\) before observing \(D\). Let \(\omega\) denote the posted wage actually paid per staffed shift. For affected postings, the statutory wage floor binds, so \(\omega=\underline w\). Because the platform charges establishments a commission proportional to the contracted payment, the establishment's per-shift cost is a constant multiple of \(\omega\); the wage terms below can be read in units of this effective cost. The proportional fee preserves the negative sign of the wage-cost response; its effect on the magnitude also depends on the demand distribution. If more vacancies are staffed than ultimately needed, the extra staffed shifts have salvage value \(s\in[0,\omega)\). If the establishment understaffs, each unit of unmet need imposes a shortfall cost \(\rho\ge 0\), capturing rationing, expedited coverage, or lost goodwill.

\subsection{The Posting Problem and the Critical Fractile}

Conditional on opening an active spot posting, the establishment chooses \(V>0\) to maximize
\begin{equation}
\Pi(V;r,\omega)
  \;=\;
  (r-\omega+\rho)\,\mathbb{E}\!\left[\min(V,D)\right]
  -(\omega-s)\,\mathbb{E}\!\left[(V-D)^{+}\right]
  -\kappa,
\label{eq:nv_profit}
\end{equation}
where \((x)^{+}=\max\{x,0\}\) and \(\kappa\ge 0\) is the fixed cost of maintaining an active posting. This formulation is measured relative to not opening the spot posting, whose payoff is normalized to zero. Equivalently, the no-posting payoff, including the unmet-need cost \(-\rho\mathbb{E}[D]\), has been subtracted. A filled unit both produces the margin \(r-\omega\) and avoids the shortfall cost \(\rho\), while an overposted staffed unit costs \(\omega-s\). The posting is active only if the maximized value of \eqref{eq:nv_profit} is non-negative.

Since
\[
\frac{\partial}{\partial V}\mathbb{E}[\min(V,D)] = 1-G(V),
\qquad
\frac{\partial}{\partial V}\mathbb{E}[(V-D)^{+}] = G(V),
\]
an interior solution satisfies
\begin{equation}
G\bigl(V^{*}\bigr)
  \;=\;
  \frac{r-\omega+\rho}{\,r-s+\rho\,}
  \;\equiv\;
  \tau(r,\omega).
\label{eq:critical_fractile}
\end{equation}
The interior case requires \(r-\omega+\rho>0\) and \(g(V^{*})>0\); if \(r-\omega+\rho\le 0\), no active posting is optimal and we set \(V^{*}=0\). Expected employment, or filled shifts, is
\[
E^{*}(r,\omega)
  =
  \mathbb{E}[\min(V^{*},D)]
  =
  \int_0^{V^{*}}\bigl(1-G(x)\bigr)\,dx,
\]
which is increasing in \(V^{*}\).

\subsection{A Binding Wage Floor: Fewer Vacancies}

For an interior affected posting, a minimum-wage increase raises \(\omega\) one-for-one and lowers the critical fractile:
\begin{equation}
\frac{\partial \tau}{\partial \omega}
  \;=\;
  -\frac{1}{\,r-s+\rho\,}
  \;<\;0,
\qquad
\frac{\partial V^{*}}{\partial \omega}
  \;=\;
  \frac{1}{g(V^{*})}\frac{\partial \tau}{\partial \omega}
  \;<\;0 .
\label{eq:dV_dw}
\end{equation}

\begin{proposition}[Vacancy creation]\label{prop:vacancy}
For an affected posting with a binding wage floor and an interior optimum, a higher floor raises \(\omega=\underline w\) one-for-one and strictly lowers the optimal number of posted vacancies and expected employment:
\[
\frac{\partial V^{*}}{\partial\omega}<0,
\qquad
\frac{\partial E^{*}}{\partial\omega}<0 .
\]
If the wage floor pushes the posting to the corner \(V^{*}=0\), the effect is weakly negative. The response operates through the establishment's posting choice, holding the demand distribution \(G\) fixed.
\end{proposition}

The mechanism is simple. A higher floor raises the cost of committing to a staffed vacancy that may not be needed and reduces the net return from meeting demand. The establishment therefore runs leaner. This is a labor-demand comparative static holding the realized-need distribution \(G\) and worker-side behavior fixed. Lower-productivity postings are closer to the corner or exit threshold. The magnitude of their interior vacancy response also depends on the density \(g(V^{*})\). A higher shortfall cost \(\rho\) reduces the magnitude of the critical-fractile response, \(1/(r-s+\rho)\). Its effect on the magnitude of the vacancy response also depends on \(g(V^{*})\), so the model does not imply a universal elasticity ranking across markets. Mapping this prediction to prefecture-level exposure requires the empirical assumption that high-Kaitz prefectures have more mass of postings near this low-productivity, low-wage region, which is exactly the bite captured by the Kaitz index.

\subsection{The Extensive Margin: Missing Jobs and Excess Jobs}

A posting remains active only if its maximized active-posting profit is non-negative:
\[
\Pi^{*}(r,\omega)=\max_{V>0}\Pi(V;r,\omega)\ge 0 .
\]
By the envelope theorem,
\[
\frac{\partial\Pi^{*}}{\partial\omega}=-V^{*}<0,
\qquad
\frac{\partial\Pi^{*}}{\partial r}=E^{*}(r,\omega)>0 .
\]
If \(\kappa>0\) and there is a break-even productivity \(r^{*}(\omega)\) satisfying \(\Pi^{*}(r^{*},\omega)=0\) with \(V^{*}>0\), which is unique because \(\partial\Pi^{*}/\partial r>0\), then
\begin{equation}
\frac{d r^{*}}{d \omega}
  =
  \frac{V^{*}}{E^{*}(r^{*},\omega)}
  \;\ge\;1,
\label{eq:threshold}
\end{equation}
with strict inequality because \(G(V^{*})=\tau>0\) at an interior optimum, so that \(E^{*}<V^{*}\). A higher wage floor therefore raises the productivity cutoff for active postings.

To connect this cutoff logic to the excess-jobs and missing-jobs decomposition, hold fixed the set of potential postings over the short window. Let \(p^{0}(r)\) denote the posting's desired wage absent the new floor, and suppose the observed posted wage is \(\omega(r;\underline w)=\max\{p^{0}(r),\underline w\}\). Consider a wage-floor increase from \(\underline w_{0}\) to \(\underline w_{1}>\underline w_{0}\). Postings with \(p^{0}(r)<\underline w_{1}\) were located below the new floor before the reform. Among these postings, survivors are mechanically relabeled to the new floor, while postings that fail the new profitability condition exit.

Let
\[
\mathcal{B}_{0}
  =
  \{r:p^{0}(r)<\underline w_{1},\,\Pi^{*}(r,\omega(r;\underline w_{0}))\ge 0\}
\]
be the set of postings that were active below the new floor before the reform, and let
\[
\mathcal{B}_{1}
  =
  \{r:p^{0}(r)<\underline w_{1},\,\Pi^{*}(r,\underline w_{1})\ge 0\}
\]
be the subset that survives after the reform. Because \(\mathcal{B}_{1}\subseteq\mathcal{B}_{0}\) and \(E^{*}(r,\underline w_{1})\le E^{*}(r,\omega(r;\underline w_{0}))\), the employment mass missing below the new floor is at least as large as the mass reappearing just above it:
\[
\int_{\mathcal{B}_{0}} E^{*}(r,\omega(r;\underline w_{0}))\,d\Phi(r)
\;\ge\;
\int_{\mathcal{B}_{1}} E^{*}(r,\underline w_{1})\,d\Phi(r).
\]
The inequality is strict whenever a positive mass of postings exits or surviving constrained postings reduce vacancy creation. The result therefore signs the net effect---missing jobs weakly exceed excess jobs---without pinning down its magnitude, which depends on how much mass exits and how far constrained survivors cut vacancy creation. Thus, in the very short run, relabeling creates excess jobs above the new floor, but exit and vacancy reductions create unmatched missing jobs below it. This argument captures the compliance-relabeling component of excess jobs; it does not rule out additional excess jobs from new entry or from postings whose desired wages were already above the new floor. The posting-history decomposition in Section~\ref{sec:firm_entry_exit_dynamics} shows that the entry component is large: employers that did not post before the revision account for most of the excess vacancies, while incumbent employers recreate few of the vacancies they cut below the floor. In this pure-spot benchmark, we hold the buffer margin in \citet{mas2026newsvendor} fixed and do not model any offset through permanent staffing or capital. The timing of these responses is also immediate in the model: because spot postings are re-chosen each period before demand is realized and carry no adjustment costs, establishments jump to the new optimum \(V^{*}\) and the new cutoff \(r^{*}\) as soon as the floor takes effect and remain there, consistent with the daily and weekly evidence that missing and excess jobs emerge immediately after implementation and then stabilize.

\subsection{Composition: Amenities and Experienced-Worker Targeting}

The core newsvendor model explains the vacancy and missing-job margins, and it also creates a selection channel for job composition. Surviving postings satisfy \(r\ge r^{*}(\underline w_{1})\), so a higher floor prunes the lower-productivity part of the posting distribution. The direction of selection across reimbursement categories depends on how amenity provision is associated with posting productivity and costs. The negative net estimate for postings offering reimbursement is consistent with this channel when reimbursement-providing postings have lower productivity net of amenity costs; identifying the association requires match-level productivity data. Separately, if low-productivity postings are more likely to use descriptions associated with less task-specific experience, raising \(r^{*}\) can tilt the surviving pool toward experienced-worker-targeted postings. This is a conditional compositional prediction, not a claim that the disappearing descriptions reveal lower worker productivity. The Skill targeting panel of Table~\ref{tb:missing_excess_net_by_category} reports separate net estimates for beginner- and experienced-worker-targeted postings.

Whether amenities also adjust directly among surviving postings is theoretically ambiguous, and the ambiguity is itself informative. Consider a reduced-form amenity decision, separate from the newsvendor core, in which a surviving establishment chooses a recruiting or match-quality amenity \(a\ge 0\)---transportation reimbursement or description language that attracts experienced, task-matched applicants---to solve
\begin{equation}
\max_{a\ge 0}
\quad
M(r,\omega)\,q(a)-C(a;\omega),
\label{eq:amenity_problem}
\end{equation}
with \(M(r,\omega)>0\), \(q'(a)>0\), \(q''(a)<0\), \(C_{a}(a;\omega)>0\), and \(C_{aa}(a;\omega)>0\). The interior optimum satisfies
\[
M(r,\omega)\,q'(a^{*})=C_{a}(a^{*};\omega).
\]
The second-order condition makes \(C_{aa}(a^{*};\omega)-M(r,\omega)q''(a^{*})>0\), so
\begin{equation}
\operatorname{sign}\frac{\partial a^{*}}{\partial \omega}
  =
  \operatorname{sign}\left[
  M_{\omega}(r,\omega)q'(a^{*})-C_{a\omega}(a^{*};\omega)
  \right].
\label{eq:amenity_comparative_static}
\end{equation}
Here \(M\) is the gross marginal value of raising the probability of a productive, task-matched fill, while \(C_{a\omega}\) captures how the marginal resource or shadow cost of amenity provision changes when the wage floor binds. These two terms need not cancel once wages are regulated and the market is imperfectly competitive, as emphasized by \citet{mas2025non}. We isolate this amenity margin from the core for transparency: because a reimbursement amenity is itself paid on each filled shift, folding it into the core would replace \(\omega\) with \(\omega+a\) in \eqref{eq:critical_fractile}, which lowers the critical fractile and the committed vacancy level in Proposition~\ref{prop:vacancy} for a given amenity but leaves the response of the critical fractile to the wage floor in \eqref{eq:dV_dw} unchanged.

The sign expression \(M_{\omega}q'(a^{*})-C_{a\omega}\) combines three forces. A \emph{recruiting}, or match-quality, force pushes \(M_{\omega}>0\): once the establishment is committed to the higher floor, an unfilled or poorly matched shift is more costly, so raising the probability of a productive, task-matched fill becomes more valuable. A \emph{compensating-differential} force pushes \(M_{\omega}<0\): with the floor binding, the higher posted wage itself does more of the work of attracting applicants, substituting for the amenity and lowering its marginal recruiting value---the multi-margin adjustment stressed by \citet{mas2025non} and \citet{clemens2021firms}. Finally, if the higher wage bill tightens the per-shift budget or raises the shadow price of cash-like amenities, then \(C_{a\omega}>0\), a further downward force. The net marginal-value effect \(M_{\omega}\) is thus of ambiguous sign, and the cost term adds a further downward push. In the intermediate monopsony benchmark of \citet{mas2025non}, where the efficient amenity is pinned down by its own optimality condition independently of the wage markdown, these terms are absent or approximately offset, so a pure wage floor can leave the direct amenity margin essentially unchanged. The net direct amenity response is therefore not determined by theory alone.

Appendix~\ref{app:other_measured_outcome} reports positive changes in observed reimbursement amounts and provision in excess wage bins, peaking at 15.1 and 16.3 percent. These estimates do not separately identify changes within surviving postings or selective exit by amenity type.

\subsection{Discussion}

The model reduces the spot labor market to a transparent short-run decision: how many flexible vacancies to commit before demand is known. A binding wage floor lowers the critical fractile, reduces vacancy creation, and raises the productivity cutoff for active postings. When postings below the new floor are either relabeled upward or exit, the survivors generate excess jobs above the floor, but exits and lower vacancy creation generate missing jobs that have no immediate offset. The amenity discussion illustrates a possible selection channel conditional on a relationship between posting attributes and productivity; the empirical results do not identify its contribution. The main limitations are deliberate. The model signs the net employment effect---missing jobs weakly exceed excess jobs---but does not pin down its magnitude, which depends on how much mass exits and how far constrained survivors cut vacancy creation. The core also treats each needed vacancy as filled with certainty and abstracts from matching frictions in worker arrival, so it is silent on the fill-risk margin except through the amenity module's \(q(a)\). We do not model the slower buffer and capital margins that would, over longer horizons, absorb some displaced tasks and shrink the net effect. Nor do we model the thick-market externality emphasized by \citet{mas2026newsvendor}, whereby fewer participating postings may thin the shared pool and raise frictions for the firms and workers that remain.

\end{document}